\documentclass[11pt,a4wide]{article}
\pdfoutput=1
\usepackage{jheppub}
\usepackage[english]{babel}
\usepackage[babel]{csquotes}
\usepackage{amsfonts}
\usepackage{array, tabularx}
\usepackage{mathtools}
\usepackage{amsthm}
\usepackage{url}
\usepackage{multicol, multirow}
\usepackage{emptypage}
\usepackage{verbatim}
\usepackage{dsfont}
\usepackage{float}
\usepackage{hyperref}
\usepackage{amsmath}
\usepackage{amssymb,tikz}
\usepackage{verbatim}
\usepackage{xspace}
\usepackage{tensor}
\usepackage{youngtab}
\usepackage{colortbl}
\usepackage[dvipsnames]{xcolor}

\usepackage{mathrsfs}   

\usepackage[font=small,labelfont={sf,bf}]{caption}

\makeatother

\newcommand{\hb}{{\bar{h}}}
\newcommand{\zb}{{\bar{z}}}
\newcommand{\1}{{\mathds{1}}}
\renewcommand{\O}{{\mathcal O}}
\newcommand{\G}{{\mathcal G}}
\newcommand{\de}{{\partial}}

\newcommand{\bare}{{\mathrm b}}

\newcommand{\incl}[2][0pt]{\raisebox{#1}{\includegraphics{./figures/#2}}}

\newcommand{\ON}{{\mathrm O(N)}}

\newcommand{\eps}{{\varepsilon}}
\newcommand{\teps}{{\tilde \epsilon}}

\newcommand{\eord}[1]{$\eps^{#1}$} 
\newcommand{\teord}[1]{$\teps^{#1}$} 
\newcommand{\Nord}[1]{$N^{-#1}$}

\DeclareMathOperator*{\dDisc}{dDisc}
\DeclareMathOperator*{\diag}{diag}
\DeclareMathOperator*{\sgn}{sgn}

\newcommand{\hchi}{\hat\chi}

\newcommand{\graycell}{\cellcolor[gray]{0.9}}
\newcounter{localfn}
\newcommand{\makefn}{$^{\fnsymbol{localfn}}$\stepcounter{localfn}}

\title{The critical O({\em N}\hspace{1.5pt}) CFT: Methods and conformal data}

\author[1]{Johan Henriksson}
\affiliation[1]{Dipartimento di Fisica ``E.\ Fermi'', Universit\`a di Pisa \emph{and} INFN, sezione di Pisa, Largo Bruno Pontecorvo 3, 56127 Pisa, Italy}

\emailAdd{johan.henriksson@df.unipi.it}

\abstract{
The critical $\ON$ CFT in spacetime dimensions $2< d<4$ is one of the most important examples of a conformal field theory, with the Ising CFT at $N=1$, $2\leqslant d <4$, as a notable special case. Apart from numerous physical applications, it serves frequently as a concrete testing ground for new approaches and techniques based on conformal symmetry. In the perturbative limits -- the $4-\varepsilon$ expansion, the large $N$ expansion and the $2+\tilde\epsilon$ expansion -- a lot of conformal data have been computed over the years. In this report, we give an overview of the critical $\ON$ CFT, including some methods to study it, and present a large collection of conformal data. The data, extracted from the literature and supplemented by many additional computations of order $\varepsilon$ anomalous dimensions, are made available through an ancillary data file.

\phantom{\cite{ThisPaper}}

\noindent \textsl{Note added:} The data file of the Arxiv submission was last updated 21 November 2025. At the following repository we will keep an up-to-date version of the data file as well as a pdf document with  up-to-date versions of all tables:  

{\begin{center} \href{https://github.com/johhen1/ON-model}{https://github.com/johhen1/ON-model}
\end{center}
}
}

\begin{document}
\maketitle



\section{Introduction}

Conformal field theories (CFTs) play a central role in contemporary theoretical physics, with applications that range from critical phenomena to string theory, and via the holographic correspondence also to quantum gravity. A reason for the great interest in CFTs is the combination of diverse applications with a plenitude of powerful methods that rely on conformal symmetry. 

Fifty years ago, Wilson and Fisher \cite{Wilson:1971dc} proposed to study three-dimensional critical phenomena with the help of $\lambda\phi^4$ field theory defined in $d=4-\eps$ spacetime dimensions. This approach, denoted the $\eps$-expansion, provided a useful and concrete example of Wilson's Nobel-prize winning theory of renormalisation \cite{Wilson:1971bg,Wilson:1971dh,Wilson:1973jj}.\footnote{See \cite{Wilson:1979qg} for a pedestrian overview of the renormalisation group framework, and \cite{Fisher:1998kv} for a historical introduction.} For $\eps\ll1$, it is possible to analyse the renormalisation group (RG) flow of $\lambda\phi^4$ theory using well-known techniques of Feynman diagrams, and a non-trivial infrared (IR) fixed-point -- the Wilson--Fisher fixed-point -- can be found perturbatively in $\eps$. The values of critical exponents computed in the $\eps$-expansion and extrapolated to $\eps=1$ have been found to give remarkably good agreement with various experimental and theoretical estimates for the critical 3d Ising model, see table~\ref{tab:IsingComparison}. The same holds for the generalisation to the $N$-component field $\varphi^i$,\footnote{In this report we distinguish the case $N=1$ by writing the field as $\phi$ instead of $\varphi$, however this distinction in notation is not normally upheld.} such as the critical XY ($N=2$) and Heisenberg ($N=3$) models.
\begin{table}[H]
\centering
\caption{Estimates and measurements of the critical exponents $\eta$ and $\nu$ in the three-dimensional Ising CFT.}\label{tab:IsingComparison}
{
\setcounter{localfn}{1} \small
\renewcommand{\arraystretch}{1.25}
\begin{tabular}{|llll|}
\hline
&&$\eta$ & $\nu$
\\\hline
Free theory && $0$ & $0.5$
\\
Wilson--Fisher $O(\eps^3)$ truncation && $0.037209$ & $0.60750$
\\
Wilson--Fisher $O(\eps^7)$ resummation &\cite{Shalaby:2020xvv}  & $0.03653(65)$ &   $0.62977(22)$ 
\\
Uniaxial magnet $\mathrm{FeF_2}$ (1987) &\cite{Belanger:1987zz} &  & $0.64(1)$
\\
Liquid-gas transition $\mathrm{CO_2}$ (1998)  &\cite{Damay1998} & $0.042(6)$ &
\\
Liquid-gas transition $\mathrm{D_2O}$ (2000)& \cite{Sullivan2000}  & & $0.62(3)$
\\
Liquid mixture\makefn\ (1988)& \cite{Sengers2009} & $ 0.032(13)$ & $0.629(3)$
\\
Liquid mixture\makefn\ (2004)& \cite{Lytle2004} & $ 0.041(5)$ & $0.632(2)$
\\
Quantum phase transition (2020) & \cite{Ebadi:2020ldi} &    & $0.62(4)$
\\
High temperature expansion& \cite{Campostrini:2002cf} & $0.03639(15)$ & $0.63012(16)$
\\
MC simulations& \cite{Hasenbusch:2021tei} & $0.036284(40)$ & $0.62998(5)$
\\
Numerical bootstrap &\cite{Kos:2016ysd} & $0.0362978(20)$ & $0.629971(4)$
\\\hline
\end{tabular}
\flushleft
\setcounter{localfn}{1} 

\makefn Light-scattering measurement in a mixture of water and isobutyric
acid.
\\ 
\makefn Turbidity measurement in a mixture of methanol and cyclohexane.
\\
}
\end{table}

It is now believed that the critical phenomena at the second-order phase-transitions in the three-dimensional systems can be described by a conformal field theory, which is continuously connected to the Wilson--Fisher fixed-point. Specifically, the existence of a two-parameter family of conformal field theories is conjectured, which is denoted the critical $\ON$ model, the critical $\ON$ CFT, the critical vector model, or simply the $\ON$ CFT. It is defined for non-negative integer $N$ and for spacetime dimension $2<d<4$ ($2\leqslant d <4$ for $N=1$), however, as will be discussed in section~\ref{sec:loopgasmodel}, it is natural to extend the range in $N$ to any real $N\geqslant-2$.

One way to define the critical $\ON$ model is the following. Consider the Lagrangian containing $N$ free scalars $\varphi^i$, $i=1,\ldots,N$ in $d$ spacetime dimensions, perturbed by the quartic interaction term $\lambda(\varphi^i\varphi^i)^2$, preserving the global $\ON$ symmetry:
\begin{equation}
\label{eq:actionintro}
S=\int d^dx\left(\frac12(\de_\mu\varphi^i)^2+\frac12m^2\varphi^2+\frac\lambda{24}(\varphi^2)^2\right).
\end{equation}
For $d<4$, the interaction is relevant and triggers an RG flow away from the free theory. For a fine-tuned value of the mass term, the flow will reach a unique non-trivial scale-invariant infrared fixed point, the critical $\ON$ CFT. 

For $N=1$, the conventional name of the theory is the Ising CFT, since it among other systems describes the critical behaviour of the Ising model, a classical statistical spin chain/lattice model with nearest neighbour interactions, i.e.\ with a Hamiltonian of the form
\begin{equation}
\label{eq:spinaction}
H=-J\sum_{<I,J>}\sigma_I\sigma_J-h\sum_I \sigma_I.
\end{equation}
In $d=1$, the Ising model shows no phase transition \cite{Ising1925}, however in $d=2$ it was solved by Onsager in 1944 \cite{Onsager:1943jn}\footnote{See also \cite{Kaufman:1949zz} and \cite{Fisher1959}.} and the critical theory, the 2d Ising CFT, was later found to be the first member $\mathcal M_{3,4}$ of a family of unitary minimal models classified by Belavin, Polyakov and A. Zamolodchikov \cite{Belavin1984}. The exact solution of the Ising CFT in $d=3$ dimensions, on the other hand, remains an open problem, although its existence has been well corroborated through the combined efforts of experiments, Monte Carlo (MC) simulations, conformal bootstrap and field theoretical considerations.\footnote{For the last point, see for instance \cite{Glimm:1973kp,Abdesselam:2006qg}. See also \cite{Aizenman2014} for a proof of continuity of the phase transition of the lattice model in three dimensions.} 

Thanks to universality -- the fact that several microscopic theories can flow to the same IR fixed-point -- physical applications of the $3$d $\ON$ CFT are ubiquitous. Apart from statistical spin models, the $N=1$ (Ising) case describes second-order phase transitions of uniaxial antiferromagnets, liquid-gas transitions in atomic and simple molecular fluids (for instance water and carbon dioxide) at the critical point of the phase diagram, and phase transitions in binary fluid mixtures. The $N=2$ theory describes the $\lambda$-line in helium and phase transitions in XY ferromagnets and in the statistical XY model, and the $N=3$ theory describes isotropic magnets and the statistical Heisenberg model.\footnote{We leave a complete treatment of applications, including experimental references, to \cite{Pelissetto:2000ek}.} In these setups, the field $\varphi$ (denoted $\phi$ or $\sigma$ at $N=1$), sometimes called the spin field and related to the critical exponent $\eta$, denotes the order parameter operator, and $\varphi^2_S$ ($\phi^2$ or $\epsilon$ at $N=1$), related to the critical exponent $\nu$, denotes the energy density operator.\footnote{Apart from systems in three (Euclidean) dimensions, the $3$d $\ON$ CFT may also describe quantum phase transitions in $2+1$ dimensions by Wick rotation. See \cite{Ebadi:2020ldi} for the first observation of the $(2+1)$-dimensional Ising CFT in a quantum simulation, \cite{Keesling:2018ish} for the $(1+1)$-dimensional case, and \cite{Fisher:1989zza,Cha1991} for a discussion the potential application  of the $(2+1)$d $\mathrm{O}(2)$ CFT to superconductor-isolator transitions in thin films. Moreover, in the large $N$ expansion the critical $\ON$ CFT, the singlet sector has been proposed to be dual to a higher-spin Vasiliev theory \cite{Vasiliev:1995dn,Vasiliev:1999ba} with certain boundary conditions \cite{Klebanov:2002ja}.} 

\begin{figure}
\centering
\includegraphics[scale=1]{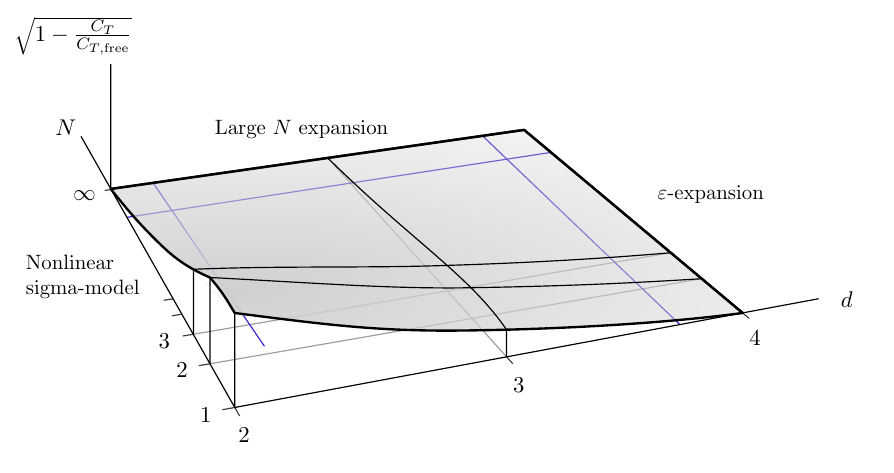}
\caption{Qualitative plot of the critical $\ON$ CFT as a continuous family of interacting conformal field theories parametrised by $d$ and $N$. Here we use $\sqrt{1-\frac{C_T}{C_{T,\mathrm{free}}}}$ for the central charge $C_T$ (see \eqref{eq:CTNscalars}--\eqref{eq:OPEwithTandJ}) as a proxy for the ``coupling strength'' and indicate the different expansion regimes. The non-linear sigma model in $2+\teps$ is a weakly coupled description of the theory using $N-1$ fields, hence in the limit $\teps\to0$, $\frac{C_T}{C_{T,\mathrm{free}}}$ approaches $\frac{N-1}N$ and not $1$ (c.f.\ \eqref{eq:CTnlsm}). For $N>2$ one must have $d>2$ strictly. The extension down to $N=-2$ is not displayed.}\label{fig:surface3d}
\end{figure}

Apart from the mentioned challenge of finding an exact solution, there are several theoretical reasons for the persistent interest in the critical $\ON$ CFT. It has a relatively simple definition, but yet a rich phenomenology. In the $(d,N)$ plane on which it is defined, it interpolates between several weakly coupled regions and a non-perturbative, ``strongly coupled'', bulk, see figure~\ref{fig:surface3d} for a visualisation.
Considering the extension to real $N\geqslant-2$, we can describe the following regions of interest:
\begin{itemize}
\item The $d=4-\eps$ expansion, or simply the $\eps$-expansion, which is weakly coupled for $\eps\ll1$. In this region the Ising or $\ON$ CFT is sometimes denoted the Wilson--Fisher CFT or Wilson--Fisher fixed-point. Results can be computed using conventional perturbation theory with Feynman diagrams.
\item The large $N$ expansion, defined for any $2<d<4$. Results can be derived in a series expansion in $1/N$ through a diagrammatic approach.
\item The non-linear sigma model for $d=2+\teps$, with $N>2$. Results can be derived in a series in small $\teps$ using Feynman diagrams.\footnote{\label{foot:Mermin}The Mermin--Wagner theorem forbids spontaneously broken symmetry when $N\geqslant2$ and $d \leqslant 2$ \cite{Mermin:1966fe,Hohenberg:1967zz,Coleman:1973ci}, so this expansion is defined for $\teps$ strictly positive.}
\item An exact solution at $d=2$ and $N=1$ (2d Ising CFT), and a family of exact results (first found by Nienhuis) for $d=2$ and $-2\leqslant N\leqslant 2$.\footnote{At $d=N=2$, the $\ON$ CFT reduces to the Kosterlitz--Thouless transition, which is an infinite-order phase transition, corresponding to a boson compactified on a circle of radius $R=1/\sqrt2$ \cite{Kosterlitz:1973xp,Dijkgraaf:1987vp}.}
\item Finally, a non-perturbative region in the bulk, where theoretical results are derived by interpolations from the perturbative regimes, numerical methods, or fixed-dimension methods such as the functional/non-perturbative RG method.
\end{itemize}
Being a well-studied and comparably simple conformal field theory, the $\ON$ CFT is often used as a toy model, or testing ground, for various methods facilitated by conformal symmetry, both perturbative or non-perturbative. In particular, inspired by the success of the numerical conformal bootstrap \cite{Rattazzi2008,Poland:2018epd}, also other methods based on conformal symmetry have been developed and included in the notion ``bootstrap'', and yielded new results for perturbative conformal data. While the perspective from experiments, statistical physics and critical phenomena is well surveyed \cite{Pelissetto:2000ek}, a comprehensive and up-to-date review of the theory from the CFT perspective is lacking. This report aims to fill that role.

A CFT perspective on the critical $\ON$ CFT takes the following point of view. A conformal field theory may be defined as a set of conformal primary operators together with their correlation functions. By an iterative application of the operator product expansion (OPE), the data needed to specify correlators of primary operators are the quantum numbers $(R,L,\Delta)$ of the individual operators (the spectrum), and the OPE coefficients $\lambda_{\O_1\O_2\O_3}$ (which also characterise three-point correlators). Here $R$ denotes an irreducible representation (irrep) of the global $\ON$ symmetry, $L$ a Lorentz (rotation) irrep, and $\Delta\geqslant0$ the scaling dimension. A conformal primary operator acting on the origin is the top component a highest-weight representation of the Euclidean conformal group $\mathrm{SO}(d+1,1)$, and subleading, ``descendant'', operators are generated by taking partial derivatives. The set of conformal primary operators with their quantum numbers and three-point functions is referred to as the conformal data, or CFT-data. Although, as experimental considerations dictate, focus has been put on computing the leading critical exponents such as $\eta$ and $\nu$, as well as the leading correction-to-scaling exponent $\omega$, more systematic determinations of larger sets of conformal data have been considered in the $4-\eps$ expansion \cite{Wegner:1972zz,Kehrein:1992fn,Kehrein:1994ff,Kehrein:1995ia} and in the large $N$ expansion \cite{Ma1976,Lang:1992pp,Lang:1994tu,Derkachov:1997qv}.

For most of this report, we will consider primary operators where the Lorentz representation $L$ takes the form of rank $\ell$ traceless-symmetric tensors, indexed by the integer label $\ell$: the spin.\footnote{In Lorentzian signature, the conformal group admits representations with a non-integer spin label $\ell$. Such operators annihilate the vacuum, but can be given an interpretation as the analytic continuation in spin of light-transforms of local operators (with integer spins) \cite{Kravchuk:2018htv}.} The spectrum, limited to such operators, can then be organised as follows: for each $(R,\ell)$, the conformal primary operators form a list organised by scaling dimensions: $\texttt{Op[}\langle R\rangle\texttt{,}l\texttt{,}i\texttt{]}$, $i=1,2,\ldots$. At higher $i$, the ordering may become ambiguous in the expansion regimes, in which case we shall refer to the ordering in the limit $\eps\ll 1/N \ll1$. Naturally, any presentation of primary operators can only contain a subset of the whole set of CFT-data. Instead of truncating all representations at the same number of operators in each representation or at a given dimension (e.g.\ by the value $\Delta^{\mathrm{4d}}$ in four dimensions), the lists in this report contain a subjectively chosen range for each representation. However, all operators with $\Delta^{\mathrm{4d}}\leqslant10$ for $N=1$, and $\Delta^{\mathrm{4d}}\leqslant6$ for general $N$ ($\Delta^{\mathrm{4d}}\leqslant8$ for $\ON$ singlet operators) are included.

The structure of this report is as follows. In section~\ref{sec:ObservablesMain}, we give an introduction to the $\ON$ CFT and present results for the main observables in the mentioned regions of the parameter space. In section~\ref{sec:presentation} we give more details on the representation theory of the global and Lorentz symmetry groups in order to clarify the notions used later on for describing the set of primary operators. In section~\ref{sec:methods}, we give a brief outline of the different perturbative methods used to study the $\ON$ CFT, and present in detail an implementation of the one-loop dilatation operator in the $\eps$-expansion based on \cite{Hogervorst:2015akt}. This implementation is then used to analyse the spectrum of primary operators which, complemented with results extracted from the literature, is presented in the tables found in section~\ref{sec:DataIsing} for $N=1$ and in section~\ref{sec:dataGenN} for generic $N$. The tables also contain references to previous determinations of the CFT-data. We conclude with a short discussion and some appendices.

Attached to this report is a data file in the form of a Mathematica package, \texttt{ONdata.m}, where the CFT-data presented in the tables are implemented in a computer-readable format. We give more details on this file at various places in the text, and in appendix~\ref{app:datafile}.

\section[\texorpdfstring{Observables in the critical $\boldsymbol{\ON}$ CFT}{Observables in the critical O(N) CFT}]{Observables in the critical $\boldsymbol{\ON}$ CFT}
\label{sec:ObservablesMain}

In this section we give a brief overview of the critical $\ON$ CFT.

\subsection{Conformal invariance}

The fundamentals of conformal field theories in spacetime dimensions $d>2$ have been laid out in \cite{Osborn:1993cr}, see also \cite{Fradkin:1997df}, and reviewed in \cite{Qualls:2015qjb,Rychkov:2016iqz,TASIBootstrap,Poland:2018epd}. Throughout this report we consider the CFT defined on flat $d$-dimensional Euclidean space $\mathbb R^d$; by Wick rotation, the conformal data is the same for the theory defined on flat Lorentzian $\mathbb R^{d-1,1}$. For results in the large $N$ expansion, we introduce the customary variable $\mu=d/2$ for aesthetic reasons. 

The conformal group $\mathrm{SO}(d+1,1)$ has generators $M_{\mu\nu}$ of rotation, $D$ of dilatation, $P_\mu$ of translation and $K_\mu$ of special conformal transformations, satisfying the algebra
\begin{align}
[M_{\mu\nu},M_{\rho\sigma}]&=-\eta_{\mu\rho}M_{\nu\sigma}+\eta_{\mu\sigma}M_{\nu\rho}+\eta_{\nu\rho}M_{\mu\sigma}-\eta_{\nu\sigma}M_{\mu\rho},\hspace{-75pt}
\\
[M_{\mu\nu},P_\rho]&=-\eta_{\mu\rho}P_\nu+\eta_{\nu\rho}P_\mu,
&
[M_{\mu\nu},K_\rho]&=-\eta_{\mu\rho}K_\nu+\eta_{\nu\rho}K_\mu,
\\
[D,P_\mu]&=P_\mu ,& 
[D,K_\mu]&=-K_\mu,
\\
[K_\mu,P_\nu]&=2\eta_{\mu\nu}D-2M_{\mu,\nu}.
\end{align}
Conformal primary operators are operators which inserted at the origin are annihilated by the generator $K_\mu$ of special conformal transformations. Alternatively, their correlation functions depend only on the insertion points and not on the local coordinate system (see e.g.\ \cite{Rychkov:2016iqz}). Moreover, they transform in irreducible representations of the commuting subgroups of the conformal group generated by dilatations $D$ and rotations $M_{\mu\nu}$ respectively,
\begin{align}
[K_\mu,\O_{M}(0)]&=0,
\\
[D,\O_{M}(0)]&=\Delta_\O \O_{MN}(0),
\label{eq:scalingdimdef}
\\
[M_{\mu\nu},\O_{M}(0)]&={(\rho_{\mu\nu})_M}^N\O_N(0).
\end{align}
Descendant operators are generated by repeated action of the generator of momentum $P_\mu$,
\begin{equation}
[P_\mu,\O_{M}(0)]=\de_{\mu}\O_M(x)|_{x=0}.
\end{equation}

Any conformal field theory has a conserved stress tensor $T^{\mu\nu}$ with fixed dimension $\Delta=d$. It is a global symmetry singlet and its correlators with local operators are fixed by conformal Ward identities \cite{Mack:1972kq}; we give precise formulas in appendix~\ref{app:normalisation}. For a CFT with a continuous global symmetry, there is also a conserved spin $1$ current $J^\mu$ of dimension $d-1$. It transforms in the (rank $2$) antisymmetric representation $A$ of $\ON$ and its correlators are fixed by conformal Ward identities. The precise form of these correlators depends on the normalisation used for operators in the different representations of the $\ON$ global symmetry, see appendix~\ref{app:normON}. The normalisations $C_T$ and $C_J$ of the two-point functions of the conserved currents are denoted the central charge and current central charge respectively, and are observables.

The spectrum in a unitary CFT satisfies the unitarity bounds \cite{Mack:1975je,Minwalla:1997ka}
\begin{equation}
\label{eq:unitaritybound}
\Delta\geqslant \frac{d-2}2 \quad\text{(scalar),}\qquad \Delta\geqslant d-2+\ell\quad \text{(spin $\ell$).}
\end{equation}
Primary operators saturating the unitarity bounds are annihilated by some specific combination of derivatives and generate short multiplets. In the interacting $\ON$ CFT for $d>2$, the only short multiplets are generated by the stress tensor, $\de_\mu T^{\mu\nu}=0$, and (for generic $N$) by the global symmetry current, $\de^\mu J^{ij}_\mu=0$. All other multiplets are long, meaning that they consist of the primary operator together with all descendants generated by repeated action of $\de_\mu$.

\subsection{Spectrum continuity and the naming of operators}
\label{sec:spectrumcontinuity}

As a full-fledged unitary conformal field theory, the critical $\ON$ CFT can only be defined for integer values of the spacetime dimension $d$ and positive integer values of the number of components $N$. Combined with the Mermin--Wagner theorem (see footnote~\ref{foot:Mermin}), this restriction limits, in principle, the scope to $d=2$, $N=1$ (2d Ising), $d=2$, $N=2$ (Kosterlitz--Thouless transition), and $d=3$, $N=1,2,3,\ldots$. At these points, the CFT-data consists of local operators with real scaling dimensions above the unitarity bound \eqref{eq:unitaritybound} and real OPE coefficients. In the $d=3$ case, this assertion has not been proved, but has been corroborated by the existence of small isolated unitary islands in the numerical conformal bootstrap \cite{Kos:2015mba,Kos:2016ysd,Chester:2019ifh,Chester:2020iyt}.	
However, as mentioned in the introduction, this report assumes that the $\ON$ CFT can be defined in a larger range, covering also non-integer values of $N$ and $d$. In the strong form of this assumption, the whole set of CFT-data -- the spectrum of operators and all their correlators -- varies continuously with $N$ and $d$. Following \cite{Hogervorst:2015akt}, we refer to this statement as ``spectrum continuity''. A weaker form of spectrum continuity assumes that the statement holds only for a limited set of observables corresponding to low-lying operators in the spectrum, such as the critical exponents. 

From an experimental or Monte Carlo point of view, (weak) spectrum continuity is a result rather than an assumption, and can be observed by simulation of the $\ON$ lattice models at various $N$ finding critical exponents that vary continuously, see e.g.\ \cite{Hasenbusch:2021rse}. In particular, the $\ON$ loop gas model, to be discussed below, lends itself to this kind of investigations \cite{Liu:2012ca}.  Likewise, the conformal bootstrap has found kinks \cite{El-Showk:2013nia} and islands \cite{Behan:2016dtz} for the $N=1$ theory at values of $d$ interpolating between $d=2$ and $d=4$. Similar observations have been made in the non-perturbative RG \cite{Chlebicki:2020pvo,DePolsi:2020pjk}. The conformal bootstrap has also reported a larger set of low-lying operator dimensions which appear to vary continuously with $d$ \cite{Cappelli:2018vir}. 

In \cite{Hogervorst:2015akt} it was noted that the $\ON$ CFT in the $4-\eps$ expansion fails to be unitary, and it was argued that on general grounds this will be the case at any non-integer $d$. One manifestation of non-unitarity is the existence of complex anomalous dimensions (see section~\ref{eq:unitaritviolating}) -- however these operators appear high in the spectrum, which may explain why the works previously mentioned (\cite{El-Showk:2013nia,Behan:2016dtz,Cappelli:2018vir}) were able to observe bootstrap features despite assuming unitarity for a non-unitary CFT. Another observation of \cite{Hogervorst:2015akt} is the existence of operators in the free theory which identically vanish for low integer $d$. We allow for the existence of such operators in the notion of spectrum continuity.

The assumption of spectrum continuity can be checked perturbatively by looking at the perturbative conformal data in the overlap between the large $N$ expansion and the $4-\eps$ and $2+\teps$ expansions. At each orders in $1/N$, all CFT-data are analytic in $d$, and likewise at each order in $\eps$ ($\teps$), all CFT-data are analytic in $N$. With no exception, the data agree in the overlapping expansion range at large $N$ and small $\eps$.\footnote{Some of the data in the $2+\teps$ expansion was computed using input from the large $N$ expansion, so the corresponding agreement does not constitute an independent check.} That indeed the different expansion regimes describe the same theory can also be argued for on the level of the Lagrangian \cite{Amit:1980bx}.

The representation theory of the rotation (Lorentz) group $\mathrm{SO}(d)$ and global symmetry group $\ON$ is different for integer values and generic values of $d$ and $N$, however the extension to non-integer values can be done in a naïve way by studying the conformal data, or be formalised using Deligne categories \cite{Binder:2019zqc}. We give more details in section~\ref{sec:presentation}.

\begin{table}[ht]
\centering
\caption{The most important operators for general $\ON$ symmetry. Symbols in brackets denote the $N=1$ case.}\label{tab:importantops}
{\small
\renewcommand{\arraystretch}{1.5}
\begin{tabular}{|llllll|}
\hline
Irrep & Impl. & $d=4-\eps$ & $1/N$ exp. & $3d$ & ass. obs.
\\\hline
$V$ & \texttt{Op[V,0,1]} &  $\varphi$ ($\phi$) &  $\varphi$ &  $\varphi$ ($\sigma$) &
\\
$S$ & \texttt{Op[S,0,1]} &  $\varphi^2_S$ ($\phi^2$)&  $\sigma$ &  $S$ ($\epsilon$) &
\\
$S$ & \texttt{Op[S,0,2]} &  $\varphi^4_S$ ($\phi^4$)&  $\sigma^2$ &  $S'$ ($\epsilon'$) &
\\
$T$ & \texttt{Op[T,0,1]} &  $\varphi^2_T$ &  $\varphi^2_T$ &  $t$ &
\\
$T_4$  & \texttt{Op[Tm[4],0,1]} &  $\varphi^4_{T_4}$ &  $\varphi^4_{T_4}$ &  $\tau$  & 
\\
$S$, $\ell=2$  & \texttt{Op[S,2,1]} &  $T^{\mu\nu}\sim\de^2\varphi^2_S$  ($\phi\de^2\phi$) &  $T^{\mu\nu}\sim\de^2\varphi^2_S$ &  $T^{\mu\nu}$ & $C_T$
\\
$A$, $\ell=1$  & \texttt{Op[A,1,1]} &  $J^\mu\sim \de\varphi^2_A$ &  $J^\mu\sim\de\varphi^2_A$ &  $J^\mu$ & $C_J$

\\\hline
\end{tabular}
}
\end{table}

Assuming spectrum continuity and level repulsion -- that energy levels with the same quantum numbers never cross \cite{vonNeumann1929}\footnote{At infinite $N$, level crossing may take place, leading to interesting features in a $1/N$ expansion \cite{Korchemsky:2015cyx}. Likewise, in two dimensions, where the conformal symmetry is enhanced to the infinite-dimensional Virasoro symmetry, crossings may occur. From truncated $\eps$-expansion data, it may appear like operators would cross at finite $\eps$, however a recent study using the conformal bootstrap found level repulsion in this case \cite{Henriksson:2022gpa}.} -- the spectrum of conformal primary operators can be organised by increasing scaling dimension for a given $\ON$ and Lorentz representation. The quantum numbers will be scaling dimension $\Delta$, $\ON$ irrep $R$ and Lorentz irrep $L$. Primarily, we shall consider operators where $L$ is a one-line Young tableau, so that we can replace $L$ by an integer $\ell\in \mathbb N=\{0,1,2,\ldots\}$. For $N=1$ ($\mathrm O(1)=\mathbb Z_2$), the options for $R$ are $ \mathbb Z_2$ even (\texttt E) and $\mathbb Z_2$ odd (\texttt O). For generic $N$ the $\ON$ representations $R$ are indexed by Young tableaux $Y_{m_1,m_2,\ldots,m_r}$ of $r$ rows of $m_i$ boxes each. At finite (integer) $N$, some Young tableaux are not present, as we will explain in section~\ref{sec:reptheory}.

Note that in the free theory, specifying $(\Delta,R,\ell)$ is not enough to uniquely label an operator.
In the interacting theory, however, we believe that this is enough.\footnote{Note that some operators are still degenerate at order $\eps$, for instance the two operators \texttt{Op[E,0,10]} and \texttt{Op[E,0,11]} in table~\ref{tab:evenscalars}. These operators have the same scaling dimension to order $\eps$ and the order $\eps^2$ result is not known, however it is expected that degeneracy is always broken at some order \cite{Kehrein:1995ia}.} We will therefore assign names to the operators of a given $(R,\ell)$ based on increasing value of $\Delta$:

\begin{equation}
\texttt{Op[}\langle R\rangle\texttt{,}l\texttt{,1]},\ 
\texttt{Op[}\langle R\rangle\texttt{,}l\texttt{,2]},\ \ldots.
\end{equation}
The identity operator $\1$ is called \texttt{Op[E,0,0]} or \texttt{Op[S,0,0]}, while all other operators \texttt{Op[$\langle R\rangle$,$l$,$i$]} have $i=1,2,\ldots$. 
We order the operators by dimension in the limit $\eps\ll1/N\ll1$. $\langle R\rangle$ denotes the symbols of the representations as implemented in the data file, 
more details on the presentation of operators are given in section~\ref{sec:presentation} and in appendix~\ref{app:datafile}.

There are traditional names for the most important operators in the various expansion regimes, as summarised in table~\ref{tab:importantops}.
We use $\phi$ for the spin field in $N=1$ and $\varphi$ for the $\ON$ vector spin field for generic $N$. In the large $N$ expansion we use $\sigma$ for the auxiliary field. Composite operators are constructed out of fields and partial derivatives $\de^\mu$. An operator constructed out of $\ell+2p$ derivatives will have spin $\ell$ if $2p$ of the derivative indices are contracted, and the remaining $\ell$ indices transform in the traceless-symmetric spin $\ell$ Lorentz representation. We write this as $\de^\ell\square^p$. We will write the ``form'' of an operator in the $\eps$-expansion as\footnote{For $N=1$, there is no need to specify the representation, since is it given by the number of fields (mod $2$), whence we write $\O=\de^\ell\square^p\phi^k$. For $k=2$, $p=0$, it is customary to write the operators as $\phi\de^\ell\phi$.}
\begin{equation}\label{eq:Oform}
\O=\de^\ell\square^p\varphi^k_R,
\end{equation}
which should be read as \emph{an operator $\O$ constructed out of a linear combination of terms with $k$ fields $\varphi$ with $\ON$ indices in the $R$ representation, each term with $2p$ contracted and $\ell$ uncontracted partial derivatives (with removal of traces), distributed in such a way that $\O$ is a conformal primary operator.} The form \eqref{eq:Oform} should be understood to be given in $d=4$. In $4-\eps$ dimensions, the dilatation operator $D$ acts on the set of operators, and the conformal primary operators are linear combinations of operators of the form \eqref{eq:Oform}. At the lowest orders in $\eps$, $D$ is block diagonal, which means that it can be diagonalised within a subset of operators of the form \eqref{eq:Oform}. We note the following:
\begin{itemize}
\item To order $\eps$, $D$ mixes only operators with the same values of $\ell$, $p$ and $k$.
\item At order $\eps^2$, there will be mixing between $\varphi^i\varphi^i$ and $\square$. 
\end{itemize}
For instance, the operator $\texttt{Op[S,0,2]}=\varphi^4_S$ will at order $\eps^2$ contain admixtures of descendant operators of the form $\square\varphi^2_S$. In writing the form as $\varphi^4_S$ it is understood that such admixtures will happen.

In the large $N$ expansion, we can write the form of the operators as
\begin{equation}
\label{eq:OformN}
\O=\de^\ell\square^p\sigma^k\varphi^m_R,
\end{equation}
Already at order $N^0$ there is admixture between $\square$ and $\sigma$, and the form \eqref{eq:OformN} is therefore not unique if $p$ and $k$ are non-zero.

\subsubsection{The loop gas model}
\label{sec:loopgasmodel}

Apart from the lattice spin model \eqref{eq:spinaction}, and its generalisation to a vector-valued spin field with (integer) $N$ components, there is another statistical model that appears to lie within the $\ON$ universality class, called the $\ON$ loop gas model \cite{Nienhuis:1982fx,Nienhuis:1984wm,Baxter1986,Baxter1987,Bloete:1989py}. This is a model of quantum closed loops on a lattice, primarily defined on two-dimensional hexagonal lattices but also extended to three-dimensional lattices \cite{Liu:2012ca}. It is a non-local theory with two couplings. The real parameter $N$ enters as a coupling constant. By tuning the other coupling constant, denoted $x$, one finds critical behaviour with exponents that agree with the usual (local) $\ON$ CFT. In two dimensions, this behaviour is seen for $N$ in the range $-2\leqslant N\leqslant 2$, parametrised by the Coulomb gas coupling\footnote{For $d=2$, there are two critical points for each $N$, denoted the dilute phase and the dense phase. Here we consider the dilute phase.}
\begin{equation}
g=1+\frac1\pi\arccos(N/2)\in [1,2],
\end{equation}
and for $d>2$ one assumes that $N\geqslant -2$. 
The limit $N\to0$ corresponds to self-avoiding random walks, found for instance in dilute solutions of polymers, and the limit $N\to-2$ to loop-erased random walks. For a discussion of applications see e.g.\ \cite{Shimada:2015gda,Peled2017}. The theory arising in the limit $N\to0$ can be understood as a logarithmic CFT \cite{Movahed:2004nr,Cardy:2013rqg,Hogervorst:2016itc}.

Assuming spectrum continuity, an interesting region to study is the vicinity of $N=d=2$.
As can be seen from the explicit formulas below, the expansion for the non-linear sigma model is defined only for $d$ and $N$ strictly greater than 2. It has been proposed that there exists a curve starting at $N=d=2$ across which critical exponents are non-analytic, denoted the ``Cardy--Hamber line'' \cite{Cardy:1980at}. However, recent studies using the non-perturbative RG indicate that the exponents are smooth across the proposed curve \cite{Chlebicki:2020pvo}.

\subsection{Main observables}

In this section, we shall give the results of the main observables in the different expansion regimes and in 2d and 3d. We focus on the scaling dimensions of the two leading singlet ($S$) operators, and the leading operators in the vector ($V$), rank two ($T$) and rank four ($T_4$) traceless-symmetric representations, as well as the central charges. In table~\ref{tab:history} we give a summary of historic determinations of these quantities.

\begin{table}[ht]
\centering
\caption{A summary of the developments of higher order results for the main observables in the $\eps$-expansion and $1/N$ expansion. The notation is ``order[ref.]''.}\label{tab:history}
\setcounter{localfn}{1}
{\small
\renewcommand{\arraystretch}{1.5}
\begin{tabular}{|cl|cl|}
\hline
 \multicolumn{2}{|c}{$d=4-\eps$ expansion}& \multicolumn{2}{|c|}{$1/N$ expansion}
\\\hline
   $\Delta_\varphi$  & \eord4\cite{Brezin:1973aa}\eord5\cite{Chetyrkin:1981jq,Kleinert:1991rg}\makefn\eord6\cite{Batkovich:2016jus}\eord7\cite{Schnetz:2016fhy}\eord8\cite{SchnetzUnp} &  $\Delta_\varphi$ & 
    \Nord2(3d)\cite{Abe1973}\Nord2\cite{SymanzikUnp,Kondor:1980zh}\Nord3\cite{Vasiliev:1982dc}\makefn
   \\
  $\Delta_{\varphi^2_S}$ & \eord4\cite{Kazakov:1979ik}\eord5\cite{Kleinert:1991rg}\eord6\cite{Kompaniets:2017yct}\eord7\cite{Schnetz:2016fhy} &  $\Delta_\sigma$ & 
  \Nord2(3d)\cite{Okabe:1978mp}\Nord2\cite{Vasiliev:1981dg}
\\
  $\Delta_{\varphi^4_S}$ & \eord4\cite{Kazakov:1979ik}\eord5\cite{Kleinert:1991rg}\eord6\cite{Kompaniets:2017yct}\eord7\cite{Schnetz:2016fhy} & $\Delta_{\sigma^2}$ & 
  \Nord1\cite{Ma:1974qh}\makefn\Nord2\cite{Broadhurst:1996ur,Gracey:1996ub}
\\
  $\Delta_{\varphi^2_T}$ & \eord2\cite{Wilson:1971vs}\eord3\cite{Yamazaki1974}\eord4\cite{Kirkham:1981pu}\eord6\cite{Kompaniets:2019zes}\makefn  &$\Delta_{\varphi^2_T}$ &  \Nord1\cite{Ma:1974qh}\Nord2\cite{Gracey:2002qa} 
  \\
  $\Delta_{\varphi^4_{T_4}}$ &  \eord5\cite{Kleinert:1994td,Calabrese:2002bm}\eord6\cite{Bednyakov:2021ojn}\makefn &  $\Delta_{\varphi^4_{T_4}}$ &   \Nord1\cite{Ma1976}\Nord2\cite{Derkachov:1997ch} 
\\
  $C_T$ & \eord2\cite{Cappelli:1990yc}\makefn\eord3\cite{Dey:2016mcs}\eord4\cite{Henriksson:2018myn} &$C_T$ & \Nord1\cite{Lang:1993ct}
\\
   $C_J$ &\eord2\cite{Petkou:1994ad}\makefn\eord3\cite{Dey:2016mcs}\eord4\cite{Henriksson:2018myn}&  $C_J$ & \Nord1(3d)\cite{Cha1991}\Nord1\cite{Lang:1992pp}

\\\hline
\end{tabular}
\flushleft
\setcounter{localfn}{1} 

\makefn The result in \cite{Chetyrkin:1981jq} contained several errors, see chapter~15 of \cite{Kleinert:2001hn}.
\\
\makefn As noted in \cite{Vasiliev:1993ux}, the expression in \cite{Vasiliev:1982dc} contains a misprint in equation (22). See also \cite{Broadhurst:1996yc,Fei:2014yja}.
\\
\makefn Ref. \cite{Brezin1976} (page 221) cites \cite{Ma:1974qh} together with Bervillier, Girardi, Brezin; Saclay preprint (1974). 
\\
\makefn Note that their $\zeta_{3,5}$ equals our $\zeta_{5,3}$, c.f \eqref{eq:multizetavalues}.
\\
\makefn This result was extracted from the more general results of \cite{Bednyakov:2021ojn}. I thank the authors of \cite{Bednyakov:2021ojn} for useful discussions and for sharing the precise form of the result.
\\
\makefn$^,$\makefn Can be extracted from \cite{Jack:1983sk}, see comment in \cite{Petkou:1994ad}.
\\
}
\end{table}

\subsubsection{Scaling dimensions}

The scaling dimension of a conformal primary operator is defined by \eqref{eq:scalingdimdef} and controls the scaling of the two-point function. For scalar operators $\O_I$, conformal symmetry dictates that
\begin{equation}
\label{eq:twopointmain}
\langle\O_I(x_1)\O_J(x_2)\rangle=c_I\frac{\delta_{IJ}}{|x_{12}|^{2\Delta_{\O_I}}},
\end{equation}
where $x_{12}=x_1-x_2$.
For spinning operators, we give precise formulas in appendix~\ref{app:normalisation}. It is customary to normalise the operators so that the two-point function \eqref{eq:twopointmain} has unit normalisation, $c_I=1$.

In general, for a weakly coupled theory with marginal coupling constants $g_i$, the eigenvalues of the matrix $\frac{\de\beta_i}{\de g_j}$ of derivatives of the beta functions give the scaling dimensions of the approximately marginal operators of the theories:
\begin{equation}
\label{eq:betaderivsmarginal}
\Delta_{\O_{\text{marginal},k}}=d+\omega_k, \qquad \text{$\omega_k$ eigenvalue of }\frac{\de\beta_i}{\de g_j}
\end{equation}
For this reason, $\frac{\de\beta_i}{\de g_j}$ is called the stability matrix, as a perturbative fixed-point with a positive-definite stability matrix is stable under marginal perturbations.
For $\phi^4$ theories in $4-\eps$ dimensions of arbitrary global symmetry, it follows from a leading order calculation that for any interacting fixed-point there is always one irrelevant operator with $\Delta_{\O}=4+O(\eps^2)$.

Sometimes $x$ denotes the dimension, and $y$ the corresponding ``exponent'', $d-\Delta_\O$, i.e.
\begin{equation}
x_\O=\Delta_\O,\qquad y_\O=d-\Delta_\O, \qquad \phi_\O=\nu y_\O ,\qquad \eta_\O=\Delta_\O-[\O].
\end{equation}
We do not employ this terminology in this report. In this notation, $[\O]$ denotes the engineering dimension of $\O$, and we have for the critical exponent $\eta=2\eta_\phi$.

\subsubsection{Critical exponents}

The most easily accessible observables in experiments or simulations are the critical exponents, conventionally denoted by Greek letters and referred to by the name of that letter. The main critical exponents can be computed from the scaling dimensions of relevant or weakly irrelevant operators in the spectrum according to the equations given in table~\ref{tab:criticalexponents}. Note that six of the critical exponents -- $\alpha$, $\beta$, $\gamma$, $\delta$, $\eta$ and $\nu$ -- are all determined in terms of the two parameters $\Delta_\varphi$ and $\Delta_{\varphi^2_S}$ \cite{Brezin:1973jc}.

\begin{table}[ht]
\centering
\caption{Conventional definitions of the critical exponents. Here $\tau=\frac{T-T_c}{T_c}$, and we use the following expressions: Two-point function of the order parameter $G^{(2)}(r)$, specific heat capacity $C$, deviation from critical density $\rho-\rho_c$, magnetic susceptibility $\chi$, isothermal compressibility $\kappa_T$, magnetisation $\mathbf M$, applied magnetic field $\mathbf H$, correlation length $\xi$, and $g$ is an externally controlled anisotropy. An alternative name for $\beta$ is the order-parameter exponent.}\label{tab:criticalexponents}
{\small
\renewcommand{\arraystretch}{1.75}
\begin{tabular}{|llll|}
\hline
Exponent &  & Conventional name & Physical definition
\\\hline
$\alpha= 2-\dfrac{d}{d-\Delta_{\varphi^2_S}}$ &\texttt{alpha}& Specific heat exponent & $C\sim |\tau|^{-\alpha}$
\\
$\beta=\dfrac{\Delta_\varphi}{d-\Delta_{\varphi^2_S}}$ &\texttt{beta}& Magnetisation exponent & $|\mathbf M|,\,\rho-\rho_c,\sim|\tau|^\beta$
\\
$\gamma=\dfrac{d-2\Delta_\varphi}{d-\Delta_{\varphi^2_S}}$ &\texttt{gamma}& Susceptibility exponent &$\chi,\,\kappa_T\sim |\tau|^{-\gamma}$
\\
$\delta=\dfrac{d-\Delta_\varphi}{\Delta_\varphi}$ &\texttt{delta}&

& $|\mathbf M|\sim|\mathbf H|^{1/\delta}$
\\
$\eta=2\Delta_{\varphi}-(d-2) $ &\texttt{eta}&  & $G^{(2)}(r)\sim\dfrac1{r^{2\Delta_\varphi}}$
\\
$\nu=\dfrac1{d-\Delta_{\varphi^2_S}}$ &\texttt{nu}&Correlation length exponent& $\xi\sim |\tau|^{-\nu}$
\\
$\phi_c=\dfrac{d-\Delta_{\varphi^2_T}}{d-\Delta_{\varphi^2_S}}$ &\texttt{phic}&Cross-over exponent& $\xi\sim |g|^{-\phi_c}$
\\
$\omega=\Delta_{\varphi^4_S}-d$ &\texttt{omega}& Correction-to-scaling exponent & see text
\\\hline
\end{tabular}
}
\end{table}

The critical exponents parametrise the scaling (often divergence) of certain physical quantities as the critical point is approached. For instance, the specific heat exponent $\alpha$ determines the behaviour of the specific heat capacity $C$ of a system as the temperature $T$ approaches its critical value $T_c$:
\begin{equation}
\label{eq:scalingCpre}
C=A_\pm |\tau|^{-\alpha}+\ldots, \qquad \tau=\frac{T-T_c}{T_c}.
\end{equation}
The amplitudes $A_\pm$ are not universal and depend on the specific material or simulation, however both $\alpha$ and the amplitude ratio $A_+/A_-$ are universal (for the latter, see section~\ref{sec:amplituderatios}). Some critical exponents are only defined above or below the fixed-point, for instance the exponent $\beta$ is only defined for $T<T_c$ in magnetic systems.

From the ``scaling hypothesis'', that physical quantities at criticality depend on a few dimensionless ratios, relations between the exponents were found, and are referred to as ``scaling laws'' \cite{Essam1963,Widom1965,Widom1965b,Kadanoff1966,Fisher1967}. Conventionally
\begin{align}
\alpha+2\beta+\gamma&=2 \quad \text{(Rushbrook)}, & \gamma-\beta(\delta-1)&=0\quad \text{(Widom)},
\\
\gamma-\nu(2-\eta)&=0\quad \text{(Fisher)}, & \alpha+d\nu&=2\quad \text{(Josephson)}.
\end{align}

\paragraph{Cross-over exponent}
For a system with a controllable anisotropy $g_{\{ij\}}\varphi^i\varphi^j$,\footnote{Curly brackets denote symmetrisation and removal of traces.} one defines the cross-over exponent as the exponent that controls the divergence of the correlation length as one approaches the $\ON$ symmetric point
\cite{Riedel1969,Fisher1972}, see also \cite{Chen:1982zz,Wegner:1972zz,Fisher:1974uq},
\begin{equation}
\xi\sim |g|^{-\phi_c}.
\end{equation}
As indicated in table~\ref{tab:criticalexponents}, the cross-over exponent is related to the dimension of the operator $\varphi^{\{i}\varphi^{j\}}=\varphi^2_T$.

\paragraph{Correction to scaling}

For practical purposes, any measurement will always take place at a finite distance from the fixed-point. 
All scaling relations such as \eqref{eq:scalingCpre} contain subleading terms which parametrise the correction to scaling. Apart from subleading integer powers to the original scaling, corrections to scaling derive from the presence of irrelevant operators in the theory \cite{Wegner:1972my}. As an example, the corrections to \eqref{eq:scalingCpre} take the form
\begin{equation}
\label{eq:scalingC}
C=A_\pm |\tau|^{-\alpha}\left(1+A_1\tau+\ldots+B_{\pm,1}|\tau|^{\nu\omega_1}+\ldots\right),
\end{equation}
where we have included only the leading correction to scaling parametrised by the ``correction-to-scaling exponent'' $\omega_1$. In general, one can include several correction-to-scaling exponents given by $\omega_i=\Delta_{\O_i}-d$ for irrelevant singlet operators $\O_i$ \cite{Wegner:1972my,Brezin:1974zr}.\footnote{In a realisation where the UV description breaks $O(N)$ symmetry, one has to include correction-to-scaling exponents for all low-lying operators that have a singlet component under the UV symmetry.} If $\omega_i\approx 0$, their effect on the scaling is significant. For the case of the $\ON$ CFT, the corrections in \eqref{eq:scalingC} and similar equations are dominated by leading correction-to-scaling exponent $\omega=\omega_1=\Delta_{\varphi^4_S}-d$.
Sometimes one employs $\Delta$ or $\theta$ for the quantity 
\begin{equation}
\Delta=\nu\omega.
\end{equation}

When dealing with a system on a lattice, one may also include the exponent $\omega_{\mathrm{NR}}$, ``non-rotational correction-to-scaling exponent'', which determines the scaling corrections due to breaking of rotational symmetry on the lattice. It is related to the scaling dimension of the spin 4 $\mathbb Z_2$ even/singlet operator $C^{\mu\nu\rho\sigma}$,
\begin{equation}
\omega_{\mathrm{NR}}=\Delta_C-d.
\end{equation}
Occasionally one writes $\sigma=\omega_{\mathrm{NR}}-2$, e.g.\ \cite{Hasenbusch:2021rse}. Sometimes the notation $\omega_A$ is used for the correction deriving from the first irrelevant $\mathbb Z_2$ odd operator in the $N=1$ case (or $V$ operator at general $N$), with the definition $\omega_A=\Delta_{\phi^5}-d$ ($\omega_{\mathrm A}=\Delta_{\varphi^5_V}-d$).

\subsubsection{OPE coefficients and central charges}

For scalar operators, conformal symmetry dictates that the three-point function takes the form
\begin{equation}
\langle\O_1(x_1)\O_2(x_2)\O_3(x_3)\rangle=\frac{\lambda_{\O_1\O_2\O_3}}{|x_{12}|^{\Delta_1+\Delta_2-\Delta_3}|x_{13}|^{\Delta_1+\Delta_3-\Delta_2}|x_{23}|^{\Delta_2+\Delta_3-\Delta_1}},
\end{equation}
where $\lambda_{\O_1\O_2\O_3}$ denotes the OPE coefficient. Similar formulas exist for operators in general Lorentz irreps, but are somewhat complicated, and in general there may be several independent OPE coefficients, see e.g.\ \cite{Kravchuk:2016qvl}. In the case where two of the operators, say $\O_1$ and $\O_2$, are scalars, there is only one OPE coefficient, which is non-zero only if $\O_3$ transforms in the traceless-symmetric Lorentz representation, spin $\ell=0,1,2,\ldots$. See equation \eqref{eq:opewithspinL} in appendix~\ref{app:normalisation} for the explicit form of the three-point function in that case.

OPE coefficients in the case where $\O_3$ is a conserved current $J^{\mu}$ or $T^{\mu\nu}$ are fixed in terms of the central charges $C_J$ and $C_T$, as discussed below. Of the other OPE coefficients, the case $\O_1=\O_2=\varphi$, $\O_3=\varphi^2_S$ has received the most attention, and has been computed to order $\eps^3$ \cite{Dey:2016mcs} (for $N=1$ numerically to order $\eps^4$ \cite{Carmi:2020ekr}), to order $N^{-3/2}$\cite{Lang:1993ct},\footnote{Note that it scales as $N^{-1/2}$.} and numerically with various methods such as the conformal bootstrap \cite{Simmons-Duffin:2016wlq,Chester:2019ifh,Chester:2020iyt} and non-perturbative RG \cite{Rose:2021zdk}.

\paragraph{Central charges.}

We use conventions where the central charges $C_T$ and $C_J$ in the theory of $N$ scalar fields in $d$ dimensions take the form
\begin{equation}\label{eq:CTNscalars}
C_{T,\mathrm{free}}=\frac{Nd}{d-1} \qquad C_{J,\mathrm{free}}=\frac{2}{d-2}.
\end{equation}
$C_T$ is denoted the central charge and $C_J$ is often called the current central charge.
In our conventions for the OPE coefficients, we have
\begin{equation}
\label{eq:OPEwithTandJ}
\lambda^2_{\varphi\varphi T}=\frac{d^2\Delta^2_{\varphi}}{(d-1)^2}\frac1{4C_T}, \qquad 
\lambda_{\varphi\varphi J}^2=\mathcal N_{A}\frac1{C_J},
\end{equation}
with $\mathcal N_{A}=\sqrt{\frac{N-1}{2N}}$. It is common to use conventions where $\mathcal N_A=1$, see appendix~\ref{app:normON} for normalisation conventions for global symmetry irreps.
It is also common in the literature to give values in conventions of canonical normalisation (see appendix~\ref{app:normalisation}), in which case
\begin{equation}
c_T=\frac{C_T}{\mathrm S_d^2}\qquad
c_J=\frac{C_J}{\mathrm S_d^2},
\end{equation}
for $
\mathrm S_d=\mathrm{Vol}(S^{d-1})=\frac{2\pi^{d/2}}{\Gamma(d/2)}
$.
The two-dimensional central charge is given by $c=\frac12C_T$. For quantum-critical systems in $2+1$ dimensions, $C_J$ is equivalent to the universal conductivity $\sigma_\infty$ by
\begin{equation}
\sigma_\infty=\frac1{32}C_J\qquad \text{$d=3$},
\end{equation}
see e.g.\ \cite{Kos:2015mba}.

\subsection{Results for main observables}

In this section we present the results for the main observables, which are the five scaling dimensions and two central charges included in table~\ref{tab:history}. We cover the three perturbative expansions as well as the cases $d=2$ and $d=3$.

\subsubsection[\texorpdfstring{$\eps$-expansion}{ε-expansion}]{$\boldsymbol\eps$-expansion}
\label{sec:epsresults}

In the $d=4-\eps$ expansion, the results for $\Delta_\varphi$, $\Delta_{\varphi^2_S}$ and $\Delta_{\varphi^4_S}$ follow from a multiplicative renormalisation procedure, as we review on section~\ref{sec:multiplicativerenorm}. For other operator dimensions, one considers diagrams with insertions of the operator under investigation. For the OPE coefficients and central charges, analytic conformal bootstrap methods have produced the highest-order results.
In the ancillary data file, the operator dimensions are given by \texttt{DeltaE[$\langle \O\rangle$]} and other quantities by \texttt{ValueE[$\langle qty\rangle$]}, where $\langle\O\rangle$ denotes the implemented expressions for the operators according to table~\ref{tab:importantops}, and $\langle qty\rangle$ is the implemented symbol of the quantity.\footnote{For critical exponents, these symbols are given in table~\ref{tab:criticalexponents}. For the central charges, \texttt{CT} denotes the ratio $\frac{C_T}{C_{T,\mathrm{free}}}$ and likewise \texttt{CJ} denotes $\frac{C_J}{C_{J,\mathrm{free}}}$.} The results, computed according to the references in table~\ref{tab:history}, read
\begin{align}
\Delta_\varphi &=1-\frac\eps2+\frac{ N+2}{4 (N+8)^2}\eps^2
+\ldots &&+ O(\eps^9)
,\\
\Delta_{\varphi^2_S} &=2-\eps+\frac{N+2}{N+8}\eps+\frac{
   (N+2) (13 N+44)}{2 (N+8)^3}\eps^2+\ldots&&+O(\eps^8),
\\
\label{eq:gammaphi4eps}
\Delta_{\varphi^4_S} &=4-\frac{3  (3
   N+14)}{(N+8)^2}\eps^2 +\ldots&&+O(\eps^8),
\\
\Delta_{\varphi^2_T} &=2-\eps+\frac2{N+8}\eps+\frac{(22-N) (N+4)}{2 (N+8)^3}\eps^2+\ldots&&+O(\eps^7),
\\
\Delta_{\varphi^4_{T_4}}&=4-2\eps+\frac{12}{N+8}\eps-\frac{5N^2+14N+152}{(N+8)^3}\eps^2+\ldots&&+O(\eps^7),
\\
\frac{C_T}{C_{T,\mathrm{free}}} &=1-\frac{5(N+2)}{12(N+8)^2}\eps^2+\ldots&&+O(\eps^5),
\\
\frac{C_J}{C_{J,\mathrm{free}}} &=1-\frac{3(N+2)}{4(N+8)^2}\eps^2
+\ldots&&+O(\eps^5).
\end{align}

For $N=1$ they reduce to
\begin{align}
\Delta_\phi &=1-\frac{\eps}2+\frac{1}{108}\eps^2+\frac{109}{11664}\eps^3+
   \bigg[\frac{7217}{1259712}-\frac{2 \zeta_3}{243}\bigg]\eps^4 +\ldots&&+O(\eps^9),
\\
\Delta_{\phi^2} &=2-\eps+\frac\eps3+\frac{19}{162}\eps^2+\bigg[\frac{937}{17496}-\frac{4 \zeta_3}{27}\bigg]\eps^3
\\&
\quad +\frac{24857-154224 \zeta _3-209952 \zeta_4+933120 \zeta _5}{1889568}\eps^4 +\ldots&&+O(\eps^8),
\\\nonumber
\Delta_{\phi^4} &=4-\frac{17}{27}\eps^2+\bigg[\frac{1603}{2916}+\frac{8 \zeta
   _3}{9}\bigg]\eps^3
   \\&  \quad 
   -\frac{178417+204768 \zeta _3-209952 \zeta _4+1399680 \zeta
   _5}{314928}\eps^4  +\ldots&&+O(\eps^8),
\\
\label{eq:CTIsing}
\frac{C_T}{C_{T,\mathrm{free}}} &=1-\frac{5}{324}\eps^2-\frac{233}{8748}\eps^3-\bigg[\frac{100651}{3779136}-\frac{55 \zeta_3}{2916}\bigg]\eps^4&&+O(\eps^5).
\end{align}

In general, the results in the $\eps$-expansion form an asymptotic series (zero radius of convergence),\footnote{The specific growth of the coefficients in the asymptotic series describing the main critical exponents has been discussed in \cite{Lipatov:1976ny,Brezin:1976vw,McKane1978,McKane:1984eq,Komarova:2001nw}; see \cite{McKane:2018ocs} for a recent overview.} which produces coefficients that first decrease in absolute value and then increase. For instance, the anomalous dimension of operators $\phi^k$ evaluate to
\begin{align}\label{eq:phinum}
\gamma_{\phi}&= 
0.00926 \eps^2 + 0.00934 \eps^3 - 0.00416 \eps^4 + 
 0.01283 \eps^5 - 0.04064 \eps^6 + 0.15738 \eps^7+0.68909 \eps^8
 ,
\\\label{eq:phi2num}
\gamma_{\phi^2}&=
0.33333\eps + 0.11728 \eps^2 - 0.12453 \eps^3 + 0.30685 \eps^4 - 
 0.95124 \eps^5 + 3.57258 \eps^6 - 15.28686 \eps^7,
\\\label{eq:phi4num}
\gamma_{\phi^4}&=2\eps-0.62963 \eps^2 + 1.61822 \eps^3 - 5.23514 \eps^4 + 20.74984 \eps^5 -  93.11128 \eps^6 + 458.74239 \eps^7
,
\end{align}
where we defined $\gamma_{\phi^k}$ by $\Delta_{\phi^k}=k[1-\frac\eps2]+\gamma_{\phi^k}$. The expansions have been resummed with various techniques to obtain estimates for the critical exponents in three dimensions, see \cite{Shalaby:2020xvv,Abhignan:2020xcj} for the most recent updates using the $O(\eps^7)$ values.

It is interesting to study the transcendental numbers that appear in these expansions, see e.g.\ \cite{Schnetz:2016fhy,Panzer:2016snt}. The types of numbers that appear in Feynman integrals are called periods. For $\phi^4$ theory, up to order $\eps^6$, all periods can be expressed in multiple zeta values, defined as 
\begin{equation}
\label{eq:multizetavalues}
\zeta_{k_1,k_2,\ldots,k_p}=\sum_{n_1=1}^\infty\sum_{n_2=1}^{n_1-1}\cdots\sum_{n_p=1}^{n_{p-1}-1}\frac{1}{n_1^{k_1}n_2^{k_2}\cdots n_p^{k_p}}.
\end{equation}
We implement them symbolically as $\texttt{mz[}k_1\texttt,k_2\texttt,\ldots\texttt]$. With only one index, the multiple zeta values reduce to the normal Riemann zeta values $\zeta_k=\zeta(k)$, implemented as $\texttt{z[}k\texttt]$.

At order $\eps^7$, there is a period, $P_{7,11}=200.357566429\ldots$, which cannot be written as a linear combination of multiple zeta values \cite{Panzer:2015ida,Panzer:2016snt}. In the ancillary data file, the command \texttt{multiZetaSub} can be used to make numerical substitutions for $P_{7,11}$ and the multiple zeta values.

\subsubsection[Large $N$ expansion]{Large $\boldsymbol N$ expansion}
In the large $N$ expansion for $2<d=2\mu\leqslant4$, composite operators are constructed out of the fields $\varphi$ and $\sigma$. Specifically,
$\varphi=\texttt{Op[V,0,1]}$, $\sigma=\texttt{Op[S,0,1]}$, $\sigma^2=\texttt{Op[S,0,2]}$, etc. Operator dimensions are implemented in the ancillary data file by \texttt{DeltaN} and other quantities by \texttt{ValueN}.
To get more readable expressions, we display all quantities at large $N$ in terms of $\gamma_\varphi^{(1)}=\eta_1/2$, i.e.\ the order $1/N$ anomalous dimension of $\varphi$, defined by $\Delta_\varphi=\mu-1+\gamma_\varphi^{(1)}/N+O(N^{-2})$. It takes the value
\begin{equation}
\label{eq:eta1half}
\frac{\eta_1/2}{N}=\frac{\texttt{eta1half}}N=\frac{(\mu-2)\Gamma(2\mu-1)}{\Gamma(1-\mu)\Gamma(\mu)^2\Gamma(\mu+1)}\frac1N,
\end{equation}
where \texttt{eta1half} denotes the implementation in the data file.

The results for the main observables, computed according to the references in table~\ref{tab:history}, are as follows:
\begin{align}
\label{eq:deltaNphi}
\Delta_\varphi &=\mu-1+\frac{\eta_1/2}N +\ldots &&+O(N^{-4}),
\\
\Delta_{\sigma} &=2-\frac{4(\mu-1)(2\mu-1)}{2-\mu}\frac{\eta_1/2}N+\ldots &&+O(N^{-3}),
\\
\Delta_{\sigma^2} &=4-4(2\mu-1)^2\frac{\eta_1/2}N+\ldots &&+O(N^{-3}),
\\
\Delta_{\varphi^2_T} &=2(\mu-1)+\frac{4}{\mu-2}\frac{\eta_1/2}N+\ldots &&+O(N^{-3}),
\\
\Delta_{\varphi^4_{T_4}}&=4(\mu-1)+\frac{8(\mu+1)}{2-\mu}\frac{\eta_1/2}N+\ldots &&+O(N^{-3}),
\\
\frac{C_T}{C_{T,\mathrm{free}}} &=1-\frac{\frac4{\mu(2-\mu)}+1+ 2\pi\,\cot(\pi \mu)+2\,S_1(2\mu-2)}{\mu+1}\frac{\eta_1/2}N\hspace{-8pt}& &+O(N^{-2}),
\\
\frac{C_J}{C_{J,\mathrm{free}}} &=1-\frac{2(2\mu-1)}{\mu(\mu-1)}\frac{\eta_1/2}N& &+O(N^{-2}),
\end{align}
where $S_1(a)$ denotes the analytic continuation of the harmonic numbers.
The anomalous dimension of $\varphi$ at order $N^{-3}$ is extremely complicated and it has not been implemented in closed form in the data file. Instead it is given as $\texttt{eta3[mu]}/2$, where $\texttt{eta3[$\mu$]}$ is implemented for $\mu=1,\frac32,2,\frac52,3$.

In three dimensions, $\mu=3/2$, the results take the form
\begin{align}\nonumber
\Delta_\varphi &=\frac12+\frac4{3\pi^2}\frac1N-\frac{256}{27\pi^4}\frac1{N^2}\\&\quad+\bigg[
\frac{32(108\ln 2-61)}{81\pi^4}-\frac{64(1594+1701\zeta_3)}{243\pi^6}
\bigg]\frac1{N^3}&&+O(N^{-4}),
\\
\Delta_{\sigma} &=2-\frac{32}{3\pi^2}\frac1N-\bigg[\frac{32}{\pi^2}-\frac{512}{27\pi^4}\bigg]\frac1{N^2}&&+O(N^{-3}),
\\
\Delta_{\sigma^2} &=4-\frac{64}{3\pi^2}\frac1N-\bigg[\frac{64}{\pi^2}-\frac{13312}{27\pi^4}\bigg]\frac1{N^2}&&+O(N^{-3}),
\\
\Delta_{\varphi^2_T} &=1+\frac{32}{3\pi^2}\frac1N-\frac{512}{27\pi^4}\frac1{N^2}&&+O(N^{-3}),
\\
\Delta_{\varphi^4_{T_4}} &=2+\frac{160}{3\pi^2}\frac1N-\bigg[-\frac{128}{\pi^2}+\frac{42496}{27\pi^4}\bigg]\frac1{N^2}&&+O(N^{-3}),
\\
\frac{C_T}{C_{T,\mathrm{free}}} &=1-\frac{40}{9\pi^2}\frac1N&&+O(N^{-2}),
\\
\frac{C_J}{C_{J,\mathrm{free}}} &=1-\frac{64}{9\pi^2}\frac1N+\frac{1.14230(2)}{N^2}&&+O(N^{-3}),
\end{align}
where \cite{Alday:2019clp} only reported a numerical expression for $C_J$ at $O(N^{-2})$, computed via a Pad\'e approximant of a large spin expansion.

\subsubsection{Non-linear sigma model}

Results for the critical $\ON$ CFT can be derived in $2+\teps$ dimensions by studying a non-linear sigma model \cite{Wegner:1972my}.
We leave some details of this expansion to section~\ref{sec:NLSMmethod}, and give here only the main results, which are valid for $\teps>0$ and $N>2$ strictly.
In the ancillary data file, operators are implemented by \texttt{DeltaZ} and other quantities by \texttt{ValueZ}.
We have $\texttt{Op[V,0,1]}=\pi$, \teord4\cite{Wegner:1987gu}, $\texttt{Op[S,0,1]}=\de\pi\cdot\de\pi$, \teord4\cite{Bernreuther:1986js}, $\texttt{Op[S,0,2]}=(\de\pi\cdot\de\pi)^2$, \teord1\cite{Brezin:1976an,Wegner1990}, $\texttt{Op[T,0,1]}=\pi^2_T$, \teord4\cite{Wegner:1987gu}, $\texttt{Op[Tm[4],0,1]}=\pi^4_{T_4}$, \teord4\cite{Wegner:1987gu}, with
\begin{align}
\Delta_\pi&=\frac\teps2+\frac{\teps}{2(N-2)}-\frac{N-1}{2(N-2)^2}\teps^2
+\ldots+
&&+O(\teps^5),
\\
\Delta_{\de\pi \cdot\de\pi}&=2-\frac{\teps^2}{N-2}-\frac{\teps^3}{2(N-2)}+\ldots&&+O(\teps^5),
\\
\Delta_{(\de\pi \cdot\de\pi)^2}&=4-\frac{2}{N-2}\teps&&+O(\teps^2),
\\
\Delta_{\pi_T^2}&=\teps+\frac2{N-2}\teps-\frac{N}{(N-2)^2}\teps^2+\ldots&&+O(\teps^5),
\\
\Delta_{\pi^4_{T_4}}&=2\teps+\frac8{N-2}\teps-\frac{2(N+2)}{(N-2)^2}\teps^2+\ldots&&+O(\teps^5).
\end{align}
Moreover, we have \cite{Diab:2016spb}
\begin{align}
\label{eq:CTnlsm}
\frac{C_T}{C_{T,\mathrm{free}}} &=\frac{N-1}{N}+\frac{3(N-1)}{4N(N-2)}\teps^2+O(\teps^3),
\\
\frac{C_J}{C_{J,\mathrm{free}}} &=\frac{N-2}N+\frac{\teps}N+O(\teps^2).
\end{align}

\subsubsection{Exact results in two dimensions}

These exists a family of exact results in $d=2$ indexed by $N\in[-2,2]$, following from computation of the partition function. These results were first been derived by Nienhuis \cite{Nienhuis:1982fx,Nienhuis:1984wm}, and later by di Francesco, Saleur and Zuber who wrote down the torus partition function
\cite{diFrancesco:1987qf,DiFrancesco:1987gwq}.
Results are given as analytic functions of $N$, and in the limit where $N$ approaches an integer they provide results for a logarithmic CFT which is an extension of the usual integer-$N$ CFT, see \cite{Gorbenko:2020xya} for a recent discussion. 
We express all quantities in terms of the Coulomb gas coupling
\begin{equation}
g=1+\frac1\pi\arccos(N/2).
\end{equation}
The values $N=\{-2,-1,0,1,2\}$ correspond to $g=\left\{2,\frac{5}{3},\frac{3}{2},\frac{4}{3},1\right\}$. 
Note that while the partition function provides the spectrum, most OPE coefficients are not known for general $g$. For a discussion of the organisation of the spectrum into $\ON$ multiplets, see \cite{Grans-Samuelsson:2021uor}.

The results for the main observables are \cite{Nienhuis:1982fx}
\begin{align}
\Delta_\varphi&=1-\frac{3 g}{8}-\frac{1}{2 g},
\\
\Delta_{\varphi^2_S}&=\frac4g-2,
\end{align}
and \cite{diFrancesco:1987qf}
\begin{equation}
\Delta_{\varphi^2_T}=1-\frac1{2g}.
\end{equation}
Moreover we have results for $c$, equivalent to $C_T$, 
\begin{equation}
\label{eq:cCTrel}
c=\frac12C_T=1-6\frac{(g-1)^2}g,
\end{equation}
and for $C_J$ \cite{Cardy1994}\footnote{As pointed out in \cite{Cardy2002}, the formula in \cite{Cardy1994} contained a typo. We have multiplied the expression in \cite{Cardy1994,Cardy2002} by a factor $2/N$, to match the conventions given in appendix~\ref{app:normalisation}.}
\begin{equation}
C_J=
\frac{2(1-g)}{g\pi\sin(\pi g)}.
\end{equation}
Note that $C_{J,\mathrm{free}}$, as defined in \eqref{eq:CTNscalars}, blows up in two dimensions.

\subsubsection[Numerical values in three dimensions for small $N$]{Numerical values in three dimensions for small $\boldsymbol N$} 

For the $\ON$ CFT in $d=3$ dimensions, many different methods can be used to give estimates of the main observables. In table~\ref{tab:IsingComparison} in the introduction, we presented a sample of results from experimental and theoretical approaches in the case $N=1$. 
Here we shall complement that list for the other experimentally relevant cases of low $N$, focussing on the most precise theoretical estimates. A comprehensive and up-to-date survey of different results for $N=1,2,3,4,5,10$ is given in \cite{DePolsi:2020pjk}. The tables of that paper also include estimates from the non-perturbative RG \cite{Balog:2019rrg} and combined MC and high-temperature expansion \cite{Campostrini:2002cf,Campostrini:2002ky,Campostrini:2006ms}.

\begin{table}[ht]
\centering
\caption{Results in three dimensions from the conformal bootstrap. The lower part of the table contains the critical exponents, computed from the values in the upper part. For comparison, we have added the exact results at $N=\infty$ (sometimes denoted the spherical model).}\label{tab:numboot}
{
\setcounter{localfn}{1} \small
\renewcommand{\arraystretch}{1.25}
\begin{tabular}{|c|ll|ll|ll|c|}
\hline
Quantity & \multicolumn{2}{c|}{ $N=1$ }& \multicolumn{2}{c|}{ $N=2$ }& \multicolumn{2}{c|}{ $N=3$ } & $N=\infty$
\\\hline
$\Delta_\varphi $ ($\Delta_\sigma$) & $0.5181489(10)$ &\cite{Kos:2016ysd}& $0.519088(22)$ &\cite{Chester:2019ifh}& $0.518942(51)$ &\cite{Chester:2020iyt} & $0.5$
\\
$\Delta_{\varphi^2_S} $ ($\Delta_\epsilon$) & $1.412625(10)$ &\cite{Kos:2016ysd}& $1.51136(22)$ &\cite{Chester:2019ifh} & $1.59489(59)$ &\cite{Chester:2020iyt} & $2$
\\
$\Delta_{\varphi^4_S} $ ($\Delta_{\epsilon'}$) & 
$3.82951(61)$ &\cite{Reehorst:2021hmp}
 & $3.794(8)$ &\cite{Liu:2020tpf} & $3.7668(100)$ & \cite{Chester2022}\makefn & $4$
\\
$\Delta_{\varphi^2_T} $ ($\Delta_{t}$) & \graycell & \graycell & $1.23629(11)$ &\cite{Chester:2019ifh}& $1.20954(32)$ &\cite{Chester:2020iyt} & $1$
\\
$\Delta_{\varphi^4_{T_4}}$ ($\Delta_{t_4}$) & \graycell& \graycell & $3.11535(73)$ &\cite{Chester:2019ifh} & $< 2.99056$\makefn  &\cite{Chester:2020iyt} & $2$
\\[4pt]
$\dfrac{C_T}{C_{T,\mathrm{free}}}$ & 
$0.946543(42)$& \cite{Reehorst:2021hmp}
& $0.944056(15)$ &\cite{Chester:2019ifh} & $0.944524(28)$ &\cite{Chester:2020iyt} & $1$
\\[8pt]
$\dfrac{C_J}{C_{J,\mathrm{free}}}$ & \graycell& \graycell & $0.904395(28)$  &\cite{Chester:2019ifh}& $0.90632(16)$ &\cite{Chester:2020iyt} & $1$
\\[4pt]\hline
$\alpha$ & $0.110087(12)$&& $-0.01526(30)$&& $-0.1350(12)$ &  & $-1$
\\
$\beta$ & $0.326418(2)$&& $0.348699(54)$&& $0.36932(22)$ & & $0.5$
\\
$\gamma$  & $1.237075(8)$&& $1.31786(20)$&& $1.39641(81)$ & & $2$
\\
$\delta$ & $4.789841(11)$&& $4.77937(24)$&& $4.78106(75)$ & & $5$
\\
$\eta$ & $0.0362978(20)$&& $0.038176(44)$& &$0.03787(13)$ & & $0$
\\
$\nu$ & $0.629971(4)$&\cite{Kos:2016ysd}& $0.671754(99)$&\cite{Chester:2019ifh}& $0.71168(41)$&\cite{Chester:2020iyt}  & $1$
\\
$\phi_c$ &  \graycell& \graycell& $1.18478(19)$ &\cite{Chester:2019ifh}& $1.27424(83)$ &\cite{Chester:2020iyt}  & $2$
\\
$\omega$& 
$0.82951(61)$ &\cite{Reehorst:2021hmp}
& $0.794(8)$ &\cite{Liu:2020tpf}& $0.7668(100)$ &\cite{Chester2022} & $1$
\\\hline
\end{tabular}
\flushleft
\setcounter{localfn}{1} 

\makefn Computed using $\Lambda=35$ data from $42$ sampling points. I thank J. Liu for sharing this value.
\\
\makefn This value is a rigorous upper bound and the true operator dimension is expected to lie not far from this value.
\\
}
\end{table}

In table~\ref{tab:numboot} we present results from the conformal bootstrap, which give high-precision estimates for operator dimensions and central charges. For completeness we also include the critical exponents.\footnote{The errors for the critical exponents have been estimated using standard error propagation, however, from the shape of the ``islands'' in the bootstrap exclusion plots, it is obvious that the errors in $\Delta_{\varphi}$ and $\Delta_{\varphi^2_S}$ are correlated and more suitable error bars could in principle be computed from the raw data.} In table~\ref{tab:numMC} we give Monte Carlo results for the critical exponents for $N=0,1,2,3$, which have statistical error bars that are of similar order of magnitude as those in the numerical bootstrap results. 
For other small values of $N$, we refer to table~VI of \cite{Hasenbusch:2021rse} for a summary of results for critical exponents, and to \cite{Kos:2013tga} for the central charges.

\begin{table}[ht]
\centering
\caption{Results in three dimensions from Monte Carlo simulations. For $N=0$ the exponent $\nu$ and the Hausdorff dimension $D_H=d-\Delta_{\varphi^2_T}$ are directly related, see section~\ref{sec:Hausdorff}.}\label{tab:numMC}
{
\setcounter{localfn}{1} \small
\renewcommand{\arraystretch}{1.25}
\begin{tabular}{|c|ll|ll|ll|ll|}
\hline
Quantity & \multicolumn{2}{c|}{ $N=0$ }& \multicolumn{2}{c|}{ $N=1$ }& \multicolumn{2}{c|}{ $N=2$ }& \multicolumn{2}{c|}{ $N=3$ }
\\\hline
$\gamma$ &  $1.15695300(95)$&\cite{Clisby:2017dah}  && &&  & &
\\
$\eta$  & 
$0.0250(14)$& \cite{Liu:2012ca}
 & $0.036284(40)$ &  \cite{Hasenbusch:2021tei} & $0.03810(8)$ & \cite{Hasenbusch:2019jkj} &  $0.03784(5)$ &\cite{Hasenbusch:2020pwj}
\\
$\nu$ & $0.58759700(40)$ &\cite{Clisby2016} & $0.62998(5)$& \cite{Hasenbusch:2021tei} & $0.67169(7)$ & \cite{Hasenbusch:2019jkj} & $0.71164(10)$ &\cite{Hasenbusch:2020pwj}
\\
$\omega$ & $0.9037(56)$&  \cite{Belohorec1997} & $0.832(6)$ & \cite{Hasenbusch:2011yya} & $0.789(4)$ &  \cite{Hasenbusch:2019jkj} &  $0.759(2)$ & \cite{Hasenbusch:2020pwj}
\\
$d_H$ & $1.7018467(16)$\makefn  & \cite{Clisby2016} & $1.7349(65)$& \cite{Winter2008} & $1.7626(66)$& \cite{Winter2008} & $1.7906(3) $ & \cite{Hasenbusch2011} 
\\\hline
\end{tabular}
\flushleft
\setcounter{localfn}{1} 

\makefn Computed from $\Delta_{\varphi^2_T}$ using \eqref{eq:Hausdorffdef}.
\\
}
\end{table}

The (numerical) conformal bootstrap gives rigorous error bars on some operator dimensions by scanning over allowed values. Once a small isolated allowed region has been found, the extremal functional method \cite{Poland:2010wg,ElShowk2012} can be used to find approximate operator dimensions for larger parts of the spectrum. These values are estimates with no error bars, however, non-rigorous error bars can be assigned by applying the extremal functional method to a collection of points on the boundary of the allowed region in the parameter space \cite{Simmons-Duffin:2016wlq}. 
We also note that in the numerical bootstrap, it is customary to denote the scalar operators in the various representations as $X$, $X'$, $X''$ etc, ordered by increasing scaling dimension. Specifically
\begin{equation}
\text{$\mathbb Z_2$ even:} \quad \epsilon,\, \epsilon',\ldots, \qquad 
\text{$\mathbb Z_2$ odd:} \quad \sigma, \,\sigma',\ldots,
\end{equation}
for $N=1$ and
\begin{equation}
\text{$S$:}\quad S,\,S',\ldots, \qquad \text{$V$:} \quad \varphi,\, \varphi'
,\ldots, \qquad \text{$T$:} \quad t,\,t',\ldots,\qquad \text{$T_4$:}\quad \tau,\,\tau',\ldots,
\end{equation}
for general $N$. 
The symbols $T,\,T',\,T'',\ldots$ are used for the singlet ($\mathbb Z_2$ even) operators of spin $2$, with $T=T^{\mu\nu}$ being the stress-tensor.

For the current central charge $C_J$, the Monte Carlo result $2\pi\sigma_\infty=0.3605(3)$, given in \cite{Katz:2014rla} with the error bars that do not include systematic effects, translates to $\frac{C_J}{C_{J,\mathrm{free}}}=0.9181(8)$, a value which is not compatible with $0.904395(28)$ from the conformal bootstrap \cite{Chester:2019ifh}.

\subsection{Other observables}

\subsubsection{Four-point correlators} We give conventions for conformal four-point correlators in section~\ref{sec:fourpoint}. In the $\eps$-expansion, the correlator $\langle\phi\phi\phi\phi\rangle$ (for $N=1$) has been computed to order $\eps^2$ \cite{Bissi:2019kkx} and $\langle\varphi^2_S\varphi^2_S\varphi^2_S\varphi^2_S\rangle$ to order $\eps$ \cite{Thesis}. More correlators involving the operators $\varphi$, $\varphi^2_S$ and $\varphi^2_T$ appeared in \cite{Bertucci:2022ptt}.
At large $N$, the correlators $\langle\varphi\varphi\varphi\varphi\rangle$ \cite{Lang:1991kp} (see \cite{Giombi:2018vtc} for an expression in $\bar D$ functions \cite{Dolan:2000ut}) and $\langle\varphi\varphi\sigma\sigma\rangle$ \cite{Lang:1992pp} (see also \cite{Alday:2015ewa,Alday:2019clp}) have been computed to order $1/N$.

In the two-dimensional Ising CFT, the non-trivial correlators are given by
\begin{align}
\G_{\sigma\sigma\sigma\sigma}(z,\zb)&=\frac{\sqrt{1+\sqrt{1-z}}\sqrt{1+\sqrt{1-\zb}}+\sqrt{1-\sqrt{1-z}}\sqrt{1-\sqrt{1-\zb}}}{2(1-z)^{1/8}(1-\zb)^{1/8}},
\\
\G_{\sigma\sigma\epsilon\epsilon}(z,\zb)&=\frac{\left(1-\frac z2\right)\left(1-\frac\zb2\right)}{(1-z)^{1/2}(1-\zb)^{1/2}},
\\
\G_{\epsilon\epsilon\epsilon\epsilon}(z,\zb)&=\frac{(1-z+z^2)(1-\zb+\zb^2)}{(1-z)(1-\zb)}.
\end{align}
These correlators can be written in terms of Virasoro conformal blocks according to the Virasoro fusion rules $\sigma\times\sigma=\mathbb I+\epsilon$, $\sigma\times\epsilon=\sigma$ and $\epsilon\times \epsilon=\mathbb I$, where $\mathbb I$ denotes the Virasoro identity multiplet. Explicit expressions for the Virasoro conformal blocks can be found for instance in \cite{AlvarezGaume:1989vk}.

\subsubsection{Three-point correlators of conserved currents}\label{sec:threepoints}
 
In general, for three-point correlators involving more than one non-scalar operator, there is more than one independent OPE coefficient. The most interesting case is three-point functions involving the conserved currents $T^{\mu\nu}$ and $J^\mu$. In canonical normalisation (see appendix~\ref{app:normalisation}), the three-point functions can be parametrised in terms of the tensor structures associated to a theory of free scalars, free fermions and free vectors,
\begin{align}
\langle  \hat J(\zeta_1,x_1)\hat J(\zeta_2,x_2)\hat T(\zeta_3,x_3)\rangle&=n_s\langle  \hat J\hat J\hat T\rangle_s+n_f\langle  \hat J\hat J\hat T\rangle_f,
\\
\langle  \hat T(\zeta_1,x_1)\hat T(\zeta_2,x_2)\hat T(\zeta_3,x_3)\rangle&=n_s\langle  \hat T\hat T\hat T\rangle_s+n_f\langle  \hat T\hat T\hat T\rangle_f+n_v\langle  \hat T\hat T\hat T\rangle_v.
\end{align}
The structure $\langle  \hat T\hat T\hat T\rangle_v$ vanishes in three dimensions. 
Explicit expressions for the free-theory correlators are given for instance in \cite{Hartman:2016dxc}, and can be found in different various conventions in many places in the literature \cite{Osborn:1993cr,Erdmenger:1996yc,Chowdhury:2012km,Zhiboedov:2012bm,Zhiboedov:2013opa,Katz:2014rla,Li:2015itl,Hofman:2016awc,Dymarsky:2017xzb,Dymarsky:2017yzx}, typically written using the embedding space formalism \cite{Costa:2011dw,Costa:2011mg}. One choice of normalisation is
\begin{equation}
\label{eq:nsnfdefn}
n_s=N_s, \qquad n_f=2^{\lfloor\frac d2\rfloor}N_f,
\end{equation}
where $N_s$ and $N_f$ are the number of complex scalars and Dirac fermions respectively, and $2^{\lfloor\frac d2\rfloor}$ corresponds to the dimension of the gamma matrices used to describe the fermion.

Taking into account the overall normalisation fixed by the central charges, the three-point function $\langle JJT\rangle$ of two global symmetry currents $J^\mu$ and the stress tensor $T^{\mu\nu}$ contains in one free parameter, denoted $\gamma$. For a $\mathrm U(1)$ current, in CFT normalisation with unit normalisation of conserved currents (see appendix~\ref{app:normalisation}), one writes \cite{Hartman:2016dxc,Dymarsky:2017xzb},
\begin{align}
&\left\langle  J(\zeta_1,x_1)J(\zeta_2,x_2)T(\zeta_3,x_3)\right\rangle=\frac{d(d-2)}{2(d-1)^2\sqrt{C_T}}\bigg(
(2-3d-4d^3\gamma)V_1V_2V_3^2
\nonumber
\\
&\quad+(1-2d-4d^2\gamma)H_{12}V_3^2
-2d(1+4\gamma)(H_{13}V_2+H_{23}V_1)V_3+2\left(\frac1{d-2}-4d\gamma\right)H_{13}H_{23}
\bigg),
\end{align}
for embedding space  tensor structures given in \cite{Dymarsky:2017xzb}. 
In terms of $n_s$ and $n_f$ in \eqref{eq:nsnfdefn}, the parameter $\gamma$ is given by
\begin{equation}
\gamma=\frac{n_f-n_s}{4d(n_s+(d-2)n_f)},
\end{equation} 
and ranges in the interval \cite{Hofman:2008ar,Hofman:2016awc}
\begin{equation}
-\frac1{4d}\leqslant \gamma\leqslant \frac1{4d(d-2)}.
\end{equation}
The lower limit corresponds to the theory of free bosons, the upper limit to the theory of free fermions. For a holographic CFT with Abelian symmetry, the parameter $\gamma$ has the interpretation of the coefficient of $W_{\mu\nu\rho\sigma}F^{\mu\nu}F^{\rho\sigma}$ in the flat-space limit of the bulk Einstein--Maxwell (effective) theory. No $1/N$ correction is known for $\gamma$ in the $\ON$ CFT: $\gamma=-\frac1{4d}+O(1/N)$ \cite{Chowdhury:2012km,Katz:2014rla}. For $N=2$ in three dimensions, \cite{Reehorst:2019pzi} found numerically
\begin{equation}
\gamma=-0.0808(5)=\frac1{12}(-0.970(6)).
\end{equation}

For the three-point function $\langle TTT\rangle$ of three stress-tensors, there are in general two free parameters when taking into account the overall scaling given by $C_T$. We are not aware of any results for these parameters in the $\ON$ CFT in the perturbative expansions.
In three dimensions, the number of tensor structures reduces to two (for parity preserving theories), meaning that in addition to $C_T$ there is one free parameter. We describe the result found in \cite{Dymarsky:2017yzx}, which introduces an angle $\theta$ by $\tan\theta=\frac{n_f}{n_s}$. $\theta=0$ corresponds to a free boson and $\theta=\frac\pi2$ to a free fermion. In terms of $\theta$, \cite{Dymarsky:2017yzx} found rigorously (assuming a set of assumptions compatible with Ising values, and that leading parity-odd scalar is irrelevant)
\begin{equation}
0.01<\theta<0.05.
\end{equation}
Assuming a larger gap in the parity-odd scalar sector, they found $0.010<\theta<0.019$.\footnote{An upper bound was found for the leading parity-odd scalar in the Ising model in \cite{Dymarsky:2017yzx}, $\Delta_{\O_{\mathrm{odd,min}}}<11.2$, see section~\ref{eq:three-rowYT}.}

\subsubsection{Universal amplitude ratios}
\label{sec:amplituderatios}

An important class of observables, directly measurable in experiments and Monte Carlo simulations, are universal amplitude ratios  \cite{Brezin1974}. Since they are not defined in terms of conformal data, they generally fall outside the scope of this report, and for a complete treatment we refer to the review \cite{Pelissetto:2000ek} and to the more recent paper \cite{DePolsi:2021cmi} from the point of view of the non-perturbative RG.
Here we focus on one particular amplitude ratio as a showcase, namely $U_0$, defined by the divergence of the heat capacity as the critical point is approached:
\begin{equation}
U_0=\frac{A_+}{A_-},
\end{equation}
where the amplitudes $A_+$ and $A_-$ were defined in \eqref{eq:scalingCpre}. 
In the $\eps$-expansion, it has been computed to order $\eps^2$ \cite{Bervillier:1986zz},
\begin{align}
\nonumber
U_0=\frac{2^\alpha N}4\bigg[1+\eps+\frac{\eps^2}{2(N+8)^2}\bigg(&3N^2+26N+100+(4-N)(N-1)\frac{\pi^2}6
\\&
-6(5N+22)\zeta_3+9(N-4)\lambda\bigg)+O(\eps^3)\bigg],
\end{align}
where $\alpha$ is the critical exponent $\alpha$ and $\lambda=\frac13\psi'(\frac13)-\frac29\pi^2=1.171953$, $\psi(x)=\frac{\Gamma'(x)}{\Gamma(x)}$ being the digamma function.\footnote{In \cite{Bervillier:1986zz}, $\lambda$ was only given numerically, but the exact form appeared in \cite{Guida:1996ep,ZinnJustin:1999bf}, and in other amplitude ratios in e.g.\ \cite{Brezin1974,Brezin1976}. See however \cite{Nicoll:1985zz}, which gives $\lambda$ with the alternative numerical expression $\lambda=1.171854\ldots$.} In three dimensions for small $N$, there are several theoretical estimates reported in \cite{Pelissetto:2000ek}, some of which are incompatible.

For more detailed tables of values of $U_0$ computed using different methods, see \cite{Gordillo-Guerrero:2011xya,Hasenbusch:2011yya} for $N=1$ and \cite{DePolsi:2021cmi} for $N=2,3,4,5$. For $N=1$, an experiment on succinonitrile--water mixture gave $U_0=0.536(5)$ \cite{Nowicki:2001gm}.
In the 2d Ising model, $U_0=1$ \cite{Onsager:1943jn}.

\subsubsection{Regge intercept}

The Regge intercept $\ell^*$ determines the growth in the Regge limit of the singlet channel of the four-point correlator of (pairwise) identical operators. Introduce the variables $\frac{x_{12}^2x_{34}^2}{x_{13}^2x_{24}^2}=u=z\zb=\sigma^2$, $\frac{x_{14}^2x_{23}^2}{x_{13}^2x_{24}^2}=v= (1-z)(1-\zb)=(1-\sigma e^\rho)(1-\sigma e^{-\rho})$. Following \cite{Costa:2012cb}, the Regge limit of a CFT four-point function is then defined as the limit $\sigma\to0$, with $\rho$ fixed and $\zb$ is evaluated on the second sheet, i.e.\ analytically continued around the branch cut starting at $\zb=1$. With this definition, the four-point function in the Regge limit $\sigma\to0$ scales as
\begin{equation}
\G(u,v)\sim \sigma^{1-\ell^*}.
\end{equation}
More precisely, the Regge intercept $\ell^*$ is the maximum over real $\nu$ of a function, ``Reggeon spin'' $\ell(\nu)$, for $\nu$ defined by $\Delta=\frac d2+i\nu$. The maximum is attained at $\nu=0$: $\ell^*=\ell(0)$.
In weakly coupled gauge theories, the intercept is given by the Pomeron exchange and the value of $\ell^*$ can be computed perturbatively using BFKL methods \cite{Kuraev:1977fs,Balitsky:1978ic,Kotikov:2002ab}. 

In a general CFT, the intercept can be understood from the point of view of light-ray operators \cite{Kravchuk:2018htv}. Consider the family of singlet operators of leading twist and even spin. The conformal data of this family can be extended to non-integer spin as describing the data of a continuous family of light-ray operators, defined for complex $\ell$. The scaling dimension of the family of singlet light-ray operators is given by a curve $f(\Delta,\ell)=0$ in the $(\Delta,\ell)$ plane, usually visualised in a Chew--Frautschi plot. Then $\ell^*$ is then given a non-perturbative definition as the intersection point (with largest $\ell$) of the curve with the vertical axis $\Delta=\frac d2$: $f(\frac d2,\ell^*)=0$. The continuation to imaginary $\Delta-\frac d2$ gives the Reggeon spin $\ell(\nu)$.

The leading perturbative corrections to $\ell^*$ can be computed using the anomalous dimension of weakly broken singlet currents $\mathcal J_{S,\ell}$ (family \texttt{ONF4[$l$]} below), which gives
\begin{align}\label{eq:interceptS}
\texttt{ValueE[intercept[S]]}&=\left(\frac12+\frac{\sqrt{6(N+2)}}{N+8}\right)\eps+\ldots+O(\eps^4),
\\
\texttt{ValueN[intercept[S]]}&=2-\mu+\sqrt{4\mu\left(\mu-1+\frac{\Gamma(2\mu-1)\Gamma(3-\mu)}{\Gamma(\mu-1)}\right)\frac{\eta_1/2}N}+O(N^{-1}).
\end{align}
For $N=1$, a prediction for the intercept to order $\varepsilon^4$ will be given in \cite{Caron-Huot:2022eqs}, encoded as \texttt{ValueE[intercept[E]]}.\footnote{I thank Murat Kolo\u{g}lu for useful discussions, and for sharing this value.} In three dimensions, the intercept at large $N$ has been computed to order $1/N$ \cite{Caron-Huot:2020ouj},\footnote{In principle, it is possible to perform this computation also for general $d$ using the $1/N^2$ anomalous dimension of $\mathcal J_{S,\ell}$. Unfortunately, no suitable identity to simplify the complicated hypergeometric functions involved in the computation has been found by the author.}
\begin{equation}
\ell^*|_{\mu=\frac32}=\frac12+\sqrt{\frac8{\pi^2N}}-\frac8{3\pi^2N}+O(N^{-3/2}).
\end{equation}
Similar to \eqref{eq:interceptS}, the intercept in the $T$ and $A$ representations can be computed perturbatively to order $\eps^3$ and to order $1/N$.
These results are in the data file as \texttt{ValueE[intercept[$\langle R\rangle$]} and \texttt{ValueN[intercept[$\langle R\rangle$]} for $\langle R\rangle = \texttt T,\texttt A$.

In $d=3$ dimensions, estimates for the Regge intercepts can be made by a numerical evaluation of the Lorentzian inversion formula \cite{Caron-Huot:2017vep} using high-precision numerical data computed from the conformal bootstrap \cite{Simmons-Duffin:2016wlq,Chester:2019ifh,Liu:2020tpf}. For $N=1$ \cite{Caron-Huot:2020ouj}, reports a value $\ell^*\approx 0.8$, and for $N=2$ \cite{Liu:2020tpf} reports $\ell^*\approx 0.82$, $\ell^*_T\approx 0.75$, $\ell^*_A\approx 0.69$. None of these values were assigned any error bars.

\subsubsection{Free energy} The rescaled free energy $\tilde F$ \cite{Giombi:2014xxa,Fei:2015oha} is defined by
\begin{equation}
\tilde F=\sin\left(\frac{\pi d}2\right)\ln Z_{S^d}=-\sin\left(\frac{\pi d}2\right)F,
\end{equation}
and conjecturally satisfies $\tilde F_{\mathrm{UV}}>\tilde F_{\mathrm{IR}}$ \cite{Giombi:2014xxa}. It was computed to order $\eps^5$ in \cite{Fei:2015oha}:
\begin{align}\nonumber
\tilde F&=N\tilde F_s(4-\eps)-\frac{\pi N(N+2)}{576(N+8)^2}\eps^3-\frac{\pi N(N+2)(13N^2+370N+1588)}{6912(N+8)^4}\eps^4
\\&\quad -\frac{\pi N (N+2) }{40 (N+8)^2}\bigg(-\frac{5 \pi ^2}{1728}-\frac{5 N+22}{(N+8)^3}\zeta _3
\\
\nonumber
&\qquad\qquad\qquad\qquad+\frac{647 N^4+32152 N^3+606576 N^2+3939520 N+8451008}{10368
   (N+8)^4} \bigg)\eps^5+O(\eps^5),
\end{align}
where $\tilde F_s(d)=\frac1{\Gamma(d+1)}\int_0^1du\sin(\pi u)\Gamma(d/2+u)\Gamma(d/2-u)$ is the free energy of a free scalar.
References \cite{Giombi:2014xxa,Fei:2015oha} also give the expression for large $N$ for $d=3$, see also \cite{Klebanov:2011gs}. At large $N$ for generic $d=2\mu$, \cite{Tarnopolsky:2016vvd} found
\begin{equation}
\tilde F=N\tilde F_s(2\mu)+\delta \tilde F\left(2\mu;2\mu-2\right)-\frac{(4 \mu ^2-6 \mu +3)\sin(\pi\mu)}{6 (\mu -1) \mu  (2 \mu -1)}\frac{\eta_1/2}N+O(N^{-2}),
\end{equation}
where $\delta\tilde F(d;\Delta)=\frac1{\Gamma(d+1)}\int_0^{\Delta-\frac d2}du\sin(\pi u)\Gamma(d/2+u)\Gamma(d/2-u)$. For $d=3$ it evaluates to
\begin{equation}
\tilde F=-\frac{N}{8}\left(\ln 2-\frac{3\zeta_3}{2\pi^2}\right)-\frac{\zeta_3}{8\pi^2}-\frac4{9\pi^2}\frac1N+O(N^{-2}).
\end{equation}

\subsubsection[Critical $N$ for cubic perturbation]{Critical $\boldsymbol N$ for cubic perturbation}

The $\ON$ CFT can be perturbed by the adding operator $\varphi^4_{T_4}$ to the action, with a coupling
\begin{equation}
\delta\mathcal L=g_{ijkl}\varphi^{\{i}\varphi^j\varphi^k\varphi^{l\}}.
\end{equation}
This coupling breaks $\ON$ symmetry, and if this perturbation is relevant, the theory flows to another fixed-point with hypercubic symmetry. The stability of the $\ON$ CFT under the perturbation is determined by the sign of $\Delta_{\varphi^4_{T_4}}-d$ and therefore depends on the values of $N$ and $d$. At infinite $N$, the $\ON$ model is unstable under the hypercubic perturbation, and for each $d$ one can define a quantity $N_{\mathrm{crit.}}$ above which this is the case. 
$N_{\mathrm{crit.}}$ was determined to order $\eps^4$ in \cite{Kleinert:1994td}, by studying the hypercubic fixed-point,
\begin{align}
N_{\mathrm{crit.}}&=4-2\eps+\frac{5(6 \zeta _3-1)}{12}\eps^2-\frac{2-90 \zeta _3+1200 \zeta _5-3 \pi ^4}{144} \eps^3
\nonumber
\\ &\quad +\frac{52731 \zeta _3-115416 \zeta _3^2-132510 \zeta _5+2176335 \zeta_7+378 \pi ^4-800 \pi ^6-189}{72576}\eps^4+O(\eps^5),
\end{align}
see also \cite{Varnashev:1999ze}. In \cite{Chester:2020iyt} it was shown numerically that for $d=3$, $N_{\mathrm{crit.}}<3$. 

\subsubsection{Hausdorff dimension}
\label{sec:Hausdorff}
An observable from the loop gas perspective is the Hausdorff, or fractal, dimension $D_H$. For generic $N$, it can be computed from $\Delta_{\varphi^2_T}$ \cite{Winter2008,Shimada:2015gda},
\begin{equation}
\label{eq:Hausdorffdef}
D_H=d-\Delta_{\varphi^2_T}.
\end{equation}
However, it is also a meaningful observable in the loop gas at $N=1$, with the value $D_H=1.7349 (65)$ reported in \cite{Winter2008}. Note that at $N=0$, $\Delta_{\varphi^2_S}=\Delta_{\varphi^2_T}$, so $D_H=1/\nu$ for the critical exponent $\nu$.

\subsubsection{Large charge semiclassics}

In \cite{Hellerman:2015nra} it was shown that in the limit of large charge under global symmetry, the scaling dimension of the lowest-dimension operator of a given charge takes a universal form independent on the specific CFT,\footnote{This happens for ``generic'' CFTs, which excludes free theories and theories with a sufficient amount of supersymmetry.} and computable in a semiclassical expansion. Consider a theory with $U(1)$ symmetry and operators with charge $m$. For the $\ON$ CFT, this setup corresponds to $N=2$ and the leading scalar operators in the $T_m$ representation -- the family \texttt{ONF3[$m$]}. The authors of \cite{Hellerman:2015nra} showed that
\begin{equation}\label{eq:semiclassics}
\Delta_m=c_0 m^{\frac{d}{d-1}}+c_1 m^{\frac{d}{d-1}-1}+\ldots+\tilde c_0+\frac{\tilde c_1}m+\ldots.
\end{equation}
where the values of the $c_i$ and $\tilde c_i$ are theory-dependent, but the scaling is universal. Moreover, in three dimensions, they showed that $\tilde c_0$ is universal and can be computed to arbitrary precision. The value reads $\tilde c_0=-0.0937256$ \cite{Monin:2016jmo}. 

In \cite{Badel:2019oxl} a function was found that interpolates between the large charge regime and the perturbative regime for the $O(2)$ CFT. This was generalised to general $N$ in \cite{Antipin:2020abu}, and the expansion takes the form
\begin{align}
\label{eq:deltaMlargecharge}
\Delta_{\varphi^m_{T_m}}=\frac{m}4\left(F_{-1}(6g^*m)+\frac1mF_0(6g^*m)+\ldots\right),
\end{align}
with
\begin{equation}
F_{-1}(x)=\frac{3^{\frac23}\left(x+\sqrt{x^2-3}\right)^{\frac13}}{3^{\frac13}+\left(x+\sqrt{x^2-3}\right)^{\frac23}}+\frac{3^{\frac13}\left[3^{\frac13}+\left(x+\sqrt{x^2-3}\right)^{\frac23}\right]}{\left(x+\sqrt{x^2-3}\right)^{\frac13}},
\end{equation}
and $F_0$ only determined in expansion around small and large $x$.
For large $x$, i.e.\ for $m\gg g^*$, \eqref{eq:deltaMlargecharge} reduces to the semiclassical regime \eqref{eq:semiclassics} with $\frac m4F_{-1}(6g^*m)\sim m^{\frac43}$, and for small $x$, with $g^*$ being the solution to $\beta(g^*)=0$ (c.f.\ \eqref{eq:betafunctionzero}), it agrees with the $\epsilon$ expansion. In fact, knowledge of $F_{-1}(x)$ and $F_0(x)$ from the computations of \cite{Badel:2019oxl,Antipin:2020abu}, combined with previous results for the cases $m=1,2,4$, facilitated the determination of \texttt{DeltaE[ONF3[$m$]]} to order $\eps^4$.\footnote{Using the more recent results of \cite{Bednyakov:2021ojn} for $m=3$ at order $\eps^6$, this can in fact be extended to order $\eps^5$.}
The relation between large charge semiclassics and other expansion has also been discussed, see \cite{Giombi:2020enj} for the large $N$ expansion, and \cite{Antipin:2021jiw} for the cubic theory in $6-\epsilon$ dimensions, discussed in section~\ref{sec:cubicsixeps}.

In three dimensions using Monte Carlo simulations, numerical values for the family of leading scalar operators in the $T_m$ representation were found in the $O(2)$ \cite{Banerjee:2017fcx} and $O(4)$  \cite{Banerjee:2019jpw} case, matching with the large charge expansion also for low values of $m$. In \cite{Banerjee:2021bbw}, this computation was extended to leading spin 1 operators in the $Y_{m,1}=H_{m+1}$ representation in the 3d $O(4)$ CFT (\texttt{ONF11[$m$,1]}).
 See also \cite{Hasenbusch2011} for a collection of Monte Carlo values for $\Delta_{\varphi^m_{T_m}}$ with $m=1,2,3,4$ and small $N$.

\subsubsection{Quantum-critical conductivity}
\label{sec:conductivity}

The study of conductivity in theories describing the quantum-critical phase transitions at finite temperature $T$ and at frequency $\omega$ (analytic continuation of Matsubara frequencies $\omega_n$), introduces a number of additional parameters \cite{Katz:2014rla}, see also \cite{Reehorst:2019pzi}. In $2+1$ dimensions, the expansion of the conductivity reads
\begin{equation}
\label{eq:conductivity}
\frac{\sigma(i\omega)}{\sigma_Q}=\sigma_\infty+b_1\left(\frac T\omega\right)^{\Delta_{\mathcal S}}+b_2\left(\frac T\omega\right)^{3}+\ldots,
\end{equation}
where $\Delta_{\mathcal S}$ is the dimension of the (only) relevant scalar singlet operator (for the $\ON$ CFT, $\varphi^2_S$) and $\sigma_Q$ is the quantum unit conductance. The constants appearing in \eqref{eq:conductivity} are $\sigma_\infty=\frac{C_J}{32}$, $b_1=\frac{C_J\lambda_{JJ\mathcal S}}{4\pi}\frac{\Gamma(\Delta_{\mathcal S})\sin(\frac{\pi\Delta_{\mathcal S}}2)}{2-\Delta_{\mathcal S}}\frac1\Upsilon$ and $b_2=-72\frac{C_J\gamma}{C_T}H_{xx}$, where $\gamma$ was discussed in section~\ref{sec:threepoints}. In the large $N$ limit, we have \cite{Katz:2014rla}
\begin{equation}
\Upsilon=\frac{1}{\pi\ln(\frac{1+\sqrt5}2)^2}\left(\frac1{\sqrt N}+\frac{0.8941}{N^{3/2}}+\ldots\right),\qquad H_{xx}=\frac{\zeta_3}{2\pi}\left(\frac{4N}5-0.3344+\ldots\right).
\end{equation}
For the 3d $N=2$ case, Monte Carlo simulations give $\Upsilon=1.18(13)$, $b_1=1.43(5)$, $b_2=-0.4(1)$ \cite{Katz:2014rla}. Combined with the bootstrap results from \cite{Reehorst:2019pzi}, one finds $\Upsilon=1.257(60)$, $H_{xx}=0.105(30)$.\footnote{As discussed in \cite{Katz:2014rla}, the Monte Carlo results for $b_2$ differ quite substanially from large-$N$ estimates that give $b_2=-0.97$ and $H_{xx}=0.24$. A Monte Carlo study gave $\Delta_P=-H_{xx}=-0.2993(7)$ \cite{Vasilyev2009}. For $N=1$ the results are $\Delta_P=-0.1526(16)$ from Monte Carlo \cite{Krech1996} and $\Delta_P=-0.143(3)$ from the thermal bootstrap \cite{Iliesiu:2018zlz}.}

\subsubsection{Observables away from flat space}
In this report, we exclusively focus on flat-space observables. However, results have been derived for the case of a compact direction (finite temperature) \cite{Iliesiu:2018fao,Iliesiu:2018zlz} and in real projective space \cite{Nakayama:2016cim}. Moreover, the Ising and $\ON$ CFT has been considered in the presence of defects \cite{Billo:2013jda,Gaiotto:2013nva,Allais:2014fqa,Billo:2016cpy,Soderberg:2017oaa,Cuomo:2021kfm} and boundaries \cite{Liendo:2012hy,Carmi:2018qzm,Dey:2020lwp}, and there are also results available for the entanglement entropy \cite{Chubukov:1993aau,Sachdev:1993pr,Metlitski:2009iyg,Whitsitt:2016irx}.

\section{Presentation of conformal data}\label{sec:presentation}

The main result of this report is a large collection of CFT-data for the critical $\ON$ CFT, presented in the tables of sections \ref{sec:DataIsing} and \ref{sec:dataGenN}, and in computer-readable format in a Mathematica package \texttt{ONdata.m}. In this section we will briefly describe the content of the data file and the notation used, leaving more details to appendix~\ref{app:datafile}. This requires giving a brief overview of the representation theory of the involved symmetry groups $\ON$ and $\mathrm{SO}(d)$. The data presented is a compilation of results from the literature, complemented by additional computations, primarily in the form of order $\eps^1$ anomalous dimensions. We summarise the additional computations in section~\ref{sec:computations}.

\subsection{Notation}

The conformal primary operators will be denoted by
\begin{equation}
\langle\O\rangle=\texttt{Op[}\langle R\rangle\texttt{,}l\texttt{,}i\texttt{]},
\end{equation}
where $\langle R\rangle$ denotes the global symmetry representation, $l$ the spin and $i$ an additional label that organises the operators of equal $R$ and $l$ by increasing scaling dimension. For $N=1$ we have $\langle R\rangle=\texttt{E},\texttt{O}$ ($\mathbb Z_2$ even and $\mathbb Z_2$ odd), and for generic $N$ we have $\langle R\rangle=\texttt{S},\texttt{V},\texttt{T},\texttt{A}$, etc. More details of the $\ON$ representation theory are provided in section~\ref{sec:reptheory} below. 
In addition to individual operators, we have implemented some families of operators, written as \texttt{IsingF1}, \texttt{IsingF2} etc. (``Ising family'') and \texttt{ONF1}, \texttt{ONF2} etc. (``$\ON$ family''). 

The scaling dimension of the operator $\O$ is implemented as
\begin{equation}
\Delta_\O=\texttt{Delta}\langle limit\rangle\texttt{[}\langle\O\rangle\texttt{]},
\end{equation}
where $\langle limit\rangle$ ranges over the different expansions, according to
\begin{itemize}
\item \texttt{DeltaE}: $d=4-\eps$ expansion, expansion parameter \texttt{e}.
\item \texttt{DeltaN}: Large $N$ expansion, expansion parameter $1/\texttt{n}$.
\item \texttt{DeltaZ}: $d=2+\teps$ expansion, expansion parameter \texttt{e}.
\end{itemize}
For quantities such as the central charges $\frac{C_T}{C_{T,\mathrm{free}}}=\texttt{CT}$ and $\frac{C_J}{C_{J,\mathrm{free}}}=\texttt{CJ}$ and the critical exponents, we have implemented $\texttt{Value}\langle limit\rangle\texttt{[}\langle qty\rangle\texttt{]}$ in the obvious way. 

The perturbative expressions will be known to some order in the expansion parameter $\langle x\rangle$, and the ``order'' symbol is implemented as $O(x^k)=\texttt{ord*}\langle x\rangle ^k$. For instance, scaling dimension of the $O(N)$ singlet operator with spin $2$ and third lowest scaling dimension is implemented as
\begin{align}
\texttt{DeltaE[Op[S,2,3]]}&=6-2\texttt{e}+\frac{44+9\,\texttt{n}+\sqrt{624-8\,\texttt{n}+9\,\texttt{n}^2}}{6(8+\texttt{n})}\texttt{e}+\texttt{ord}\ \texttt{e}^2,
\label{eq:exampleDeltaE}
\\\label{eq:exampleDeltaN}
\texttt{DeltaN[Op[S,2,3]]}&=6-\frac{2(-42+\,\texttt{mu}\,(47+\texttt{mu}\,(7+4\,\texttt{mu}\,(\,\texttt{mu}-7))))\Gamma(2\,\texttt{mu}-1)}{3(1+\texttt{mu}\,)\Gamma(1-\texttt{mu}\,)\Gamma(\,\texttt{mu}\,)^3\,\Gamma(1+\texttt{mu}\,)\,\texttt{n}}+\texttt{ord}/\texttt{n}^2.
\end{align}
For expressions in the large $N$ expansion, \texttt{mu} denotes the quantity $\mu=d/2$. In the $2+\teps$ expansion of the non-linear sigma model, we have only implemented a limited set of operators. For instance, for the operator in \eqref{eq:exampleDeltaE}--\eqref{eq:exampleDeltaN}, $\texttt{DeltaZ[Op[S,2,3]]}$ remains un-evaluated.

In the tables, we indicate by \eord k[ref.] and \Nord k[ref.] a reference to the first computation of the order $\eps^k$ and $N^{-k}$ result in the respective expansions. This reference is omitted for order $\eps^1$ results such as \eqref{eq:exampleDeltaE}, which will be derived by the method explained in section~\ref{sec:orderepssystematics}. For instance, in the presentation of the operator $\texttt{Op[S,2,3]}$, found in table~\ref{tab:singletspinning}, no reference is given to the anomalous dimension in the $\eps$-expansion. Note that, as expected, to the overlap of the orders, the expressions \eqref{eq:exampleDeltaE} and \eqref{eq:exampleDeltaN} agree: for $\mu=2-\eps/2$, both expressions expand as $6-\frac{80\eps}{9N}+O(\eps^2)+O(N^{-2})$.

Squared OPE coefficients $\lambda^2_{\varphi\varphi\O}$ will be denoted by $\texttt{Ope}\langle limit\rangle\texttt{[}\langle\O\rangle\texttt{]}$, and squared three-point functions $\lambda^2_{\O_1\O_2\O_3}$ of generic operators will be denoted $\texttt{Ope}\langle limit\rangle\texttt{[}\langle\O_1\rangle,\langle\O_2\rangle,\langle\O_3\rangle\texttt{]}$. We use conventions where the squared OPE coefficients in the theory of a free scalar in $d$ dimensions are
\begin{equation}
\label{eq:freeOPE}
\left.\lambda^2_{\phi\phi\mathcal J_\ell}\right|_{\mathrm{free}}=\frac{2 \Gamma (\ell+\mu -1)^2 \Gamma (\ell+2 \mu -3)}{\Gamma (2 \ell+2 \mu -3)\Gamma (\mu -1)^2 \Gamma (\ell+1) }
\end{equation}
for $\ell$ even.
In four dimensions this expression reduces to $\frac{2\Gamma(\ell+1)^2}{\Gamma(2\ell+1)}$. Normalisation conventions for $\ON$ irreps are discussed in appendix~\ref{app:normON}

\subsection[Representation theory for $\ON$ global symmetry]{Representation theory for $\boldsymbol{\ON}$ global symmetry}\label{sec:reptheory}

Irreducible representations of the global $\ON$ symmetry are labelled by Young tableaux.
Specifically, we define the $\ON$ representation $Y_{m_1,m_2,\ldots m_r}$ with $m_1\geqslant m_2\geqslant\ldots\geqslant m_r$ as the representation corresponding to the Young tableau of $r$ rows with row $i$ containing $m_i$ boxes. We introduce special notation for three families of representations: $T_m=Y_m$ (traceless-symmetric), $H_m=Y_{m-1,1}$ (hook), $A_m=Y_{1,\ldots,1}$ (antisymmetric), and the specific representations $S=Y_{\varnothing}$ (singlet), $V=Y_1$ (vector), $T=T_2=Y_2$, $A=A_2=Y_{1,1}$, and $B_4=Y_{2,2}$ (box). 
In table~\ref{tab:irrepsON} we list the irreps of rank $\leqslant4$ for low values of $N$, together with their symbols in the ancillary data file and their dimensionality.
We have not implemented any operators in the representations $A_4$ or $Y_{2,1,1}$. 
\begin{table}
\centering
\caption{Dimensions of the irreducible representations of $\ON$ symmetry with rank up to $4$. For $N=2$ and $N=3$ we also give additional symbols for the irreps, corresponding to those used in \cite{Reehorst:2019pzi} and \cite{Chester:2020iyt}. For $N=3$, the parity corresponds to the parity of the number of boxes in the Young tableau.}\label{tab:irrepsON}
{\small
\renewcommand{\arraystretch}{1.25}
\begin{tabular}{|cll|cc|cc|c|c|}
\hline 
\multicolumn{3}{|c|}{Irrep}& \multicolumn{2}{c|}{$N=2$} &\multicolumn{2}{c|}{$N=3$} &$N=4$ & $N\geqslant5$ \\
\hline
$S$  & \raisebox{-0pt}{$\bullet$}&\texttt{S} & $q^p=0^+$ & $1$& $\mathbf q^p=0^+$ &$1$& $1$& $1$
\\
\hline
$V$ &  \raisebox{-0pt}{\tiny\yng(1)}&\texttt{V}& $q=1$ & $2$ & $\mathbf q^p=1^-$ &$3$ & $4$ & $N$
\\
\hline
$A$ & \raisebox{-3pt}{\tiny \yng(1,1)} &\texttt{A}&$q^p=0^-$ &$1$ & $\mathbf q^p=1^+$ &$3$ & $3+\overline 3$ & $N(N-1)/2$ 
\\
\hline
$T$&\raisebox{-0pt}{\tiny \yng(2)}&\texttt{T}  &  $q=2$ &$2$ &$\mathbf q^p=2^+$ & $5$ & $9$ & $(N-1)(N+2)/2$ 
\\
\hline
$A_3$&\raisebox{-3pt}{\tiny \yng(1,1,1)}&\texttt{A3}&  \multicolumn{2}{c|}{\graycell}& $\mathbf q^p=0^-$ &$1$ & $4^-$ & $N(N-1)(N-2)/6$ 
\\
\hline
$T_3$&\raisebox{-0pt}{\tiny \yng(3)} &\texttt{Tm[3]}&  $q=3$ &$2$ &$\mathbf q^p=3^-$ & $7$ & $16$ &$N(N+4)(N-1)/6$
\\
\hline
$H_3$&\raisebox{-3pt}{\tiny \yng(2,1)}&\texttt{Hm[3]} & \multicolumn{2}{c|}{\graycell} & $\mathbf q^p=2^-$ &$5$ & $8+\overline 8$ & $N(N+2)(N-2)/3$ 
\\
\hline
$A_4$& \raisebox{-3pt}{\tiny \yng(1,1,1,1)}&---   & \multicolumn{2}{c|}{\graycell}&\multicolumn{2}{c|}{\graycell}& $1^-$ & $N(N-1)(N-2)(N-3)/24$
\\
\hline
$T_4$& \raisebox{-0pt}{\tiny \yng(4)}&\texttt{Tm[4]}  &  $q=4$ & $2$ & $\mathbf q^p=4^+$ &$9$ & $25$ & $N(N+1)(N+6)(N-1)/24$ 
\\
\hline
$H_4$ &  \raisebox{-3pt}{\tiny \yng(3,1)} &\texttt{Hm[4]}&   \multicolumn{2}{c|}{\graycell} &$\mathbf q^p=3^+$ & $7$ & $15+\overline{15}$ & $(N+1)(N+4)(N-1)(N-2)/8$
\\
\hline
$B_4$ &  \raisebox{-3pt}{\tiny \yng(2,2)}&\texttt{B4} & \multicolumn{2}{c|}{\graycell}&\multicolumn{2}{c|}{\graycell} & $5+\overline 5$ & $N(N+1)(N+2)(N-3)/12$
\\
\hline
$Y_{2,1,1}$ & \raisebox{-3pt}{\tiny \yng(2,1,1)}&--- &\multicolumn{2}{c|}{\graycell}&\multicolumn{2}{c|}{\graycell} & $9^-$ & $N(N+2)(N-1)(N-3)/8$
\\
\hline
\end{tabular}
}
\end{table}

In table~\ref{tab:irrepsON} we also include the special cases at low values of $N$, where the general theory degenerates. For finite integer $N$, the set of tensors that can be used to construct $\ON$ invariant objects includes, in addition to the Kronecker delta $\delta_{ij}$, the rank $N$ antisymmetric tensor $\epsilon_{i_1\cdots i_N}$. The antisymmetric tensor can be used to reduce representations with $\geqslant\lfloor \frac N2\rfloor$ rows, and to completely annihilate representations with $>N$ rows. The representations resulting from the use of an odd number of antisymmetric tensors are distinguished from those with an even number of antisymmetric tensors by a label parity label ``$-$'', not to be confused with spacetime parity.

For $N=2$, the representations can be labelled by charge $q$ and, for $q=0$, parity. The tensor products are $q\otimes q'=q-q'\oplus q+q'$ for $q'<q$ and $q\otimes q=0^+\oplus 0^-\oplus 2q$. For $N=3$, usual $\mathrm{SO}(3)$ addition rules apply, supplemented with an additional parity label that is multiplicative. For $N=4$, the representations can be labelled by two spin labels and a parity label, similar to the Lorentz $\mathrm{SO}(4)$ case considered in the next subsection.

The non-trivial tensor product decompositions involving the low-lying $\ON$ representations are
\begin{align}
\label{eq:tensorprod1}
V\otimes V&=S\oplus T\oplus A,
\\
V\otimes A&=V\oplus A_3\oplus H_3,
\\
V\otimes T&=V\oplus T_3\oplus H_3,
\label{eq:tensorprod3}
\\
A\otimes A&=S\oplus T\oplus A\oplus A_4\oplus B_4\oplus Y_{2,1,1},
\\
A\otimes T&=T\oplus A\oplus H_4\oplus Y_{2,1,1},
\\
T\otimes T &= S\oplus T\oplus A\oplus T_4\oplus H_4\oplus B_4.
\label{eq:tensorprod6}
\end{align}
The general tensor products~\eqref{eq:tensorprod1}--\eqref{eq:tensorprod6} reduce to the special cases at $N=3$ and $N=4$ by removal of the irreps that do not exist for these values of $N$. For the $N=2$ theory, the tensor product $T\otimes T$ does not contain $T$. 

For non-integer values of $N$, the representation of the $\ON$ symmetry can be formalised for instance using Deligne categories \cite{Binder:2019zqc}. In \cite{Grans-Samuelsson:2021uor}, the representation theory for non-integer $N$ was discussed with focus on the 2d loop gas model.

\subsection[Representation theory for $\mathrm{SO}(d)$ Lorentz symmetry]{Representation theory for $\boldsymbol{\mathrm{SO}(d)}$ Lorentz symmetry}
\label{sec:reprSOd}
In section~\ref{sec:ObservablesMain} we stated that the conformal primary operators have a fixed scaling dimension and transform in an irreducible representation of the $\mathrm{SO}(d)$ rotational symmetry, which in Lorentzian signature becomes $\mathrm{SO}(d-1,1)$ Lorentz symmetry. For the purposes of the representation theory, we consider the case of Euclidean signature, however we sometimes use the notation ``$\mathrm{SO}(d)$ Lorentz symmetry'' to distinguish it from internal $\ON$ rotations.

The $\ON$ CFT respects parity invariance in addition to the conformal symmetry, implying that conformal primary operators in the $\ON$ CFT have definite parity. It is useful to combine the effect of Lorentz symmetry and parity by considering $\mathrm{O}(d)$ representation theory. Since we mostly consider generic values of $d$, we will think of the integer values as degenerate cases of a general $\mathrm{O}(d)$ representation theory, which can for instance be formalised using Deligne categories \cite{Binder:2019zqc}, something that is not possible for $\mathrm{SO}(d)$. For simplicity however, we will in other parts of the report keep referring to the combined Lorentz and parity symmetry as ``$\mathrm{SO}(d)$ Lorentz symmetry''.

The description of irreducible representations is now analogous to the previous subsection. General $\mathrm{O}(d)$ representations are indexed by
Young tableaux $y_{r_1,r_2,\ldots}$ with rows of length $r_1,r_2,\ldots$. In order to use the assumption of spectrum continuity, we define such irreps for generic $d$, however at integer values of $d$ some irreps vanish identically.
The one-row Young tableaux $y_\ell$ exist for any $d$ and represent traceless-symmetric tensors corresponding to spin-$\ell$ operators.

In the case $d=3$, the irreps are labelled by a spin $\ell=0,1,2\ldots$ and a parity label. In the case $d=4$, the irreps are labelled by a pair of spins $(j,\bar j)$ taking half-integer values with integer differences, and a parity label. We summarise what happens to the one- two- and three-row Young tableaux in $d=3$ and $d=4$:
\begin{description}
\item[One-row Young tableaux] exist for all $d$ and correspond to spin $\ell$ parity-even operators. They are completely traceless-symmetric tensors with $\ell$ Lorentz indices.
\item[Two-row Young tableaux] are denoted $y_{r_1,r_2}$. For $d=3$, one can use $r_2$ antisymmetric tensors to convert such a representation to a representation of spin $r_1$ and parity $(-1)^{r_2}$. For $d=4$, the irrep $y_{r_1,r_2}$ becomes a parity-even representation that decomposes into a sum $\left(\frac{r_1+r_2}2,\frac{r_1-r_2}2\right)+\left(\frac{r_1-r_2}2,\frac{r_1+r_2}2\right)$, using the notation in \cite{Kehrein:1994ff}.
\item[Three-row Young tableaux] are denoted $y_{r_1,r_2,r_3}$. For $d=3$, one can use $r_3$ antisymmetric tensors to remove the third row completely, and an additional $r_2-r_3$ antisymmetric tensors to reduce to a spin $r_1-r_3$ operator of parity $(-1)^{r_2}$. For $d=4$ one can use $r_3$ antisymmetric tensors to reduce to a $y_{r_1,r_2-r_3}$ representation with parity $(-1)^{r_3}$. 
\end{description}

In general, we refer to Young tableaux with more than one row as Lorentz non-traceless-symmetric representations. They do not appear in the OPE of two scalar operators. Two-row Young tableaux appear for instance in the OPE of a current or stress-tensor with a scalar, and three-row Young tableaux appear in the OPE of a current or stress-tensor with itself. 

Note that it is the three-row Young tableaux that can produce parity odd scalars in three dimensions, the existence of which was shown in \cite{Dymarsky:2017yzx} by conformal bootstrap of the stress-tensor four-point function. These operators are pseudovectors in $d=4$. We discuss results for such operators in sections~\ref{sec:non-TS-Ising} and \ref{sec:nonTSLorentz} below.

It is known that the spectrum of the free scalar contains aditional states that vanish identically for low integer spacetime dimensions $d$, called ``evanescent operators'' \cite{Hogervorst:2015akt}. For instance, Lorentz representations with more than three rows belong to this category, since they vanish identically for $d=3$. In this report, we limit to operators that exist for $d=4$, where the spectrum of the $\ON$ CFT reduces to that of $N$ free scalars. To index such operators, we have performed a decomposition in characters of the four-dimensional conformal group and of the $\ON$ group. This decomposition will not detect any evanescent operators that vanish identically at $d=4$.

\subsection{Computations}
\label{sec:computations}

While most of the results presented in this report are compiled from the literature, a number of additional computations have been made. In summary, they include
\begin{enumerate}
\item A character decomposition to index all primary operators up to a certain scaling dimension. We give more details in appendix~\ref{app:characters}.
\item In the $\eps$-expansion, the order $\eps^1$ anomalous dimensions of all operators included in the tables below have been computed. Specifically, the approach of \cite{Hogervorst:2015akt} was used, which is equivalent to the approach in \cite{Kehrein:1992fn,Kehrein:1994ff}. We give more details on this computation in section~\ref{sec:orderepssystematics}. Most of the results found for $N=1$ were already tabulated in \cite{Kehrein:1994ff}, while most of the results found for general $N$ have not been computed before.
\item For operators containing $m$ fields $\varphi$ transforming in the $T_m$ representations, we have computed the order $\eps^2$ anomalous dimension using the approach of \cite{Kehrein:1995ia} and the order $1/N$ anomalous dimension using the method of \cite{Derkachov:1997qv}.
\item A conformal block decomposition of the large $N$ correlator $\langle\varphi\varphi\varphi\varphi\rangle$, computed in \cite{Lang:1991kp}.
\item In the perturbative expansions, when there are two degenerate operators and total OPE coefficients $\langle\lambda^2_{\varphi\varphi\O}\rangle$ and average anomalous dimensions $\langle\lambda^2_{\varphi\varphi\O}\gamma_\O\rangle$ are available, the individual OPE coefficients can be extracted knowing the anomalous dimensions $\gamma_\O$ of the individual operators. This process is sometimes denoted ``unmixing'' and has been performed in a few cases. We give more details and an explicit example in section~\ref{sec:unmixing} immediately below.
\end{enumerate}
In general, in the tables we indicate by \cite{ThisPaper} results that were derived for the first time in this work.
To avoid cluttering the tables, we do not give any references next to the operators for which only the order $\eps^1$ anomalous dimension is known. In the cases where we used the approach of \cite{Kehrein:1995ia} or \cite{Derkachov:1997qv} to find data not listed in those papers, we give the reference as \cite{Kehrein:1995ia,ThisPaper} or \cite{Derkachov:1997qv,ThisPaper} respectively.

\subsubsection{Unmixing of operators with low degeneracy}
\label{sec:unmixing}

Consider the $\eps$-expansion of the correlator $\langle\varphi\varphi\varphi\varphi\rangle$ for general $N$. The conformal data appearing in the OPE decomposition of this correlator have been computed to order $\eps^3$ \cite{Dey:2016mcs} and $\eps^4$ \cite{Henriksson:2018myn} using analytic bootstrap methods, see section~\ref{sec:analyticbootstrapmethods}. Such computation can only resolve OPE coefficients and anomalous dimensions for operators that are non-degenerate in the limit $\eps\to0$, such as the weakly broken currents $\mathcal J_{R,\ell}$, which have twists $\tau^{\mathrm{4d}}=\Delta^{\mathrm{4d}}-\ell=2$. However, results for operators at twist $\tau^{\mathrm{4d}}=4$ were also found, which may be interpreted as weighted averages. For instance, in the singlet representation, \cite{Henriksson:2018myn} gives
\begin{align}
\label{eq:avera}
\left\langle\lambda^2_{\varphi\varphi \O_{4,\ell}}\right\rangle&=\frac{\Gamma(\ell+2)^2}{\Gamma(2\ell+3)}\frac{(N+2)}{4N(N+8)^2}\left(1+\frac6{(\ell+1)(\ell+2)}\right)\eps^2+O(\eps^3),
\\
\label{eq:averagamma}
\left\langle\lambda^2_{\varphi\varphi \O_{4,\ell}}\gamma_{\O_{4,\ell}}\right\rangle&=\frac{\Gamma(\ell+2)^2}{\Gamma(2\ell+3)}\frac{(N+2)}{4N(N+8)^2}\left(1+\frac{14}{(\ell+1)(\ell+2)}\right)\eps^3+O(\eps^4),
\end{align}
where the bracket indicates a sum over individual operators $\O_{4,\ell,i}$, of the form $\de^\ell\varphi^4_S$ with dimensions $4+\ell-2\eps+\gamma_{\O,\ell,i}$. One can immediately confirm that for $\ell=0$, equations \eqref{eq:avera}--\eqref{eq:averagamma} are consistent with $\lambda_{\varphi\varphi\varphi^4_S}=\frac{N+2}{2N(N+8)^2}\eps^2+O(\eps^3)$, c.f.\ \eqref{eq:opephi2khyp}, and $\gamma_{\varphi^4_S}=2\eps+O(\eps^2)$, c.f.\ \eqref{eq:gammaphi4eps}.

The interesting case happens at $\ell=2$, where there are two degenerate operators of the form $\de^2\varphi^4_S$, $\O_{4,2,1}= \texttt{Op[S,2,2]}$ and $\O_{4,2,2}=\texttt{Op[S,2,3]}$. In section~\ref{sec:orderepssystematics} below we will find that their anomalous dimensions take the form 
\begin{equation}
\label{eq:opS22S23}
\gamma_{\O_{4,2,1}},
\gamma_{\O_{4,2,2}}=\frac{9 N+44\mp\sqrt{9 N^2-8 N+624}}{6 (N+8)}\eps+O(\eps^2).
\end{equation}
Solving two equations, \eqref{eq:avera} and \eqref{eq:averagamma} with $\ell=2$, for two unknowns, $\lambda^2_{\varphi\varphi\O_{4,2,1}}$, $\lambda^2_{\varphi\varphi\O_{4,2,2}}$, we find
\begin{equation}
\label{eq:OPEs22}
\lambda^2_{\varphi\varphi\O_{4,2,1}},\lambda^2_{\varphi\varphi\O_{4,2,2}}=\frac{N+2}{320N(N+8)^2}\frac{\pm N\mp 76+3\sqrt{9N^2-8N+624}}{\sqrt{9N^2-8N+624}}\eps^2+O(\eps^3).
\end{equation}
The resulting expressions are implemented in the ancillary data file as \texttt{OpeE[Op[S,2,2]]} and \texttt{OpeE[Op[S,2,3]]} for the upper and lower sign respectively.

A similar computation can be done in all cases where there are two degenerate operators and where the averages $\langle\lambda^2_{\O_1\O_2 \O_{i}}\rangle$ and $\langle\lambda^2_{\O_1\O_2 \O_{i}}\gamma_{\O_i}\rangle$ are known. The results for performing such an unmixing are reported with a reference to joint reference to \cite{ThisPaper} (this paper) and to the paper where the average was found. 

\section{Methods} \label{sec:methods}

In this section we outline some of the methods that can be used to compute conformal data of the critical $\ON$ CFT. We limit the discussion to methods used to find perturbative data in the various expansion limits, and therefore omit methods such as the numerical conformal bootstrap \cite{Poland:2018epd}, Monte Carlo simulations (e.g.\ \cite{Hasenbusch:2019jkj,Hasenbusch:2020pwj}), high temperature expansions \cite{Campostrini:2002cf}, non-perturbative/functional RG methods \cite{Berges:2000ew,Delamotte:2007pf,Gurau:2014vwa,Dupuis:2020fhh,Balog:2019rrg,DePolsi:2020pjk,DePolsi:2021cmi} and fixed-dimension expansions \cite{Guida:1998bx,Jasch2001,Pogorelov:2008zz}.

\subsection{Feynman diagrams in dimensional expansions}\label{sec:Feynmandiagrams}

The $d=4-\eps$ expansion was first developed since it reduces the computation of conformal data, such as critical exponents, to the familiar task of evaluating Feynman diagrams.
The computation follows the standard procedure of multiplicative renormalisation, where one computes the two- and four-point correlators of the field $\varphi$ to high loop-order, giving as the result the scaling dimensions for the operators $\varphi$, $\varphi_S^2$ and $\varphi_S^4$. There is no conceptual limit to the order of the computation -- the obstacle is the evaluation of the diagrams where the state-of-the-art is order $\eps^8$ results for $\Delta_\varphi$ and order $\eps^7$ for $\Delta_{\varphi_S^2}$ and $\Delta_{\varphi_S^4}$ \cite{Batkovich:2016jus,Kompaniets:2016hct,Kompaniets:2017yct,Schnetz:2016fhy,SchnetzUnp}.\footnote{Six-loop expressions for general $\phi^4$ theory can be found in \cite{Bednyakov:2021ojn}.}

The anomalous dimensions of composite operators can be found by considering diagrams with additional insertions. In general, one has to consider diagrams for the insertion of all operators with the same value of the engineering dimension and other quantum numbers, see e.g.\ \cite{Brezin1976}, and the procedure can in principle be performed to arbitrary order.\footnote{For a nicely worked out example, however from $\phi^3$ theory, see \cite{Amit1977}.}
To leading order in the $4-\eps$ expansion, the mixing is simple to analyse, and we present in section~\ref{sec:orderepssystematics} the systematic procedure introduced in \cite{Hogervorst:2015akt}.

\subsubsection[\texorpdfstring{Multiplicative renormalisation in the $\eps$-expansion}{Multiplicative renormalisation in the ε-expansion}]{Multiplicative renormalisation in the $\boldsymbol \eps$-expansion}
\label{sec:multiplicativerenorm}

The standard multiplicative renormalisation in $d=4-\eps$ expansion gives the scaling dimension of the leading operators,
\begin{align}
\label{eq:deltaphifromRG}
\Delta_\varphi&=1-\frac\eps2+\gamma_\varphi(g^*),
\\
\label{eq:deltaphi2fromRG}
\Delta_{\varphi^2_S}&=2-\eps+\gamma_{m^2}(g^*),
\\
\label{eq:deltaphi4fromRG}
\Delta_{\varphi^4_S}&=4-\eps+\frac{\de\beta}{\de g}(g^*),
\end{align}
in terms of the coupling constant $g$ evaluated at the fixed-point $\beta(g^*)=0$.
The RG functions $\gamma_\phi$, $\gamma_{m^2}$ and $\beta$ will be defined below. An early introduction and computation up to order $\eps^3$ ($\eps^4$ for $\varphi$) is given in \cite{Brezin1976}, and a very comprehensive presentation can be found in \cite{Kleinert:2001hn}.

We present the method of counterterms in the minimal subtraction (MS) scheme.
Start from a bare action in Euclidean signature of the form
\begin{equation}
S=\int \mathrm d^dx \left(
\frac12\de^\mu\varphi^i_\bare(x)\de_\mu\varphi^i_\bare(x)+\frac{m_\bare^2}2\varphi^i_\bare(x)\varphi^i_\bare(x)+\frac{\lambda_\bare}{4!}\left(\varphi^i_\bare(x)\varphi^i_\bare(x)\right)^2
\right),
\end{equation}
where the subscript denotes bare quantities. 
The normalisation of the interaction term is chosen such that it reduces to the usual $\lambda\phi^4$ interaction for $N=1$. Here $\lambda_\bare$ is dimensionful, so one introduces $g_\bare=M^{-\eps}\lambda_\bare$ where $M$ is a mass renormalisation scale.

Introduce the field and coupling constant redefinitions 
\begin{align}
\varphi^i_\bare(x)&=Z^{1/2}\varphi^i(x),
\\
m^2_\bare&=Z_{m^2}Z^{-1}m^2,
\\
g_\bare&=Z_gZ^{-2}g,
\end{align}
and define the counterterms $Z=1+c_Z$ and $Z_i=1+c_i$ to get a renormalised action of the form
\begin{equation}
\label{eq:Srenorm}
S=\int \mathrm d^dx\left(
\frac12(\de\varphi)^2+c_Z\frac12(\de\varphi)^2+\frac{m^2}2\varphi^2+c_{m^2}\frac{m^2}2\varphi^2+\frac{M^\eps g}{4!}(\varphi^2)^2+c_g\frac{M^\eps g}{4!}(\varphi^2)^2
\right).
\end{equation}
The counterterms are chosen to absorb the divergences that appear in the limit $\eps\to0$.
Moreover, define amputated correlators (vertex functions) for the one-point-irreducible correlators $\langle \varphi\cdots\varphi \rangle$ in momentum space by
\begin{equation}
\Gamma_\bare^{(n)}(k_i)=\prod_{j=1}^{n}\frac1{G_\bare^{(2)}(k_j)}G_\bare^{(n)}(k_i)_{\mathrm{1PI}},
\end{equation} and renormalised vertex functions by $\Gamma^{(n)}(p_1,\ldots,p_{n},M)=Z^{n/2}\Gamma_\bare^{(n)}(p_1,\ldots,p_{n},M)$. 
With the definitions above, the Callan--Symanzik (CS) equations take the following form in the MS scheme,
\begin{equation}\label{eq:CSmixing}
\left(
M\frac\de{\de M}+\beta(g)\frac\de{\de g}+\gamma_{m^2}(g)m^2\frac\de{\de m^2}-n\gamma_\phi(g)
\right)\Gamma^{(n)}(k_1,\ldots,k_{n},M)=0.
\end{equation}
where 
\begin{align}
\gamma_\phi(g)&=M\left.\frac{\de \ln Z^{1/2}}{\de M}\right|_{m^2_\bare,\lambda_\bare},
\label{eq:gammaphiRGdef}
\\
\beta(g)&=M\left.\frac{\de g}{\de M}\right|_{m^2_\bare,\lambda_\bare},
\label{eq:betadef}
\\
\label{eq:gammam2def}
\gamma_{m^2}(g)&=\frac M{m^2}\left.\frac{\de m^2}{\de M}\right|_{m^2_\bare,\lambda_\bare}.
\end{align}

We will now evaluate the diagrams needed to find the leading order terms in the RG functions \eqref{eq:gammaphiRGdef}--\eqref{eq:gammam2def}. From the action~\eqref{eq:Srenorm} we read off the Feynman rules, which with the definition  $T_{ijkl}=\frac{\delta_{ij}\delta_{kl}+\delta_{ik}\delta_{jl}+\delta_{il}\delta_{jk}}{3}$ take the form\footnote{Since we are ultimately interested in the theory at the fixed-point, a simpler scheme can be achieved by putting the renormalised mass $m^2$ to zero in all computations. However keeping track of the $m^2$ dependence gives a direct access to $\gamma_{m^2}$, and thus to $\Delta_{\varphi^2_S}$.}
\begin{itemize}
\item $i\incl[-2pt]{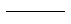}j=\dfrac{\delta_{ij}}{k^2+m^2}$ ,
\item $i\incl[-2pt]{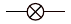}j=-k^2c_Z\delta_{ij}-m^2c_{m^2}\delta_{ij}$,
\item $\begin{matrix}i\\j\end{matrix}\incl[-5pt]{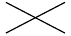}\begin{matrix}k\\l\end{matrix}=-gM^\eps\, T_{ijkl}$,
\item $\begin{matrix}i\\j\end{matrix}\incl[-5pt]{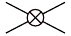}\begin{matrix}k\\l\end{matrix}=-gc_gM^\eps\,T_{ijkl}$.
\end{itemize}
First we compute the inverse propagator
\begin{equation}
\Gamma^{(2)}_{ij}=\delta_{ij}\left(k^2+m^2+\frac12 \mathsf S_1 \incl[-5pt]{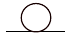}-k^2 c_Z-m^2 c_{m^2}+O(g^2)\right),
\end{equation}
where $\frac12$ is the symmetry factor for $N=1$ and $\mathsf S_1=\frac{N+2}3$ is additional symmetry factor arising from $\ON$ index contractions, computed by solving $T_{ijkk}=\mathsf S_1\delta_{ij}$. 
The self-energy diagram evaluates to
\begin{align}\nonumber
\incl[-5pt]{bubble}&=-g\mu^\eps\int \frac{\mathrm d^dp}{(2\pi)^d}\frac1{p^2+m^2}=-gM^\eps\frac{(m^2)^{\frac{d-2}2}}{(4\pi)^{\frac d2}}\Gamma\left(\frac{2-d}2\right)
\\&=\frac{m^2g}{(4\pi)^2}\left(
\frac2\eps+1-\gamma_{\mathrm{E}}+\ln\frac{4\pi M^2}{m^2}+O(\eps),
\right)
\end{align}
where $\gamma_{\mathrm{E}}=-\psi(1)$ is the Euler--Mascheroni constant and $\psi(x)=\frac{\Gamma'(x)}{\Gamma(x)}$ denotes the digamma function. 
So we get
\begin{align}
\label{eq:Zleading}
Z&=1+O(g^2),
\\
\label{eq:Zm2leading}
Z_{m^2}&=1+\frac{g}{(4\pi)^2}\frac{N+2}3\frac1\eps+O(\eps^0,g^2).
\end{align}
Note that there is no field redefinition to this order.

Next we compute the four-point vertex, which in principle depends on the two momentum combinations $(k_1+k_2)^2$ and $(k_1+k_3)^2$. However, the divergent part is independent of these momenta, and we can write
\begin{equation}\label{eq:Gamma4div}
M^{-\eps}\Gamma^{(4)}_{ijkl}=\frac{\delta_{ij}\delta_{kl}+\delta_{ik}\delta_{jl}+\delta_{il}\delta_{jk}}{3}\left(-g+\frac32\mathsf S_2\incl[-5pt]{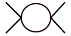}-g c_g+O(g^3)\right),
\end{equation}
where $\mathsf S_2=\frac{N+8}9$, which follows from solving $\frac13(T_{ijmn}T_{mnkl}+T_{ikmn}T_{mnjl}+T_{ilmn}T_{mnjk})=\mathsf S_2T_{ijkl}$.
The diagram evaluates to\footnote{In the massless renormalisation, this diagram can in fact be evaluated exactly for arbitrary $d$, see \eqref{eq:Jintegralresult} in appendix~\ref{app:fouriertransform}.}
\begin{align}
\incl[-5pt]{crossbubble}&=M^\eps g^2\int \frac{d^dk}{(2\pi)^d}\frac1{k^2+m^2}\frac{1}{(k-p)^2+m^2}
\nonumber
\\&=M^\eps \frac{g^2}{(4\pi)^2}\left(\frac2\eps-\gamma_{\mathrm{E}}+\int_0^1dx\ln\left(\frac{4\pi^2}{k^2x(1-x)+m^2}\right)+O(\eps)\right),
\label{eq:loopdiagramcomp}
\end{align}
which gives
\begin{equation}
\label{eq:Zgleading}
Z_g=1+\frac{N+8}3\frac{g}{(4\pi)^2}\frac1\eps+O(\eps^0,g^2).
\end{equation}
Equations \eqref{eq:Zleading}, \eqref{eq:Zm2leading} and \eqref{eq:Zgleading} give the leading expansions of the RG functions by \eqref{eq:gammaphiRGdef}--\eqref{eq:gammam2def},
\begin{align}
\beta(g)=-\eps g+\frac{N+8}3\frac{g^2}{(4\pi)^2}+O(g^3),\quad \gamma_\phi(g)=0+O(g^2),\quad \gamma_{m^2}(g)=\frac{N+2}3\frac{g}{(4\pi)^2}+O(g^2).
\end{align}
The complete expressions, using the results from the seven-loop computation in \cite{Schnetz:2016fhy,SchnetzUnp}, read
\begin{align}
\texttt{ValueE[RGbeta[$\bar g$]]}&=\beta(\bar g)=-\bar g\eps+\frac{N+8}3 \bar g^2+\ldots+O(\bar g^8),
\label{eq:RGbeta}
\\
 \texttt{ValueE[RGgamma[$\bar g$]]}&=\gamma_\phi(\bar g)=\frac{N+2}{36}\bar g^2+\ldots+O(\bar g^9),
\\
 \texttt{ValueE[RGm2[$\bar g$]]}&=\gamma_{m^2}(\bar g)=\frac{N+2}3 \bar g+\ldots+O(\bar g^8),
\end{align}
where we have defined $\bar g=\frac g{(4\pi)^2}$ and rescaled $\beta(g)$.
Solving $\beta(\bar g^*)=0$ in \eqref{eq:RGbeta} gives
\begin{equation}\label{eq:betafunctionzero}
\bar g^*=\frac{3}{N+8}\eps+\frac{9(3N+14)}{(N+8)^3}\eps^2+\ldots,
\end{equation}
and the scaling dimensions of the operators $\varphi$, $\varphi^2_S$ and $\varphi^4_S$ follow from \eqref{eq:deltaphifromRG}--\eqref{eq:deltaphi4fromRG}:
\begin{align}
\Delta_\varphi&=[1-\tfrac\eps2]+O(\eps^2),
\\
\Delta_{\varphi^2_S}&=[2-\eps]+\frac{N+2}{N+8}\eps+O(\eps^2),
\label{eq:deltaphi2SRG}
\\
\Delta_{\varphi^4_S}&=[4-2\eps]+2\eps+O(\eps^2).
\end{align}
To find the leading non-zero term in $\gamma_\phi(g)$, one has to go to order $g^2$ and evaluate the ``sunset'' diagram.\footnote{For the mass and coupling constant renormalisation, additional diagrams have to be evaluated to find complete results at this order.} For simplicity, continue in massless renormalisation, with the renormalised mass $m^2=0$. Then, for the term proportional to $k^2$,
\begin{equation}
\Gamma_{ij}^{(2)}|_{k^2}=\delta_{ij}\left(k^2-k^2c_\phi+\frac16\mathsf S_3\incl[-5pt]{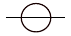}\big|_{k^2}+O(g^3)\right),
\end{equation}
where $\mathsf S_3=\frac{N+2}3$ follows from solving $T_{iklm}T_{jklm}=\mathsf S_3\delta_{ij}$. 
The evaluation of the sunset diagram is somewhat complicated but doable by hand (see e.g.\ section~4.4 of \cite{Ramond:1981pw} or section~8.3.2 of \cite{Kleinert:2001hn}), and gives (for $m^2=0$)
\begin{equation}
\incl[-5pt]{sunset}\big|_{k^2}=-k^2\frac{g^2}{(4\pi)^4}\frac1{2\eps}+O(\eps^0).
\end{equation}
This gives
\begin{equation}
Z=1-\frac{g^2}{(4\pi)^4}\frac{N+2}{36}\frac1\eps+O( g^2),
\end{equation}
 and ultimately
\begin{equation}
\Delta_{\varphi}=[1+\tfrac\eps2]+\frac{N+2}{4(N+8)^2}\eps^2+O(\eps^3).
\end{equation}

\subsubsection[\texorpdfstring{Anomalous dimension of composite operators in the $\eps$-expansion}{Anomalous dimension of composite operators in the ε-expansion}]{Anomalous dimension of composite operators in the $\boldsymbol \eps$-expansion}

To find the anomalous dimension of more complicated operators in the theory, we consider the renormalisation of composite operators. From this point, we continue in massless renormalisation. Define the vertex function of $n$ insertions of $\varphi$ and one insertion of a composite operator $\O$,
\begin{equation}
\Gamma_\bare^{(n,\O)}(k_i)=\prod_{i=1}^{n}\frac1{G_\bare^{(2)}(k_i)}G_\bare^{(n,\O)}(k_i)_{\mathrm{1PI}},
\end{equation}
where the Fourier transform to momentum space has been performed on all independent momenta, which we can take to be the momenta of the $n$ insertions of $\varphi$. In this expression, the operator $\O=\O_\bare$ is a normal-ordered combination of fields $\varphi$ and gradients $\de^\mu$. The renormalised operator $\O$ is defined in terms of the bare composite operator $\O_\bare$ by $\O_\bare=Z_\O\O$
The renormalised vertex function with insertions is then given by
\begin{equation}
\label{eq:vertexfunctionO}
\Gamma^{(n,\O)}(p_1,\ldots,p_{n},M)=Z^{n/2}Z_\O^{-1}\Gamma_\bare^{(n,\O)}(p_1,\ldots,p_{n},M),
\end{equation}
where $Z_\O=1+c_\O$. The term $c_\O$ is called an ``emerging field counterterm'' and is chosen to cancel divergences appearing in \eqref{eq:vertexfunctionO}.
The corresponding CS equation takes the form
\begin{equation}\label{eq:CSnomixing}
\left[
M\frac\de{\de M}+\beta(g)\frac\de{\de g}-n\gamma_\phi(g)+\gamma_\O(g)
\right]\Gamma^{(n,\O)}(p_1,\ldots,p_{n},M)=0,
\end{equation}
where 
\begin{equation}
\gamma_\O=M\left.\frac{\de\ln Z_\O}{\de M}\right|_{\lambda_\bare}.
\end{equation}

Before proceeding, we will briefly explain why the anomalous dimension that enters in the CS equation indeed corresponds to the CFT anomalous dimension of the renormalised operator at the fixed-point, i.e. to show that $\Delta_\O=\Delta^{(0)}_\O+\gamma_\O(g^*)$. Look at the CS equation for $n=0$ with two insertions of $\O$, which at the fixed-point $\beta(g^*)=0$ takes the form
\begin{equation}
\label{eq:CDtwopoint}
\left[
M\frac\de{\de M}+2\gamma_\O(g)
\right]\Gamma^{(\O,\O)}(p,g^*,M)=0.
\end{equation}
From dimensional analysis (in the bare theory) we must have that $\Gamma^{(\O,\O)}$ scales with momentum as $p^\kappa$ for  $\kappa=2\Delta_\O^{(0)}-d$. This means that $\Gamma^{(\O,\O)}\sim p^\kappa f(p/M)$, and we can use the Euler equation $M \frac\de{\de M}+p\frac\de{\de p}=\kappa$ in \eqref{eq:CDtwopoint}. The solution to the resulting differential equation is $\Gamma^{(\O,\O)}(p,g^*,M)\sim p^\kappa (p/M)^{2\gamma_\O(g^*)}$, which by the Fourier transform to position space gives
\begin{equation}
\langle\O(x_1)\O(x_2)\rangle\sim x^{d-\kappa-2\gamma_\O}.
\end{equation}
Recalling $\kappa=2\Delta_\O^{(0)}-d$, we get the result, $\langle\O(x_1)\O(x_2)\rangle\sim x^{-2\Delta_\O}$, which indeed matches the expected form of a two-point function of conformal operator, c.f.\ \eqref{eq:twopointmain}.

In general, to find anomalous dimensions of composite operators, one has to consider a collection of operators $\O_I$ with identical $\Delta_{\O_I}^{(0)}$. Such operators mix in diagrams, and we write, to leading order,\footnote{A simplifying feature in $\phi^4$ theory is that there is no field redefinition at leading order, or equivalently that $\phi$ gets no anomalous dimension at leading order.}
\begin{equation}
\label{eq:barerenormops}
\O_{\bare,I}(x)=Z_{IJ} \O_J(x),
\end{equation}
with $Z_{IJ}=\delta_{IJ}+c_{IJ}$. The $c_{IJ}$ are emerging field counterterms corresponding to the insertion of $\O_I$ in order to keep the diagrams $\Gamma^{(n,\O_J)}$ finite. One gets a CS equation of the form
\begin{equation}\label{eq:CSmixing}
\left[
\left(M\frac\de{\de M}+\beta(g)\frac\de{\de g}-n\gamma_\phi(g)\right)\delta_{IJ}+\gamma_{IJ}(g)
\right]\Gamma^{(n,\O_J)}(p_1,\ldots,p_{n},M)=0,
\end{equation}
where $\gamma_{IJ}=(Z^{-1})_{IK}M\frac{\de}{\de M} Z_{KJ}$. The anomalous dimensions of the individual operators is then found by computing the eigenvalues of the matrix $\gamma_{IJ}$.

\paragraph{Example: anomalous dimensions of bilinears} 
As an example, we look at the anomalous dimensions of the bilinear scalars. The more general case with mixing will be considered in section~\ref{sec:orderepssystematics}. Consider the operator $\varphi^2_R=R_{ij}\varphi^i\varphi^j$, where $R_{ij}=\delta_{ij}$ for the singlet $S$ representation and $R_{ij}=t_it_j$ (with $t_it_i=0$) for the traceless-symmetric $T$ representation.

We introduce an emerging field counterterm to the diagrammatic rules, and write down an expression for the relevant vertex function,
\begin{equation}
\Gamma^{(2,\varphi_R^2)}=1+\frac12 \mathsf S_R\incl[-5pt]{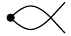}-c_{\varphi_R^2}+O(g^2).
\end{equation}
Evaluating the diagram gives
\begin{equation}
Z_{\varphi^2_R}=1+c_{\phi^2}=1-\mathsf S_R\frac g{(4\pi)^2}\frac1\eps +O(\eps^0),
\end{equation}
from which it follows that
\begin{equation}
\gamma_{\varphi_R^2}=\mathsf S_R\frac g{(4\pi)^2}+O(g^2).
\end{equation}
The symmetry factors are found by solving $R_{ij}T_{ijkl}=\mathsf S_RR_{kl}$, giving $\mathsf S_S=\frac{N+2}3$ and $\mathsf S_T=\frac23$.
Taking these results into account and inserting the value of the coupling constant at the fixed-point we get
\begin{equation}\label{eq:gammaphi2SandT}
\Delta_{\varphi^2_S}=2-\eps+\frac{N+2}{N+8}\eps+O(\eps^2),
\qquad
\Delta_{\varphi^2_T}=2-\eps+\frac{2}{N+8}\eps+O(\eps^2).
\end{equation}
The value for $\varphi^2_S$ agrees with what we found above, in \eqref{eq:deltaphi2SRG}.

\subsubsection{Non-linear sigma model}
\label{sec:NLSMmethod}

For $N>2$, it is possible to write down a perturbative expansion for the $\ON$ CFT starting from the action of a non-linear sigma model in $d=2+\teps$ dimensions \cite{Polyakov:1975rr,Brezin:1975sq,Bardeen:1976zh,Brezin:1976ap,Brezin:1976qa,Hikami:1977vr}. Here we will give a somewhat schematic presentation following \cite{Brezin:1976qa,Hikami:1977vr}. 

Start from the action of $N$ constrained fields,
\begin{equation}
S=\int d^{d}x\left(\frac12\de_\mu\varphi^i\de^\mu\varphi^i+\alpha(\varphi^2-f_\pi^2)\right),
\end{equation}
where the auxiliary field $\alpha$ introduces the constraint $\varphi^2=f_\pi^2$ for some coupling $f_\pi$. Solving the constraint (spontaneous symmetry breaking) we can write the action in terms of the Goldstone fields $\pi^a=\varphi^a$, $a=1,\ldots,N-1$, $\varphi^N=f_\pi^2-\pi^a\pi^a$,
\begin{equation}
S=\int d^dx\left(\frac12\de_\mu\pi^a\de^\mu \pi^a+\frac t2\frac{(\pi^a\de_\mu \pi^a)^2}{1-t\pi^a\pi^a}\right),
\end{equation}
where $t=\frac1{f_\pi^2}$. We now put $d=2+\teps $ and write down a renormalised action. Absorb the coupling $t$ in the field and rewrite in terms of bare quantities: $\sqrt t\pi^a\rightsquigarrow \pi^a_\bare$, $t\rightsquigarrow t_\bare$. Then introduce a renormalised field and coupling by $\pi^a_\bare=\sqrt Z\pi^a$, $t_b=M^{-\teps}Z_1t$ to get
\begin{equation}
S=\frac{M^{\teps}}{2Z_1t}\int d^{2-\eps}x\left(Z\de_\mu\pi^a\de^\mu\pi^a+\frac{Z^2(\pi^a\de_\mu\pi^a)^2}{1-Z\pi^a\pi^a}\right).
\end{equation}
The renormalisation constants can be found by massless dimensional regularisation at $d=2+\teps$,
\begin{align}
\zeta(t)&=M\left.\frac{\de \ln Z}{\de M}\right|_{t_\bare},
\\
\beta(t)&=M\left.\frac{\de t}{\de M}\right|_{t_\bare}.
\end{align}
The leading and subleading order computations give \cite{Hikami:1977vr}
\begin{align}
\zeta(\bar t)&=(N-1)\bar t+O(\bar t^3),
\\
\beta(\bar t)&=\teps\bar t-(N-2)\bar t^2-(N-2)\bar t^3+O(\bar t^4),
\end{align}
for $\bar t=t\frac{S_d}{(2\pi)^d}=t\frac{2\pi^{d/2}}{(2\pi)^d\Gamma(d/2)}$ ($S_d$ is defined in \eqref{eq:Sddef}). Solving for $\beta(\bar t)=0$ perturbatively in $\teps$ we find a UV fixed-point for $\bar t^*=\frac1{N-2}\teps-\frac{1}{(N-2)^2}\teps^2+O(\teps^3)$. The leading vector and singlet operator dimensions are then given by\footnote{$\frac12\zeta(t)$ almost plays the same role as $\gamma_\varphi(g)$ in the $4-\eps$ expansion, but is defined to be the anomalous dimension of $\pi$ with respect to $\Delta_\pi^{(0)}=0$ and not to $\Delta_\pi^{(0)}=\frac{d-2}2$, explaining the precise form \eqref{eq:deltapi}. Equation~\eqref{eq:deltasZ} follows from \eqref{eq:betaderivsmarginal}.}
\begin{align}
\label{eq:deltapi}
\Delta_\pi&=\frac12\zeta(\bar t^*)=\frac\teps2+\frac{\teps}{2(N-2)}-\frac{N-1}{(N-2)^2}\teps^2+\ldots,
\\
\label{eq:deltasZ}
\Delta_{\de\pi\cdot\de\pi}&=d+\beta'(\bar t^*)=2-\frac{\teps^2}{N-2}+\ldots.
\end{align}
The form of the higher order terms were guessed in \cite{Hikami:1977vr} and could be fixed using matching with large $N$ results, giving $\Delta_\pi$ and $\Delta_{\de\pi\cdot\de\pi}$ to order $\teps^3$. The computation was extended to order $\teps^4$ in \cite{Wegner:1987gu} for $\Delta_\pi$ and in \cite{Bernreuther:1986js} for $\Delta_{\de\pi\cdot\de\pi}$. Moreover, 
the result to order $\teps^4$ in \cite{Wegner:1987gu} was extended to all operators $\pi^m_{T_m}$ in the rank $m$ representation of $\ON$, i.e. the family \texttt{ONF3[$m$]} defined below.

In \cite{Brezin:1976an} the study of composite operators with derivatives was initiated, and \cite{Wegner1990} identified the family of conformal primaries corresponding to $\sigma^k$ at large $N$, with dimensions
\begin{equation}
\texttt{DeltaZ[ONF1[$k$]]}=\Delta_{(\de\pi^i\de\pi^i)^k}=2k-\frac{k(k-1)}{N-2}\teps+O(\teps^2).
\end{equation}
The negative sign in the order $\teps$ correction triggered a discussion about the stability of the fixed-point in the non-linear sigma model, but the agreement with $\Delta_{\sigma^k}$ to order $1/N^2$ indicated that the stability problem might be cured by higher orders or non-perturbative effects \cite{Derkachov:1997gc}. The complete mapping of operators between the non-linear sigma model and the large $N$ expansion has not been worked out.

In \cite{Giombi:2016hkj} the order $\teps^2$ anomalous dimensions of singlet weakly broken currents $\mathcal J_{S,\ell}=\pi^i\de^\ell\pi^i$ were found,
\begin{equation}
\texttt{DeltaZ[ONF4[$l$]]}=l+\teps+\frac{1}{N-2}\left(S_1(l-2)+\frac1l-\frac12\right)\teps^2+O(\teps^3),
\end{equation}
where $S_1$ denotes the harmonic numbers.

Finally we would like to stress the fact that the fixed-point in the non-linear sigma model is a UV fixed-point. This is consistent with the central charge being larger than that for $N-1$ free fields,\footnote{In two dimensions, $C_T$ is proportional to $c$, which decreases under RG flow \cite{Zamolodchikov:1986gt}. We would therefore expect that in the vicinity of $d=2$, $C_T$ also decreases under RG flow.}
\begin{equation}
\label{eq:CTnlsm2}
\frac{C_T}{\frac{N-1}NC_{T,\mathrm{free}}}=1+\frac{3}{4(N-2)}\teps^2+O(\teps^3),
\end{equation}
where we used the result from \cite{Diab:2016spb} quoted in \eqref{eq:CTnlsm} above.

\subsubsection{Cubic model near six dimensions}
\label{sec:cubicsixeps}

The work of \cite{Fei:2014yja} proposed that the critical $\ON$ CFT is connected, via the large $N$ expansion, to a cubic model of $N+1$ fields $\varphi^i,\sigma$ near the corresponding upper critical dimension, $d=6-\epsilon$, with results computable in an $\epsilon$-expansion. They proposed an action of the form
\begin{equation}
\label{eq:actioncubic}
S=\int d^{6-\epsilon}x\left(
\frac12(\de_\mu\varphi^i)^2+\frac12(\de_\mu\sigma)^2+\frac{g_1}2\sigma\varphi^i\varphi^i+\frac{g_2}6\sigma^3
\right),
\end{equation}
containing $N$ fields $\varphi^i$ preserving $\ON$ symmetry, and a single field $\sigma$ with no $\mathbb Z_2$ symmetry. In $d=6$ dimensions, $\varphi^i$ and $\sigma$ reduce to free fields, and the action~\eqref{eq:actioncubic} contains the two marginal perturbations compatible with these symmetries. Standard multiplicative renormalisation finds several fixed-points, one of which gives operator dimensions that agree in the large $N$ limit with the large $N$ results in the $\ON$ CFT continued to $d=6-\epsilon$. In the leading order analysis, this fixed-point only gives real operator dimensions for $N>\tilde N_{\mathrm{crit}}$ with $\tilde N_{\mathrm{crit}}=1038$. 

The initial multiplicative renormalisation analysis of \cite{Fei:2014yja} has been extended to order $\epsilon^3$ in \cite{Fei:2014xta}, to order $\epsilon^4$ in \cite{Gracey:2015tta} and to order $\epsilon^5$ in \cite{Kompaniets:2021hwg,Borinsky:2021jdb} giving operator dimensions of the form
\begin{align}
\Delta_{\varphi}&=2-\frac{\epsilon}2+\left(\frac1N+\frac{44}{N^2}+\frac{1936}{N^3}+\ldots\right)\epsilon-\left(\frac{11}{12N}+\frac{835}{6N^2}-\frac{16352}{N^3}+\ldots\right)\epsilon^2+\ldots+O(\epsilon^6),
\\
\label{eq:sigmacubic}
\Delta_\sigma&=2+\left(\frac{40}N+\frac{6800}{N^2}+\frac{2637760}{N^3}+\ldots\right)\epsilon-\left(\frac{104}{3 N}+\frac{34190}{3 N^2}+\frac{5912792}{N^3}+\ldots\right)\epsilon^2+\ldots+O(\epsilon^6),
\end{align}
which are to be identified with $\varphi$ and $\sigma $ at large $N$. For bilinear operators, a mixing problem has to be resolved and two operators are found as linear combinations of $\sigma^2$ and $\varphi^i\varphi^i$ with dimensions $\Delta_\pm$. The lower eigenvalue satisfies $\Delta_-+\Delta_\sigma=d$, so the corresponding operator is a descendant,\footnote{Strictly speaking the computation finds the RG dimension of the equations-of-motion operator, and not the dimension of the corresponding descendant state.} while \cite{Kompaniets:2021hwg}
\begin{align}
\Delta_+&=4-\left(\frac{100}{N}+\frac{49760}{N^2}+\frac{27470080}{N^3}+\ldots\right)\epsilon+\left(\frac{395}{3 N}+\frac{237476}{3 N^2}+\frac{56985896}{N^3}+\ldots\right)\epsilon^2\nonumber\\&\quad +\ldots+O(\eps^6).
\end{align}
This expansion agrees with the dimension of the operator $\texttt{Op[S,0,2]}=\sigma^2$.

In \cite{Fei:2014xta} two operators at $\Delta=6+O(\epsilon)$ were considered by analysing the eigenvalues $\lambda_i$ of the $2\times2$ matrix $\de_j\beta_i$ (c.f.\ \eqref{eq:betaderivsmarginal}), matching with $\texttt{Op[S,0,4]}=\sigma^3$ and $\texttt{Op[S,0,3]}=\square\sigma^2$, with $\Delta_i=6-\epsilon+\lambda_i$. We get from the result of \cite{Kompaniets:2021hwg}
\begin{align}
\Delta_1&=6-\left(\frac{420}{N}+\frac{159600}{N^2}+\frac{83791680}{N^3}+\ldots\right)\epsilon+\left(\frac{499}{N}+\frac{164910}{N^2}+\frac{114980496}{N^3}+\ldots\right
)\epsilon^2\nonumber\\&\quad+\ldots+O(\epsilon^6).
\\
\Delta_2&=6+\left(\frac{155}{3 N}+\frac{12700}{3 N^2}+\frac{718400}{3 N^3}+\ldots\right)\epsilon^2+\ldots+O(\epsilon^6),
\end{align}
Note that in $d=6-\epsilon$ dimensions, these operators are in the opposite order compared to $d=4-\eps$.

In \cite{Giombi:2016hkj}, the two singlet twist families $[\varphi,\varphi]_{S,0,\ell}$ and $[\sigma,\sigma]_{0,\ell}$ were considered. Again a mixing problem had to be solved, and the resulting expressions agree with the families \texttt{ONF4[$l$]} and \texttt{ONF7[$l$]} at large $N$,\footnote{As in the other expressions, any order in $1/N$ in the $O(\epsilon)$ term can be extracted from the results of \cite{Giombi:2016hkj}. The terms at $\epsilon/N^2$ read $8 (\ell-2) (\ell+5) (11 \ell^6+99 \ell^5+429 \ell^4+1089 \ell^3+2196 \ell^2+3024 \ell+1360)/((\ell+1)^4 (\ell+2)^4)$ and $32  (425 \ell^8+5100 \ell^7+21976 \ell^6+37134 \ell^5-7872 \ell^4-122292 \ell^3-180809 \ell^2-110454 \ell-24880)/((\ell+1)^4 (\ell+2)^4) $.}
\begin{align}
\label{eq:JScubic}
\Delta_{\mathcal J_{S,\ell}}&=4-\epsilon+\ell+\left(\frac{2(\ell-2)(\ell+5)(\ell^2+3\ell+8)}{(\ell+1)^2(\ell+2)^2}\frac{1}{N}
+\ldots\right)\eps+O(\eps^2),
\\
\Delta_{[\sigma,\sigma]_{0,\ell}}&=4-\epsilon+\ell+\left(1+\frac{16(5\ell^4+30\ell^3+38\ell^2-21\ell-25)}{(\ell+1)^2(\ell+2)^2}\frac1{N}+\ldots\right)\eps+O(\epsilon^2).
\end{align}
We note that $\Delta_{[\varphi,\varphi]_{S,0,\ell}}+\Delta_\sigma=d-\epsilon$. The order $\epsilon$ anomalous dimensions for non-singlet broken currents were also found in \cite{Giombi:2016hkj}, taking the form
\begin{equation}
\Delta_{\mathcal J_{T,\ell}}=\Delta_{\mathcal J_{A,\ell}}+O(\epsilon^2)=4-\epsilon+\ell+\left(\frac{2(\ell-1)(\ell+4)}{(\ell+1)(\ell+2)}\frac1N+\ldots\right)\epsilon+O(\epsilon^2),
\end{equation}
where the spin $\ell$ is even for $T$ and odd for $A$.
Evaluating $\Delta_{[\varphi,\varphi]_{T,0,\ell}}$ at spin $\ell=0$ gives
\begin{equation}
\Delta_{\varphi^2_T}=4-\epsilon-\left(\frac4N+\frac{176}{N^2}+\frac{1936}{N^3}+\ldots\right)\epsilon+O(\epsilon^2).
\end{equation}
We are not aware of any direct computation of this quantity, or of any computation at all of $\Delta_{\varphi^4_{T_4}}$ in this expansion.
For the central charge corrections, \cite{Diab:2016spb} found
\begin{align}
\frac{C_T}{C_{T,\mathrm{free}}} &=\frac{N+1}N-\left(\frac7{4N}+\frac{98}{N^2}+\frac{10192}{N^3}+\ldots\right)\epsilon+O(\epsilon^2),
\\
\frac{C_J}{C_{J,\mathrm{free}}} &=1-\left(\frac5{3N}+\frac{220}{3N^2}+\frac{9680}{3N^3}+\ldots\right)\epsilon+O(\epsilon^2).
\end{align}

A perturbative determination of $\tilde N_{\mathrm{crit.}}$ was done in \cite{Kompaniets:2021hwg}, giving
\begin{equation}
\tilde N_{\mathrm{crit.}}=1038.2661 -1219.6796\epsilon -1456.6933\epsilon^2 + 3621.6847\epsilon^3 + 986.2232\epsilon^4 .
\end{equation}
In \cite{Li:2016wdp} a numerical bootstrap study was performed in five dimensions. However, the work \cite{Giombi:2019upv} indicated that the theory remains non-unitary in $d=5$ even at large $N$.
 
Finally, let us mention the $N=0$ version of the action \eqref{eq:actioncubic}. The (non-unitary) fixed-point found in that case is not connected to the critical $\ON$ CFT but instead describes the Yang--Lee edge singularity \cite{Fisher:1978pf,Cardy:1985yy}, connecting in 2d to the minimal model $\mathcal M_{2,5}$ with central charge $c=-\frac{22}5$. In the $d=6-\epsilon$ expansion of that theory, the anomalous dimensions of the operators $\sigma$ and $\sigma^3$ ($\sigma^2$ is a descendant) have been determined to order $\epsilon^5$ \cite{Kompaniets:2021hwg,Borinsky:2021jdb}.

\subsection[\texorpdfstring{One-loop dilatation operator in the $\eps$-expansion}{One-loop dilatation operator in the ε-expansion}]{One-loop dilatation operator in the $\boldsymbol\eps$-expansion}
\label{sec:orderepssystematics}

In this section, we will describe in detail how to implement the one-loop dilatation operator in the $\eps$-expansion, which was first done systematically in \cite{Kehrein:1992fn,Kehrein:1994ff,Kehrein:1995ia}. These papers take the starting point in diagrams, and give an implementation of $D$ in terms of creation and annihilation operators acting on the factors $\de_{\mu_1}\cdots\de_{\mu_r}\varphi^i$ that build up composite operators.\footnote{See also \cite{Derkachov:1995wg} for a similar approach.} Equivalent formulations were later described in \cite{Hogervorst:2015akt} based on conformal perturbation theory, and in \cite{Liendo:2017wsn} starting from the idea of multiplet recombination described in \cite{Rychkov:2015naa}, c.f.\ section~\ref{sec:multipletrecombination}. After reviewing the connection to the multiplicative renormalisation discussed in section~\ref{sec:Feynmandiagrams}, we will describe the implementation of \cite{Hogervorst:2015akt}.

In \eqref{eq:barerenormops}, renormalised operators were introduced by $\O_I=(Z^{-1})_{IJ}\O_{\bare,J}$, which to leading order in $\eps$ reads
\begin{equation}
\label{eq:renormofcomposites}
\O_I=\O_{\bare,I}-c_{IJ}\O_{\bare,J},
\end{equation}
where $c_{IJ}$ are the emerging field counterterms introduced to keep diagrams with insertions of the $\O_I$ finite. To leading order the anomalous dimension matrix $\gamma_{IJ}$ of \eqref{eq:CSmixing} takes the form 
\begin{equation}
\label{eq:gammaIJmatrix}
\gamma_{IJ}=M\frac\de{\de M}c_{IJ} =\varGamma_{IJ}\eps+O(\eps^2),
\end{equation}
where we defined the one-loop anomalous dimension matrix $\varGamma_{IJ}$ as acting on the space of renormalised operators. 

Look for a set of operators that diagonalise \eqref{eq:gammaIJmatrix}, $\tilde \O_I=A_{IJ}\O_I$, where $A_{IJ}\varGamma_{JK}(A^{-1})_{KL}$ is diagonal. Such operators are egenstates of the one-loop part of the dilatation operator $D$, satisfying
\begin{equation}
D\O_I=(\Delta^{(0)}\delta_{IJ}+\varGamma_{IJ}\eps)\O_J+O(\eps^2). 
\end{equation}
Here we think of the operators as inserted at the origin. The eigenstates $\tilde \O_I$ thus satisfy
\begin{equation}
D\tilde\O_I(0)|0\rangle=\Delta_{\tilde \O_I}\tilde \O_I(0)|0\rangle=\left(\Delta^{(0)}_{\O_I}+\gamma_{O_I}^{(1)}\eps\right)\tilde \O_I(0)|0\rangle,
\end{equation}
and one can see that for fixed $I$, the vector $A_{IJ}$ is a left eigenvector of the one-loop anomalous dimension matrix, $
A_{IJ}\varGamma_{JK}=\gamma_{\O_I}^{(1)}A_{IK}$.

At order $\eps$, the field redefinition does not change the number of fields $\varphi$ and derivatives $\de^\mu$ in a composite operator.\footnote{In the subset of operators that are completely traceless-symmetric in the fields, $\mathbf R_{i_1\cdots i_p}=t_{i_1}\cdots t_{i_p}$ with $t_it^i=0$, this statement can be extended to two-loops \cite{Kehrein:1995ia}.} The dilatation operator is also block-diagonal with respect to global $\ON$ and Lorentz $\mathrm{SO}( d)$ symmetries, which we describe with the help of tensor structures $\mathbf R$ and $\mathbf L$ respectively. The most general form of a composite operator built out of $p$ fields and $L=\sum_{k=1}^p r_k$ gradients is therefore a linear combination of operators of the form\footnote{In writing the ansatz \eqref{eq:operatorRLform}, we can remove all operators with a factor $\square\varphi$, i.e.\ a pair of contracted derivatives acting on a single field. Such operators are redundant and have non-vanishing correlators only at coincident points, see for instance the comment on page $5$ in \cite{Hogervorst:2015akt}.}
\begin{equation}\label{eq:operatorRLform}
\O_I=\mathbf R_{i_1\ldots i_p}\mathbf L^{\mu_{11}\cdots \mu_{1 r_1}\cdots \mu_{p1}\cdots\mu_{pr_p}}\,(\de_{\mu_{11}}\cdots \de_{\mu_{1r_1}}\varphi^{i_1})\cdots 
(\de_{\mu_{p1}}\cdots \de_{\mu_{pr_p}}\varphi^{i_p}),
\end{equation}
 and has engineering dimension $\Delta^{(0)}_\O=p(1-\frac12\eps)+L$. For Lorentz traceless-symmetric representations, the spin is given by $L$ minus the number of contractions $\delta^{\mu\nu}$ in the tensor $\mathbf L$. 
The task to find the eigenstates of the one-loop dilatation operator is then reduced to diagonalising the matrix of anomalous dimensions within the subspace of operators of the form \eqref{eq:operatorRLform}. The eigenstates will either correspond to primary operators, or descendants of primary operators with fewer derivatives and the same global symmetry.

We now describe the implementation of the one-loop dilatation operator given in \cite{Hogervorst:2015akt},\footnote{Some additional details can be found in appendix~C of \cite{Hogervorst:2015tka}.} which starts from the formula for shift in scaling dimension in conformal perturbation theory (see \cite{Komargodski:2016auf} and \cite{Cardy:1996xt}). In this formalism, one considers the perturbation of a CFT by the operator $\alpha\mathcal X$ for some small parameter $\alpha$, and finds the leading order anomalous dimension of $\O$ by $\gamma_\O\propto \alpha\lambda_{\O\O  \mathcal X}$. We will have $ \mathcal X=\frac14(\varphi^2)^2$. 

The one-loop anomalous dimension operator $\varGamma$ on the subspace of operators of the form \eqref{eq:operatorRLform} with $p$ fields and $L$ derivatives will be implemented as\footnote{In \eqref{eq:dil1}, the specification $|_{pL}$ means that we should compute the contractions between $(\varphi(x)^2)^2$ and $\O(0)$, and only keep the resulting operators with $p$ fields and $L$ derivatives. This implies that two of the four factors in $(\varphi(x)^2)^2$ must be consumed by the Wick contractions.}
\begin{equation}\label{eq:dil1}
\varGamma\O(0)|0\rangle =\mathcal N\left.\int_{|x|=1} d^{d-1} x (\varphi(x)^2)^2 \O(0)|0\rangle \right|_{pL},
\end{equation}
where, since we are already working at order $\eps$, we can evaluate the integral at order $\eps^0$. Using the normalisation convention where $\varphi$, but not composite operators, have unit two-point function (the intermediate unit normalisation described in appendix~\ref{app:normalisation}), the two-point function including derivatives on one field is given by
\begin{equation}\label{eq:twopointderiv}
\left\langle \varphi^i(x)\de_{\mu_1}\cdots\de_{\mu_r}\varphi^j(0)\right\rangle=2^r\left(\tfrac{d-2}2\right)_r\frac{x_{\mu_1}\cdots x_{\mu_r}-\text{traces}}{|x|^{d-2+2r}}\delta^{ij},
\end{equation}
where $(a)_n=\frac{\Gamma(a+n)}{\Gamma(a)}$ denotes the (rising) Pochhammer symbol.
In these conventions $\mathcal N$ has the value
\begin{equation}
\mathcal N=\frac1{S_d}\frac1{16(N+8)}.
\end{equation}

The procedure to find $\varGamma$ within a given subspace of operators is the following:
\begin{enumerate}
\item Perform Wick contractions between two of the four fields in $(\varphi(x)^2)^2$ and two of the fields in $\O( 0)$, using the two-point function \eqref{eq:twopointderiv}.
\item Expand the remaining two $x$-dependent fields around the origin, using 
\begin{equation}\label{eq:expandaroundorigin}
\varphi^i(x)\varphi^j(x)=\varphi^i(0)\varphi^j(0)+x^\mu\de_\mu\left(\varphi^i(0)\varphi^j(0)\right)+x^\mu x^\nu\frac{\de_\mu\de_\nu}{2}\left(\varphi^i(0)\varphi^j(0)\right)+\ldots.
\end{equation}
\item Perform the integration over $x$ using 
\begin{equation}\label{eq:integratexpowers}
\frac1{S_d}\int\limits_{|x|=1}d^{d-1}x \, x^{\mu_1}\cdots x^{\mu_{2m}}=\frac{\delta_{\mu_1\mu_2}\cdots\delta_{\mu_{2m-1}\mu_{2m}}+\text{permutations}}{2^m\left(\tfrac d2\right)_m},
\end{equation}
where there are $(2m-1)!!$ permutations and we get zero of we have an odd number of $x^\mu$.
\end{enumerate}

We will consider three examples. As a warm-up, we recompute the anomalous dimension of the operators $\varphi^2_S$ and $\varphi^2_T$, which correspond to $L=0$ and $p=2$ with $R_{ij}=\delta_{ij}$ and $R_{ij}=t_it_j$ respectively, for a null vector $t^it_i=0$. Under the first step we write 
\begin{equation}
(\varphi(x)^2)^2=\delta_{ab}\delta_{cd}\varphi^a(x)\varphi^b(x)\varphi^c(x)\varphi^d(x).
\end{equation}
Since $\varphi^2_R$ contains no derivatives, the spacetime dependence of all Wick contractions becomes the trivial factor $|x|^{-d+2}$. For the $\ON$ indices, the $2\times 4!=48$ possible contractions group in three inequivalent terms, giving 
\begin{equation}\label{eq:ONcontractionsphi2T}
48R_{ij}T_{ijkl}\varphi^k(x)\varphi^l(x)=16\mathsf S_RR_{ij}\varphi^i(x)\varphi^j(x),
\end{equation}
for $T_{ijkl}=\frac{\delta_{ij}\delta_{kl}+\delta_{ik}\delta_{jl}+\delta_{il}\delta_{jk}}{3}$.
Finally, there is no need for going beyond the leading term in the expansion \eqref{eq:expandaroundorigin}, and the integral \eqref{eq:integratexpowers} trivialises. We get
\begin{equation}
\gamma_{\varphi^2_R}=\frac{1}{16(N+8)}\times 16\mathsf S_R=\frac{\mathsf S_R}{N+8}.
\end{equation}
Computing the index contractions in \eqref{eq:ONcontractionsphi2T} gives for the singlet representation $ \mathsf S_S=N+2$ and for the rank $2$ traceless-symmetric representation $\mathsf S_T=2$, leading to $\gamma_{\varphi^2_S}=\frac{N+2}{N+8}$  and $\gamma_{\varphi^2_T}=\frac{2}{N+8}$, in agreement with \eqref{eq:gammaphi2SandT}.

Consider now a more complicated example. We look at spin $\ell=2$ operators of the schematic form $\de^2\varphi^4_S$, which corresponds to a tensor structure
\begin{equation}
\mathbf R_{ijkl}\mathbf L^{\mu\nu}=\delta_{ij}\delta_{kl}l^\mu l^\nu,
\end{equation}
with $l^{\mu}l_\mu=0$. For this choice of tensor structure, there are three linearly independent operators of the form \eqref{eq:operatorRLform}:
\begin{align}
\O_1 &=\mathbf R_{ijkl}\mathbf L^{\mu\nu}\, \varphi^i\varphi^j\varphi^k(\de_\mu\de_\nu\varphi^l),
\\
\O_2 &=\mathbf R_{ijkl}\mathbf L^{\mu\nu} \,\varphi^i\varphi^j(\de_\mu\varphi^k)(\de_\nu\varphi^l),
\\
\O_3 &=\mathbf R_{ijkl}\mathbf L^{\mu\nu} \,\varphi^i(\de_\mu\varphi^j)\varphi^k(\de_\nu\varphi^l).
\end{align}
In this basis we find
\begin{equation}
\varGamma =\frac1{3(N+8)}\begin{pmatrix}5 (N+8) & 2(N+4) & 8\\
 N+2 & 4 (N+5) & 24 \\
 3 & 10 & 2 (3 N+16) 
 \end{pmatrix}.
\end{equation}
One left eigenvector is $(1,1,2)$, with eigenvalue $2$, which corresponds to the operator $\O_1+\O_2+2\O_3$. By inspection, this is a total derivative of the operator $\varphi^4_S=\varphi^i\varphi^i\varphi^j\varphi^j$ and is therefore a descendant, and indeed we can see that the eigenvalue $2$ gives a scaling dimension $[6-2\eps]+2\eps+O(\eps^2)=\Delta_{\varphi^4_S}+2$, which is consistent with \eqref{eq:betaderivsmarginal}. The other two operators are primary operators with dimensions
\begin{equation}\label{eq:eigenvaluesS2}
\Delta_\mp=[6-2\eps]+\frac{9 N+44\mp\sqrt{9 N^2-8 N+624}}{6 (N+8)}\eps+O(\eps^2).
\end{equation}
They are the operators \texttt{Op[S,2,2]} and \texttt{Op[S,2,3]}, discussed already in section~\ref{sec:unmixing} and appearing in table~\ref{tab:singletspinning} below.

Let us make a couple of remarks on the example just studied. The case $N=1$ needs to be considered separately, since in this case $\O_{2}$ and $\O_{3}$ are identical. We get a $2\times 2$ matrix $\varGamma$ with eigenvalues $\{2,\frac{13}9\}$. Again, the first is a descendant and the second is a new operator, denoted \texttt{Op[E,2,2]}. We can compare with the eigenvalues in \eqref{eq:eigenvaluesS2} and note that as $N\to 1$, the lower value must decouple,\footnote{Indeed, as can be seen from \eqref{eq:OPEs22}, the OPE coefficient of this operator in the $\varphi\times\varphi$ OPE vanishes when $N=1$.} while the upper becomes $\frac{13}9$, showing that $\texttt{Op[S,2,3]}\to \texttt{Op[E,2,2]}$.
It is also instructing to study the dependence of the dimensions $\Delta_\mp$ in \eqref{eq:eigenvaluesS2} as $N\to\infty$. We note that
\begin{equation}
\Delta_-\to 6-\eps+O(N^{-1})=4+2\frac{d-2}2+O(N^{-1}),\qquad \Delta_+\to 6 +O(N^{-1}),
\end{equation}
which means that we can make a connection with the description of operators in the large $N $ expansion. At large $N$ and generic $d$, there is only one operator of spin $2$ and dimension $6+O(N^{-1})$, namely the operator $\sigma\de^2\sigma$, which is a member of the infinite family of bilinears in $\sigma$ of the form $[\sigma,\sigma]_{0,\ell}\sim\sigma\de^\ell\sigma$. The order $1/N$ anomalous dimensions of the operators $\sigma\de^\ell\sigma$ have been computed (family \texttt{ONF7[$l $]} below), and for $\ell=2$ it agrees with $\Delta_+$ in the overlap of expansions. The large $N$ interpretation of the lower operator is somewhat obscured, and without invoking a more careful analysis we may view the operator as a mixture of $[\varphi,\varphi]_{S,1,2}=\de^2\square\varphi^2_S$ and $\de^2\sigma\varphi^2_S$, due to the mixing between $\sigma $ and $\square$ in the large $N$ expansion.

\begin{figure}
\centering
\includegraphics[scale=1]{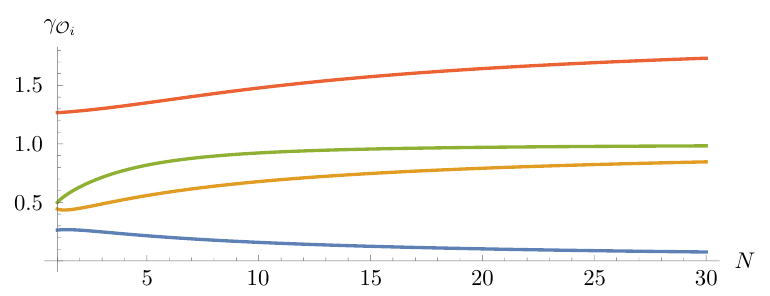}
\caption{Anomalous dimensions of operators \texttt{Op[S,4,$i$]}, $i=2,3,4,5$ as functions of $N$.}\label{fig:delta4}
\end{figure}

At spin three, one new eigenvalue, $1$, is found, but the corresponding operator of the form $\de^3\varphi^4_S$ is a descendant of the broken current $\mathcal J_{S,4}$ by the multiplet recombination effect, c.f.\ section~\ref{sec:multipletrecombination}. At spin four, we find four new eigenvalues corresponding to four new primary operators, and in figure~\ref{fig:delta4} we plot them for a range of $N$. We observe that as $N\to\infty$, the lowest operator dimension approaches $4+4\frac{d-2}2$, the two middle ones approach $6+2\frac{d-2}2$ and the highest operator dimension approaches $8$. The first and last operators therefore take the forms $\de^4\varphi^4_S$ and $\sigma\de^4\sigma$ respectively in the large $N$ limit. The interpretation of the middle two operators is again a bit obscured. Operators one and three from the top survive at $N=1$.

As a third and final example, consider operators of the form $\square\varphi^{2k}_S$ for $k\leqslant2$, where both Lorentz indices and all $2k$ $\ON$ indices are contracted: $\mathbf L_{\mu\nu}=\eta_{\mu\nu}$, $\mathbf R_{i_1\cdots i_{2k}}=\delta_{i_1i_2}\cdots\delta_{i_{2k-1}i_{2k}}$. There are two linearly independent operators of this form:
\begin{align}
\O_1 &=(\varphi^i\varphi^i)^{k-1}\de_\mu\varphi^j\de^\mu\varphi^j,
\\
\O_2 &=(\varphi^i\varphi^i)^{k-2}\varphi^j\de_\mu\varphi^j\varphi^l\de^\mu\varphi^l.
\end{align}
In this basis we find the matrix
\begin{equation}
\varGamma=\frac1{N+8}\begin{pmatrix}
6k^2+(N-12)k+8 & 8(k-1) \\
4 & 6k^2+(N-4)k-4
\end{pmatrix},
\end{equation}
which has the eigenvalues $\gamma_1=k\frac{6k+N-4}{N+8}$, $\gamma_2=\frac{6k^2+(N-12)k+4}{N+8}$. The first value agrees with the anomalous dimension of the operator $\varphi^{2k}_S$ \cite{Wegner:1972zz,Derkachov:1997gc}, and indeed the corresponding left eigenvector is a descendant,
\begin{equation}
\square\big((\varphi^i\varphi^i)^k\big)=2k\O_1+4k(k-1)\O_2.
\end{equation}
The other eigenvalue represents a family of new conformal primary operators, which we schematically denote $\square\varphi^{2k}_S$. We include this result as the anomalous dimension of the family \texttt{ONF9[$k$]} below.

Finally, let us remark that for $N=1$, all the eigenvalues in the cases of operators with up to five derivatives come out as rational numbers, meaning that all operators in the $N=1$ theory constructed with up to five derivatives have a rational anomalous dimension at order $\eps$ \cite{Kehrein:1994ff}. A complete list of such operators can be found in that paper.

\subsection[Large $N$ expansion]{Large $\boldsymbol N$ expansion}
\label{sec:largeN}

For any spacetime dimension $2< d<4$, the critical $\ON$ CFT can be described through an expansion in powers of $1/N$, with propagating degrees of freedom mediated by the vector field $\varphi^i$, with $\Delta_{\varphi}=\frac{d-2}2+O(1/N)$, and an auxiliary field $\sigma$ with $\Delta_\sigma=2+O(1/N)$. A framework for the leading-order computations was developed in \cite{Abe1972,Ma:1972zz,Brezin:1972se,Suzuki1972,Ferrell:1972zz,Ma:1973cmn} and a  systematic framework for computing higher-order corrections, denoted skeleton expansion, or conformal bootstrap, was developed in \cite{Vasiliev:1981yc,Vasiliev:1981dg,Vasiliev:1982dc}.\footnote{The order $1/N^2$ correction to $\Delta_\varphi$ was computed already in \cite{Abe1973}. The ``old'' conformal bootstrap, sometimes known as the Polyakov--Migdal bootstrap dates back to the 1960's \cite{Patashinskii1964,Polyakov1968,Migdal1969,Polyakov:1970xd,Parisi:1971zza,Mack:1972kq}.} A systematic determination of leading anomalous dimensions of a larger set of operators was undertaken in \cite{Lang:1990ni,Lang:1990re,Lang:1992pp,Lang:1992zw,Vasiliev:1993ux,Lang:1994tu,Derkachov:1997qv}. In particular \cite{Derkachov:1997qv} gave a simple formalism for computing the order $1/N$ anomalous dimension for any operator with $m$ fields and arbitrary spin transforming in the rank $m$ traceless-symmetric representation $T_m$ of the $\ON$ symmetry. Higher-order computations of anomalous dimensions for several operators were performed in \cite{Lang:1991kp,Derkachov:1997ch} and of OPE coefficients in \cite{Lang:1993ct,Goykhman:2019kcj}.

An action for the large $N$ $O(N)$ model can be written by coupling $\varphi$ to an auxiliary field $\sigma$, sometimes denoted a Hubbard--Stratonovich transformation,
\begin{equation}\label{eq:HSaction}
S=\int d^dx\left(\frac12(\de_\mu\varphi^i)^2+\frac1{2\sqrt N}\varphi^i\varphi^i\sigma-\frac32\frac{\sigma^2}{N\lambda}\right).
\end{equation}
The field $\sigma $ has no kinetic term in \eqref{eq:HSaction}, but a propagator for $\sigma$ is dynamically generated by a sum over bubble diagrams.
Integrating out the auxiliary field $\sigma$ gives back the usual $\lambda(\varphi^2)^2$ interaction \eqref{eq:actionintro}, given in the introduction. In the IR limit, on the other hand, the $\sigma^2$ term becomes irrelevant and decouples from the action.

When constructing diagrams using the action \eqref{eq:HSaction}, $1/N$ cannot be directly used as a loop counting parameter, since a closed $\varphi$ loop gives a factor $N$. In particular, it is the balancing of these two powers introduces a dynamically generated propagator for the field $\sigma$. Typically, at each order in $1/N$, diagrams at different loop orders will contribute.

The action \eqref{eq:HSaction} can be used as the starting point for a computational framework, see e.g.\ \cite{Goykhman:2019kcj}.
Here we follow instead the framework used in e.g.\ \cite{Vasiliev1983,Derkachov:1997ch}, which starts from a Euclidean action of the fields $\varphi^i$ and $\psi\sim\frac\sigma{\sqrt N}$:
\begin{equation}\label{eq:actionVasiliev}
S=\int d^dx\left(\frac12(\de_\mu\varphi^i)^2-\frac12\psi\varphi^a\varphi^a\right).
\end{equation}
The action \eqref{eq:actionVasiliev} is now extended with a complicated zero, and split into two terms, $S=S_0+S_{\mathrm I}$ with
\begin{align}\label{eq:S0largeN}
S_0&=\int d^dx\left(\frac12(\de_\mu\varphi^i)^2+\frac12\psi M^{-2\delta}K_\delta^{-1}\psi\right),
\end{align}and\begin{align}\label{eq:SIlargeN}
S_{\mathrm I}&=-\int d^dx\left(\frac12\psi\varphi^i\varphi^i+\frac12 \psi K^{-1} \psi\right).
\end{align}
The term $S_{\mathrm I}$ will be treated as an interaction to $S_0$, and $K^{-1}$ corresponds to the dynamically generated inverse propagator. 
In \eqref{eq:S0largeN}, $\delta$ is introduced as a regularisation parameter, $K^{-1}_{\delta}(p)=C(\delta)^{-1}K^{-1}(p)p^{2\delta}$, with expressions regular in the limit $\delta\to0$ (with $C(0)=0$). 

We first work out the normalisation of the propagators. In this computation one may put $\delta=0$ from the beginning. Define the inverse propagators for $\phi$ and $\psi$ by
\begin{align}
\delta^{ij}G_\varphi(x)=\langle\varphi^i(x)\varphi^j(0)\rangle=\frac{C_\varphi}{x^{2\Delta_\varphi}}\delta^{ij}
,\qquad
G_\psi(x)=\langle\psi(x)\psi(0)\rangle = \frac{C_\psi}{x^{2\Delta_\psi}}.
\end{align}
The normalisation $C_\varphi$ of the inverse $\varphi$ propagator agrees to leading order with the one of a free scalar, which follows directly from the Fourier transform of $1/p^2$ (see appendix~\ref{app:fouriertransform})
\begin{equation}
C_\varphi=\frac{\Gamma(\mu-1)}{4\pi^\mu}=\frac1{(d-2)S_d},
\end{equation}
where $S_d=\mathrm{Vol}(S^{d-1})=2\pi^{d/2}/\Gamma(d/2)$ is the volume of the $d-1$ dimensional sphere.

For the $\psi$ propagator, define in momentum space
\begin{equation}
\label{eq:Kpdef}
K(p)=\frac1{G_\psi(p)}=\frac{\Gamma(\Delta_\psi)}{\pi^\mu 2^{2(\mu-\Delta_\psi)}\Gamma(\mu-\Delta_\psi)}\frac{p^{2(\mu-\Delta_\psi)}}{C_\psi}.
\end{equation}
Now we have
\begin{equation}
G_\psi(p)=\incl[-2.5pt]{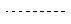}+\frac N2\incl[-5pt]{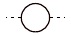}+\frac{N^2}4\incl[-5pt]{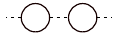}+\ldots,
\end{equation}
where the factor $N$ comes from the $\ON$ index contraction and $\frac12$ is a symmetry factor.
Summing the geometric series gives 
\begin{equation}\label{eq:Kinvexpr}
K(p)=1-\frac N2\incl[-5pt]{sigmabubble}=1-\frac {N}2\int \frac{d^dk}{(2\pi)^d}\frac{\Lambda^{4-d}}{p^{2}(k-p)^2},
\end{equation}
where $\Lambda$ is a UV cutoff. 
In the limit $\Lambda\to\infty$, the contribution from the loop diagram dominates over the constant $1$, which can be dropped \cite{Ma1976}.
The integral in \eqref{eq:Kinvexpr} defines a convolution, and we can evaluate the answer in position space, giving an expression for $K(x)$ as a product (see appendix~\ref{app:fouriertransform}). Fourier-transforming back to momentum space gives
\begin{equation}\label{eq:Kinvposeq}
K(p)=-\frac{N}{2}\frac{\pi^\mu\Gamma(\mu-1)^2\Gamma(2-\mu)}{\Gamma(2\mu-2)}\frac{\Lambda^{4-2\mu}}{p^{4-2\mu}}.
\end{equation}
Comparing with \eqref{eq:Kpdef} we see that we have, to leading order, $\Delta_\psi=2$, and 
\begin{equation}\label{eq:Cpsires}
C_\psi=-\frac{32 \Gamma (2 \mu -2)\Lambda^{2\mu-4}}{N\Gamma (2-\mu ) \Gamma (\mu -2) \Gamma (\mu -1)^2}.
\end{equation}
Note that, despite the appearance, this constant is positive for $1<\mu<2$. As already advertised, the scaling of $C_\psi$ with $N$ motivates the introduction of $\sigma\sim \psi/\sqrt{C_\psi}\sim \sqrt N\psi$, connecting to the heuristically introduced action \eqref{eq:HSaction}.

The action $S_0+S_{\mathrm I}$ in \eqref{eq:S0largeN}+\eqref{eq:SIlargeN} is not multiplicatively renormalisable in the usual sense. 
To get a multiplicatively renormalisable theory, \cite{Vasiliev1983} proposed to introduce an ``intermediate theory'' or ``enhanced model'' with additional couplings $u$ and $v$. We follow conventions in \cite{Derkachov:1997ch}, see also \cite{Derkachov:1998js} for a nice presentation. 
Introduce auxiliary couplings $u_\bare$ and $v_\bare$ and write
\begin{equation}\label{eq:enhancedmodel}
S_0+S_{\mathrm{I}}=\int d^dx\left(\frac12(\de_\mu\varphi_\bare)^2+\frac{u_\bare}2M^{-2\delta}\psi_\bare K_\delta\psi_\bare
-\frac12\psi_\bare\varphi_\bare^2-\frac{v_\bare}2\psi_\bare K\psi_\bare
\right).
\end{equation}
Then introducing renormalised quantities by $\varphi_\bare=\sqrt{Z}\varphi$, $\psi_\bare=Z_g/Z\psi$, $u_\bare=uZ^2/Z_g^2$ and $v_\bare=vZ^2/Z_g^2$, where the renormalisation constants depend on $u$ and $v$, the bare action \eqref{eq:enhancedmodel} becomes
\begin{align}\label{eq:enhancedmodelren}
S&=\int d^dx\left(
Z\frac12(\de_\mu\varphi)^2+u\frac{M^{-2\delta}}2\psi K_\delta\psi-Z_g\frac{1}2\psi\varphi^2-v\frac12\psi K\psi
\right).
\end{align}
The action \eqref{eq:enhancedmodelren} of the enhanced model is then used to perform multiplicative renormalisation, treating $u$ and $v$ as coupling constants. An advantage of the enhanced model is that facilitates the computation of higher-order anomalous dimensions of composite operators \cite{Derkachov:1997ch}.

We conclude this section by giving a sample computation, namely the evaluation of the anomalous dimension $\gamma_\varphi$ at order $1/N$. For this computation, the original action with $u=v=1$ suffices.
 We get for $G_\varphi(p)$\footnote{In principle, also tadpole diagrams can also be written down at this order, however they do not contribute to the field renormalisation.}
\begin{equation}
G_\varphi(p)=\incl[-2.5pt]{line}+\incl{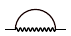}+\incl[-2.5pt]{lineCT}+\ldots,
\end{equation}
where the wavy line refers to the effective $\psi$ propagator.
The diagram can be written as
\begin{align}
\incl{bubblemix-new}=M^{2\delta}\int \frac{d^dk}{(2\pi)^d}\frac{\pi^\mu 2^{2\mu-4}\Gamma(\mu-2)}{\Gamma(2)}\frac{C_\psi}{k^{4-2\mu+2\Delta}(p-k)^2}.
\end{align}
and evaluated using the result \eqref{eq:Jintegralresult} for $\alpha=1$ and $\beta=\mu-2+\Delta$. It is divergent in the limit $\delta\to0$,
\begin{equation}
\incl{bubblemix-new}+\incl[-2.5pt]{lineCT}=-\frac{(\mu-2)\Gamma(2\mu-1)}{\Gamma(1-\mu)\Gamma(\mu)^2\Gamma(\mu+1)}\frac{p^2M^{2\delta}}{N\delta}+c_Zp^2+O(\Delta^0),
\end{equation}
and the divergence can be cancelled by the counterterm to get a finite result as $\delta\to0$. Then, from $Z=1+c_Z$ it follows that
\begin{equation}
\gamma_\varphi=\lim_{\delta\to0}M\frac{\de}{\de M} \sqrt Z=\frac{(\mu-2)\Gamma(2\mu-1)}{\Gamma(1-\mu)\Gamma(\mu)^2\Gamma(\mu+1)N},
\end{equation}
which is finite, and agrees with \eqref{eq:deltaNphi} for the order $1/N$ anomalous dimension of $\varphi$.

Other anomalous dimensions, even at order $1/N$, are more difficult to find and require the evaluation of several diagrams. For instance, to find $\gamma_\sigma $ one needs to evaluate
\begin{equation}
\incl[-5pt]{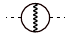}+\incl[-5pt]{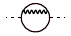}+\incl[-5pt]{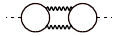}+\text{counterterms}=\text{finite}.
\end{equation}

\subsection{CFT methods}

While the diagrammatic methods have a long history, and remain very valuable for finding perturbative data, more recently some alternative methods have been derived which make direct use of conformal symmetry. These methods, sometimes unified under the term ``analytic bootstrap'', will, when applied in the perturbative regimes, give results that agree with those computed by the diagrammatic methods. However, they may provide new insights into the implications of the constraints from conformal symmetry, and in some cases produce results to higher orders than what were available with the  diagrammatic expansions.

\subsubsection{Conformal four-point functions and the OPE}
\label{sec:fourpoint}

We begin by reviewing some expressions from conformal field theory, continuing the discussion from section~\ref{sec:ObservablesMain}.
Define the four-point function of conformal primary scalar operators by
\begin{equation}
\left\langle\O_1(x_1)\O_2(x_2)\O_3(x_3)\O_4(x_4)\right\rangle=\mathbf K_{\O_1\O_2\O_3\O_4}(x_1,x_2,x_3,x_4)\,\G_{\O_1\O_2\O_3\O_4}(u,v),
\end{equation}
where $\frac{x_{12}^2x_{34}^2}{x_{13}^2x_{24}^2}=u=z\zb$ and $\frac{x_{14}^2x_{23}^2}{x_{13}^2x_{24}^2}=v=(1-z)(1-\zb)$ are the conformal cross-ratios and 
where
\begin{equation}\label{eq:Kkinematic}
\mathbf K_{\O_1\O_2\O_3\O_4}(x_1,x_2,x_3,x_4)=\left(\frac{x_{24}^2}{x_{14}^2}\right)^{\frac{\Delta_1-\Delta_2}2}\left(\frac{x_{14}^2}{x_{13}^2}\right)^{\frac{\Delta_3-\Delta_4}2}\frac{1}{(x_{12}^2)^{\frac{\Delta_1+\Delta_2}2}(x_{34}^2)^{\frac{\Delta_3+\Delta_4}2}}.
\end{equation}

In the presence of a global symmetry with $\O_i$ transforming in the representation $R_i$, we write
\begin{equation}
\G_{\O_1\O_2\O_3\O_4}(u,v)=\sum_R \mathbf P^R_{R_1R_2;R_3R_4}\G^R_{\O_1\O_2\O_3\O_4}(u,v),
\end{equation}
where the sum over $R$ is over irreducible representations in both tensor products $R_1\otimes R_2$ and $R_3\otimes R_4$. Here $\mathbf P^R$ are called projectors and are carry the index structure of the expression, implying that the correlators $\G^R_{\O_1\O_2\O_3\O_4}(u,v)$ are scalar functions of the cross-ratios. The crossing equation, following from interchanging the roles of the operators at $x_1$ and $x_3$, takes the form
\begin{equation}
\G^R_{\O_1\O_2\O_3\O_4}(u,v)=\frac{u^{\frac{\Delta_1+\Delta_2}2}}{v^{\frac{\Delta_2+\Delta_3}2}}\sum_{R'}\mathbf M^{RR'}\G^{R'}_{\O_3\O_2\O_1\O_4}(v,u),
\end{equation}
for a crossing matrix $\mathbf M^{RR'}$ defined in appendix~\ref{app:normON}. 

The OPE of two scalar operators is a sum over primary operators transforming in traceless-symmetric Lorentz representations, where each term is the OPE coefficient times a primary operator plus a tower of descendants in a combination dictated by conformal symmetry. 
For a scalar $\O_3$ in the $\O_1\times\O_2$ OPE, the contribution takes the form
\begin{align}
\nonumber
\O_1(x)\O_2(0)&=\frac{\lambda_{\O_1\O_2\O_3}}{|x|^{\Delta_1+\Delta_2-\Delta_3}}\left(1+c_1x^\mu\de_\mu+c_2 x^\mu x^\nu\de_\mu\de_\nu+c_3 x^2\square+\ldots\right)\O_3(0)\\
&\quad+\text{other operators},
\label{eq:OPEscalar}
\end{align}
where $c_i$ are fixed by conformal symmetry. Specifically, in \eqref{eq:OPEscalar}, $c_1=\frac{\Delta_{12}+\Delta_3}{2\Delta_3}$, $c_2=\frac{(\Delta_{12}+\Delta_3)(\Delta_{12}+\Delta_3+2)}{8\Delta_3(\Delta_3+1)}$, $c_3=\frac{(\Delta_{12}+\Delta_3)(\Delta_{12}-\Delta_3)}{16\Delta_3(\Delta_3+1)(\Delta_3+1-d/2)}$, with $\Delta_{ij}=\Delta_i-\Delta_j$, which follows from a repeated action on \eqref{eq:OPEscalar} with the generator $K_\rho$ of special conformal transformations.

An application of the OPE in the four-point function of scalars leads to the conformal block decomposition,
\begin{equation}
\G^R_{\O_1\O_2\O_3\O_4}(u,v)=\sum_{\O}\lambda_{\O_1\O_2\O}\lambda_{O_3\O_4\O}G_{\Delta_{\O},\ell_\O}(u,v),
\end{equation}
where the functions $G_{\Delta_{\O},\ell_\O}(u,v)$ are the conformal blocks, which depend only on the crossratios and on the dimensions of the involved operators and the spin of the exchanged operator.
We choose conventions for the conformal blocks such that the OPE coefficients for the operators $[\O,\O]_{n,\ell}$ in the generalised free field correlator,
\begin{equation}
\label{eq:GFFcorr}
\G(u,v)=1+u^{\Delta_\O}+\left(\frac uv\right)^{\Delta_\O},
\end{equation}
take the form
\begin{align}
\lambda^2_{\O\O[\O,\O]_{n,\ell}}&=
a_{n,\ell}^{\mathrm{GFF}}[\Delta_\O]
\nonumber\\&:=\frac{(1+(-1)^\ell)(\Delta_\O+1-\mu)_n^2(\Delta_\O)_{n+\ell}^2}{\ell !\, n!\, (\ell+\mu)_n(2\Delta_\O+n+1-2\mu)_n(2\Delta_\O+\ell+n-\mu)_n(2\Delta_\O+2n+\ell-1)_\ell}
\label{eq:GFFOPE}
,
\end{align}
where $(a)_n=\frac{\Gamma(a+n)}{\Gamma(a)}$ is the (rising) Pochhammer symbol.
These conventions correspond to writing the conformal blocks in $d=4$, computed in \cite{Dolan:2000ut}, as
\begin{equation}
\label{eq:CB}
G_{\Delta,\ell}=(-1)^{\ell}\frac{z \zb}{z-\zb}\left(k^{[\Delta_{12},\Delta_{34}]}_{\frac{\Delta+\ell}2}(z)k^{[\Delta_{12},\Delta_{34}]}_{\frac{\Delta-\ell}2-1}(\zb)-k^{[\Delta_{12},\Delta_{34}]}_{\frac{\Delta+\ell}2}(\zb)k^{[\Delta_{12},\Delta_{34}]}_{\frac{\Delta-\ell}2-1}(z)\right),
\end{equation}
with $k_{\beta}^{[\Delta_{12},\Delta_{34}]}(x)=x^\beta{_2F_1}\left(\beta-\frac{\Delta_{12}}2,\beta+\frac{\Delta_{34}}2;2\beta;x\right)$ and $\Delta_{ij}=\Delta_i-\Delta_j$. Our conventions agree with row six of table~I in \cite{Poland:2018epd}.
In addition to the closed-form expressions, in arbitrary spacetime dimension there exist an expansion in subcollinear blocks (see equation~(5.16) of \cite{Simmons-Duffin:2016wlq}), which can be used for the conformal block decomposition of the four-point functions away from even integer dimensions.

\subsubsection{Multiplet recombination}\label{sec:multipletrecombination}
In the free $\ON$ symmetric CFT, the conformal multiplets corresponding to the field $\varphi$ and to the currents $\mathcal J_{R,\ell}\sim (\varphi\de^\ell\varphi)_R$ are short, which can be seen from the equations of motion that set certain descendants to zero: $\square\varphi^i=0$, $\de_{\mu_1}\mathcal J_{R,\ell}^{\mu_1\mu_2\cdots\mu_{\ell}}=0$. In the interacting theory, the only short multiplets are the stress-tensor $T^{\mu\nu}=\mathcal J_{S,2}^{\mu\nu}$ and the global symmetry current $J^\mu=\mathcal J_{A,1}^\mu$. The other operators, $\varphi^i$ and the broken currents $\mathcal J_{R,\ell}$, form long multiplets, which are constructed by a recombination of the corresponding short multiplet and another long multiplet, which goes in at a descendant level. This implies that compared to the free theory, the spectrum of conformal primaries of the critical $\ON$ CFT in the $\eps$-expansion does not contain certain multiplets, specifically those corresponding to the operator $\varphi^3_V$ and to one operator of the form $\de^{\ell-1}\varphi^4_R$ for each $\ell=4,6,\ldots$ ($S$), $\ell=2,4,\ldots$ ($T$) and $\ell=3,5,\ldots$ ($A$). 

In \cite{Rychkov:2015naa}, the multiplet recombination effect was used to determine some leading anomalous dimensions in the $\eps$-expansion of the Ising and $\ON$ CFTs. The method has been extended to spinning operators \cite{Skvortsov:2015pea,Giombi:2016hkj,Roumpedakis:2016qcg}, to the cubic model in $6-\epsilon$ dimensions \cite{Nii:2016lpa} and the large $N$ expansion \cite{Giombi:2017rhm}.\footnote{In a similar spirit is the work \cite{Liendo:2017wsn}, which assumes that the one-loop dilatation acts like a two-site Hamiltonian, in agreement with section~\ref{sec:orderepssystematics}, and fixes the normalisation using the multiplet recombination idea.}
In this section we outline the computations of \cite{Rychkov:2015naa}, and a more detailed presentation can be found there.
We denote $\phi_0^k$ operators in the free theory, and $\phi^k$ operators in the Wilson--Fisher fixed-point. 

The main assumption of \cite{Rychkov:2015naa} is that $\phi^3$ is a descendant of $\phi$, such that
\begin{equation}
\label{eq:multipletrecombinationEOM}
\square\phi=\alpha\phi^3,
\end{equation}
for some constant $\alpha$ that will be determined later. We will take a little shortcut, and assume from the beginning that $\phi$ gets no anomalous dimension at order $\eps$, so that we take as an ansatz
\begin{align}
\Delta_\phi=\frac{d-2}2+\eps^2\delta+O(\eps^3)\label{eq:deltaphiformRecomb}
,\qquad
\Delta_{\phi^k}=k\frac{d-2}2+\eps\gamma_k+O(\eps^2),
\end{align}
with $\gamma_1=0$. 

Normalising the two-point function of $\phi$ as
\begin{equation}
\langle\phi(x)\phi(0)\rangle=\frac1{|x|^{2\Delta_\phi}},
\end{equation}
a direct computation shows that
\begin{equation}
\langle\square_x\phi(x)\square_y\phi(y)\rangle|_{y=0}=\frac{32\delta}{|x|^{2\Delta_\phi+4}}.
\end{equation}
To leading order, we impose that this agrees with the computation from the free theory using the assumption \eqref{eq:multipletrecombinationEOM}. The free theory two-point function of $\alpha\phi_0(x)^3$ evaluates to
\begin{equation}
\alpha^2\langle\phi_0(x)^3\phi_0(0)^3\rangle=\alpha^2\frac{6}{|x|^6},
\end{equation}
and we find
\begin{equation}\label{eq:deltasolRecomb}
\delta=\frac{3\alpha^2}{16}.
\end{equation}

Next, consider the OPE of operators of consecutive powers of $\phi$. In the free theory, Wick contractions give
\begin{equation}\label{eq:freeOPEkkp1}
\phi_0(x)^k \phi_0(0)^{k+1}=\frac{\Gamma(k)}{|x|^{2k}}\left(\phi_0(0)+\frac k2x^2\phi_0(0)^3+\ldots\right).
\end{equation}
In the interacting theory we will have by \eqref{eq:OPEscalar}\footnote{\emph{A priori} it is not clear that the equations below hold for the cases $k=2$ and $k=3$, since the descendant operator $\phi^3$ is involved. However, this turns out to be the case, for details see \cite{Rychkov:2015naa}.}
\begin{equation}
\phi^k(x)\phi^{k+1}(0)=\frac C{ |x|^{\Delta_{\phi^k}+\Delta_{\phi^{k+1}}-\Delta_\phi}}\left(
1+\ldots+c_3\square +\ldots\right)\phi(0)+\ldots \sim C|x|^{-2k}(\phi+c_3\alpha \phi^3).
\end{equation}
Matching with \eqref{eq:freeOPEkkp1} gives to leading order $C=\Gamma(k)$ and \begin{equation}
\label{eq:c3constant}
\frac k2=c_3\alpha=\frac{\gamma_{k+1}-\gamma_k}{16\delta}\alpha,
\end{equation}
where we used the expression for $c_3$ given after \eqref{eq:OPEscalar}.\footnote{Note that when using $\Delta_\phi$ in \eqref{eq:deltaphiformRecomb} for $\Delta_3$ in \eqref{eq:c3constant}, the denominator scales as $\delta\sim\eps^2$ in the limit $\eps\to0$.}
Using \eqref{eq:deltasolRecomb} for $\delta$, \eqref{eq:c3constant} becomes the recursion relation
\begin{equation}
\gamma_{k+1}-\gamma_k=\frac{3\alpha}2k,
\end{equation}
which, for $\gamma_1=0$, has the solution $\gamma_k=\frac34\alpha k(k-1)$. Finally, $\alpha$ can be fixed by assuming that $\gamma_3=1$ as required by the assumption that $\phi^3$ is a descendant of $\phi$ in the interacting theory. This gives $\alpha=\frac29\eps$, and therefore
\begin{align}
\Delta_\phi&=1-\frac\eps2+\frac{\eps^2}{108}+O(\eps^3),
\\
\Delta_{\phi^k}&=k\left[1-\frac\eps2\right]+\frac{k(k-1)}6\eps+O(\eps^2),
\end{align}
which agree with the know values. 
The latter result forms the order $\eps$ anomalous dimension of operators in the family \texttt{IsingF1[$k$]} defined below.

In the argument above, we phrased the multiplet recombination effect as $\phi^3$ becoming a descendant of $\phi$, however 
from the point-of-view of renormalisation of composite operators the picture is somewhat different. When considering the mixing of $\phi^3$ and $\square\phi$, two eigenvalues are found: $\Delta_\pm=3-\frac\eps2\pm \gamma_\varphi$. The lower eigenvalue corresponds to the equation-of-motion operator\footnote{In the renormalisation group picture, the theory contains, in addition to primaries and descendants, ``redundant operators'', of which the equation-of-motion operator is the simplest. The correlators of redundant operators vanish at non-coincident points, and their dimensions are fixed in terms of scaling dimensions of the conformal primaries \cite{Wegner:1976bn}.} and satisfies $\Delta_-+\Delta_\phi=4-\eps=d$, with the corresponding eigenvector proportional to $\square\phi-\alpha\phi^3$. The upper eigenvalue corresponds to the descendant state with $\Delta_+=\Delta_\phi+2$, and the eigenvector is proportional to $\square\phi$. This means that the renormalised descendant eigenoperator has no component along $\phi^3$.\footnote{I thank Andreas Stergiou for useful discussions.} Therefore one may say that the states in the $\phi^3$ multiplet disappear since they go into the equation-of-motion operator, and that the $\phi$ multiplet gains the corresponding set of states since its descendant $\square\phi$ is no longer identically zero.
Despite this conflicting picture, in the rest of the report we will still describe the multiplet recombination effect as $\phi^3$ becoming a descendant of $\phi$, and similar for the broken currents.

\subsubsection{Analytic bootstrap methods}\label{sec:analyticbootstrapmethods}

The methods that fall under the notion ``analytic bootstrap'' have in common that they study the constraints imposed by conformal symmetry on four-point functions of primary operators. Several of the analytic bootstrap methods rely in simplifications of the conformal data at large spin, and of the four-point function when it is similar to the generalised free field correlator~\eqref{eq:GFFcorr}. 

The spectrum of conformal primaries at large spin satisfies ``twist additivity'', which is the statement that for any two operators $\O_1$ and $\O_2$ of twists $\tau_1=\Delta_1-\ell_1$ and $\tau_2=\Delta_2-\ell_2$, there is an infinite sequence of operators $\O_\ell$ of increasing spin with twists $\tau_\ell$ approaching $\tau_1+\tau_2$. This structure was argued for already by Parisi \cite{Parisi:1973xn} and Callan \emph{et.\ al.}\ \cite{Callan:1973pu} and directly observed in the $\eps$-expansion in \cite{Kehrein:1995ia,Derkachov:1996ph}. Twist additivity was derived for a general CFT in \cite{Fitzpatrick:2012yx,Komargodski:2012ek} and follows from studying the crossing equation of the $\langle\O_1\O_2\O_2\O_1\rangle$ correlator.\footnote{See \cite{Qiao:2017xif} for a rigorous analysis of a 1d toy model for \cite{Fitzpatrick:2012yx,Komargodski:2012ek}, based on Tauberian theorems. I thank Slava Rychkov for useful discussions.} In the generic case, the existence if the identity operator in the crossed channel OPE implies the existence of operators $[\O_1,\O_2]_{n,\ell}$ of the schematic form $\O_1\square^n\de^\ell\O_2$. 
These considerations were further quantified by the Lorentzian inversion formula \cite{Caron-Huot:2017vep}, which showed that the conformal data of such operators are analytic functions in spin.

In the $\ON$ CFT, the leading twist families are the weakly broken currents $\mathcal J_{R,\ell}=[\varphi,\varphi]_{R,0,\ell}=(\varphi\de^\ell\varphi)_R$. They have scaling dimensions of the form
\begin{equation}
\Delta_{R,\ell}=2\Delta_\varphi+\ell+\gamma_{R,\ell},
\end{equation}
where $\gamma_{R,\ell}\to0$ as $\ell\to\infty$. 
In the large $N$ limit, it is possible to identify additional twist families, like $[\sigma,\sigma]_{n,\ell}$.
The considerations of \cite{Fitzpatrick:2012yx,Komargodski:2012ek} can be used to determine the conformal data for these operators \cite{Alday:2015ota,Alday:2015ewa,Alday2016b}, however, a more systematic method uses the mentioned Lorentzian inversion formula \cite{Caron-Huot:2017vep}. Applied to the $\ON$ CFT, the Lorentzian inversion formula can be used both perturbatively in $\eps$ \cite{Alday:2017zzv,Henriksson:2018myn} and $1/N$ \cite{Alday:2019clp}, and non-perturbatively for the three-dimensional theories at finite $N$ \cite{Albayrak:2019gnz,Caron-Huot:2020ouj,Liu:2020tpf}.\footnote{See also \cite{Simmons-Duffin:2016wlq} for a similar computation without the Lorentzian inversion formula.}

\paragraph{CFT-data from the Lorentzian inversion formula}

Let us briefly review how the Lorentzian inversion formula can be used to determine the order $\eps^2$ anomalous dimensions and OPE coefficients of the weakly broken currents $\mathcal J_\ell$ in the $4-\eps$ expansion. For simplicity we keep $N=1$.

Consider the four-point function $\langle\phi\phi\phi\phi\rangle\sim\G(z,\zb)$, and focus on the operators $\mathcal J_\ell=\phi\de^\ell\phi$ with dimensions $\Delta_\ell=2\Delta_\phi+\ell+\gamma_\ell$ and OPE coefficients $a_\ell=\lambda^2_{\phi\phi\mathcal J_\ell}$, using normalisations consistent with \eqref{eq:GFFOPE}. Working in perturbation theory, where $\gamma_\ell$ is small, the Lorentzian inversion formula implies that the conformal data of $\mathcal J_\ell$ can be computed using \cite{Alday:2017vkk} (see also \cite{Thesis})
\begin{equation}\label{eq:agammafromT}
(\gamma_\ell)^ka_\ell=T_\hb^{(k)}+\frac12\de_\hb T^{(k+1)}_\hb+\ldots\Big|_{\hb=\Delta_\phi+\ell},
\end{equation}
where the functions $T^{(k)}_\hb$ are computed from the double-discontinuity of the correlator using
\begin{equation}
\label{eq:T01def}
T^{(0)}_\hb+\frac12 T^{(1)}_\hb\ln z+\ldots = \kappa\int\limits_0^1\frac{d\zb}{\zb^2}k_\hb(\zb)\dDisc[\G(z,\zb)]\Big|_{z^{\Delta_\phi}},
\end{equation}
where $\kappa=\frac{\Gamma(\hb)^4}{\pi^2\Gamma(2\hb)\Gamma(2\hb-1)}$ and $k_{\hb}(x)=x^{\hb}{_2F_1}(\hb,\hb;2\hb;x)$.
Here $\dDisc[\G(z,\zb)]=\G(z,\zb)- \frac12(\G^\circlearrowleft(z,\zb)+ \G^\circlearrowright(z,\zb))$ is the difference between the correlator and its two analytic continuations in $\zb$ around the branch cut starting at $\zb=1$. 

In general, the double-discontinuity can be computed by evaluating the contribution from operators appearing in the crossed-channel OPE,
\begin{equation}
\dDisc[\G(z,\zb)]=\dDisc\left[\left(\frac{z \zb}{(1-z)(1-\zb)}\right)^{\Delta_\phi}\sum_{\O'}\lambda^2_{\phi\phi\O'}G_{\Delta_{\O'},\ell_{\O'}}(1-\zb,1-z))\right].
\end{equation}
The double-discontinuity deriving from an operator of twist $\tau'=\Delta'-\ell'$ is proportional to the factor $\sin^2(\pi\frac{\tau'-2\Delta_\phi}2)$.
To order $\eps^3$, the entire double-discontinuity can be found by considering as crossed-channel operators the identity $\O'=\1$ and the bilinear scalar $\O'=\phi^2$. This is because all operators in the $\eps$-expansion, apart from the identity, have a twist of the form $\tau'=2\Delta_\phi+2n+\gamma_{\O'}$ for integer $\ell$ and small $\gamma_{\O'}$. The contribution of a crossed-channel operator $\O'$ is therefore proportional to 
\begin{equation}
\lambda_{\phi\phi\O'}^2\sin^2\left(\pi\frac{\tau'-2\Delta_\phi}2\right)\sim\frac14\lambda_{\phi\phi\O'}^2\gamma_{\O'}^2
\end{equation}
which is $O(\eps^4)$ for all operators except $\1$ and $\phi^2$.\footnote{For operators containing more than two factors $\phi$, the OPE coefficient $\lambda^2_{\phi\phi\O'}$ is suppressed and is of $O(\eps^2)$, and the factor $\gamma^2_{\O'}$ introduces two additional factors $\eps$, giving $O(\eps^4)$. For operators with two factors $\phi$, $\lambda^2_{\phi\phi\O'}=O(1)$, but for the weakly broken currents, $\gamma_{\O'}^2=O(\eps^4)$. Left is only the operator $\phi^2$. Note that for general $N$, one needs both $\varphi^2_S$ and $\varphi^2_T$.}

For $\O'=\1$, the conformal block is just the constant $1$ to all orders. We then turn to $\phi^2$.
Denote by $g$ its anomalous dimension, which we will take as a free parameter, to be determined later on: $\Delta_{\phi^2}=2\Delta_\phi+g+O(\eps^2)$.

Working only to order $\eps^2$, we can evaluate the conformal block of $\O'=\phi^2$ using the $d=4$ conformal block \eqref{eq:CB}. One finds that
\begin{align}
\nonumber
\dDisc[\G(z,\zb)]&=\dDisc\left[\left(\frac{z \zb}{(1-z)(1-\zb)}\right)^{\Delta_\phi}+\lambda^2_{\phi\phi\phi^2}\left(\frac{z\zb}{1-z}\right)^{\Delta_\phi}\frac{g^2}8\ln^2(1-\zb)\frac{\ln \zb-\ln z}\zb
\right]\\&\quad+O(\eps^3),
\end{align}
where $\lambda^2_{\phi\phi\phi^2}=2+O(\eps)$, which follows from the OPE decomposition of the free theory correlator, c.f.\ \eqref{eq:freeOPE}.
Evaluating the double-discontinuity and integral in \eqref{eq:T01def}, for instance using the formulas in \cite{Alday:2017zzv}, one finds that
\begin{align}
T^{(0)}&=
T_\hb^{\mathrm{GFF}}[\Delta_\phi]-\frac{g^2}{\hb^2(\hb-1)^2}+O(\eps^3),
\\
T^{(1)}&=-\frac{g^2}{\hb(\hb-1)}+O(\eps^3),
\end{align}
and $T^{(k)}_\hb=O(\eps^3)$ for $k\geqslant2$, where $T^{\mathrm{GFF}}_\hb[\Delta_\phi]|_{\hb=\Delta_\phi+\ell}=a_{0,\ell}^{\mathrm{GFF}}[\Delta_\phi]$, defined in \eqref{eq:GFFOPE}.\footnote{In the Lorentzian inversion formula, the identity operator always gives a contribution that corresponds to the generalised free field OPE coefficients.} For the anomalous dimensions, \eqref{eq:agammafromT} gives
\begin{equation}\label{eq:Deltaellansg2}
\Delta_\ell=2\Delta_\phi+\ell-\frac{g^2}{\ell(\ell+1)}+O(\eps^2),
\end{equation} 
and by demanding conservation of the stress-tensor, $\Delta_2=d$, we find
\begin{equation}
\Delta_\phi=1-\frac\eps2+\frac{g^2}{12}+O(\eps^3).
\end{equation}
The leading order expression for $g$ in terms of $\eps$ can be fixed by an analytic continuation in spin to spin zero by promoting $\ell(\ell+1)\to\frac{\Delta+\ell-2}2\frac{\Delta+\ell}2$, the validity of which we will discuss shortly. This gives
\begin{equation}
\label{eq:tospinzero}
2\Delta_\phi+g=2\Delta_\phi-\frac{g^2}{\frac{2-\varepsilon+g}2\frac{-\varepsilon+g}2}+O(\varepsilon^2),
\end{equation}
from which follows $
g(3g-\varepsilon)=0$. 
With the non-trivial solution $g=\eps/3$, one finds the known results $
\Delta_{\phi^2}=2\Delta_\phi+\frac{\varepsilon}3+\ldots,
$ and $\Delta_{\phi}=\frac{d-2}2+\frac{\varepsilon^2}{108}+\ldots$. Moreover, for the family of broken currents one finds
\begin{align}
\texttt{DeltaE[IsingF2[$l$]]}=2+l-\eps+\left(\frac1{54}-\frac{1}{9l(l+1)}\right)\eps^2+\ldots+O(\eps^5).
\end{align}
Likewise, for the OPE coefficients,
\begin{equation}
\lambda_{\phi\phi\mathcal J_\ell}^2=a^{\mathrm{GFF}}_{0,\ell}[\Delta_\phi]+\frac{2\Gamma(\ell+1)^2}{\Gamma(2\ell+1)}\frac{\eps^2}{9\ell(\ell+1)}\left(\frac1{\ell+1}+S_1(2\ell)-S_1(\ell)\right)+O(\epsilon^3),
\end{equation}
where $a_\ell^{\mathrm{GFF}}[\Delta_\phi]$ is the expansion of \eqref{eq:GFFOPE} for $\Delta_\phi=1-\frac\eps2+\frac{\eps^2}{108}+\ldots$. Explicitly,
\begin{align}
\texttt{OpeE[IsingF2[$l$]]}&=\frac{2\Gamma(l+1)^2}{\Gamma(2l+1)}\bigg(
1+\eps\left(S_1(2l)-2S_1(l)\right)+\eps^2\Big(
\frac1{9l(l+1)^2}+2S_1(l)^2+
\nonumber
\\&\qquad\qquad+\frac12S_1(2l)^2+\frac12S_2(2l)-\frac34S_2(l)+\Big(\frac1{27}-\frac1{9l(l+1)}\Big)S_1(l)
\nonumber
\\&\qquad\qquad-2S_1(l)S_1(2l)+\Big(-\frac1{54}+\frac1{9l(l+1)}\Big)S_1(2l)
\Big)
\bigg)+\ldots+O(\eps^5).
\end{align}
Using the relation \eqref{eq:OPEwithTandJ}, the correction to the central charge can be computed from the value at $\ell=2$, giving \eqref{eq:CTIsing}.

\paragraph{Analytic continuation to spin zero}
In equation~\eqref{eq:tospinzero} above, we presented the analytic continuation in spin to spin zero made in \cite{Alday:2017zzv}. There is no rigorous argument that motivates such continuation, since the Lorentzian inversion formula is only guaranteed to correctly reproduce the CFT-data for spins $>1$. Instead, we may view \eqref{eq:tospinzero} as an interesting observation that an analytic continuation to spin zero correctly reproduces the order $\eps$ anomalous dimension of bilinear scalar operators, an assumption which appears to be valid for general $\phi^4$ theories \cite{Henriksson:2020fqi}. 

Other similar relations have been observed. In the large $N$ expansion it was noted, for instance in \cite{Alday:2019clp}, that an analytic continuation in spin of $\Delta_{\mathcal J_{T,\ell}}$ and $\Delta_{\mathcal J_{S,\ell}}$ gives the dimension $\Delta_{\varphi^2_T}$ and the shadow dimension $d-\Delta_\sigma$ at order $1/N$. From the explicit expressions for $\Delta_{\mathcal J_{S,\ell}}$ \cite{Manashov:2017xtt} and $\Delta_\sigma$ \cite{Vasiliev:1981dg}, the singlet ($S$) shadow relation is broken at order $1/N^2$. 
Likewise, in the $6-\epsilon$ expansion, the dimensions \eqref{eq:JScubic} of broken currents satisfy the shadow relation $\Delta_{\mathcal J_{S,0}}=d-\Delta_\sigma$; a similar relation was found for $N=0$ \cite{Goncalves:2018nlv}.

\paragraph{Mellin space bootstrap}

We also make a brief comment on the analytic bootstrap approach of Mellin space bootstrap \cite{Gopakumar:2016wkt,Gopakumar:2016cpb,Dey:2016mcs}, connected to dispersive functionals in \cite{Carmi:2020ekr}. The fundamental ingredient in the Mellin space bootstrap is the Mellin amplitudes, defined (for identical external operators $\O$) by \cite{Mack:2009mi,Penedones:2010ue}\footnote{Note that different conventions for the Mellin amplitude exists in the literature.}
\begin{equation}
\G(u,v)=1+u^{\Delta_\O}+\left(\frac uv\right)^{\Delta_\O}+\int\frac{ds}{2\pi i}\frac{dt}{2\pi i}u^sv^{t-\Delta_\O}\Gamma(\Delta_\O-s)^2\Gamma(\Delta_\O-t)^2\Gamma(\Delta_\O-\tilde u)^2\mathcal M(s,t),
\end{equation}
where $s+t+\tilde u=\Delta_\O$.
Here the generalised free field part of the correlator has not been included in the Mellin amplitude, which represents the ``connected'' part of a correlator, borrowing terminology from holographic CFT. The Mellin amplitude has poles at the location of operators exchanged in the $\O\times\O$ OPE, plus poles at integer-shifted values, with residues that can be written as Mack polynomials \cite{Mack:2009mi}, see also \cite{Costa:2012cb,Gopakumar:2016cpb}. In the Mellin space bootstrap, crossing symmetry is build in, and consistency with the OPE gives constraints on the conformal data through the cancellation of spurious poles.

The results from the Mellin space bootstrap do not require analyticity in spin to spin zero, but cancellations appearing in the computations mean that, like in the application of the Lorentzian inversion formula, the spinning operators can be determined to higher order than the operator at spin zero. There is however is an additional relation between the OPE coefficient $\lambda^2_{\phi\phi\phi^2}$ and anomalous dimension $\gamma_{\phi^2}$ at spin zero. At leading non-trivial order, this relation gives
\begin{equation}
\lambda^2_{\phi\phi\phi^2}=2\left(1-g\right)+O(\eps^2)=2\left(1-\frac\eps3\right)+O(\eps^2).
\end{equation}
Using the higher order terms in the mentioned relation facilitates the determination of $\lambda^2_{\varphi\varphi\varphi^2_S}$ and $\lambda^2_{\varphi\varphi\varphi^2_S}$ to order $\eps^3$ \cite{Dey:2016mcs} and $\lambda^2_{\phi\phi\phi^2}$ numerically to order $\eps^4$ for $N=1$ \cite{Carmi:2020ekr}, by using the anomalous dimension at the same order as input.

\section[\texorpdfstring{Conformal data for $\boldsymbol{N=1}$}{Conformal data for N=1}]{Conformal data for $\boldsymbol{N=1}$}
\label{sec:DataIsing}

This section presents the conformal data for the Ising CFT ($N=1$). We focus on perturbative data in the $\eps$-expansion, but for individual operators we also give numerical values in the three-dimensional Ising CFT computed by the numerical conformal bootstrap. The numerical values are primarily those extracted in \cite{Simmons-Duffin:2016wlq} using the extremal functional method \cite{Poland:2010wg,ElShowk2012}, and the error bars are therefore non-rigorous. 
The operators are organised by global symmetry representation, $\mathbb Z_2$ even (\texttt E) or $\mathbb Z_2$ odd (\texttt O), and by spin $\ell$.

\subsection{Families of operators}

We give a non-exhaustive list of 13 different families of operators, where the scaling dimensions, and in a few cases the OPE coefficients, are given as closed-form functions of some parameter. We limit ourselves to families containing operators in traceless-symmetric Lorentz representations. Several of these families were listed in table~6 of \cite{Kehrein:1994ff}, who also gave some families of operators in non-traceless-symmetric Lorentz representations, see section~\ref{sec:two-rowIsing}.

\paragraph{\texttt{IsingF1[$k$]}} Scalar operators of the form $\phi^k$, $k\geqslant1$, $k\neq3$. We have \eord2\cite{Derkachov:1997gc}\footnote{We took their general result for $\varphi^{2s}_S$, put $N=1$, and generalised to odd values of $k=2s$.} 
\begin{equation}
\label{eq:isingF1}
\texttt{DeltaE[IsingF1[$k$]]}=k\left[1-\tfrac \eps2\right]+\frac{k(k-1)}6\eps-\frac{k(17k^2-67k+47)}{324}\eps^2+O(\eps^3).
\end{equation}
There is no primary operator at $k=3$.\footnote{A previous version stated that the operator at $k=3$ is a descendant, however the formula \eqref{eq:isingF1} is not applicable for $k=3$.} The computation of the order $\eps$ anomalous dimension was discussed in section~\ref{sec:multipletrecombination}. 
\paragraph{\texttt{IsingF2[$l$]}} Broken currents of the form $\mathcal J_\ell$, $\ell\in2\mathbb Z_+$. We have \eord2\cite{Wilson:1973jj}\eord4\cite{Derkachov:1997pf},\footnote{
As pointed out in \cite{Carmi:2020ekr}, equation~(3.21) of \cite{Derkachov:1997pf} contains a typo, where $\frac{65}{81}$ should read $\frac{65}{96}$.}
\begin{equation}
\texttt{DeltaE[IsingF2[$l$]]}=2+l-\eps+\left(\frac{1}{54}-\frac{1}{9l(l+1)}\right)\eps^2+\ldots+O(\eps^5).
\end{equation}
The computation of the order $\eps^2$ correction was reviewed in section~\ref{sec:analyticbootstrapmethods}.
\paragraph{\texttt{IsingF3[$l$]}} Spinning operators of the form $\de^\ell \phi^3$, $\ell\geqslant2$, \eord1\cite{Kehrein:1992fn}\eord2\cite{Bertucci:2022ptt},
\begin{equation}
\texttt{DeltaE[IsingF3[$l$]]}=3+l-\frac{3\eps}2+\frac{\eps}3\left(1+\frac{2(-1)^l}{l+1}\right)+\ldots+O(\eps^3).
\end{equation}
Starting from $\ell=6$, there are additional operators of the type $\de^\ell\phi^3$, however they all have vanishing first-order anomalous dimensions. The family \texttt{IsingF3} is identified by the non-vanishing of the anomalous dimension to order $\eps$. 
\paragraph{\texttt{IsingF4[$k$]}} Spin-two operators of the form $\de^2\phi^k$, $k\geqslant2$, \eord1\cite{Kehrein:1994ff},
\begin{equation}
\texttt{DeltaE[IsingF4[$k$]]}=2+k\left[1-\tfrac \eps2\right]+\frac{(k-2)(3k+1)}{18}\eps+O(\eps^2).
\end{equation}
\paragraph{\texttt{IsingF5[$k$]}} Spin-three operators of the form $\de^3\phi^k$, $k\geqslant3$, $k\neq4$, \eord1\cite{Kehrein:1994ff},
\begin{equation}
\texttt{DeltaE[IsingF5[$k$]]}=3+k\left[1-\tfrac \eps2\right]+\frac{k^2-2k-2}{6}\eps+O(\eps^2).
\end{equation}
\paragraph{\texttt{IsingF6[$k$]}} Spin-four operators of the form $\de^4\phi^k$, $k\geqslant2$, \eord1\cite{Kehrein:1994ff},
\begin{equation}
\texttt{DeltaE[IsingF6[$k$]]}=4+k\left[1-\tfrac \eps2\right]+\frac{(k-2)(5k-1)}{30}\eps+O(\eps^2).
\end{equation}
\paragraph{\texttt{IsingF7[$k$]}} Spin-four operators of the form $\de^4\phi^k$, $k\geqslant4$, \eord1\cite{Kehrein:1994ff},
\begin{equation}
\texttt{DeltaE[IsingF7[$k$]]}=4+k\left[1-\tfrac \eps2\right]+\frac{3k^2-7k-12}{18}\eps+O(\eps^2).
\end{equation}
\paragraph{\texttt{IsingF8[$k$]}} Spin-five operators of the form $\de^5\phi^k$, $k\geqslant3$, $k\neq4$, \eord1\cite{Kehrein:1994ff},
\begin{equation}
\texttt{DeltaE[IsingF8[$k$]]}=5+k\left[1-\tfrac \eps2\right]+\frac{3k^2-7k-2}{18}\eps+O(\eps^2).
\end{equation}
\paragraph{\texttt{IsingF9[$k$]}} Spin-five operators of the form $\de^5\phi^k$, $k\geqslant5$, \eord1\cite{Kehrein:1994ff},
\begin{equation}
\texttt{DeltaE[IsingF9[$k$]]}=5+k\left[1-\tfrac \eps2\right]+\frac{3k^2-8k-20}{18}\eps+O(\eps^2).
\end{equation}
\paragraph{\texttt{IsingF10[$k$]}} Scalar operators of the form $\square^2\phi^k$, $k\geqslant4$, \eord1\cite{Kehrein:1992fn},
\begin{equation}
\texttt{DeltaE[IsingF10[$k$]]}=4+k\left[1-\tfrac \eps2\right]+\frac{k(3k-7)}{18}\eps +O(\eps^2).\end{equation}

\paragraph{\texttt{IsingF11[$k$]}} Spin-one operators of the form $\de\square^2\phi^k$, $k\geqslant5$, \eord1\cite{Kehrein:1994ff},
\begin{equation}
\texttt{DeltaE[IsingF11[$k$]]}=5+k\left[1-\tfrac \eps2\right]+\frac{3k^2-8k-4}{18}\eps +O(\eps^2)\end{equation}
\paragraph{\texttt{IsingF12[$k$]}} Spin-two operators of the form $\de^2\square\phi^k$, $k\geqslant4$, \eord1\cite{Kehrein:1994ff},
\begin{equation}
\texttt{DeltaE[IsingF12[$k$]]}=4+k\left[1-\tfrac \eps2\right]+\frac{3k^2-7k-4}{18}\eps +O(\eps^2).\end{equation}
\paragraph{\texttt{IsingF13[$k$]}} Spin-three operators of the form $\de^3\square\phi^k$, $k\geqslant5$, \eord1\cite{Kehrein:1994ff},
\begin{equation}
\texttt{DeltaE[IsingF13[$k$]]}=5+k\left[1-\tfrac \eps2\right]+\frac{3k^2-8k-10}{18}\eps +O(\eps^2).
\end{equation}

\subsection{Scalar operators}

In table~\ref{tab:evenscalars} we present the spectrum of $\mathbb Z_2$ even scalar operators for $\Delta^{\mathrm{4d}}\leqslant 12$. Likewise, in table~\ref{tab:Z2oddscalars} we present the spectrum of $\mathbb Z_2$ odd scalar operators of $\Delta^{\mathrm{4d}}\leqslant 11$.

\begin{table}[ht]
\centering
\caption{$\mathbb Z_2$ even scalar operators for $N=1$. The table includes operators with $\Delta^{\mathrm{4d}}\leqslant 12$. The fact that there are two operators with anomalous dimension $\frac83$ of the form $\square^3\phi^6$ is not a typo, for instance both operators are reported in table~5 of \cite{Kehrein:1994ff}.}\label{tab:evenscalars}
{\small
\renewcommand{\arraystretch}{1.25}
\begin{tabular}{ccllll}

$\O$ &$\O|\eps$ & $\Delta^{(1)}_{4-\eps}$ & $\Delta(\eps)$ & $\Delta_{3\mathrm d}$ & Family
\\\hline
\texttt{Op[E,0,0]} & $\1$ & $0$ & exact & $0$ &
\\
\texttt{Op[E,0,1]} & $\phi^2$ & $[2-\eps]+\frac{1}{3}\eps$ & 
\eord7\cite{Schnetz:2016fhy}
&
$1.412625(10)$\cite{Kos:2016ysd} &  $\texttt 1_2$ 
\\
\texttt{Op[E,0,2]} & $\phi^4$ & $[4-2\eps]+2\eps$ & 
\eord7\cite{Schnetz:2016fhy}
& $3.82951(61)$\cite{Reehorst:2021hmp}
& $\texttt 1_4$ 
\\
\texttt{Op[E,0,3]} & $\phi^6$ & $[6-3\eps]+5\eps$ 
& \eord2\cite{Derkachov:1997gc}  &$6.8956(43)$\cite{Simmons-Duffin:2016wlq}& $\texttt 1_6$
\\
\texttt{Op[E,0,4]} & $\square^2\phi^4$ & $[8-2\eps]+\frac{10}{9}\eps$  
& \eord1
&$7.2535(51)$\cite{Simmons-Duffin:2016wlq}&  $\texttt{10}_4$
\\
\texttt{Op[E,0,5]} & $\phi^8$ & $[8-4\eps]+\frac{28}{3}\eps$ 
& \eord2\cite{Derkachov:1997gc}  &&  $\texttt 1_8$
\\
\texttt{Op[E,0,6]} & $\square^3\phi^4$ & $[10-2\eps]+\frac{1}{3}\eps$  
& \eord1
&&  
\\
\texttt{Op[E,0,7]} & $\square^2\phi^6$ & $[10-3\eps]+\frac{11}{3}\eps$ 
& \eord1&& $\texttt{10}_6$
\\
\texttt{Op[E,0,8]} & $\phi^{10}$ & $[10-5\eps]+15\eps$ 
& \eord2\cite{Derkachov:1997gc}    &&  $\texttt 1_{10}$
\\
\texttt{Op[E,0,9]} & $\square^4\phi^4$ & $[12-2\eps]+\frac{14}{15}\eps$ 
& \eord1&& 
\\
\texttt{Op[E,0,10]} & $\square^3\phi^6$ & $[12-3\eps]+\frac83\eps$ 
& \eord1&& 
\\
\texttt{Op[E,0,11]} & $\square^3\phi^6$ & $[12-3\eps]+\frac83\eps$ 
& \eord1&& 
\\
\texttt{Op[E,0,12]} & $\square^2\phi^8$ & $[12-4\eps]+\frac{68}9\eps$ 
& \eord1&&  $\texttt{10}_8$
\\
\texttt{Op[E,0,13]} & $\phi^{12}$ & $[12-6\eps]+22\eps$  &\eord2\cite{Derkachov:1997gc} & 
& $\texttt 1_{12}$
\\\hline
\end{tabular}
}
\end{table}

We remark that in some places in the literature, an additional scalar operator is reported with dimension (for $d=3$) near $4.67$ \cite{Newman1984,ElShowk2012,Gliozzi:2014jsa}, which does not match the spectrum of \cite{Simmons-Duffin:2016wlq} or the expectations from the $\eps$-expansion. \cite{Hasenbusch:2021tei} notes that it is likely that ``$\omega' = 1.67(11)$ is an artifact of the scaling field method.'' It is interesting to note that the operator \texttt{Op[S,0,3]}, when setting $N=\eps=1$, gives a prediction near this value, however this operator does not exist for $N=1$.

\begin{table}[ht]
\centering
\caption{$\mathbb Z_2$ odd scalar operators for $N=1$. The table includes operators with $\Delta^{\mathrm{4d}}\leqslant11$.}\label{tab:Z2oddscalars}
{\small
\setcounter{localfn}{1} 
\renewcommand{\arraystretch}{1.25}
\begin{tabular}{ccllll}

$\O$ &$\O|\eps$ & $\Delta^{(1)}_{4-\eps}$ & $\Delta(\eps)$ & $\Delta_{3\mathrm d}$ & Family
\\\hline
\texttt{Op[O,0,1]} & $\phi$ & $[1-\frac12\eps]+0\eps$ & 
\eord8\cite{SchnetzUnp}  &  $0.5181489(10)$\cite{Kos:2016ysd} &   $\texttt 1_1$
\\
\texttt{Op[O,0,2]} & $\phi^5$ & $[5-\frac52\eps]+\frac{10}3\eps$ & \eord3\cite{Zhang1982}\makefn &  
$5.2906(11)$\cite{Simmons-Duffin:2016wlq}\makefn &   $\texttt 1_5$
\\
\texttt{Op[O,0,3]} & $\phi^7$ & $[7-\frac72\eps]+7\eps$ & \eord2\cite{Derkachov:1997gc} && $\texttt 1_7$
\\
\texttt{Op[O,0,4]} & $\square^2\phi^5$ & $[9-\frac52\eps]+\frac{20}9\eps$ & \eord1 && $\texttt{10}_5$
\\
\texttt{Op[O,0,5]} & $\phi^9$ & $[9-\frac95\eps]+12\eps$ & \eord2\cite{Derkachov:1997gc} &&  $\texttt 1_9$
\\
\texttt{Op[O,0,6]} & $\square^3\phi^5$ & $[11-\frac52\eps]+\frac43\eps$ & \eord1 && 
\\
\texttt{Op[O,0,7]} & $\square^2\phi^7$ & $[11-\frac72\eps]+\frac{49}9$ & \eord1 && $\texttt{10}_7$
\\
\texttt{Op[O,0,8]} & $\phi^{11}$ & $[11-\frac{11}2\eps]+\frac{55}3\eps$ & \eord2\cite{Derkachov:1997gc} &&  $\texttt 1_{11}$
\\\hline
\end{tabular}
\flushleft
\setcounter{localfn}{1} 

\makefn I thank Hugh Osborn for useful discussions regarding this value.
\\
\makefn A value with rigorous error bars, $5.262(89)$, was given in \cite{Reehorst:2021hmp}.
\\
}
\end{table}

\subsection{Spinning operators}
\label{sec:spinningIsing}

For conformal primary operators of non-zero spin, consider first the $\mathbb Z_2$ even case, where we present operators of even and odd spin separately. For even spin, the operators are presented in table~\ref{tab:evenspintwoising} for spin $2$, in table~\ref{tab:evenspinfourising} for spin $4$ and in table~\ref{tab:evenHSising} for spin $6,8,\ldots,16$. The leading operator is always the weakly broken current $\mathcal J_\ell$. For spins up to $8$ we also include subleading operators. In table~\ref{tab:evenOpoddspinIsing} we present $\mathbb Z_2$ even operators of odd spin, including spin up to $9$. Altogether, all $\mathbb Z_2$ even operators of $\Delta^{\mathrm{4d}}\leqslant10$ are included in the tables~\ref{tab:evenspintwoising}--\ref{tab:evenOpoddspinIsing}.

\begin{table}[ht]
\centering
\caption{$\mathbb Z_2$ even operators of spin two, for $N=1$. The table includes operators with $\Delta_{4d}\leqslant 10$.}\label{tab:evenspintwoising}
{\small
`\setcounter{localfn}{1} 
\renewcommand{\arraystretch}{1.25}
\begin{tabular}{ccllll}

$\O$ &$\O|\eps$ & $\Delta^{(1)}_{4-\eps}$ & $\Delta(\eps)$ & $\Delta_{3\mathrm d}$ & Family
\\\hline
\texttt{Op[E,2,1]} & $T^{\mu\nu}$ & $[4-\eps]$ & exact & $3$&  $\texttt 2_2$, $\texttt 4_2$
\\
\texttt{Op[E,2,2]} & $\de^2\phi^4$ & $[6-2\eps]+\frac{13}{9}\eps$ &  \eord1
&$5.50915(44)$\cite{Simmons-Duffin:2016wlq}\makefn &  $\texttt 4_4$
\\
\texttt{Op[E,2,3]} & $\de^2\square\phi^4$ & $[8-2\eps]+\frac{8}{9}\eps$ &   \eord1
&$7.0758(58)$\cite{Simmons-Duffin:2016wlq}& $\texttt{12}_4$
\\
\texttt{Op[E,2,4]} & $\de^2\phi^6$ & $[8-3\eps]+\frac{38}{9}\eps$ & \eord1
&& $\texttt 4_6$
\\
\texttt{Op[E,2,5]} & $\de^2\square^2\phi^4$ & $[10-2\eps]+\frac19\eps$ & \eord1
&&
\\
\texttt{Op[E,2,6]} & $\de^2\square^2\phi^4$ & $[10-2\eps]+\frac{14}{15}\eps$ & \eord1
&&
\\
\texttt{Op[E,2,7]} & $\de^2\square\phi^6$ & $[10-3\eps]+\frac{31}{9}\eps$ & \eord1
&&  $\texttt{12}_6$
\\
\texttt{Op[E,2,8]} & $\de^2\phi^8$ & $[10-4\eps]+\frac{25}{3}\eps$ & \eord1
&& $\texttt 4_8$
\\
\hline
\end{tabular}
\flushleft
\setcounter{localfn}{1} 

\makefn A value with rigorous error bars, $5.499(17)$, was given in \cite{Reehorst:2021hmp}.
\\
}
\end{table}

\begin{table}[ht]
\centering
\caption{$\mathbb Z_2$ even operators of spin four, for $N=1$. The table includes operators with $\Delta^{\mathrm{4d}}\leqslant 10$.}\label{tab:evenspinfourising}
{\small
\renewcommand{\arraystretch}{1.25}
\begin{tabular}{ccllll}

$\O$ &$\O|\eps$ & $\Delta^{(1)}_{4-\eps}$ & $\Delta(\eps)$ & $\Delta_{3\mathrm d}$ & Family
\\\hline
\texttt{Op[E,4,1]} & $C^{\mu\nu\rho\sigma}$ & $[6-\eps]+0\eps$ & \eord4\cite{Derkachov:1997pf} & $5.022665(28)$\cite{Simmons-Duffin:2016wlq} &  $\texttt 2_4$, $\texttt 6_2$
\\
\texttt{Op[E,4,2]} & $\de^4\phi^4$ & $[8-2\eps]+\frac{4}{9}\eps$ &\eord1  & $6.42065(64)$\cite{Simmons-Duffin:2016wlq} &  $\texttt 7_4$
\\
\texttt{Op[E,4,3]} & $\de^4\phi^4$ & $[8-2\eps]+\frac{19}{15}\eps$ & \eord1 & $7.38568(28)$\cite{Simmons-Duffin:2016wlq} &  $\texttt 6_4$
\\
\texttt{Op[E,4,4]} & $\de^4\square\phi^4$ & $[10-2\eps]+\tfrac{49-\sqrt{697}}{90}\eps$ &  \eord1& $8.9410(99)$\cite{Simmons-Duffin:2016wlq} &  
\\
\texttt{Op[E,4,5]} & $\de^4\square\phi^4$ & $[10-2\eps]+\tfrac{49+\sqrt{697}}{90}\eps$ & \eord1 &&  
\\
\texttt{Op[E,4,6]} & $\de^4\phi^6$ & $[10-3\eps]+\frac{58}{15}\eps$ &\eord1  &&  $\texttt 6_6$ 
\\
\texttt{Op[E,4,7]} & $\de^4\phi^6$ & $[10-3\eps]+3\eps$ &\eord1  &&  $\texttt 7_6$
\\
\hline
\end{tabular}
}
\end{table}

\begin{table}[ht]
\centering
\caption{$\mathbb Z_2$ even operators for $N=1$ of leading twist for even spin. The identification with numerical data can only be done for the smallest operator at each spin. For spin $6$ and $8$ we also include operators of subleading twist $\tau^{\mathrm{4d}}=4$.}\label{tab:evenHSising}
{\small
\renewcommand{\arraystretch}{1.25}
\begin{tabular}{ccllll}

$\O$ &$\O|\eps$ & $\Delta^{(1)}_{4-\eps}$ & $\Delta(\eps)$ & $\Delta_{3\mathrm d}$ & Family
\\\hline
\texttt{Op[E,6,1]} & $\mathcal J_6$ & $[8-\eps]+0\eps$ & \eord4\cite{Derkachov:1997pf} & $7.028488(16)$\cite{Simmons-Duffin:2016wlq} & $\texttt 2_7$
\\
\texttt{Op[E,6,2]} & $\de^6\phi^4$ & $[10-2\eps]+0.4966\eps$ & \eord1 & &  
\\
\texttt{Op[E,6,3]} & $\de^6\phi^4$ & $[10-2\eps]+0.6707\eps$ & \eord1 & &  
\\
\texttt{Op[E,6,4]} & $\de^6\phi^4$ & $[10-2\eps]+1.1787\eps$ &\eord1  & &  
\\
\hline
\texttt{Op[E,8,1]} & $\mathcal J_8$ & $[10-\eps]+0\eps$ & \eord4\cite{Derkachov:1997pf} & $9.031023(30)$\cite{Simmons-Duffin:2016wlq} & $\texttt 2_8$
\\
\texttt{Op[E,8,2]} & $\de^8\phi^4$ & $[12-2\eps]+0.1142\eps$ &  \eord1& &  
\\
\texttt{Op[E,8,3]} & $\de^8\phi^4$ & $[12-2\eps]+0.5217\eps$ &\eord1  & &  
\\
\texttt{Op[E,8,4]} & $\de^8\phi^4$ & $[12-2\eps]+0.6752\eps$ & \eord1 & &  
\\
\texttt{Op[E,8,5]} & $\de^8\phi^4$ & $[12-2\eps]+1.1276\eps$ &  \eord1& &  
\\
\hline
\texttt{Op[E,10,1]} & $\mathcal J_{10}$ & $[12-\eps]+0\eps$ & \eord4\cite{Derkachov:1997pf} & $11.0324141(99)$\cite{Simmons-Duffin:2016wlq} & $\texttt 2_{10}$
\\
\hline
\texttt{Op[E,12,1]} & $\mathcal J_{12}$ & $[14-\eps]+0\eps$ & \eord4\cite{Derkachov:1997pf} & $13.033286(12)$\cite{Simmons-Duffin:2016wlq} & $\texttt 2_{12}$
\\
\hline
\texttt{Op[E,14,1]} & $\mathcal J_{14}$ & $[16-\eps]+0\eps$ & \eord4\cite{Derkachov:1997pf} &   $15.033838(15)$\cite{Simmons-Duffin:2016wlq} & $\texttt 2_{14}$
\\
\hline
\texttt{Op[E,16,1]} & $\mathcal J_{16}$ & $[18-\eps]+0\eps$ & \eord4\cite{Derkachov:1997pf} & $17.034258(34)$\cite{Simmons-Duffin:2016wlq} & $\texttt 2_{16}$
\\
\end{tabular}
}
\end{table}

\begin{table}[ht]
\centering
\caption{$\mathbb Z_2$ even operators for $N=1$ of leading twist for odd spin.}\label{tab:evenOpoddspinIsing}
{\small
\renewcommand{\arraystretch}{1.25}
\begin{tabular}{ccllll}

$\O$ &$\O|\eps$ & $\Delta^{(1)}_{4-\eps}$ & $\Delta(\eps)$ & $\Delta_{3\mathrm d}$ & Family
\\\hline
\texttt{Op[E,1,1]} &  $\de\square^2\phi^6$ & $[11-3\eps]+\frac{28}9\eps $  & \eord1 & &  $\texttt{11}_6$
\\\hline
\texttt{Op[E,3,1]} & $\de^3\phi^6$ & $[9-3\eps]+\frac{11}3\eps$ &\eord1  & & $\texttt 5_6$
\\\hline
\texttt{Op[E,5,1]} & $\de^5\square\phi^4$  & $[11-2\eps]+\frac{29}{45}\eps$  & \eord1& 
\\
\texttt{Op[E,5,2]} &  $\de^5\phi^6$ & $[11-3\eps]+\frac{20}9\eps$ & \eord1   & &  $\texttt 9_6$ 
\\
\texttt{Op[E,5,3]} &  $\de^5\phi^6$ & $[11-3\eps]+\frac{32}9\eps$ & \eord1   & &  $\texttt 8_6$ 
\\\hline
\texttt{Op[E,7,1]} &    $\de^7\phi^4$ & $[11-2\eps]+\frac13\eps$     &   \eord1 &  &
\\\hline
\texttt{Op[E,9,1]} & $\de^9\phi^4$ & $[13-2\eps]+\frac13\eps$   &   \eord1 &  &
\\
\texttt{Op[E,9,2]} &$\de^9\phi^4$ & $[13-2\eps]+\frac59\eps$ & \eord1  &  &
\\\hline
\end{tabular}
}
\end{table}

Global symmetry singlet (i.e.\ $\mathbb Z_2$ even for $N=1$) operators of spin $\ell=1$ are particularly interesting, since they may be virial current candidates. The existence of a virial current, i.e.\ a conserved spin-$1$ singlet operator with $\Delta=d-1$, is a necessary condition for a theory that is scale invariant but not conformally invariant \cite{Polchinski:1987dy,Nakayama:2013is}. From table~\ref{tab:evenOpoddspinIsing}, we note that the leading candidate for the virial current is of the form $\de\square^2\phi^6$ with dimension $\Delta=11+\frac\eps9+O(\eps^2)$,\footnote{This value was given already in table~4 of \cite{Kehrein:1994ff}.} far above the value needed to be a virial current. 
A conformal bootstrap study including two different $\mathbb Z_2$ even operators (or two different $\mathbb Z_2$ odd operators) as external operators could be used to find a numerical lower bound in three dimensions. Other papers discussing virial current candidates are \cite{Paulos:2015jfa,DePolsi:2018vxc,Meneses:2018xpu,Delamotte:2018fnz,DePolsi:2019owi}. 

Spinning $\mathbb Z_2$ odd operators are presented in table~\ref{tab:Z2oddspinning}. This table contains all such operators with $\Delta^{\mathrm{4d}}\leqslant10$. Note that for operators with three factors $\phi$, there is for each spin $\ell\geqslant2$ only one operator with a non-vanishing anomalous dimension at order $\eps$, agreeing with \texttt{IsingF3[$\ell$]}. The identification with the numerical values of \cite{Simmons-Duffin:2016wlq} has only been done when there is no degeneracy.

\begin{table}[ht]
\centering
\caption{A selection of spinning $\mathbb Z_2$ odd operators for $N=1$. The table includes all such operators with $\Delta^{\mathrm{4d}}\leqslant10$.}\label{tab:Z2oddspinning}
{\small
\renewcommand{\arraystretch}{1.25}
\begin{tabular}{ccllll}
$\O$ &$\O|\eps$ & $\Delta^{(1)}_{4-\eps}$ & $\Delta(\eps)$ & $\Delta_{3\mathrm d}$ & Family
\\\hline
\texttt{Op[O,1,1]} &  $\de\square^2\phi^5$  &  $[10-\frac52\eps]+\frac{31}{18}\eps$ & \eord1 & & $\texttt{11}_5$
\\\hline
\texttt{Op[O,2,1]} & $\de^2\phi^3$ & $[5-\frac32\eps]+\frac59\eps$ & \eord2\cite{Bertucci:2022ptt} & $4.180305(18)$\cite{Simmons-Duffin:2016wlq}  & $\texttt 3_2$, $\texttt 4_3$
\\
\texttt{Op[O,2,2]} & $\de^2\phi^5$ & $[7-\frac52\eps]+\frac83\eps$ & \eord1 & $6.9873(53)$\cite{Simmons-Duffin:2016wlq}  &  $\texttt 4_5$
\\
\texttt{Op[O,2,3]} & $\de^2\square\phi^5$ & $[9-\frac52\eps]+2\eps$ & \eord1 &   &  $\texttt {12}_5$
\\
\texttt{Op[O,2,4]} & $\de^2\phi^7$ & $[9-\frac72\eps]+\frac{55}9\eps$ & \eord1 &  &  $\texttt 4_7$
\\\hline
\texttt{Op[O,3,1]} & $\de^3\phi^3$& $[6-\frac32\eps]+\frac16\eps$ &\eord2\cite{Bertucci:2022ptt}  &  $4.63804(88)$\cite{Simmons-Duffin:2016wlq} &  $\texttt 3_3$, $\texttt 5_3$
\\
\texttt{Op[O,3,2]} & $\de^3\phi^5$ & $[8-\frac52\eps]+\frac{13}6\eps$ &  \eord1&  &   $\texttt 5_5$
\\
\texttt{Op[O,3,3]} & $\de^3\square\phi^5$ & $[10-\frac52\eps]+\frac{25}{18}\eps$ &  \eord1&  &   $\texttt{13}_5$
\\
\texttt{Op[O,3,4]} & $\de^3\phi^7$ & $[10-\frac72\eps]+\frac{11}2\eps$ &  \eord1&  &   $\texttt 5_7$
\\\hline
\texttt{Op[O,4,1]} & $\de^4\phi^3$& $[7-\frac32\eps]+\frac7{15}\eps$ &\eord2\cite{Bertucci:2022ptt}&  $6.112674(19)$\cite{Simmons-Duffin:2016wlq}&  $\texttt 3_4$, $\texttt 6_3$
\\
\texttt{Op[O,4,2]} & $\de^4\phi^5$& $[9-\frac52\eps]+\frac{14}9\eps$ &\eord1 &  &  $\texttt 7_5$
\\
\texttt{Op[O,4,3]} & $\de^4\phi^5$& $[9-\frac52\eps]+\frac{12}5\eps$ &\eord1 &  &   $\texttt 6_5$
\\\hline
\texttt{Op[O,5,1]} & $\de^5\phi^3$& $[8-\frac32\eps]+\frac29\eps$ &\eord2\cite{Bertucci:2022ptt} & $6.709778(27)$\cite{Simmons-Duffin:2016wlq}  &  $\texttt 3_5$, $\texttt 8_3$
\\
\texttt{Op[O,5,2]} & $\de^5\phi^5$& $[10-\frac52\eps]+\frac56\eps$ &\eord1 &  &  $\texttt 9_5$
\\
\texttt{Op[O,5,3]} & $\de^5\phi^5$& $[10-\frac52\eps]+\frac{19}9\eps$ &\eord1 &  &  $\texttt 8_5$
\\\hline
\texttt{Op[O,6,1]} & $\de^6\phi^3$& $[9-\frac32\eps]+0\eps$ & \eord1  &  &   
\\
\texttt{Op[O,6,2]} & $\de^6\phi^3$& $[9-\frac32\eps]+\frac37\eps$ & \eord2\cite{Bertucci:2022ptt} &
 &  $\texttt 3_6$
\\\hline
\texttt{Op[O,7,1]} &$\de^7\phi^3$ & $[10-\frac32\eps]+\frac14\eps$ & \eord2\cite{Bertucci:2022ptt} & $8.747293(56)$\cite{Simmons-Duffin:2016wlq}
  &  $\texttt 3_7$
\\\hline
\texttt{Op[O,8,1]} & $\de^8\phi^3$& $[11-\frac32\eps]+0\eps$ &\eord1  &   &  
\\
\texttt{Op[O,8,2]} & $\de^8\phi^3$& $[11-\frac32\eps]+\frac{11}{27}\eps$ & \eord2\cite{Bertucci:2022ptt}  &
&  $\texttt 3_8$
\\\hline
\texttt{Op[O,9,1]} & $\de^9\phi^3$& $[12-\frac32\eps]+0\eps$ &  \eord1&  &   
\\
\texttt{Op[O,9,2]} &$\de^9\phi^3$ & $[12-\frac32\eps]+\frac4{15}\eps$ & \eord2\cite{Bertucci:2022ptt} &
 &  $\texttt 3_9$
\\\hline
\end{tabular}
}
\end{table}

\subsection{Operators in non-traceless-symmetric Lorentz representations}
\label{sec:non-TS-Ising}

As explained in section~\ref{sec:reprSOd}, the representation theory of the Lorentz symmetry (rotation) group $\mathrm{SO}(d)$ depends on the value of $d$, but when taking parity into account it can be continued to generic $d$. The combined Lorentz and parity representations are labelled by Young tableaux, and here we will discuss results for operators in two-row and three-row Young tableaux. Representations with more than three rows contain operators with rather high scaling dimensions, and vanish identically in the physically relevant case $d=3$.

\subsubsection{Two-row Lorentz Young tableaux}
\label{sec:two-rowIsing}

In table~\ref{tab:IsingNonTS} we present all operators in two-row Young tableaux $y_{r_1,r_2}$ in the Ising CFT with $\Delta^{\mathrm{4d}}\leqslant10$. 
In three dimensions, the representations with $r_2>1$ are equivalent to representations with $r_2=0$ or $1$, and parity $(-1)^{r_2}$. The representations $y_{r_1,1}$ become in three dimensions parity-odd spin $\ell=r_1$ representations. 

Reference \cite{Kehrein:1994ff} also gives some families for operators in two-row Young tableaux. The simplest case is the family of operators with $k$ fields and ${\tiny \yng(2,2)}$ Lorentz representation, for $k\geqslant3$. For this family, $\Delta=k[1-\frac\eps2]+4+ \frac13(1/2k(k-1)-2/3k-1)\eps+O(\eps^2)$.

\begin{table}
\centering
\caption{Lorentz non-traceless-symmetric operators with $\Delta^{\mathrm{4d}}\leqslant 10$. The 4d notation for the representations is the one used in \cite{Kehrein:1994ff}.}\label{tab:IsingNonTS}
{\small
\renewcommand{\arraystretch}{1.25}
\begin{tabular}{|cccc|} 
\hline
Lorentz YT & 4d representation  & Field content & $\Delta^{(1)}_{4-\eps}$   \\
\hline
 \raisebox{-3pt}{\tiny\yng(2,2)}&  $(2,0)+(0,2)$ &  $\de^4\phi^3$ & $[7-\frac32\eps]+0\eps$ 
\\\hline
 \raisebox{-3pt}{\tiny\yng(4,1)}&  $(\frac52,\frac32)+(\frac32,\frac52)$ &  $\de^5\phi^3$ & $[8-\frac32\eps]+0\eps$ 
\\
 \raisebox{-3pt}{\tiny\yng(2,2)}&  $(2,0)+(0,2)$ &  $\de^4\phi^4$ & $[8-2\eps]+\frac79\eps$ 
\\\hline
 \raisebox{-3pt}{\tiny\yng(4,2)}&  $(3,1)+(1,3)$ &  $\de^6\phi^3$ & $[9-\frac32\eps]+0\eps$ 
\\
 \raisebox{-3pt}{\tiny\yng(3,2)}&  $(\frac52,\frac12)+(\frac12,\frac52)$ &  $\de^5\phi^4$ & $[9-2\eps]+\frac29\eps$ 
\\
 \raisebox{-3pt}{\tiny\yng(2,1)}&  $(\frac32,\frac12)+(\frac12,\frac32)$ &  $\de^5\phi^4$  & $[9-2\eps]+\frac12\eps$ 
 \\
 \raisebox{-3pt}{\tiny\yng(4,1)}&  $(\frac52,\frac32)+(\frac32,\frac52)$ &  $\de^5\phi^4$ & $[9-2\eps]+\frac{13}{18}\eps$ 
 \\
 \raisebox{-3pt}{\tiny\yng(2,2)}&  $(2,0)+(0,2)$ &  $\de^4\phi^5$ & $[9-\frac52\eps]+\frac{17}9\eps$ 
\\\hline
 \raisebox{-3pt}{\tiny\yng(5,2)}&  $(\frac72,\frac32)+(\frac32,\frac72)$ &  $\de^7\phi^3$ & $[10-\frac32\eps]+0\eps$ 
\\
 \raisebox{-3pt}{\tiny\yng(6,1)}&  $(\frac72,\frac52)+(\frac52,\frac72)$ &  $\de^7\phi^3$ & $[10-\frac32\eps]+0\eps$ 
\\
 \raisebox{-3pt}{\tiny\yng(3,1)}&  $(2,1)+(1,2)$ &  $\de^6\phi^4$ & $[10-2\eps]+\frac{28}{45}\eps$ 
 \\
 \raisebox{-3pt}{\tiny\yng(4,2)}&  $(3,1)+(1,3)$ &  $\de^6\phi^4$ & $[10-2\eps]+\frac{37-\sqrt{649}}{90}\eps$ 
 \\
 \raisebox{-3pt}{\tiny\yng(4,2)}&  $(3,1)+(1,3)$ &  $\de^6\phi^4$ & $[10-2\eps]+\frac{37+\sqrt{649}}{90}\eps$ 
 \\
 \raisebox{-3pt}{\tiny\yng(5,1)}&  $(3,2)+(2,3)$ &  $\de^6\phi^4$ & $[10-2\eps]+\frac{13}{45}\eps$ 
 \\
  \raisebox{-3pt}{\tiny\yng(2,1)}&  $(\frac32,\frac12)+(\frac12,\frac32)$ &  $\de^5\phi^5$ & $[10-\frac52\eps]+\frac{14}9\eps$ 
 \\
 \raisebox{-3pt}{\tiny\yng(4,1)}&  $(\frac52,\frac32)+(\frac32,\frac52)$ &  $\de^5\phi^5$ & $[10-\frac52\eps]+\frac{16}9\eps$ 
\\
 \raisebox{-3pt}{\tiny\yng(3,2)}&  $(\frac52,\frac12)+(\frac12,\frac52)$ &  $\de^5\phi^5$ & $[10-\frac52\eps]+\frac{23}{18}\eps$ 
 \\
 \raisebox{-3pt}{\tiny\yng(2,2)}&  $(2,0)+(0,2)$ &  $\de^4\phi^6$ & $[10-3\eps]+\frac{10}3\eps$ 
 \\\hline
\end{tabular}
}
\end{table}

\subsubsection{Three-row Lorentz Young tableaux}
\label{eq:three-rowYT}

Consider now operators transforming in representations corresponding to three-row Young tableaux. In the Ising CFT, there is no such operator with $\Delta^{\mathrm{4d}}\leqslant10$. A particularly interesting case is operators in the representation given by the Lorentz Young tableau $y_{1,1,1}={\tiny \yng(1,1,1)}$, since they become parity-odd scalars in three dimensions. 
For $N=1$ and $d=3$, the theory of a free scalar field has an operator the form $\epsilon^{\mu\nu\rho}\de_\mu\de_\nu\de_\rho \square^3\phi^4$. 
For the interacting theory, the corresponding operator is the leading operator in the $y_{1,1,1}$ Lorentz representation, which has dimension $[13-2\eps]+0\eps+O(\eps^2)$. In \cite{Dymarsky:2017yzx}, the numerical conformal bootstrap studied the four-point function of stress-tensors in three dimensions. A general upper bound for the parity-odd scalar was found: $\Delta_{\O_{\mathrm{odd,min}}}\lesssim 11.5$. When inserting some input to single out the Ising model, a somewhat lower bound was found: $\Delta_{\O_{\mathrm{odd,min}}}<11.2$.

\subsubsection{Unitarity-violating operators}
\label{eq:unitaritviolating}

In \cite{Hogervorst:2015akt}, unitarity of the free scalar CFT and the Ising CFT was investigated away from integer values of the spacetime dimension $d$. They found that in the free theory, there exist operators $R_n$, denoted ``evanescent operators'', corresponding to states which has negative norm for non-integer $d$ in the range $n-2<d<n-1$.\footnote{The operators $R_n$ are descendants and in general no form was given for the corresponding primary in the free theory. The primary with negative norm below three dimensions corresponding to $R_4$ is for $d=3$ of the form $\de^2\square^2\phi^4$.} Moreover, in the Wilson--Fisher fixed-point in $4-\eps$ dimensions, operators were found with non-real anomalous dimensions at order $\eps$. Specifically, four scalar descendant operators were found of the form $\square^8\phi^7$ with dimensions
\begin{equation}
\Delta=23-\frac72\eps+(0.470381\pm0.155408i)\eps+O(\eps^2),\quad 23-\frac72\eps+(1.191261\pm 0.029877i)\eps+O(\eps)^2.
\end{equation}
It was not stated which primary operators they descend from.

\subsection[\texorpdfstring{OPE coefficients in the $\phi\times\phi$ OPE}{OPE coefficients in the φ*φ OPE}]{OPE coefficients in the $\boldsymbol{\phi\times\phi}$ OPE}

In table~\ref{tab:OPEIsing} we present OPE coefficients $\lambda_{\phi\phi\O}^2$ for a selection of $\mathbb Z_2 $ even operators. Specifying only the square of the OPE coefficients leaves the sign of the actual OPE coefficient undetermined, however the numerical values from \cite{Simmons-Duffin:2016wlq} may be used to infer the sign.

\begin{table}[ht]
\centering
\caption{Squared OPE coefficients in $\phi\times\phi$ for operators for $N=1$. Numerical values are taken from \cite{Simmons-Duffin:2016wlq} and denote non-squared OPE coefficients.}
\label{tab:OPEIsing}
{\small
\setcounter{localfn}{1} 
\renewcommand{\arraystretch}{1.25}
\begin{tabular}{cclll}

$\phi,\phi,\O$ &$\O|\eps$ & $\lambda_{\phi\phi\O}^2(\eps)$ & $\lambda_{\phi\phi\O}^2(\eps)$  & $\lambda_{\phi\phi\O}^{3d}$
\\\hline
$\phi,\phi,\texttt{Op[E,0,0]}$ & $\1$ & $1$ & exact  &  $1$
\\
$\phi,\phi,\texttt{Op[E,0,1]} $ & $\phi^2$ & $2$ & \eord3\cite{Gopakumar:2016cpb}\eord4(num)\cite{Carmi:2020ekr} & $1.0518537(41)$ 
\\
$\phi,\phi,\texttt{Op[E,0,2]}$ & $\phi^4$ & $\frac1{54}\eps^2$ & \eord2\cite{Gliozzi:2017hni}\eord3\cite{Carmi:2020ekr} 
&  
$0.053012(55) $\makefn
\\
$\phi,\phi,\texttt{Op[E,0,3]}$ & $\phi^6$ & $\frac5{104976}\eps^4$ & \eord4\cite{Codello:2017qek} 
&  $0.0007338(31) $
\\
$\phi,\phi,\texttt{Op[E,0,4]}$ & $\square^2\phi^4$ &  &
&  $0.000162(12)  $
\\
\hline
$\phi,\phi,\texttt{Op[E,2,1]}$ & $T^{\mu\nu}$ & $\frac13 $ & \eord3\cite{Gopakumar:2016cpb}\eord4\cite{Alday:2017zzv}
&  $0.32613776(45) $\makefn
\\
$\phi,\phi,\texttt{Op[E,2,2]}$ & $\de^2\phi^4$ & $\frac1{1440}\eps^2 $ & \eord2\cite{Alday:2017zzv}\eord3(num.)\cite{Carmi:2020ekr}
&  $0.0105745(42)$\makefn
\\ $\phi,\phi,\texttt{Op[E,2,3]}$ & $\de^2\square\phi^4$ &  $\frac{\eps^4}{2592000}$ & \eord4\cite{Bertucci:2022ptt} & $0.0004773(62)$
\\\hline
$\phi,\phi,\texttt{Op[E,4,1]}$ & $C^{\mu\nu\rho\sigma}$ & $\frac1{35} $ & \eord3\cite{Gopakumar:2016cpb}\eord4\cite{Alday:2017zzv}
&  $0.069076(43)$
\\
$\phi,\phi,\texttt{Op[E,4,2]}$ & $\de^4\phi^4$ & $\frac{1}{419580} \eps^2$ &\eord2\cite{Alday:2017zzv,ThisPaper}
&  $0.0019552(12)$
\\
$\phi,\phi,\texttt{Op[E,4,3]}$ & $\de^4\phi^4$ & $\frac{1}{23976}\eps^2$ & \eord2\cite{Alday:2017zzv,ThisPaper}
&  $0.00237745(44)$
\\\hline
$\phi,\phi,\texttt{Op[E,6,1]}$ & $\mathcal J_6$ & $\frac1{462} $ & \eord3\cite{Gopakumar:2016cpb}\eord4\cite{Alday:2017zzv}
&  $0.0157416(41)$
\\
$\phi,\phi,\texttt{Op[E,6,2]}$ & $\de^6\phi^4$ & $O( \eps^2)$ & \eord2\cite{Alday:2017zzv,ThisPaper}
\\
$\phi,\phi,\texttt{Op[E,6,3]}$ & $\de^6\phi^4$ & $O( \eps^2)$ & \eord2\cite{Alday:2017zzv,ThisPaper}
\\
$\phi,\phi,\texttt{Op[E,6,4]}$ & $\de^6\phi^4$ & $O( \eps^2)$ & \eord2\cite{Alday:2017zzv,ThisPaper}
\\\hline
$\phi,\phi,\texttt{Op[E,8,1]}$ & $\mathcal J_8$ & $\frac1{6435} $ & \eord3\cite{Gopakumar:2016cpb}\eord4\cite{Alday:2017zzv}
&  $0.0036850(54)$
\\\hline
$\phi,\phi,\texttt{Op[E,10,1]}$ & $\mathcal J_{10}$ & $\frac1{92378} $ & \eord3\cite{Gopakumar:2016cpb}\eord4\cite{Alday:2017zzv}
&  $0.00087562(13) $
\\
\hline
\end{tabular}
\flushleft
\setcounter{localfn}{1} 

\makefn A value with rigorous error bars, $0.05304(16)$, was given in \cite{Reehorst:2021hmp}.
\\
\makefn Related to the central charge $C_T$.
\\
\makefn A value with rigorous error bars, $0.01054(10)$, was given in \cite{Reehorst:2021hmp}.
\\
}
\end{table}

For the family \texttt{IsingF2[$l$]}, the OPE coefficients were determined to order $\eps^3$ in \cite{Gopakumar:2016cpb} and $\eps^4$ in \cite{Alday:2017zzv}. For no other family of operators the OPE coefficients  $\lambda_{\phi\phi\O}^2$ have been determined on closed form.
See however \eqref{eq:opephi2khyp} below for a prediction of the OPE coefficient $\lambda_{\phi\phi\phi^{2k}}$, given for general $N$, which reduces to the values quoted here for $N=1$ and $k=2$ and $k=3$.

\subsection[\texorpdfstring{OPE coefficients in the $\phi\times\phi^2$ OPE}{OPE coefficients in the φ*φ^2 OPE}]{OPE coefficients in the $\boldsymbol{\phi\times\phi^2}$ OPE}
In table~\ref{tab:OPEIsing12} we present OPE coefficients $\lambda_{\phi\phi^2\O}^2$ for a selection of $\mathbb Z_2 $ odd operators. This OPE coefficient is also implemented for operators in the family \texttt{IsingF3[$l$]}, determined to order $\eps$ in \cite{Bertucci:2022ptt}.

\begin{table}[ht]
\centering
\caption{Squared OPE coefficients in $\phi\times\phi^2$ for operators for $N=1$. Numerical values are from \cite{Simmons-Duffin:2016wlq} and denote non-squared OPE coefficients.}\label{tab:OPEIsing12}
{\small
\setcounter{localfn}{1} 
\renewcommand{\arraystretch}{1.25}
\begin{tabular}{cclll}
$\phi,\phi^2,\O$ &$\O|\eps$ & $\lambda_{\phi\phi^2\O}^2(\eps)$ & $\lambda_{\phi\phi^2\O}^2(\eps)$  & $\lambda_{\phi\phi^2\O}^{3d}$
\\\hline
$\phi,\phi^2,\texttt{Op[O,0,1]}$ & $\phi$ & $2$ & \eord3\cite{Gopakumar:2016cpb}\eord4(num)\cite{Carmi:2020ekr} & $1.0518537(41)$
\\
$\phi,\phi^2,\texttt{Op[O,0,2]}$ & $\phi^5$ &    $ \frac5{108}\eps^2$ & \eord2\cite{Codello:2017qek} & $0.057235(20)$\makefn
\\
\hline
$\phi,\phi^2,\texttt{Op[O,2,1]}$ & $\de^2\phi^3$ & $\frac12$ & \eord1\cite{Bertucci:2022ptt} & 
 $0.38915941(81)$
\\
$\phi,\phi^2,\texttt{Op[O,2,2]}$ & $\de^2\phi^5$ & $O(\eps^2)$  &  & 
 $0.017413(73)$
\\
\hline
$\phi,\phi^2,\texttt{Op[O,3,1]}$ & $\de^3\phi^3$ & $\frac2{35}$ &\eord1\cite{Bertucci:2022ptt}  & 
 $0.1385(34)$
\\
\hline
$\phi,\phi^2,\texttt{Op[O,4,1]}$ & $\de^4\phi^3$ & $\frac1{18}$  &\eord1\cite{Bertucci:2022ptt}  & 
 $0.1077052(16)$
\\
\hline
$\phi,\phi^2,\texttt{Op[O,5,1]}$ & $\de^5\phi^3$ & $\frac2{231}$ &\eord1\cite{Bertucci:2022ptt}  & 
 $0.04191549(88)$
\\
\hline
$\phi,\phi^2,\texttt{Op[O,6,1]}$ & $\de^6\phi^3$ & $O(\eps^2)$ &  & 
 
\\
$\phi,\phi^2,\texttt{Op[O,6,2]}$ & $\de^6\phi^3$ &  $\frac3{572}$  & \eord1\cite{Bertucci:2022ptt} & 
 $$
\\
\hline
$\phi,\phi^2,\texttt{Op[O,7,1]}$ & $\de^7\phi^3$ &  $\frac2{2145}$   & \eord1\cite{Bertucci:2022ptt} & 
 $0.01161255(13)$
\\\hline
$\phi,\phi^2,\texttt{Op[O,8,1]}$ & $\de^8\phi^3$ & $O(\eps^2)$ &  & 
 $$
\\
$\phi,\phi^2,\texttt{Op[O,8,2]}$ & $\de^8\phi^3$ & $\frac1{2210}$  & \eord1\cite{Bertucci:2022ptt} & 
 $$
\\\hline
$\phi,\phi^2,\texttt{Op[O,9,1]}$ & $\de^9\phi^3$ & $O(\eps^2)$ &  & 
 $$
\\
$\phi,\phi^2,\texttt{Op[O,9,2]}$ & $\de^9\phi^3$ & $\frac{4}{46189}$ & \eord1\cite{Bertucci:2022ptt} & 
 $$
\\\hline
\end{tabular}
\flushleft
\setcounter{localfn}{1} 

\makefn A value with rigorous error bars, $0.0565(15)$, was given in \cite{Reehorst:2021hmp}.
\\
}
\end{table}

\subsection[\texorpdfstring{OPE coefficients in the $\phi^2\times\phi^2$ OPE}{OPE coefficients in the φ^2*φ^2 OPE}]{OPE coefficients in the $\boldsymbol{\phi^2\times\phi^2}$ OPE}

\begin{table}[ht]
\centering
\caption{Squared OPE coefficients in $\phi^2\times\phi^2$ for operators for $N=1$. Numerical values are from \cite{Simmons-Duffin:2016wlq} and denote non-squared OPE coefficients.}\label{tab:OPEIsing22}
{\small
\setcounter{localfn}{1} 
\renewcommand{\arraystretch}{1.25}
\begin{tabular}{cclll}
$\phi^2,\phi^2,\O$ &$\O|\eps$ & $\lambda_{\phi^2\phi^2\O}^2(\eps)$ & $\lambda_{\phi^2\phi^2\O}^2(\eps)$  & ${\lambda}_{\phi^2\phi^2\O}^{3d}$
\\\hline
$\phi^2,\phi^2,\texttt{Op[E,0,0]}$ & $\1$ & $1$  & exact &  $1$
\\
$\phi^2,\phi^2,\texttt{Op[E,0,1]} $ & $\phi^2$ & $8$ & \eord1\cite{Thesis,ThisPaper} & $1.532435(19)$
\\
$\phi^2,\phi^2,\texttt{Op[E,0,2]}$ & $\phi^4$ &  $6$ &\eord1\cite{Thesis,ThisPaper}  & $1.5360(16)$\makefn
\\
$\phi^2,\phi^2,\texttt{Op[E,0,3]}$ & $\phi^6$ &  &  & $0.1279(17)$
\\
$\phi^2,\phi^2,\texttt{Op[E,0,4]}$ & $\square^2\phi^4$ & $\frac16$ & \eord1\cite{Thesis,ThisPaper} & $0.1874(31)$
\\
\hline
$\phi^2,\phi^2,\texttt{Op[E,2,1]}$ & $T^{\mu\nu}$ & $\frac43$ & \eord4\cite{Alday:2017zzv} & $0.8891471(40)$
\\
$\phi^2,\phi^2,\texttt{Op[E,2,2]}$ & $\de^2\phi^4$ &$\frac85$ &  \eord1\cite{Thesis,ThisPaper}  & $0.69023(49)$\makefn
\\
$\phi^2,\phi^2,\texttt{Op[E,2,3]}$ & $\de^2\square\phi^4$ &$\frac15$ & \eord1\cite{Thesis,ThisPaper} & $0.21882(73)$
\\
$\phi^2,\phi^2,\texttt{Op[E,2,4]}$ & $\de^2\phi^6$ &  & & $$
\\
$\phi^2,\phi^2,\texttt{Op[E,2,5]}$ & $\de^2\square^2\phi^4$ &$\frac1{999}$ & \eord1\cite{Thesis,ThisPaper,Bertucci:2022ptt} & $$
\\
$\phi^2,\phi^2,\texttt{Op[E,2,6]}$ & $\de^2\square^2\phi^4$ &$\frac4{111}$ & \eord1\cite{Thesis,ThisPaper,Bertucci:2022ptt} & $$
\\\hline
$\phi^2,\phi^2,\texttt{Op[E,4,1]}$ & $C^{\mu\nu\rho\sigma}$ &$\frac4{35}$ & \eord1\cite{Thesis,ThisPaper} & 0.24792(20)$$
\\
$\phi^2,\phi^2,\texttt{Op[E,4,2]}$ & $\de^4\phi^4$ & $\frac{125}{2331}$ &\eord1\cite{Thesis,ThisPaper}  & $-0.110247(54)$
\\
$\phi^2,\phi^2,\texttt{Op[E,4,3]}$ & $\de^4\phi^4$ & $\frac8{37}$ &\eord1\cite{Thesis,ThisPaper}  & $0.22975(10)$
\\
$\phi^2,\phi^2,\texttt{Op[E,4,4]}$ & $\de^4\square\phi^4$ & $\frac{2091- 71   \sqrt{697}}{107338}$ &\eord1\cite{Thesis,ThisPaper,Bertucci:2022ptt}  & $0.08635(18)$
\\
$\phi^2,\phi^2,\texttt{Op[E,4,5]}$ & $\de^4\square\phi^4$ & $\frac{2091+ 71   \sqrt{697}}{107338}$ &\eord1\cite{Thesis,ThisPaper,Bertucci:2022ptt}  & $$
\\\hline
$\phi^2,\phi^2,\texttt{Op[E,6,1]}$ & $\mathcal J_6$ & $\frac2{231}$ & \eord1\cite{Thesis,ThisPaper} & $0.066136(36)$
\\\hline
$\phi^2,\phi^2,\texttt{Op[E,8,1]}$ & $\mathcal J_8$ & $\frac4{6435}$ & \eord1\cite{Thesis,ThisPaper} & $0.017318(30)$
\\\hline
$\phi^2,\phi^2,\texttt{Op[E,10,1]}$ & $\mathcal J_{10}$ & $\frac2{46189}$ & \eord1\cite{Thesis,ThisPaper} & $0.0044811(15)$
\\
\hline
\end{tabular}
\flushleft
\setcounter{localfn}{1} 

\makefn A value with rigorous error bars, $1.5362(12)$, was given in \cite{Reehorst:2021hmp}.
\\
\makefn A value with rigorous error bars, $0.678(15)$, was given in \cite{Reehorst:2021hmp}.
\\
}
\end{table}

In table~\ref{tab:OPEIsing22} we present OPE coefficients $\lambda_{\phi^2\phi^2\O}^2$ for a selection of $\mathbb Z_2 $ even operators. The OPE coefficient for the family \texttt{IsingF2[$l$]} as reported in \cite{Thesis}, has also been implemented.

\section[\texorpdfstring{Conformal data for general $\boldsymbol N$}{Conformal data for general N}]{Conformal data for general $\boldsymbol N$}
\label{sec:dataGenN}

This section presents the conformal data for general $N$. We include perturbative data both in the $4-\eps$ expansion and in the large $N$ expansion.
Operators are organised by global symmetry representation $R$ and spin $\ell$. For each $R$ and $\ell$, the operators are organised by increasing scaling dimension in the limit $\eps\ll1/N\ll 1$.

\subsection{Families of operators}
\label{sec:familiesON}

For general $N$ we introduce a number of families \texttt{ONF$\langle i\rangle$}. We express the $1/N$ part of the large $N$ operator dimension in terms of the anomalous dimension of $\varphi$: $\Delta_{\varphi}=\mu-1+\frac{\eta_1/2}{N}+O(N^{-2})$.

\paragraph{\texttt{ONF1[$k$]}} Scalar singlet operators of the form $\varphi^{2k}_S\sim \sigma^k$, $k\geqslant1$. We have \eord1\cite{Wegner:1972zz}\eord2\cite{Derkachov:1997gc} and \Nord1\cite{Vasiliev:1993ux,Lang:1992zw}\Nord2\cite{Derkachov:1997gc},
\begin{align}
\texttt{DeltaE[ONF1[$k$]]}&=k[2-\eps]+\frac{k(6k+N-4)}{N+8}\eps+\ldots+O(\eps^3),
\\
\texttt{DeltaN[ONF1[$k$]]}&=2k+\frac{2k(2\mu-1)}{\mu-2}\left(k\mu(2\mu-3)-\mu^2+\mu+2\right)\frac{\eta_1/2}{N}+\ldots+O(1/N^3).
\end{align}

\paragraph{\texttt{ONF2[$k$]}} Scalar fundamental operators of the form $\varphi^{2k+1}_V\sim \sigma^k\varphi$, $k\geqslant0$, $k\neq1$. We have \eord1\cite{Wegner:1972zz} and \Nord1\cite{Lang:1993ct},
\begin{align}
\texttt{DeltaE[ONF2[$k$]]}&=(2k+1)[1-\tfrac\eps2]+\frac{k(6k+N+2)}{N+8}\eps+O(\eps^2),
\\
\texttt{DeltaN[ONF2[$k$]]}&=2k+\mu-1+\left(\frac{2k(k-1)\mu(2\mu-3)(2\mu-1)}{2-\mu}+1\right)\frac{\eta_1/2}{N}+O(N^{-2}).
\end{align}
In fact \texttt{ONF1} and \texttt{ONF2} have the largest order $\eps$ anomalous dimensions for operator with $2k$ and $2k+1$ fields respectively \cite{Kehrein:1992fn}.

\paragraph{\texttt{ONF3[$m$]}} Scalar completely symmetric traceless tensors of the form $\varphi^{m}_{T_m}$, $m\geqslant 1$. 
We have \eord2\cite{Houghton1974}\eord3\cite{Wallace1975}\footnote{
As pointed out in \cite{Derkachov:1997ch}, the result presented in \cite{Wallace1975} is incorrect.}\eord4\cite{Antipin:2020abu}\footnote{For $N=2$, some results also appeared in \cite{Badel:2019oxl}, however with an error propagating from a typo in \cite{Calabrese:2002bm}.}\eord5\cite{ThisPaper,Bednyakov:2021ojn,Antipin:2020abu}\footnote{This was found by combining the order $\eps^5$ result for $\varphi^3_{T_3}$ of \cite{Bednyakov:2021ojn} with the general formula at large $m$ from \cite{Antipin:2020abu}. I thank the authors of both papers for useful discussions.} and \Nord1\cite{Lang:1990ni,Lang:1990re}\Nord2\cite{Derkachov:1997ch},
\begin{align}
\texttt{DeltaE[ONF3[$m$]]}&=m\left[1-\frac 12\eps\right]+\frac{m(m-1)}{N+8}\eps
+\ldots+O(\eps^6) ,
\\
\texttt{DeltaN[ONF3[$m$]]}&=m(\mu-1)+\frac{m(m\mu+2-2\mu)}{\mu-2}\frac{\eta_1/2}{N}+\ldots+O(N^{-3}).
\end{align}
\paragraph{\texttt{ONF4[$l$]}} Singlet broken currents of the form $\mathcal J_{S,\ell}=[\varphi,\varphi]_{S,0,\ell}$, $\ell\in 2\mathbb Z_+$. We have \eord2\cite{Wilson:1973jj}\eord4\cite{Manashov:2017xtt}\footnote{It was determined by reconstruction of $O(N)$ symmetry factors in the computation of the much earlier work \cite{Derkachov:1997pf}.} and \Nord1\cite{Lang:1992zw}\Nord2\cite{Manashov:2017xtt},\footnote{In the ancillary data file, we have rewritten the expression given in \cite{Manashov:2017xtt} using the identity (checked numerically)
\begin{align*}
&{_3}F_2(1,\mu-1,\ell+2\mu-3;\ell+\mu,\ell+2\mu-2;1)\\&\qquad\qquad\qquad=\frac{(\ell+\mu-1)(\ell+2\mu-3)}{\mu-2}\left(
\ln 2+S_1(\ell/2)-S_1(\ell)+\frac12S_1\left(\tfrac{\ell-5}2+\mu\right)-\frac12S_1\left(\tfrac{\ell-4}2+\mu\right)
\right).
\end{align*}
Notice also that $_3F_2(1,\tfrac32,\ell+\tfrac12;\ell+\tfrac32,\ell+2;1)=2(\ell+1)\left(
\frac{3+4\ell^2}{3+2\ell}+\ell(2\ell+1)\left(S_1(\tfrac \ell2+\tfrac14)-S_1(\tfrac\ell2+\frac34)\right)
\right)
$, which is useful when evaluating the expression in three dimensions ($\mu=\frac32$).}
\begin{align}
\hspace{-20pt}\texttt{DeltaE[ONF4[$l$]]}&=2-\eps+l+\frac{N+2}{2(N+8)^2}\left(1-\frac6{l(l+1)}\right)\eps^2+\ldots+O(\eps^5),
\\
\hspace{-20pt}\texttt{DeltaN[ONF4[$l$]]}&=2(\mu-1)+l+\bigg(2 -\frac{2 \mu\left((\mu\! -\!1)\!+\!\frac{ \Gamma (2\mu -1) \Gamma (l+1)}{\Gamma (l+2 \mu -3)}\right)}{(l+\mu-1)(l+\mu-2)}  
   \bigg)\frac{\eta_1/2}{N}+\!\ldots\!+O(N^{-3}).
\end{align}
\paragraph{\texttt{ONF5[$l$]}} Traceless-symmetric broken currents of the form $\mathcal J_{T,\ell}=[\varphi,\varphi]_{T,0,\ell}$, $\ell\in 2\mathbb Z_+$. We have \eord2\cite{Wilson:1973jj}\eord3\cite{Braun:2013tva}\eord4\cite{Henriksson:2018myn} and \Nord1\cite{Lang:1992zw}\Nord2\cite{Derkachov:1997ch},
\begin{align}
\texttt{DeltaE[ONF5[$l$]]}&=2-\eps+l+\frac1{2(N+8)^2}\left(N+2-\frac{2(N+6)}{l(l+1)}\right)\eps^2+\ldots+O(\eps^5),
\\
\texttt{DeltaN[ONF5[$l$]]}&=2(\mu-1)+l+\left(2-\frac{2\mu(\mu-1)}{(l+\mu-1)(l+\mu-2)}\right)\frac{\eta_1/2}{N}+\ldots+O(N^{-3}).
\end{align}
\paragraph{\texttt{ONF6[$l$]}} Antisymmetric broken currents of the form $\mathcal J_{A,\ell}=[\varphi,\varphi]_{A,0,\ell}$, $\ell\in 2\mathbb N+1$. We have \eord3\cite{Braun:2013tva}\eord4\cite{Henriksson:2018myn} and \Nord1\cite{Lang:1992zw}\Nord2\cite{Derkachov:1997ch},
\begin{align}
\texttt{DeltaE[ONF6[$l$]]}&=2-\eps+l+\frac{N+2}{2(N+8)^2}\left(1-\frac2{l(l+1)}\right)\eps^2+\ldots+O(\eps^5),
\\
\texttt{DeltaN[ONF6[$l$]]}&=2(\mu-1)+l+\left(2-\frac{2\mu(\mu-1)}{(l+\mu-1)(l+\mu-2)}\right)\frac{\eta_1/2}{N}+\ldots+O(N^{-3}).
\end{align}
\paragraph{\texttt{ONF7[$l$]}} Spinning singlet operators of the form $[\sigma,\sigma]_{0,\ell}$, $\ell\in 2\mathbb N$. We have \Nord1\cite{Lang:1992zw},
\begin{align}
\texttt{DeltaN[ONF7[$l$]]}&=4+l-\frac{8}{(2-\mu)^2}\bigg((\mu-1)(2-\mu)(2\mu-1)
\\\nonumber & \quad+
\frac{\mu(2\mu-3)}{(l+1)(l+2)}
\left((\mu-1)(2\mu-3)-\frac{(\mu-l-3)_l}{(\mu)_l}\right)
\!\!\bigg)\frac{\eta_1/2}{N}+O(N^{-2}),
\end{align}
where $(a)_k=\frac{\Gamma(a+k)}{\Gamma(a)}$ is the (rising) Pochhammer symbol. There is no simple expression for these operators in the $\eps$-expansion, see discussion in section~\ref{sec:orderepssystematics}.
\paragraph{\texttt{ONF8[$l$]}} Spinning fundamental operators of the form $[\varphi,\sigma]_{0,\ell}$, $\ell\geqslant1$. We have \eord1\cite{Kehrein:1992fn} and \Nord1\cite{Lang:1992zw},
\begin{align}\label{eq:ONF8eps}
\texttt{DeltaE[ONF8[}l\texttt{]]}&=[3-\tfrac32\eps]+l\\\nonumber&\hspace{-55pt}+\frac{4(-1)^l+(N+4)(l+1)+\sqrt{16(l+1)(-1)^l+48+16N+(l+1)^2N^2}}{2(N+8)(l+1)}\eps+O(\eps^2),
\\
\texttt{DeltaN[ONF8[}l\texttt{]]}&=\mu+1+l+\bigg(
\frac{4\mu(\mu-1)(2\mu-3)}{(l+1)(l+\mu-1)(2-\mu)}
 \\  \nonumber &\quad
-\frac{8\mu^2-11\mu+2}{2-\mu}+\frac{2(-1)^l\Gamma(l+1)\Gamma(\mu+1)}{(2-\mu)\Gamma(l+\mu)}
\bigg)\frac{\eta_1/2}{N}+O(N^{-2}).
\end{align}
\paragraph{\texttt{ONF9[$k$]}} Scalar singlet operators of the form $\square\varphi^{2k}_S\sim \square\sigma^k$,\footnote{Although in general, there is a mixing between $\sigma$ and $\square$ when relating operators between the $\eps$-expansion and the large $N$ expansion, the writing of this family of operators is consistent, see \cite{Derkachov:1997gc}.} $k\geqslant2$. We have \eord1\cite{ThisPaper} and \Nord1\cite{Vasiliev:1993ux},
\begin{align}
\texttt{DeltaE[ONF9[$k$]]}&=2+k(2-\eps)+\frac{6k^2+(N-12)k+4}{N+8}\eps+O(\eps^2),
\\
\texttt{DeltaN[ONF9[$k$]]}&=2k+2+\frac{2(2\mu-1)}{3(2-\mu)}\big(3k^2\mu(2\mu-3)-k(10\mu^2-\mu-18)\nonumber\\&\qquad\qquad\qquad\qquad\qquad\quad+2\mu^3-\mu^2-7\mu+6\big)\frac{\eta_1/2}N+O(N^{-2}).
\end{align}

\paragraph{\texttt{ONF10[$k$,$m$]}} Scalar operators of the form $\varphi^{2k+m}_{T_m}\sim\sigma^k\varphi^m_{T_m}$, $k,m\geqslant0$. We have \eord1\cite{Wegner:1972zz}
 and \Nord1\cite{Lang:1993ct},
\begin{align}
\texttt{DeltaE[ONF10[}k\texttt,m\texttt{]]}&=(2k+m)[1-\tfrac\eps2]+\frac{6k^2+k(6m+N-4)+m(m-1)}{N+8}\eps+O(\eps^2),
\\
\texttt{DeltaN[ONF10[}k\texttt,m\texttt{]]}&=2k+m(\mu-1)+\frac1{2-\mu}\big[2k^2\mu(2\mu-1)(2\mu-3)+m (m \mu  +2-2 \mu)
\nonumber
\\
&\qquad -2 k (2 \mu -1) (2 \mu ^2-\mu -2-2   m(\mu-1))\big]\frac{\eta_1/2}N+O(N^{-2}).
\end{align}
For $m=0$ the family reduces to $\texttt{ONF1[}k\texttt{]}$, for $m=1$ to $\texttt{ONF2[}k\texttt{]}$ and for $k=0$ to $\texttt{ONF3[}m\texttt{]}$.

\paragraph{\texttt{ONF11[$m$,$l$]}} Spin-$l$ operators of the form $\de^l\varphi^{m+l}_{Y_{m,l}}$ transforming in the $\ON$ representation $Y_{m,l}$. We have \eord1\cite{ThisPaper} and \Nord1\cite{Lang:1990re}.
\begin{align}
\texttt{DeltaE[ONF11[}m\texttt,l\texttt{]]}&=(m+l)[1-\tfrac\eps2]+l+\frac{3 m^2+3 m l-3 m+l^2-4 l}{3(N+8)}\eps+O(\eps^2),
\\
\texttt{DeltaN[ONF11[}m\texttt,l\texttt{]]}&=(m+l)(\mu-1)+l\\&\quad\nonumber+\left[(m+l)(m+l-2)-4(l-1)\left(\tfrac{m}{2\mu}+\tfrac{l}{2\mu+2}\right)\right]\frac{\mu }{2-\mu}\frac{\eta_1/2}N+O(N^{-2}).
\end{align}
For $l=0$ the family reduces to \texttt{ONF3[$m$]}. Note that $Y_{m,0}=T_m$, and $Y_{m,1}=H_{m+1}$, and that $Y_{2,2}=B_4$. Note also that operators with $m+l$ fields in the $\ON$ representation $Y_{m,l}$ and with no contracted derivatives must have spin $\geqslant l$.

Apart from the eleven families of operators for generic $N$ listed above, \cite{Kehrein:1992fn} gives three additional families in the $\eps$-expansion. These families are only defined by the fact that they acquire an order $\eps$ anomalous dimension, which singles them out from additional operators of the same type. This property is not visible in the large $N$ expansion, whence we do not include them in our list of families.
\begin{itemize}
\item $\de^\ell\varphi^3_V$ with non-zero order $\eps$ dimension \eord1\cite{Kehrein:1992fn}. This is the sister family to \texttt{ONF8[$\ell$]}. Dimension given by \eqref{eq:ONF8eps}, but with a minus sign in front of the square root.
\item $\de^\ell\varphi^3_{T_3}$ with non-zero order $\eps$ dimension \eord1\cite{Kehrein:1992fn}: $[3-\frac32\eps]+\ell+\frac{2}{N+8}\left(1+\frac{2(-1)^\ell}{\ell+1}\right)\eps+O(\eps^2)$, $\ell\neq1$.
\item $\de^\ell\varphi^3_{H_3}$ with non-zero order $\eps$ dimension \eord1\cite{Kehrein:1992fn}: $[3-\frac32\eps]+\ell+\frac{1}{N+8}\left(2-\frac{2(-1)^\ell}{\ell+1}\right)\eps+O(\eps^2)$, $\ell\neq0$.
\end{itemize}
Moreover, in \cite{Lang:1993ct}, a closed-form expression at large $N$ is given for the anomalous dimension of spin-one operators of the form $\de\sigma^k\varphi$.

\subsection{Singlet operators}
For singlet operators, we include all conformal primary operators with $\Delta^{\mathrm{4d}}\leqslant8$. The dimensions in the $\eps$-expansion are often quite cumbersome, being roots of some polynomial, and in the tables we only display the leading behaviour at large $N$. For instance, for the operator \texttt{Op[S,2,3]}, the dimension given explicitly in \eqref{eq:exampleDeltaE} above will be abbreviated as $[6-2\eps]+2\eps N^0_{\cdots}$, where $N^0_{\cdots}$ is short for $N^0+O(N^{-1})$. Full results are implemented in the ancillary data file.

\subsubsection{Scalar singlet operators}

In table~\ref{tab:singletscalars}, we give a presentation of the $\ON$ singlet and Lorentz scalar operators of dimension $\Delta^{\mathrm{4d}}\leqslant10$.

\begin{table}[ht]
\centering
\caption{Singlet scalar operators for $N>1$. $N^0_{\cdots}$ denotes $N^0+O(N^{-1})$, where there is an exact expression. Potentially extend up to $\Delta=10$.}\label{tab:singletscalars}
{\small
\renewcommand{\arraystretch}{1.25}
\begin{tabular}{cccllll}
$\O$ &$\O|\eps$&$\O|N$  & $\Delta^{(1)}_{4-\eps}$& $\Delta(\eps)$   & $\Delta(N)$& Family
\\\hline
\texttt{Op[S,0,0]} & $\1$&$\1$ & $0$ & exact   & exact &
\\
\texttt{Op[S,0,1]} & $\varphi_S^2$ &$\sigma$ & $[2-\eps]+\frac{N+2}{N+8}\eps$ & \eord7\cite{Schnetz:2016fhy}   &
\Nord2\cite{Vasiliev:1981dg} & $\texttt 1_2$, $\texttt{10}_{1,0}$
\\
\texttt{Op[S,0,2]} & $\varphi_S^4$ &$\sigma^2$  & $[4-2\eps]+2\eps$ & \eord7\cite{Schnetz:2016fhy}  & 
\Nord2\cite{Broadhurst:1996ur} & $\texttt 1_2$, $\texttt 7_0$, $\texttt{10}_{2,0}$
\\
\texttt{Op[S,0,3]} & $\square\varphi_S^4$ &$\square\sigma^2$& $[6-2\eps]+\frac{2N+4}{N+8}\eps $ &\eord1   &\Nord1\cite{Vasiliev:1993ux} & $\texttt 9_2$
\\
\texttt{Op[S,0,4]} & $\varphi_S^6$ &$\sigma^3$  & $[6-3\eps]+\frac{3N+42}{N+8}$ & \eord2\cite{Derkachov:1997gc} & \Nord2\cite{Derkachov:1997gc}& $\texttt 1_3$, $\texttt{10}_{3,0}$
\\
\texttt{Op[S,0,5]} &$\square^2\varphi^4_S$&$\square^2\varphi^4_S$&$[8-2\eps]+0\eps N^0_{\cdots}$&\eord1& &
\\
\texttt{Op[S,0,6]} &$\square^2\varphi^4_S$&$\square^2\sigma^2$&$[8-2\eps]+2\eps N^0_{\cdots}$&\eord1& &
\\
\texttt{Op[S,0,7]} &$\square\varphi^6_S$&$\square \sigma^3$&$[8-3\eps]+\frac{3N+22}{N+8}\eps$&\eord1&  \Nord1\cite{Vasiliev:1993ux} & $\texttt 9_3$
\\
\texttt{Op[S,0,8]} &$\varphi^8_S$&$\sigma^4$&$[8-4\eps]+\frac{4N+80}{N+8}\eps$&\eord2\cite{Derkachov:1997gc} & 
\Nord2\cite{Derkachov:1997gc}&  $\texttt 1_4$, $\texttt{10}_{4,0}$
\\ 
\texttt{Op[S,0,9]} &$\square^3\varphi^4_S$&$$&$[10-2\eps]+0\eps N^0_{\cdots}$&\eord1 &
\\
\texttt{Op[S,0,10]} &$\square^2\varphi^6_S$&$$&$[10-3\eps]+\eps N^0_{\cdots}$&\eord1& &
\\
\texttt{Op[S,0,11]} &$\square^2\varphi^6_S$&$$&$[10-3\eps]+2\eps N^0_{\cdots}$& \eord1&
\\
\texttt{Op[S,0,12]} &$\square^2\varphi^6_S$&$$&$[10-3\eps]+3\eps N^0_{\cdots}$& \eord1&
\\
\texttt{Op[S,0,13]} &$\square^3\varphi^4_S$&$$&$[10-2\eps]+2\eps N^0_{\cdots}$&\eord1& &
\\
\texttt{Op[S,0,14]} &$\square^2\varphi^6_S$&$$&$[10-3\eps]+3\eps N^0_{\cdots}$&\eord1& &
\\
\texttt{Op[S,0,15]} &$\square\varphi^8_S$&$\square\sigma^4$&$[10-4\eps]+\frac{4N+13}{N+8}\eps$&\eord1 & \Nord1\cite{Vasiliev:1993ux} & $\texttt 9_4$
\\
\texttt{Op[S,0,16]} &$\varphi^{10}_S$&$\sigma^5$&$[10-5\eps]+\frac{5N+130}{N+8}\eps$&\eord2\cite{Derkachov:1997gc} & 
\Nord2\cite{Derkachov:1997gc}&   $\texttt 1_5$, $\texttt{10}_{5,0}$
\\\hline
\end{tabular}
}
\end{table}

\subsubsection{Spinning singlet operators}

In table~\ref{tab:singletspinning}, we present a selection of singlet operators with non-zero spin. The list is complete up to $\Delta^{\mathrm{4d}}\leqslant8$, but for some spins it contains additional operators.

\begin{table}[ht]
\centering
\caption{Singlet spinning operators for $N>1$.}\label{tab:singletspinning}
{\small
\renewcommand{\arraystretch}{1.25}
\begin{tabular}{cccllll}
$\O$ &$\O|\eps$&$\O|N$  & $\Delta^{(1)}_{4-\eps}$& $\Delta(\eps)$   & $\Delta(N)$& Family
\\\hline
\texttt{Op[S,1,1]} & $\de\square\varphi^6_S$& $\de\square\sigma^3$&$[9-3\eps]+\frac{9N+58}{3(N+8)}\eps$  &    &  &
\\\hline
\texttt{Op[S,2,1]} & $T^{\mu\nu}$&$T^{\mu\nu}$ & $[4-\eps]$ & exact   & exact & $\texttt 4_2$
\\
\texttt{Op[S,2,2]} & $\de^2\varphi^4_S$ & $[\varphi,\varphi]_{S,1,2}$  &  $[6-2\eps]+\eps N^0_{\cdots}$ &\eord1&&
\\
\texttt{Op[S,2,3]} & $\de^2\varphi^4_S$  & $[\sigma,\sigma]_{0,2}$ & $[6-2\eps]+2\eps N^0_{\cdots}$ &\eord1& \Nord1\cite{Lang:1992zw}  &  $\texttt 7_2$
\\
\texttt{Op[S,2,4]} &$\de^2\square\varphi^4_S$&&$[8-2\eps]+0\eps N^0_{\cdots}$&\eord1&&
\\
\texttt{Op[S,2,5]} &$\de^2\square\varphi^4_S$&&$[8-2\eps]+\eps N^0_{\cdots}$&\eord1&&
\\
\texttt{Op[S,2,6]} &$\de^2\varphi^6_S$&&$[8-3\eps]+2\eps N^0_{\cdots}$&\eord1&&
\\
\texttt{Op[S,2,7]} &$\de^2\square\varphi^4_S$&&$[8-2\eps]+2\eps N^0_{\cdots}$&\eord1&&
\\
\texttt{Op[S,2,8]} &$\de^2\varphi^6_S$&&$[8-3\eps]+3\eps N^0_{\cdots}$&\eord1&&
\\\hline
\texttt{Op[S,4,1]} & $\mathcal J_{S,4}$ &$\mathcal J_{S,4}$& $[6-\eps]$    &\eord4\cite{Manashov:2017xtt}& 
\Nord2\cite{Manashov:2017xtt}& $\texttt 4_4$
\\
\texttt{Op[S,4,2]} & $\de^4\varphi^4_S$ & $\de^4\varphi^4_S$  & $[8-2\eps]+0\eps N^0_{\cdots}$ &\eord1&&
\\
\texttt{Op[S,4,3]} & $\de^4\varphi^4_S$   & & $[8-2\eps]+\eps N^0_{\cdots}$ &\eord1&&
\\
\texttt{Op[S,4,4]} & $\de^4\varphi^4_S$  &  & $[8-2\eps]+\eps N^0_{\cdots}$ &\eord1&&
\\
\texttt{Op[S,4,5]} & $\de^4\varphi^4_S$  & $[\sigma,\sigma]_{0,4}$ & $[8-2\eps]+2\eps N^0_{\cdots}$&\eord1& \Nord1\cite{Lang:1992zw} & $\texttt 7_4$
\\\hline
\texttt{Op[S,5,1]} &$\de^5\varphi^4_S$ &$\de^5\sigma\varphi^2_S$  &$[9-2\eps]+\frac{N+2}{N+8}\eps$&\eord1&&
\\\hline
\texttt{Op[S,6,1]} &$\mathcal J_{S,6}$&$\mathcal J_{S,6}$& $[8-\eps]$ &\eord4\cite{Manashov:2017xtt}&  
\Nord2\cite{Manashov:2017xtt}& $\texttt 4_6$
\\
\texttt{Op[S,6,2]} & $\de^6\varphi^4_S$ & $\de^6\varphi^4_S$  & $[10-2\eps]+0\eps N^0_{\cdots}$ &\eord1&&
\\
\texttt{Op[S,6,3]} & $\de^6\varphi^4_S$ & $\de^6\varphi^4_S$  & $[10-2\eps]+0\eps N^0_{\cdots}$ &\eord1&&
\\
\texttt{Op[S,6,4]} & $\de^6\varphi^4_S$ & & $[10-2\eps]+\eps N^0_{\cdots}$ &\eord1&&
\\
\texttt{Op[S,6,5]} & $\de^6\varphi^4_S$ &   & $[10-2\eps]+\eps N^0_{\cdots}$ &\eord1&&
\\
\texttt{Op[S,6,6]} & $\de^6\varphi^4_S$ & & $[10-2\eps]+\eps N^0_{\cdots}$ &\eord1&&
\\
\texttt{Op[S,6,7]} & $\de^6\varphi^4_S$ & $[\sigma,\sigma]_{0,6}$  & $[10-2\eps]+2\eps N^0_{\cdots}$ &\eord1&\Nord1\cite{Lang:1992zw}& $\texttt 7_6$
\\\hline
\texttt{Op[S,8,1]} &$\mathcal J_{S,8}$&$\mathcal J_{S,8}$ &$[10-\eps]$ &\eord4\cite{Manashov:2017xtt}&
\Nord2\cite{Manashov:2017xtt}& $\texttt 4_8$
\\\hline
\texttt{Op[S,10,1]} &$\mathcal J_{S,10}$&$\mathcal J_{S,10}$ &$[12-\eps]$ &\eord4\cite{Manashov:2017xtt}&
\Nord2\cite{Manashov:2017xtt}& $\texttt 4_{10}$
\\\hline
\end{tabular}
}
\end{table}

Singlet operators of spin one are virial current candidates, c.f.\ section~\ref{sec:spinningIsing}. As can be seen in table~\ref{tab:singletspinning}, the leading operator of this type has dimension $\Delta=9+O(\eps)$, which is well above the value for a virial current.\footnote{The existence of this operator has also been confirmed by a character counting by the methods described in appendix~\ref{app:characters}. A character decomposition using $d=4$ conformal characters finds also a spin-$1$ operator of the form $\de\square^2\varphi^4_S$, however this operator is a pseudovector, i.e.\ an operator in the $y_{1,1,1}$ Lorentz representation. See section~\ref{sec:reprSOd} for a discussion on the Lorentz representation theory and section~\ref{sec:nonTSLorentz} where this operator is discussed.} 

Not included in our tables are some results from \cite{Lang:1994tu}, which computed at large $N$ the individual anomalous dimensions of the two operators of the form $\de^6\sigma^3$, which are degenerate at infinite $N$. It is, in principle, possible to use their method to find large $N$ anomalous dimensions of more operators than those included in this report.

\subsection[Operators in the $\ON$ representations $V$, $T$ and $A$]{Operators in the $\boldsymbol{\ON}$ representations $\boldsymbol V$, $\boldsymbol T$ and $\boldsymbol A$}

In table~\ref{tab:ONV} we present a selection of conformal primary operators in the fundamental (vector, $V$) representation. Likewise, in tables~\ref{tab:Toperators} and \ref{tab:Aoperators} we present a selection of operators in the rank two traceless-symmetric ($T$) and antisymmetric ($A$) representations.

\begin{table}[ht]
\centering
\caption{Operators in the vector representation $V$.}\label{tab:ONV}
{\small
\renewcommand{\arraystretch}{1.25}
\begin{tabular}{cccllll}
$\O$ &$\O|\eps$&$\O|N$  & $\Delta^{(1)}_{4-\eps}$& $\Delta(\eps)$   & $\Delta(N)$& Family
\\\hline
\texttt{Op[V,0,1]} & $\varphi$& $\varphi $ & $[1-\frac12\eps]+0\eps$&
\eord8\cite{SchnetzUnp} & 
 \Nord3\cite{Vasiliev:1982dc}* & $\texttt 2_0,\texttt 3_1,\texttt{11}_{1,0},\texttt{10}_{0,1}$
\\
\texttt{Op[V,0,2]} & $\varphi^5_V$ & $\sigma^2\varphi$ & $[5-\frac52\eps]+\frac{2N+28}{N+8}\eps$ &\eord1 &  \Nord1\cite{Lang:1993ct}&   $\texttt 2_2$, $\texttt{10}_{2,1}$
\\
\texttt{Op[V,0,3]} & $\square\varphi^5_V$ & $\square\sigma^2\varphi$ &  $[7-\frac52\eps]+\frac{2N+12}{N+8}\eps$ & \eord1 
\\
\texttt{Op[V,0,4]} & $\varphi^7_V$ & $\sigma^3\varphi $ & $[7-\frac72\eps]+\frac{3N+60}{N+8}\eps$ & \eord1 & \Nord1\cite{Lang:1993ct} &    $\texttt 2_3$, $\texttt{10}_{3,1}$
\\\hline
\texttt{Op[V,1,1]} & $\de\varphi^3_V$ & $\de\sigma\varphi$ & $[4-\frac32\eps]+\frac{N+2}{N+8}\eps$ & \eord1& \Nord1\cite{Lang:1992pp} & $\texttt 8_1$, $\texttt{12}_1$
\\
\texttt{Op[V,1,2]} & $\de\varphi^5_V$ & $\de\sigma^2\varphi$ & $[6-\frac52\eps]+\frac{2N+18}{N+8}\eps$ & \eord1&  \Nord1\cite{Lang:1993ct}
\\\hline
\texttt{Op[V,2,1]} & $\de^2\varphi^3_V$ &$$ & $[5-\frac32\eps]+0\eps N^0_{\cdots}$ & \eord1
\\
\texttt{Op[V,2,2]} & $\de^2\varphi^3_V$ &$\de^2\sigma\varphi$ & $[5-\frac32\eps]+\eps N^0_{\cdots}$& \eord1 & \Nord1\cite{Lang:1992zw} & $\texttt 8_2$
\\
\texttt{Op[V,2,3]} & $\de^2\varphi^5_V$ && $[7-\frac52\eps]+\eps N^0_{\cdots}$ &\eord1
\\
\texttt{Op[V,2,4]} & $\de^2\varphi^5_V$ && $[7-\frac52\eps]+2\eps N^0_{\cdots}$ & \eord1
\\
\texttt{Op[V,2,5]} & $\de^2\varphi^5_V$ && $[7-\frac52\eps]+2\eps N^0_{\cdots}$ & \eord1
\\\hline
\texttt{Op[V,3,1]} & $\de^3\varphi^3_V$ &$$ & $[6-\frac32\eps]+0\eps N^0_{\cdots}$ & \eord1
\\
\texttt{Op[V,3,2]} & $\de^3\varphi^3_V$ &$\de^3\sigma\varphi$ & $[6-\frac32\eps]+\eps N^0_{\cdots}$& \eord1 & \Nord1\cite{Lang:1992zw} & $\texttt 8_3$
\\\hline
\texttt{Op[V,4,1]} & $\de^4\varphi^3_V$ &$$ & $[7-\frac32\eps]+0\eps$ & \eord1
\\
\texttt{Op[V,4,2]} & $\de^4\varphi^3_V$ &$$ & $[7-\frac32\eps]+0\eps N^0_{\cdots}$ & \eord1
\\
\texttt{Op[V,4,3]} & $\de^4\varphi^3_V$ &$\de^4\sigma\varphi$ & $[7-\frac32\eps]+\eps N^0_{\cdots}$& \eord1 & \Nord1\cite{Lang:1992zw} & $\texttt 8_4$
\\\hline
\texttt{Op[V,5,1]} & $\de^5\varphi^3_V$ &$$ & $[8-\frac32\eps]+0\eps$ & \eord1
\\
\texttt{Op[V,5,2]} &$\de^5\varphi^3_V$ &$$ & $[8-\frac32\eps]+0\eps N^0_{\cdots}$ & \eord1
\\
\texttt{Op[V,5,3]} &$\de^5\varphi^3_V$ &$\de^5\sigma\varphi$ & $[8-\frac32\eps]+\eps N^0_{\cdots}$ & \eord1 & \Nord1\cite{Lang:1992zw} & $\texttt 8_5$
\\\hline
\texttt{Op[V,6,1]} & $\de^6\varphi^3_V$ &$$ & $[9-\frac32\eps]+0\eps$ & \eord1
\\
\texttt{Op[V,6,2]} & $\de^6\varphi^3_V$ &$$ & $[9-\frac32\eps]+0\eps$ & \eord1
\\
\texttt{Op[V,6,3]} &$\de^6\varphi^3_V$ &$$ & $[9-\frac32\eps]+0\eps N^0_{\cdots}$ & \eord1
\\
\texttt{Op[V,6,4]} &$\de^6\varphi^3_V$ &$\de^6\sigma\varphi$ & $[9-\frac32\eps]+\eps N^0_{\cdots}$ & \eord1 & \Nord1\cite{Lang:1992zw} & $\texttt 8_6$
\\\hline
\texttt{Op[V,7,1]} & $\de^7\varphi^3_V$ &$$ & $[10-\frac32\eps]+0\eps$ & \eord1
\\
\texttt{Op[V,7,2]} & $\de^7\varphi^3_V$ &$$ & $[10-\frac32\eps]+0\eps$ & \eord1
\\
\texttt{Op[V,7,3]} &$\de^7\varphi^3_V$ &$$ & $[10-\frac32\eps]+0\eps N^0_{\cdots}$ & \eord1
\\
\texttt{Op[V,7,4]} &$\de^7\varphi^3_V$ &$$ & $[10-\frac32\eps]+\eps N^0_{\cdots}$ & \eord1 & \Nord1\cite{Lang:1992zw} & $\texttt 8_7$
\\\hline
\multicolumn{7}{l}{* As noted in \cite{Vasiliev:1993ux}, the expression in \cite{Vasiliev:1982dc} contains a misprint in equation (22).}
\end{tabular}
}
\end{table}

\begin{table}[ht]
\centering
\caption{Operators in the $T$ representation. Note the double eigenvalue at \texttt{Op[T,0,7]} and \texttt{Op[T,0,8]}.} \label{tab:Toperators}
{\small
\renewcommand{\arraystretch}{1.25}
\begin{tabular}{cccllll}
$\O$ &$\O|\eps$&$\O|N$  & $\Delta^{(1)}_{4-\eps}$& $\Delta(\eps)$   & $\Delta(N)$& Family
\\\hline
\texttt{Op[T,0,1]} & $\varphi_T^2$ &$\varphi_T^2$ & $[2-\eps]+\frac2{N+8}\eps$ & 
\eord6\cite{Kompaniets:2019zes}   &
\Nord2\cite{Gracey:2002qa} & $\texttt 3_2$, $\texttt{10}_{0,2}$, $\texttt{11}_{2,0}$
\\
\texttt{Op[T,0,2]} & $\varphi_T^4$ &$\sigma\varphi^2_T$ & $[4-2\eps]+\frac{N+16}{N+8}\eps$ &  \eord6\cite{Bednyakov:2021ojn}    & \Nord1\cite{Ma:1974qh} &$\texttt{10}_{1,2}$
\\
\texttt{Op[T,0,3]} & $\square\varphi_T^4$  & 
 & $[6-2\eps]+\frac{N+4}{N+8}\eps$& \eord1   & &
\\
\texttt{Op[T,0,4]} & $\varphi_T^6$   &  $\sigma^2\varphi^2_T$
&$[6-3\eps]+\frac{2N+42}{N+8}\eps$&  \eord1 &  \Nord1\cite{Lang:1993ct}  & $\texttt{10}_{2,2}$
\\
\texttt{Op[T,0,5]} &  $\square^2\varphi^4_T$ & 
&  $[8-2\eps]+0\eps N^0_{\cdots}$ &\eord1& &
\\
\texttt{Op[T,0,6]} &  $\square^2\varphi^4_T$ &    
&$[8-2\eps]+\eps N^0_{\cdots}$ &\eord1& &
\\
\texttt{Op[T,0,7]} &  $\square\varphi^6_T$ & &  $[8-3\eps]+\frac{2N+22}{N+8}\eps$ &\eord1& &
\\
\texttt{Op[T,0,8]} &  $\square\varphi^6_T$ & &  $[8-3\eps]+\frac{2N+22}{N+8}\eps$ &\eord1& & 
\\
\texttt{Op[T,0,9]} &  $\varphi^8_T$ &  $\sigma^3\varphi^2_T$ 
&$[8-4\eps]+\frac{3N+80}{N+8}\eps$&  \eord1& \Nord1\cite{Lang:1993ct} & $\texttt{10}_{3,2}$
\\\hline
\texttt{Op[T,1,1]} & $\de\square\varphi^4_T$  & &$[7-2\eps]+\frac{3N+16}{3(N+8)}\eps $ &\eord1& & 
\\
\texttt{Op[T,1,2]} & $\de\varphi^6_T$  & &$[7-3\eps]+\frac{2N+30}{N+8}\eps $ &\eord1& &
\\\hline
\texttt{Op[T,2,1]} & $\mathcal J_{T,2}$ &$\mathcal J_{T,2}$ & $[4-\eps]+0\eps$ & 
\eord4\cite{Henriksson:2018myn}   & 
\Nord2\cite{Derkachov:1997ch}&  $\texttt 5_2$
\\
\texttt{Op[T,2,2]} & $\de^2\varphi^4_T$& $\de^2\varphi^4_T$ & $[6-2\eps]+0\eps N^0_{\cdots}$ &\eord1& &
\\
\texttt{Op[T,2,3]} & $\de^2\varphi^4_T$& & $[6-2\eps]+\eps N^0_{\cdots}$ &\eord1& &
\\
\texttt{Op[T,2,4]} & $\de^2\varphi^4_T$& & $[6-2\eps]+\eps N^0_{\cdots}$ &\eord1& &
\\\hline
\texttt{Op[T,3,1]} & $\de^3\varphi^4_T$& & $[7-2\eps]+0\eps N^0_{\cdots}$ &\eord1& &
\\
\texttt{Op[T,3,2]} & $\de^3\varphi^4_T$& & $[7-2\eps]+\eps N^0_{\cdots}$ &\eord1& &
\\\hline
\texttt{Op[T,4,1]} & $\mathcal J_{T,4}$ &$\mathcal J_{T,4}$ & $[6-\eps]+0\eps$ & \eord4\cite{Henriksson:2018myn}   &  
\Nord2\cite{Derkachov:1997ch}& $\texttt 5_4$
\\\hline
\texttt{Op[T,6,1]} &  $\mathcal J_{T,6}$ &$\mathcal J_{T,6}$ & $[8-\eps]+0\eps$ & \eord4\cite{Henriksson:2018myn}   &  
\Nord2\cite{Derkachov:1997ch}& $\texttt 5_6$
\\\hline
\end{tabular}
}
\end{table}

\begin{table}[ht]
\centering
\caption{Operators in the $A$ representation.}\label{tab:Aoperators}
{\small
\renewcommand{\arraystretch}{1.25}
\begin{tabular}{cccllll}
$\O$ &$\O|\eps$&$\O|N$  & $\Delta^{(1)}_{4-\eps}$& $\Delta(\eps)$   & $\Delta(N)$& Family
\\\hline
\texttt{Op[A,1,1]} & $J^\mu_{ij}$ &  $J^\mu_{ij}$ & $[3-\eps] $ & exact & exact & $\texttt 6_1$
\\
\texttt{Op[A,1,2]} & $\de\varphi^4_A$ && $[5-2\eps]+\frac{N+10}{N+8}\eps$ &\eord1& &
\\
\texttt{Op[A,1,3]} & $\de\square\varphi^4_A$ && $[7-2\eps]+0\eps N^0_{\ldots}$ &\eord1& &
\\
\texttt{Op[A,1,4]} & $\de\square\varphi^4_A$ && $[7-2\eps]+\eps N^0_{\ldots}$ &\eord1& &
\\
\texttt{Op[A,1,5]} & $\de\varphi^6_A$ && $[7-3\eps]+\frac{2N+32}{N+8}\eps$  &\eord1& &
\\\hline
\texttt{Op[A,3,1]} &  $\mathcal J_{A,3}$ &$\mathcal J_{A,3}$ & $[5-\eps]$  & \eord4\cite{Henriksson:2018myn}   &
\Nord2\cite{Derkachov:1997ch}& $\texttt 6_3$
\\
\texttt{Op[A,3,2]} &$\de^3\varphi^4_A$ && $[7-2\eps]+0\eps N^0_{\ldots}$ &\eord1& &
\\
\texttt{Op[A,3,3]} &$\de^3\varphi^4_A$ && $[7-2\eps]+\eps N^0_{\ldots}$ &\eord1& &
\\
\texttt{Op[A,3,4]} &$\de^3\varphi^4_A$ && $[7-2\eps]+\eps N^0_{\ldots}$ &\eord1& &
\\\hline
\texttt{Op[A,5,1]} &  $\mathcal J_{A,5}$ &$\mathcal J_{A,5}$ & $[7-\eps]$ &  \eord4\cite{Henriksson:2018myn}   &  
\Nord2\cite{Derkachov:1997ch} & $\texttt 6_5$
\\\hline
\end{tabular}
}
\end{table}

\subsection[Operators in other $\ON$ representations]{Operators in other $\boldsymbol{\ON}$ representations}

As the number of Young tableau boxes used to construct the $\ON$ representation increases, the number of different irreps increases rapidly. For the tables of primary operators, we focus our attention some more interesting irreps, where we include a larger selection of operators compared to other irreps. These are primarily the irreps that appear in the tensor product of traceless-symmetric irreps, since these representations contain the leading scalar operators in the spectrum. By the tensor products \eqref{eq:tensorprod3} and \eqref{eq:tensorprod6}, this implies a focus on the representations $T_3$, $H_3$, $T_4$, $H_4$ and $B_4$. Moreover, for operators constructed out of $m$ fields in the rank $m$ traceless-symmetric representation $T_m$, the systematic procedures of \cite{Kehrein:1995ia} and \cite{Derkachov:1997qv} facilitate a systematic determination of the order $\eps^2$ and $1/N$ anomalous dimensions, and we therefore present a larger number of such operators.

\begin{table}[ht]
\centering
\caption{Some operators in the rank $3$ traceless-symmetric representation.}\label{tab:T3}
{\small
\renewcommand{\arraystretch}{1.25}
\begin{tabular}{cccllll}
$\O$ &$\O|\eps$&$\O|N$  & $\Delta^{(1)}_{4-\eps}$& $\Delta(\eps)$   & $\Delta(N)$& Family
\\\hline
\texttt{Op[Tm[3],0,1]} & $\varphi^3_{T_3} $& $\varphi^3_{T_3} $ &$[3-\frac32\eps]+\frac{6}{N+8}\eps$& 
\eord6\cite{Bednyakov:2021ojn} & 
\Nord2\cite{Derkachov:1997ch}&  $\texttt 3_3$, $\texttt{10}_{0,3}$, $\texttt{11}_{3,0}$
\\
\texttt{Op[Tm[3],0,2]} & $\varphi^5_{T_3} $& $\sigma\varphi^3_{T_3} $ &$[5-\frac52\eps]+\frac{N+26}{N+8}\eps$& \eord1& \Nord1\cite{Lang:1993ct}& $\texttt{10}_{1,3}$
\\\hline
\texttt{Op[Tm[3],1,1]} & $\de\varphi^5_{T_3}$& $\de\sigma\varphi^3_{T_3}$&$[6-\frac52\eps]+\frac{N+16}{N+8}\eps$  & \eord1  &  &
\\\hline
\texttt{Op[Tm[3],2,1]} & $\de^2\varphi^3_{T_3}$& $\de^2\varphi^3_{T_3}$&$[5-\frac32\eps]+\frac{10}{3(N+8)}\eps$  &  \eord2\cite{Kehrein:1995ia}  &  \Nord1\cite{Derkachov:1997qv,ThisPaper}&
\\\hline
\texttt{Op[Tm[3],3,1]} & $\de^3\varphi^3_{T_3}$&$\de^3\varphi^3_{T_3}$ & $[6-\frac32\eps]+\frac{1}{N+8}\eps$ &  \eord2\cite{Kehrein:1995ia}  &  \Nord1\cite{Derkachov:1997qv,ThisPaper}&
\\\hline
\texttt{Op[Tm[3],4,1]} & $\de^4\varphi^3_{T_3}$&$\de^4\varphi^3_{T_3}$ & $[7-\frac32\eps]+\frac{14}{5(N+8)}\eps$ &  \eord2\cite{Kehrein:1995ia}  &  \Nord1\cite{Derkachov:1997qv,ThisPaper}&
\\\hline
\texttt{Op[Tm[3],5,1]} & $\de^5\varphi^3_{T_3}$&$\de^5\varphi^3_{T_3}$ & $[8-\frac32\eps]+\frac{4}{3(N+8)}\eps$ &  \eord2\cite{Kehrein:1995ia}  &  \Nord1\cite{Derkachov:1997qv,ThisPaper}&
\\\hline
\texttt{Op[Tm[3],6,1]} & $\de^6\varphi^3_{T_3}$&$\de^6\varphi^3_{T_3}$ & $[9-\frac32\eps]+0\eps$ &  \eord2\cite{Kehrein:1995ia}  & \Nord1\cite{Derkachov:1997qv,ThisPaper} &
\\
\texttt{Op[Tm[3],6,2]} &$\de^6\varphi^3_{T_3}$&$\de^6\varphi^3_{T_3}$ & $[9-\frac32\eps]+\frac{18}{7(N+8)}\eps$ &  \eord2\cite{Kehrein:1995ia}  & \Nord1\cite{Derkachov:1997qv,ThisPaper} &
\\\hline
\texttt{Op[Tm[3],7,1]} & $\de^7\varphi^3_{T_3}$&$\de^7\varphi^3_{T_3}$ & $[10-\frac32\eps]+\frac3{2(N+8)}\eps$ &  \eord2\cite{Kehrein:1995ia,ThisPaper}  &  \Nord1\cite{Derkachov:1997qv,ThisPaper}  &
\\\hline
\end{tabular}
}
\end{table}

\begin{table}[ht]
\centering
\caption{Some operators in the rank $4$ traceless-symmetric representation. Writing ``\cite{Kehrein:1995ia,ThisPaper}'' means that this particular result was not listed in \cite{Kehrein:1995ia}, but has been computed here using the method of that paper, and likewise for ``\cite{Derkachov:1997qv,ThisPaper}''.}\label{tab:T4}
{\small
\renewcommand{\arraystretch}{1.25}
\begin{tabular}{cccllll}
$\O$ &$\O|\eps$&$\O|N$  & $\Delta^{(1)}_{4-\eps}$& $\Delta(\eps)$   & $\Delta(N)$& Family
\\\hline
\texttt{Op[Tm[4],0,1]} & $\varphi^4_{T_4} $& $\varphi^4_{T_4} $ &$[4-2\eps]+\frac{12}{N+8}\eps$&
\eord6\cite{Bednyakov:2021ojn}
 &  
\Nord2\cite{Derkachov:1997ch} & $\texttt 3_4$, $\texttt{10}_{0,4}$, $\texttt{11}_{4,0}$
\\
\texttt{Op[Tm[4],0,2]} & $\varphi^6_{T_4}$ & $\sigma\varphi^4_{T_4}$ & $[6-3\eps]+\frac{N+38}{N+8}\eps$ & \eord1 &   \Nord1\cite{Lang:1993ct}& $\texttt{10}_{1,4}$
\\\hline
\texttt{Op[Tm[4],2,1]} & $\de^2\varphi^4_{T_4}$& $\de^2\varphi^4_{T_4}$ & $[6-2\eps]+\frac{26}{3(N+8)}\eps$ &  \eord2\cite{Kehrein:1995ia}  &  \Nord1\cite{Derkachov:1997qv,ThisPaper}&
\\\hline
\texttt{Op[Tm[4],3,1]} & $\de^3\varphi^4_{T_4}$&$\de^3\varphi^4_{T_4}$ & $[7-2\eps]+\frac6{N+8}\eps$ &  \eord2\cite{Kehrein:1995ia}  &  \Nord1\cite{Derkachov:1997qv,ThisPaper}&
\\\hline
\texttt{Op[Tm[4],4,1]} & $\de^4\varphi^4_{T_4}$&$\de^4\varphi^4_{T_4}$ & $[8-2\eps]+\frac{8}{3(N+8)}\eps$ &  \eord2\cite{Kehrein:1995ia,ThisPaper}  &  \Nord1\cite{Derkachov:1997qv,ThisPaper}&
\\
\texttt{Op[Tm[4],4,2]} & $\de^4\varphi^4_{T_4}$&$\de^4\varphi^4_{T_4}$ & $[8-2\eps]+\frac{38}{5(N+8)}\eps$ &  \eord2\cite{Kehrein:1995ia,ThisPaper}  &  \Nord1\cite{Derkachov:1997qv,ThisPaper}&
\\\hline
\texttt{Op[Tm[4],5,1]} & $\de^5\varphi^4_{T_4}$&$\de^5\varphi^5_{T_4}$ & $[9-2\eps]+\frac6{N+8}\eps$ &  \eord2\cite{Kehrein:1995ia,ThisPaper}  & \Nord1\cite{Derkachov:1997qv,ThisPaper} &
\\\hline
\texttt{Op[Tm[4],6,1]} & $\de^6\varphi^4_{T_4}$&$\de^6\varphi^4_{T_4}$ & $[10-2\eps]+ 0\eps N^0_{\cdots}$ &  \eord2\cite{Kehrein:1995ia,ThisPaper}  & \Nord1\cite{Derkachov:1997qv,ThisPaper} &
\\
\texttt{Op[Tm[4],6,2]} & $\de^6\varphi^4_{T_4}$&$\de^6\varphi^4_{T_4}$ & $[10-2\eps]+ 0\eps N^0_{\cdots}$ &  \eord2\cite{Kehrein:1995ia,ThisPaper}  & \Nord1\cite{Derkachov:1997qv,ThisPaper} &
\\
\texttt{Op[Tm[4],6,3]} & $\de^6\varphi^4_{T_4}$&$\de^6\varphi^4_{T_4}$ & $[10-2\eps]+ 0\eps N^0_{\cdots}$ &  \eord2\cite{Kehrein:1995ia,ThisPaper}  & \Nord1\cite{Derkachov:1997qv,ThisPaper} &
\\\hline
\end{tabular}
}
\end{table}

In tables~\ref{tab:T3} and \ref{tab:T4}, we present a selection of operators in the traceless-symmetric $T_3$ and $T_4$ representations.
To order $\eps$, the structure of these operators is the same as for operators of $\phi^m$ type for $N=1$, but the order $\eps$ anomalous dimension is multiplied by $\frac6{N+8}$. The exception is that for the $T_4$ representation there is now one additional sequence of operators, with odd spin, four fields and anomalous dimensions $\frac6{N+8}$, which were absent in the $N=1$ case, since for $N=1$ they are descendants of broken currents.

For the leading scalar $T_3$ operator \texttt{Op[Tm[3],0,1]}, we report \eord3\cite{Wallace1975}\eord4\cite{Antipin:2020abu}\eord6\cite{Bednyakov:2021ojn}\footnote{Like the order $\eps^6$ results for \texttt{Op[T,0,2]} and \texttt{Op[Tm[4],0,1]}, this result was extracted from the more general results of \cite{Bednyakov:2021ojn}. I thank the authors of \cite{Bednyakov:2021ojn} for sharing these results in computer-readable format.} and \Nord1\cite{Lang:1990ni}\Nord2\cite{Derkachov:1997ch}. Numerical values in $d=3$ are $\Delta_{\varphi^3_{T_3}}=2.1086(3)$  for $N=2$ \cite{Chester:2019ifh}, $\Delta_{\varphi^3_{T_3}}=2.0384(10)$ for $N=3$ \cite{Hasenbusch2011}, and $\Delta_{\varphi^3_{T_3}}=1.9768(10)$ for $N=4$ \cite{Hasenbusch2011}.

\begin{table}[ht]
\centering
\caption{Scalar operators in various $\ON$ representations.}\label{tab:otherONoperators}
{\small
\renewcommand{\arraystretch}{1.25}
\begin{tabular}{cccllll}
$\O$ &$\O|\eps$&$\O|N$  & $\Delta^{(1)}_{4-\eps}$& $\Delta(\eps)$   & $\Delta(N)$& Family
\\\hline
\texttt{Op[B4,0,1]} & $\square\varphi^4_{B_4}$ & $\square\varphi^4_{B_4}$ & $[6-2\eps]+\frac{6}{N+8}\eps$  & \eord1&& 
\\\hline 
\texttt{Op[Tm[5],0,1]} & $\varphi^5_{T_5} $& $\varphi^5_{T_5} $ & $[5-\tfrac52\eps]+\frac{20}{N+8}\eps$ & 
 \eord5\cite{ThisPaper,Bednyakov:2021ojn,Antipin:2020abu}
 &
\Nord2\cite{Derkachov:1997ch}& $\texttt 3_5$, $\texttt{10}_{0,5}$, $\texttt{11}_{5,0}$
\\\hline 
\texttt{Op[Tm[6],0,1]} & $\varphi^6_{T_6} $& $\varphi^6_{T_6} $ &   $[6-3\eps]+\frac{30}{N+8}\eps$  &
\eord5\cite{ThisPaper,Bednyakov:2021ojn,Antipin:2020abu}& 
\Nord2\cite{Derkachov:1997ch}& $\texttt 3_6$, $\texttt{10}_{0,6}$, $\texttt{11}_{6,0}$
\\\hline 
\end{tabular}
}
\end{table}

\begin{table}[ht] 
\centering
\caption{Spinning operators in various $\ON$ representations.}\label{tab:otherONspinning}
{\small
\renewcommand{\arraystretch}{1.25}
\begin{tabular}{cccllll}
$\O$ &$\O|\eps$&$\O|N$  & $\Delta^{(1)}_{4-\eps}$& $\Delta(\eps)$   & $\Delta(N)$& Family
\\\hline
\texttt{Op[Hm[3],1,1]} & $\de\varphi^3_{H_3}$ & $\de\varphi^3_{H_3}$ & $[4-\frac32\eps]+\frac3{N+8}\eps$ & \eord1 & \Nord1\cite{Lang:1990re} & $\texttt{11}_{2,1}$
\\
\texttt{Op[Hm[3],1,2]} & $\de\varphi^5_{H_3}$ & $\de\sigma\varphi^3_{H_3}$ & $[6-\frac52\eps]+\frac{N+19}{N+8}\eps$ & \eord1 &  & 
\\
\texttt{Op[Hm[3],2,1]} & $\de^2\varphi^3_{H_3}$ & $\de^2\varphi^3_{H_3}$ & $[5-\frac32\eps]+\frac4{3(N+8)}\eps$ & \eord1 &
\\
\texttt{Op[Hm[3],3,1]} & $\de^3\varphi^3_{H_3}$ & $\de^3\varphi^3_{H_3}$ & $[6-\frac32\eps]+\frac5{2(N+8)}\eps$ & \eord1 &
\\
\texttt{Op[Hm[3],4,1]} & $\de^4\varphi^3_{H_3}$ & $\de^4\varphi^3_{H_3}$ & $[7-\frac32\eps]+0\eps$ & \eord1 &
\\
\texttt{Op[Hm[3],4,2]} & $\de^4\varphi^3_{H_3}$ & $\de^4\varphi^3_{H_3}$ & $[7-\frac32\eps]+\frac8{5(N+8)}\eps$ & \eord1 &
\\
\texttt{Op[Hm[3],5,1]} & $\de^5\varphi^3_{H_3}$ & $\de^5\varphi^3_{H_3}$ & $[8-\frac32\eps]+0\eps$ &  \eord1 &  & 
\\
\texttt{Op[Hm[3],5,2]} & $\de^5\varphi^3_{H_3}$ & $\de^5\varphi^3_{H_3}$ & $[8-\frac32\eps]+\frac7{3(N+8)}\eps$ & \eord1 &
\\
\texttt{Op[Hm[3],6,1]} & $\de^6\varphi^3_{H_3}$ & $\de^6\varphi^3_{H_3}$ & $[9-\frac32\eps]+0\eps$ & \eord1 &
\\
\texttt{Op[Hm[3],6,2]} & $\de^6\varphi^3_{H_3}$ & $\de^6\varphi^3_{H_3}$ & $[9-\frac32\eps]+\frac{12}{7(N+8)}\eps$ & \eord1 &
\\\hline
\texttt{Op[A3,3,1]} & $\de^{3}\varphi^3_{A_3}$& $\de^{3}\varphi^3_{A_3}$ & $[6-\frac32\eps]+0\eps$ & \eord1 && 
\\
\texttt{Op[A3,5,1]} &$\de^{5}\varphi^3_{A_3}$& $\de^{5}\varphi^3_{A_3}$ & $[8-\frac32\eps]+0\eps$ & \eord1 && 
\\
\texttt{Op[A3,6,1]} &
$\de^{6}\varphi^3_{A_3}$& $\de^{6}\varphi^3_{A_3}$ & $[9-\frac32\eps]+0\eps$ & \eord1 && 
\\\hline
\texttt{Op[Hm[4],1,1]} & $\de\varphi^4_{H_4}$ &$\de\varphi^4_{H_4}$ & $[5-2\eps]+\frac{8}{N+8}\eps$ &\eord1& \Nord1\cite{Lang:1990re} & $\texttt{11}_{3,1}$
\\
\texttt{Op[Hm[4],2,1]} & $\de^2\varphi^4_{H_4}$ &$\de^2\varphi^4_{H_4}$ & $[6-2\eps]+\frac6{N+8}\eps$ &\eord1&
\\\hline
\texttt{Op[B4,2,1]} &$\de^2\varphi^4_{B_4}$ &$\de^2\varphi^4_{B_4}$ & $[6-2\eps]+\frac{14}{3(N+8)}\eps$ &\eord1& \Nord1\cite{Lang:1990re} & $\texttt{11}_{2,2}$
\\\hline
\texttt{Op[Hm[5],1,1]} &$\de\varphi^5_{H_5}$ &$\de\varphi^5_{H_5}$ & $[6-\frac52\eps]+\frac{15}{N+8}\eps$  &\eord1&   \Nord1\cite{Lang:1990re} & $\texttt{11}_{4,1}$
\\\hline
\end{tabular}
}
\end{table}

In tables~\ref{tab:otherONoperators} and \ref{tab:otherONspinning}, we give a selection of scalar and spinning operators respectively, transforming in various $\ON$ representations. With the inclusion of these operators, the tables of this report include all operators in traceless-symmetric Lorentz representations of $\Delta^{\mathrm{4d}}\leqslant6$.
In \cite{Kehrein:1992fn} it was noted that any operator constructed out of $m$ fields in the $A_m$ representation has vanishing anomalous dimension at order $\eps$.

\subsection{Operators in non-traceless-symmetric Lorentz representations}
\label{sec:nonTSLorentz}

When considering operators in non-traceless-symmetric Lorentz representation for general $N$, the complexity of the spectrum quickly increases. This can be seen for instance by performing the character counting as described in appendix~\ref{app:charactergenN}. 
Here we will just mention a few such operators, in order to give a complete presentation of the spectrum up to $\Delta^{\mathrm{4d}}\leqslant6$.

\begin{table}
\centering
\caption{$\ON$ singlet and Lorentz non-traceless-symmetric operators up to 4d dimension $8$. The 4d irrep notation is the one used in \cite{Kehrein:1994ff}.}\label{tab:SingletNonTS}
{\small
\renewcommand{\arraystretch}{1.25}
\begin{tabular}{|cccc|} 
\hline
Lorentz YT & 4d irrep  & Field content & $\Delta^{(1)}_{4-\eps}$   \\
\hline
 \raisebox{-3pt}{\tiny\yng(2,1)}&  $(\frac32,\frac12)+(\frac12,\frac32)$ &  $\de^3\varphi^4_S$ & $[7-2\eps]+\frac{N+2}{N+8}\eps$ 
\\\hline
 \raisebox{-3pt}{\tiny\yng(3,1)}&  $(2,1)+(1,2)$ &  $\de^4\varphi^4_S$ & $[8-2\eps]+\frac{N+2}{N+8}\eps$ 
\\
 \raisebox{-3pt}{\tiny\yng(2,2)}&  $(2,0)+(0,2)$ &  $\de^4\varphi^4_S$ & $[8-2\eps]+\frac{3 N+20-\sqrt{9 N^2+64 N+288}}{6 (N+8)}\eps$ 
\\
 \raisebox{-3pt}{\tiny\yng(2,2)}&  $(2,0)+(0,2)$ &  $\de^4\varphi^4_S$ & $[8-2\eps]+\frac{3 N+20+\sqrt{9 N^2+64 N+288}}{6 (N+8)}\eps$ 
\\\hline
\end{tabular}
}
\end{table}

\begin{description}
\item[Dimension $\boldsymbol{\Delta=5}$] One operator in $\ON$ irrep $A_3$, Lorentz irrep \raisebox{-3pt}{\tiny\yng(1,1)}: $\varphi^{[i}\de_{[\mu}\varphi^{j}\de_{\nu]}\varphi^{k]}$, dimension $\Delta=[5-\frac32\eps]+O(\eps^2)$.
\item[Dimension $\boldsymbol{\Delta=6}$] Two operators constructed out of three fields and three gradients: $\ON$ irreps $V$ and $H_3$, both in Lorentz irrep \raisebox{-3pt}{\tiny\yng(2,1)}, and both with dimension $\Delta=[6-\frac32\eps]+O(\eps^2)$. Moreover, two operators constructed out of four fields and two gradients, both in Lorentz irrep \raisebox{-3pt}{\tiny\yng(1,1)}: $\ON$ irrep $A$, dimension $\Delta=[6-2\eps]+\frac{N+2}{N+8}\eps+O(\eps^2)$; $\ON$ irrep $Y_{2,1,1}$, dimension $\Delta=[6-2\eps]+\frac4{N+8}\eps+O(\eps^2)$.
\end{description}
In table~\ref{tab:SingletNonTS} we give the leading $\ON$ singlet operators in non-traceless-symmetric $\ON$ representations, up to $\Delta^{\mathrm{4d}}\leqslant 8$.

Operators in the Lorentz representation labelled by the Young tableau ${\tiny \yng(1,1,1)}$ are related to parity-odd scalars in the three-dimensional theory.
In the free three-dimensional $\ON$ symmetric CFT, one can for generic $N>1$ construct a parity odd singlet scalar out of four fields and five gradients, giving (in the free theory) dimension $7$ \cite{Dymarsky:2017xzb}, equation 3.1. It has the form $\epsilon^{\mu\nu\rho}\de_\mu\de_\nu\de_\rho \square \varphi^4_S$. The corresponding operator in the $4-\eps$ expansion has dimension $[9-2\eps]+0\eps+O(\eps^2)$.

\subsection[OPE coefficients in the $\varphi\times\varphi$ OPE]{OPE coefficients in the $\boldsymbol{\varphi\times\varphi}$ OPE}

In table~\ref{tab:OPEgenN} we present OPE coefficients $\lambda^2_{\varphi\varphi\O}$ for a selection of operators $\O$. In addition to individual operators, the OPE coefficients for operators in the families of weakly broken currents $\mathcal J_{R,\ell}$ (\texttt{ONF4[$l$]},\texttt{ONF5[$l$]},\texttt{ONF6[$l$]}) have been computed to high order. For all $R=S,T,A$, the order $\eps^3$ computation was performed in \cite{Dey:2016mcs} and the order $\eps^4$ in \cite{Henriksson:2018myn}. In the large $N$ expansion, the order $1/N$ correction was found for $R=T,A$ in \cite{Dey:2016mcs}\footnote{However, they can be extracted using expressions given already in \cite{Lang:1991kp}. We have not found any reference to such computation before the work of \cite{Dey:2016mcs}, and we write \cite{Lang:1991kp,Dey:2016mcs} in the tables.} and for $R=S$ in \cite{Alday:2019clp}. In principle, the order $1/N^2$ OPE coefficients for the $T$ and $A$ irreps can be determined numerically by evaluating the computation in \cite{Alday:2019clp} to high precision. This was done only in the case $\mathcal J_{A,1}$, to find the $1/N^2$ correction to $C_J$.

\begin{table}[ht]
\centering
\caption{OPE coefficients in the $\varphi\times\varphi$ OPE for operators for general $N$.}\label{tab:OPEgenN}
{\small
\renewcommand{\arraystretch}{1.25}
\begin{tabular}{cccclcl}
$\O$ &$\O|\eps$ & $\O|N$ & $\lambda_{\varphi\varphi\O}^2(\eps)$& $\lambda_{\varphi\varphi\O}^2(\eps)$ & $\lambda_{\varphi\varphi\O}^2(N)$ & $\lambda_{\varphi\varphi\O}^2(N)$ 
\\\hline
$\varphi,\varphi,\texttt{Op[S,0,0]}$ & $\1$ & $\1$  &$1$ & exact & $1$ & exact
\\
$\varphi,\varphi,\texttt{Op[S,0,1]}$ & $\varphi^2_S$ & $\sigma$ & $O(1)$ & \eord3\cite{Dey:2016mcs} & $O(N^{-1})$ & \Nord2\cite{Lang:1993ct}
\\
$\varphi,\varphi,\texttt{Op[S,0,2]}$ & $\varphi^4_S$ & $\sigma^2$  & $O(\eps^2)$& \eord2\cite{Henriksson:2018myn} & $O(N^{-2})$ & \Nord2\cite{Alday:2019clp}
\\
$\varphi,\varphi,\texttt{Op[S,0,4]}$ & $\varphi^6_S$ & $\sigma^3$  & $O(\eps^4)$& \eord4\cite{Padayasi:2021sik} & 
\\\hline
$\varphi,\varphi,\texttt{Op[S,2,1]}$ & $T^{\mu\nu}$ & $T^{\mu\nu}$   & $O(1)$ & 
\eord4\cite{Henriksson:2018myn}& $O(N^{-1})$ & \Nord2\cite{Cappelli:1990yc}
\\
$\varphi,\varphi,\texttt{Op[S,2,2]}$ & $\de^2\varphi^4_S$ &    & $O(\eps^2)$ & \eord2\cite{ThisPaper,Henriksson:2018myn} & &
\\
$\varphi,\varphi,\texttt{Op[S,2,3]}$ & $\de^2\varphi^4_S$ & $[\sigma,\sigma]_{0,2}$   & $O(\eps^2)$ & \eord2\cite{ThisPaper,Henriksson:2018myn} & $O(N^{-2})$& \Nord2\cite{Alday:2019clp}
\\\hline
$\varphi,\varphi,\texttt{Op[S,4,1]}$ & $\mathcal J_{S,4}$ & $\mathcal J_{S,4}$ & $O(1)$& 
\eord4\cite{Henriksson:2018myn} & $O(N^{-1})$ & \Nord2\cite{Alday:2019clp}
\\\hline
$\varphi,\varphi,\texttt{Op[T,0,1]}$ & $\varphi^2_T$ &  $\varphi^2_T$  & $O(1)$&  \eord3\cite{Dey:2016mcs} & $O(1)$& \Nord1\cite{Lang:1991kp,Dey:2016mcs}
\\
$\varphi,\varphi,\texttt{Op[T,0,2]}$ & $\varphi^4_T$ &   & $O(\eps^2)$ & \eord2\cite{Henriksson:2018myn}& $O(N^{-1})$ & \Nord1\cite{Lang:1991kp,ThisPaper}
\\\hline
$\varphi,\varphi,\texttt{Op[T,2,1]}$ & $\mathcal J_{T,2}$ & $\mathcal J_{T,2}$ & $O(1)$& 
\eord4\cite{Henriksson:2018myn} & $O(1)$& \Nord1\cite{Lang:1991kp,Dey:2016mcs}
\\\hline
$\varphi,\varphi,\texttt{Op[T,4,1]}$ & $\mathcal J_{T,4}$ & $\mathcal J_{T,4}$ & $O(1)$& 
\eord4\cite{Henriksson:2018myn} & $O(1)$& \Nord1\cite{Lang:1991kp,Dey:2016mcs}
\\\hline
$\varphi,\varphi,\texttt{Op[A,1,1]}$ & $ J^\mu$ &  $ J^\mu$   & $O(1)$& 
\eord4\cite{Henriksson:2018myn} & 
 $O(1)$ &
\Nord1\cite{Lang:1992pp}\Nord2(num)\cite{Alday:2019clp}
\\
$\varphi,\varphi,\texttt{Op[A,1,2]}$ & $\de\varphi^4_A$ &  & $O(\eps^2)$  & \eord2\cite{Henriksson:2018myn} & $O(N^{-1})$ &  \Nord1\cite{Lang:1991kp,ThisPaper}
\\\hline
$\varphi,\varphi,\texttt{Op[A,3,1]}$ & $\mathcal J_{A,3}$ & $\mathcal J_{A,3}$ & $O(1)$& 
\eord4\cite{Henriksson:2018myn} & $O(1)$& \Nord1\cite{Lang:1991kp,Dey:2016mcs}
\\\hline
\end{tabular}
}
\end{table}

In \cite{Padayasi:2021sik}, a general formula was conjectured for the OPE coefficient of $\varphi^{2k}_S$ (\texttt{ONF1[$k$]}) in the $\varphi\times\varphi$ OPE,
\begin{equation}
\label{eq:opephi2khyp}
\lambda^2_{\varphi\varphi\varphi^{2k}_S}\overset ?=\frac{64k\Gamma(k)^5\left(\frac N2\right)_k}{2^{2k}\Gamma(2k-1)^2N^2(N+8)^{2k-2}}\eps^{2k-2},\quad k>1.
\end{equation}
For $k=2$ and $k=3$ the formula is known to be correct.

For operators in the $T$ and $A$ representations, the normalisation of the OPE coefficients depends on the normalisation used for the tensor structures. In appendix~\ref{app:normON} we describe the normalisations used here, which are compatible with $\lambda_{\O_1\O_2\O_3}=\lambda_{\O_{\sigma(1)}\O_{\sigma(2)}\O_{\sigma(3)}}$ for any permutation $\sigma$. They differ from those used in other parts of the literature. Our conventions are such that the OPE coefficients for operators in the theory of $N$ free scalars take the form
\begin{equation}
\lambda^2_{\varphi\varphi\mathcal J_{R,\ell}}=\mathcal N^R_{VVVV}\left.\lambda^2_{\phi\phi\mathcal J_\ell}\right|_{\mathrm{free}}
, 
\end{equation}
for $\lambda^2_{\phi\phi\mathcal J_\ell}\big|_{\mathrm{free}}$ defined in \eqref{eq:freeOPE} with spin $\ell$ even for $S$ and $T$ and odd for $A$. The normalisation constants, given by \eqref{eq:normconst}, are
\begin{equation}
\mathcal N^S_{VVVV}=\frac1N,
\qquad
\mathcal N^T_{VVVV}=\frac{\sqrt{(N+1)(N-2)/2}}N,
\qquad
\mathcal N^A_{VVVV}=\frac{\sqrt{N(N-1)/2}}N.
\end{equation}
For the results at large $N$ in the ancillary data file,  in the case of the representations $R=T$ and $R=A$, the factor $\mathcal N_{VVVV}^R$ has been factored out, and thus not been expanded in a series in $1/N$

\subsection{Other OPE coefficients}

In table~\ref{tab:OPEotherON} we present a small selection of OPE coefficients for general $N$ that do not include two $\varphi$ operators.
In the $\eps$-expansion, a collection of OPE coefficients of this type were reported in \cite{Bertucci:2022ptt}.
In \cite{Chai:2021uhv}, the OPE coefficients $\lambda_{\sigma\sigma\sigma^2}$ and $\lambda_{\sigma^2\sigma^2\sigma}$ were evaluated at large $N$ to subleading order in three dimensions.
Numerical values of several low-lying OPE coefficients in three dimensions for were determined using the conformal bootstrap for the cases $N=2$ \cite{Liu:2020tpf,Chester:2019ifh} and $N=3$ \cite{Chester:2020iyt}.

\begin{table}[ht]
\centering
\caption{Other OPE coefficients implemented for general $N$.}\label{tab:OPEotherON}
{\small
\renewcommand{\arraystretch}{1.25}
\begin{tabular}{cclll}

$\O_1,\O_2,\O_3$ &$\O_1,\O_2,\O_3|_\eps$ &$\O_1,\O_2,\O_3|_N$ & $\lambda_{\O_1\O_2\O_3}^2|_\eps$ & $\lambda^2_{\O_1\O_2\O_3}|_{N}$  
\\\hline
\texttt{Op[S,0,1],Op[S,0,1],Op[S,0,1]} &$\varphi^2_S,\varphi^2_S,\varphi^2_S$ & $\sigma,\sigma,\sigma$ & \eord1\cite{Thesis}  & \Nord2\cite{Goykhman:2019kcj}
\\
\texttt{Op[S,0,1],Op[S,0,1],Op[S,0,2]} &$\varphi^2_S,\varphi^2_S,\varphi^4_S$ & $\sigma,\sigma,\sigma^2$ &   \eord1\cite{Bertucci:2022ptt}   & \Nord1(3d)\cite{Chai:2021uhv}
\\
\hline
\texttt{Op[T,0,1],Op[T,0,1],Op[S,0,1]} & $\varphi^2_T,\varphi^2_T,\varphi^2_S$ &  $\varphi^2_T,\varphi^2_T,\sigma$ &
   \eord1\cite{Bertucci:2022ptt}   &
\\
\texttt{Op[T,0,1],Op[T,0,1],Op[S,0,2]} & $\varphi^2_T,\varphi^2_T,\varphi^4_S$& $\varphi^2_T,\varphi^2_T,\sigma^2$ & 
  \eord1\cite{Bertucci:2022ptt}   &
\\
\texttt{Op[T,0,1],Op[T,0,1],Op[T,0,1]} & $\varphi^2_T,\varphi^2_T,\varphi^2_T$ &   $\varphi^2_T,\varphi^2_T,\varphi^2_T$  &
  \eord1\cite{Bertucci:2022ptt}   &
\\ 
\texttt{Op[T,0,1],Op[T,0,1],Op[T,0,2]} & $\varphi^2_T,\varphi^2_T,\varphi^4_T$ &   $\varphi^2_T,\varphi^2_T,\sigma\varphi^2_T$  &
    \eord1\cite{Bertucci:2022ptt}   &
\\
\texttt{Op[T,0,1],Op[T,0,1],Op[Tm[4],0,1]} & $\varphi^2_T,\varphi^2_T,\varphi^4_{T_4}$ & $\varphi^2_T,\varphi^2_T,\varphi^4_{T_4}$ 
& 
    \eord1\cite{Bertucci:2022ptt}   &
\\
\texttt{Op[T,0,1],Op[T,0,1],Op[B4,0,1]} & $\varphi^2_T,\varphi^2_T,\square\varphi^4_{B_4}$ & $\varphi^2_T,\varphi^2_T,\square\varphi^4_{B_4}$ 
& 
    \eord1\cite{Bertucci:2022ptt}   &
\\
\hline
\end{tabular}
}
\end{table}

\section{Discussion}

In this report we have given a presentation of the critical $\ON$ CFT as a family of CFTs parametrised by the spacetime dimension $d$ and the number of components $N$. In the presentation, we have taken point-of-view of conformal field theory, and emphasised the results for conformal data in the various expansion limits -- in particular the $\eps$-expansion and the large $N$ expansion. The results extracted from the literature have been supplemented by the determination of order $\eps$ anomalous dimensions for a large set of operators, and the spectrum in the limit $\eps\to0$ has been checked by the character decomposition described in appendix~\ref{app:characters}.

In various places, we have aimed to articulate the often implicit assumption of spectrum continuity, introduced in section~\ref{sec:spectrumcontinuity}. In its strong form -- that the whole set of CFT-data varies continuously with $d$ and $N$ -- spectrum continuity is still far from being verified and requires further attention. The discovery of evanescent operators \cite{Hogervorst:2015akt} (section~\ref{eq:unitaritviolating}) and the formalisation of the $\ON$ representation theory in terms of Deligne categories \cite{Binder:2019zqc} (sections~\ref{sec:spectrumcontinuity} and \ref{sec:reprSOd}) are important steps towards an improved understanding. The most intriguing region of the parameter space with respect to spectrum continuity is the limit $d\to2$, which has recently been investigated by the numerical bootstrap \cite{Cappelli:2018vir}, non-perturbative RG \cite{Chlebicki:2020pvo} and direct crossing analysis \cite{Li:2021uki}. Moreover, a complete dictionary between composite operators in the non-linear sigma model and in the large $N$ expansion is missing (section~\ref{sec:NLSMmethod}).

Assuming spectrum continuity, the ordering of operators in the bulk of the $d,N$ plane is not expected to change, by level repulsion arguments. This picture can break down at large $N$, where in fact operator dimensions determined to order $1/N$ do cross and the ordering of operators may change. In such an expansion, the level repulsion may be studied perturbatively, as was done for 4d $\mathcal N=4$ super-Yang--Mills theory in \cite{Korchemsky:2015cyx}, however to look at this in the $\ON$ CFT would require the computation of additional perturbative data. A specific case that would be interesting to investigate is the mixing of $\texttt{Op[S,0,5]}=\de^4\varphi^4_S$ of dimension $2d+O(N^{-1})$ with the two operators $\texttt{Op[S,0,3]}=\square\sigma^2$ and $\texttt{Op[S,0,4]}=\sigma^3$ of dimension $6+O(N^{-1})$. All these three operators become approximately degenerate in $d=3$. Unfortunately, the order $N^{-1}$ anomalous dimension of \texttt{Op[S,0,5]} is not known. 

Over the years, a large collection of conformal data has been computed. As described in section~\ref{sec:orderepssystematics} and heavily used in this report, in the $4-\eps$ expansion there is a systematic procedure for finding the leading anomalous dimension of arbitrary operators. In the large $N$ expansion, such procedure only exists for operators constructed out of $m$ fields in the completely traceless-symmetric representation $T_m$ \cite{Derkachov:1997qv}, so other operators must be given a careful treatment \cite{Lang:1993ct,Lang:1994tu}. This fact has resulted in a state where several moderately low-lying operators have unknown anomalous dimensions at order $N^{-1}$. Apart from the mentioned operator \texttt{Op[S,0,5]}, other relatively low-lying operators for which the large $N$ anomalous dimensions remains unknown (to the best of our knowledge) are \texttt{Op[S,2,2]}, \texttt{Op[S,0,6]}, \texttt{Op[V,0,3]}, \texttt{Op[T,0,3]}, \texttt{Op[T,0,4]}, \texttt{Op[T,2,$i$]} for $i=2,3,4$, \texttt{Op[A,1,2]}, \texttt{Op[Tm[3],0,2]}, \texttt{Op[Tm[4],0,2]}, and so on. Likewise, the order $N^{-1}$ correction to the parameters $\gamma$ and $\theta$ in the correlators $\langle JJT\rangle$ and $\langle TTT\rangle$ are still unknown, and it would be desirable to determine them.

The $\ON$ CFT, with the Ising CFT as a special case, continues to be one of the most important examples of a conformal field theory, thanks to its simple formulation and rich physics. The exact solution, even in the $d=3$ case, remains an open problem. It may be that the 3d critical exponents, and more generally, the CFT-data, are expressible in terms of known mathematical constants (say in $d=3$), but a more likely scenario is that the $\ON$ CFT defines a whole new set of mathematical constants (or more generally functions of $d$ and $N$). This view, that there for a given $d$ and $N$ should be exactly one allowed value for each piece of conformal data, is a beautiful consequence of renormalisation group theory and the conformal bootstrap.


\section*{Acknowledgements}

I would like to thank the following people for useful discussions:
Oleg Antipin, 
Alexander Bednyakov, 
Francesco Bertucci, 
John Cardy, 
Gonzalo De Polsi, 
John Gracey, 
Matthijs Hogervorst, 
Stephan Kehrein,
Murat Kolo\u{g}lu,
Zohar Komargodski, 
Stefanos Kousvos, 
Wenliang Li, 
Pedro Liendo, 
Junyu Liu, 
Mark van Loon, 
Alexander Manashov, 
Brian McPeak, 
Marco Meineri, 
Carlo Meneghelli, 
Hugh Osborn,
Erik Panzer, 
Sylvain Ribault, 
Slava Rychkov, 
Oliver Schnetz, 
Jo\~ao Silva, 
Aninda Sinha, 
Andreas Stergiou, 
Ning Su, 
Ettore Vicari, 
Alessandro Vichi, 
Mats Wallin, 
Nicol\'as Wschebor,
Omar Zanusso, 
and
Bob Ziff. 
This project has received funding from the European Research Council (ERC) under the European Union's Horizon 2020 research and innovation programme (grant agreement no.~758903).

\appendix

\section{Normalisation and Fourier transforms}
In this appendix we review some normalisation conventions for conformal and global symmetry.

\subsection[Fourier transforms in $d$ spacetime dimensions]{Fourier transforms in $\boldsymbol d$ spacetime dimensions}
\label{app:fouriertransform}

In this appendix we write $x=|x|$, $p=|p|$, $px =p^\mu x_\mu$ and $(k-p)=|k-p|$, and define the Fourier transform by
\begin{equation}
f(p)=\int d^dxe^{-ipx}f(x),\qquad f(x)=\int \frac{d^dp}{(2\pi)^d}e^{ipx}f(p).
\end{equation}
Then, for a pure power,
\begin{equation}
\label{eq:fouriertransform}
\int d^dx \frac{e^{-ipx}}{x^{2\alpha}}=\frac{\pi^{\mu}2^{2(\mu-\alpha)}\Gamma(\mu-\alpha)}{\Gamma(\alpha)}\frac1{k^{2(\mu-\alpha)}}\quad \Leftrightarrow\quad \int \frac{d^dp}{(2\pi)^d}\frac{e^{ipx}}{p^{2\alpha}}=\frac{\Gamma(\mu-\alpha)}{\pi^{\mu} 2^{2\alpha}\Gamma(\alpha)}\frac1{x^{2(\mu-\alpha)}},
\end{equation}
where $d=2\mu$.

We can use the above definitions to evaluate the convolution integral,
\begin{equation}
J_{\alpha,\beta}(p)=\int \frac{d^d k}{(2\pi)^d}\frac{1}{k^{2\alpha}(p-k)^{2\beta}} = \left(p^{-2\alpha}*p^{-2\beta}\right)[p],
\end{equation}
needed in section~\ref{sec:largeN}.
Taking the inverse Fourier transform we get a simple product
\begin{equation}
J_{\alpha,\beta}(x)=\frac{\Gamma(\mu-\alpha)\Gamma(\mu-\beta)}{\pi^{2\mu}2^{2\alpha+2\beta}\Gamma(\alpha)\Gamma(\beta)}\frac1{x^{4\mu-2\alpha-2\beta}},
\end{equation}
which, through the Fourier transform, gives
\begin{equation}
\label{eq:Jintegralresult}
J_{\alpha,\beta}(p)=\frac{\Gamma(\mu-\alpha)\Gamma(\mu-\beta)\Gamma(\alpha+\beta-\mu)}{2^{2\mu}\pi^\mu\Gamma(\alpha)\Gamma(\beta)\Gamma(2\mu-\alpha-\beta)}\frac1{p^{2\alpha+2\beta-2\mu}}.
\end{equation}
See e.g.\ \cite{Kazakov:1983dyk} for more results of this type, useful for large $N$ computations.

\subsection{Normalisation conventions for operators}
\label{app:normalisation}
In this appendix, we outline some normalisation conventions for conformal primary operators.  We let
\begin{equation}
\label{eq:Sddef}
S_d=\mathrm{Vol}(S^{d-1})=\frac{2\pi^{d/2}}{\Gamma(d/2)}
\end{equation}
denote the volume of the $d-1$ dimensional sphere. We consider the theory of one free scalar field $\phi$, leaving the discussion of general $N$ to section~\ref{app:normON}. 

\paragraph{Canonical normalisation}
Consider the theory of a free scalar in $d$ dimensions, with an action
\begin{equation}
S=\int d^dx\frac12\de_\mu\hat\phi(x)\de^\mu\hat\phi(x),
\end{equation}
which fixes the normalisation of operators in the canonical normalisation, denoted by the hat.
The two-point function of $\hat \phi$ is given by the Fourier transform of the propagator $k^{-2}$, which using \eqref{eq:fouriertransform} becomes
\begin{equation}
\label{eq:twopointhat}
\langle\hat\phi(x_1)\hat\phi_j(x_2)\rangle=\frac1{(d-2)S_d}\frac{1}{(x_{12}^2)^{\frac{d-2}2}}.
\end{equation}
We employ the notation that composite operators with a hat are constructed without introducing further normalisation, so that all correlators are computed through Wick contractions using \eqref{eq:twopointhat}. For instance, the (normal ordered) operator $\hat\phi^k(x)=:\hat\phi(x)^k:$ has the two-point function
\begin{equation}
\langle\hat \phi^k(x_1)\hat\phi^k(x_2)\rangle=\frac{k!}{((d-2)S_d)^{k}}\frac{1}{(x^2_{12})^{k\frac{d-2}2}}.
\end{equation}
The stress tensor is given by the improved form \cite{Osborn:1993cr} $\hat T_{\mu\nu}=\de_\mu\hat\phi\de_\nu\hat\phi-\frac1{4(d-1)}((d-2)\de_\mu\de_\nu+\eta_{\mu\nu}\square)\hat\phi^2$, and satisfies
\begin{align}
\langle\hat\phi(x_1)\hat\phi(x_2)\hat T^{\mu\nu}(x_3)\rangle&=\frac{-d}{(d-1)S_d}\frac{d-2}2\frac{1}{(d-2)S_d}\frac{ Z_{123}^\mu Z_{123}^\nu-\eta^{\mu\nu} Z_{123}^2/d}{(x_{13}^2)^{\frac{d-2}2}(x_{23}^2)^{\frac{d-2}2}},
\\
\langle \hat T_{\mu\nu}(x_1)\hat T_{\rho\sigma}(x_2)\rangle&=
\frac d{d-1} \frac1{S_d^2}
\frac{\mathcal I_{\mu\nu\rho\sigma}(x_{12})}{x_{12}^{2d}} = \frac{C_{T,\mathrm{free},N=1}}{S_d^2}\frac{\mathcal I_{\mu\nu\rho\sigma}(x_{12})}{x_{12}^{2d}},
\label{eq:stresstensortwopoint}
\end{align}
with $\mathcal I_{\mu\nu\rho\sigma}(x)=\frac12(I_{\mu\sigma}(x)I_{\nu\rho}(x)+I_{\mu\rho}(x)I_{\nu\sigma}(x))-\frac1d\eta_{\mu\nu}\eta_{\rho\sigma}$, $I_{\mu\nu}(x)=\eta_{\mu\nu}-2\frac{x_\mu x_\nu}{x^2}$, $Z_{123}^\mu=\frac{x_{13}^\mu}{x_{13}^2}-\frac{x^\mu_{23}}{x_{23}^2}$, and $C_{T,\mathrm{free}}=\frac{Nd}{d-1}$, c.f.\ \eqref{eq:CTNscalars}. Note that it is often customary to include the factors $S_d^2$ in the central charge $C_T$.

\paragraph{Intermediate unit normalisation}
For some purposes, it is convenient to normalise $\phi$, but not composite operators, to have a unit two-point function. This convention is useful when considering CFT computations in or in the vicinity of the free theory, so that it is possible to talk about composite operators. In this normalisation, 
\begin{equation}
\langle \phi(x_1) \phi(x_2)\rangle=\frac{1}{(x_{12}^2)^{\frac{d-2}2}},
\qquad
\langle  \phi^k(x_1) \phi^k(x_2)\rangle=\frac{k!}{(x^2_{12})^{k\frac{d-2}2}}.
\end{equation}

\paragraph{CFT normalisation}
In conformal field theory, it is customary normalise all operators (except conserved currents) to have unit two-point function.
To define unit normalisation for spinning operators, we follow the review article \cite{Poland:2018epd} and introduce polarisation vectors $\zeta_\mu$ so that for operators in the spin $\ell$ traceless-symmetric Lorentz representation we have $\O_\ell(\zeta,x)=\zeta_{\mu_1}\cdots\zeta_{\mu_\ell}\O_\ell^{\mu_1\cdots\mu_\ell}$. We then normalise two-point function as
\begin{equation}\label{eq:twopointspinL}
\left\langle\O_\ell^I(\zeta_1,x_1)\O_\ell^J(\zeta_2,x_2)\right\rangle=\frac{(\zeta_1^\mu I_{\mu\nu}(x_{12})\zeta_2^\nu)^\ell-\text{traces}}{(x_{12}^{2})^{\Delta_I}}\delta^{IJ},
\end{equation}
with $I_{\mu\nu}(x)$ defined after \eqref{eq:stresstensortwopoint}. For three-point functions of two scalar primaries and one spin $\ell$ primary we write
\begin{equation}\label{eq:opewithspinL}
\left\langle\O_1(x_1)\O_2(x_2)\O_{3,\ell}(\zeta,x_3)\right\rangle=2^{\ell/2}\lambda_{\O_1\O_2\O_3}\frac{(\zeta\cdot Z_{123})^\ell-\text{traces}}{(x_{12}^2)^{\frac{\Delta_{12;3}+\ell}2}(x_{13}^2)^{\frac{\Delta_{13;2}-\ell}2}(x_{23}^2)^{\frac{\Delta_{23;1}-\ell}2}},
\end{equation}
where $Z_{123}^\mu=\frac{x_{13}^\mu}{x_{13}^2}-\frac{x_{23}^2}{x_{23}^2}$ and $\Delta_{ij;k}=\Delta_i+\Delta_j-\Delta_k$. The extra $2^{\ell/2}$ compared to \cite{Poland:2018epd} is to compensate for the absence of a factor $2^{-\ell}$ in the conformal blocks \eqref{eq:CB} compared to that review. Our conventions agree with \cite{Simmons-Duffin:2016wlq}, which is row six of table~I in \cite{Poland:2018epd}.

To rewrite the free theory in the CFT normalisation, we take $\phi(x)=\sqrt{(d-2)S_d}\hat\phi(x)$ and $\phi^k(x)=\sqrt{(d-2)^kS_d^k/k!}\,\hat\phi(x)^k$ so that for all scalar operators $\O$ we have
\begin{equation}
\langle\O(x_1)\O(x_2)\rangle=\frac1{x_{12}^{2\Delta_\O}},
\end{equation}
in agreement with \eqref{eq:twopointspinL}, $\ell=0$.

In order to keep conformal Ward identities \cite{Erdmenger:1996yc,TASIBootstrap} unchanged, it is often customary to keep the normalisation of the conserved currents unchanged. This means working with conventions where
\begin{align}\label{eq:normT}
\langle \hat T(\zeta_1,x_1)\hat T(\zeta_2;x_2)\rangle&=\frac{C_T}{S_d^2}\frac{(\zeta_1^\mu I_{\mu\nu}(x_{12})\zeta_2^\nu)^2-\zeta_1^2\zeta_2^2/d}{(x_{12}^{2})^{d}},
\\
\label{eq:opewithT}
\langle\O(x_1)\O(x_2)\hat T(\zeta,x_3)\rangle&=\frac{-d\Delta_\O}{(d-1)S_d}\frac{(\zeta\cdot Z_{123})^2-\zeta^2Z_{123}^2/d}{(x_{12}^2)^{\Delta_\O-\frac{d-2}2}(x_{13}^2)^{\frac{d-2}2}(x_{23}^2)^{\frac{d-2}2}}.
\end{align}
For the free scalar case with $C_T=C_{T,\mathrm{free},N=1}=\frac d{d-1}$, one can check that these formulas are in agreement with the intermediate unit normalisation. Note that often the factor $S_d^{-2}$ in \eqref{eq:normT} is included in the definition of $C_T$, for instance in \cite{Osborn:1993cr}.

\paragraph{CFT unit normalisation of conserved currents}
Finally, we mention the version of the CFT-normalisation convention used in this report, where the conserved currents are treated in the same way as any other spinning operator, with two-point functions of the form \eqref{eq:twopointspinL}. For the stress-tensor, this defines $T_{\mu\nu}=S_d/\sqrt{C_T}\,\hat T_{\mu\nu}$, so that
\begin{align}
\langle  T(\zeta_1,x_1) T(\zeta_2,x_2)\rangle&=\frac{(\zeta_1^\mu I_{\mu\nu}(x_{12})\zeta_2^\nu)^2-\zeta_1^2\zeta_2^2/d}{(x_{12}^{2})^{d}},
\\
\langle \O(x_1)\O(x_2)T(\zeta,x_3)\rangle &=\lambda_{\O\O T}\frac{(\zeta\cdot Z_{123})^2-\zeta^2Z_{123}^2/d}{(x_{12}^2)^{\Delta_\O-\frac{d-2}2}(x_{13}^2)^{\frac{d-2}2}(x_{23}^2)^{\frac{d-2}2}}.
\end{align}
The conformal Ward identities are then manifest in terms of a fixed value of the OPE coefficient
\begin{equation}
\lambda_{\O\O T}=\frac{-d\Delta_\O}{2(d-1)\sqrt{C_T}}.
\end{equation}
Again we warn for different conventions coming from the factors $2^{\ell/2}$ in \eqref{eq:opewithspinL} and $S^2_d$ in \eqref{eq:normT}, and reiterate that our conventions agree with \cite{Simmons-Duffin:2016wlq}.

\subsection[Normalisation conventions for $\ON$ representations]{Normalisation conventions for $\boldsymbol{\ON}$ representations}
\label{app:normON}

Consider scalar operators $\O_R^I$ in $\ON$ representations $R$ with indices $I=1,\ldots,\dim R$. We normalise by
\begin{equation}
\left\langle\O_R^I(x_1)\O_R^J(x_2)\right\rangle=\frac{\mathbf I^{IJ}_{R}}{(x_{12}^2)^{\Delta_\O}},
\end{equation}
where $\mathbf I^{KL}_{R}=\delta^{IJ}$.
For the four-point function, we write for $\O_i$ in the irrep $R_i$
\begin{equation}
\left\langle \O^I_{1}(x_1)\O^J_{2}(x_2)\O^K_{3}(x_3)\O^L_{4}(x_4)\right\rangle=\mathbf K_{\O_{1}\O_{2}\O_{3}\O_{4}}(x_i)\sum_R \mathbf P^{R,IJKL}_{R_1R_2R_3R_4} \G^R_{\O_{1}\O_{2}\O_{3}\O_{4}}(u,v),
\end{equation}
where $\mathbf K$ is the kinematic factor given in \eqref{eq:Kkinematic}. The objects $\mathbf P^{R,IJKL}_{R_1R_2;R_3R_4}$ are called projectors and satisfy
\begin{equation}\label{eq:projectoridentity}
\left(\mathcal N^R_{R_1R_2R_3R_4}\mathbf P^{R,IJKL}_{R_1R_2R_3R_4}\right)\left(\mathcal N^{R'}_{R_3R_4R_5R_6}\mathbf P^{R',KLMN}_{R_3R_4R_5R_6}\right)=\delta^{RR'} \mathcal N^R_{R_1R_2R_5R_6}\mathbf P^{R,IJMN}_{R_1R_2R_5R_6},
\end{equation}
for normalisation constants given by
\begin{equation}
\label{eq:normconst}
\mathcal N^R_{R_1R_2R_3R_4}= \frac{\sqrt{\dim R}}{(\dim R_1\dim R_2\dim R_3\dim R_4)^{1/4}}.
\end{equation}
Using the projectors, crossing matrices can be written down, so that
\begin{equation}
\mathcal G^R_{\O_{1}\O_{2}\O_{3}\O_{4}}(u,v)=
\frac{u^{\frac{\Delta_1+\Delta_2}2}}{v^{\frac{\Delta_2+\Delta_3}2}}
\sum_{R'}
\mathbf M^{RR'}_{R_1R_2;R_3R_4}
 \G_{\O_{3}\O_{2}\O_{1}\O_{4}}^{R'}(v,u).
\end{equation}
Note that one could introduce rescaled projectors that satisfy \eqref{eq:projectoridentity} with $\mathcal N=1$, however decomposing the correlator with such projectors would not give OPE coefficients that satisfy all desired properties. For instance, for the singlet representation, we would like to impose $\mathbf P^{S,IJKL}_{RR;R'R'}=\mathbf I^{IJ}_R\mathbf I^{KL}_{R'}$ to guarantee the OPE coefficient $\lambda_{\O_R\O_R\1}=1$, but such choice immediately requires that $\mathcal N_{S,RRRR}=1/\dim R$.
A general framework for computing the crossing matrices, that can be applied to $\ON$ symmetry, was worked out by the authors of \cite{He:2021sto}, see \cite{HeFuture}.\footnote{I thank Ning Su for sharing with me many unpublished results.} Their results give rise to projectors with $\mathcal N$ satisfying \eqref{eq:normconst}, and with OPE coefficients invariant under any permutation $\sigma$, i.e.\ $\lambda_{\O_1\O_2\O_3}=\lambda_{\O_{\sigma(1)}\O_{\sigma(2)}\O_{\sigma(3)}}$. In these conventions, we give the crossing matrices as \texttt{crossingM[$\langle R_1\rangle$,$\langle R_2\rangle$,$\langle R_3\rangle$,$\langle R_4\rangle$]} for $R_i=\texttt{S,V,T,A}$. For instance, 
\begin{align}\nonumber
&\texttt{crossingM[V,V,V,V]}\\&\quad=\texttt{\{\{S,T,A\},\{S,T,A\},}\begin{pmatrix}
\dfrac1N &\dfrac{\sqrt{ (N+2)(N-1)}}{\sqrt2N}  & -\dfrac{\sqrt{N-1}}{\sqrt{2N}}
\\
\dfrac{\sqrt{ (N+2)(N-1)}}{\sqrt2N}&\dfrac{N-2}{2N}&\dfrac{\sqrt{N+2}}{2\sqrt N}
\\
-\dfrac{\sqrt{N-1}}{\sqrt{2N}}&\dfrac{\sqrt{N+2}}{2\sqrt N}&\dfrac12
\end{pmatrix}\texttt{\}}.
\label{eq:crossingoutput}
\end{align}
Apart from the commonly used matrix \eqref{eq:crossingoutput}, some other crossing matrices have previously appeared in the literature: $\mathbf M_{TTTT}$ \cite{Reehorst:2020phk}, $\mathbf M_{AAAA}$ \cite{Baume:2021chx}. 
We note the general properties $\mathbf M^{RR'}_{rrrr}\mathbf M_{rrrr}^{R'R''}=\delta^{RR''}$
and $\mathbf M_{rrrr}^{RS}=\sgn(R)\frac{\sqrt{\dim R}}{\dim r}$, where $\sgn( R)$ denotes the parity of the irrep under $1\leftrightarrow2$. If at least one $R_i=S$, the crossing matrix is trivial. 

In general, for a scalar operator $\O_R$ in the $\ON$ representation $R$, the OPE coefficient with the global symmetry current $J$ takes the form
\begin{equation}
\lambda^2_{\O_R\O_RJ}=\frac{q_{\mathrm{eff},R}^2}{C_J}\mathcal N_{RRRR}^A
,
\end{equation}
where the normalisation constant was defined in \eqref{eq:normconst}.
For $R=T_m$, one has
\begin{equation}
q^2_{\mathrm{eff},T_m}=\frac{m(m+N-2)\dim T_m}{N(N-1)/2}=1,N+2,\frac{(N+1)(N+4)}2,\ldots.
\end{equation}

When considering less complicated systems of crossing, one is free to normalise the projectors in a different way than dictated by \eqref{eq:projectoridentity}. For instance, a common choice for the projectors in the $\varphi$ four-point function is
\begin{align}
\label{eq:projectorS}
\tilde{\mathbf P}^{S,ijkl}_{VVVV}=\mathbf P^{S,ijkl}_{VVVV}&=\delta^{ij}\delta^{kl},
\\
\tilde{\mathbf P}^{T,ijkl}_{VVVV}=\mathcal N^T_{VVVV}\mathbf P
^{T,ijkl}_{VVVV}&=\frac{\delta^{ik}\delta^{jl}+\delta^{il}\delta^{jk}}2-\frac1N\delta^{ij}\delta^{kl},
\\
\label{eq:projectorA}
\tilde{\mathbf P}^{A,ijkl}_{VVVV}=\mathcal N^A_{VVVV}\mathbf P
^{A,ijkl}_{VVVV}&=\frac{\delta^{ik}\delta^{jl}-\delta^{il}\delta^{jk}}2.
\end{align}
With the projectors \eqref{eq:projectorS}--\eqref{eq:projectorA}, the crossing matrix for four fundamentals reads
\begin{equation}
\label{eq:crossingmatrixON}
{\renewcommand*{\arraystretch}{1.75}
\tilde{\mathbf M}_{VVVV}=\diag(1,\mathcal N_T^{-1},\mathcal N_A^{-1})\mathbf M_{VVVV}\diag(1,\mathcal N_T,\mathcal N_A)=\begin{pmatrix}
\dfrac1N &\dfrac{ (N+2)(N-1)}{2N^2}  & \dfrac{1-N}{2N}\\
1&\dfrac{N-2}{2N}&\dfrac12\\
-1&\dfrac{N+2}{2N}&\dfrac12
\end{pmatrix}.
}
\end{equation}

\section{Operator counting using characters}
\label{app:characters}

Here we briefly review the characters for the conformal and $O(N)$ groups \cite{Dolan:2005wy,Henning:2017fpj,Meneses:2018xpu},\footnote{I thank Carlo Meneghelli for useful discussions.} see also \cite{Barabanschikov:2005ri,Liendo:2017wsn}. We first consider the $N=1$ case and focus on the conformal characters. 

\subsection{Conformal character decomposition}

We review here the characters for the four-dimensional conformal group, where the Lorentz symmetry $\mathrm{SO}(4)=\mathrm{SO}(3)\times\mathrm{SO}(3)$ can be described by two spin labels $j,\bar j=0,\frac12,1,\frac32,\ldots$. For conformal characters in arbitrary spacetime dimension $d$, we refer to \cite{Dolan:2005wy}.

The conformal character of a generic 4d conformal multiplet of scaling dimension $\Delta$ and spin $(j,\bar j)$ takes the form
\begin{equation}
\chi_{[\Delta,j,\bar j]}(q,x,y)=q^\Delta\chi_j(x)\chi_{\bar j}(y)P_4(q,x,y),
\end{equation}
with
\begin{equation}
P_4(q,x,y)=\sum_{m=0}^\infty\sum_{n=0}^\infty q^{2m+n}\chi_{n/2}(x)\chi_{n/2}(y),
\end{equation}
and the $\mathrm{SO}(3)$ character $\chi_j(x)$ is given by
\begin{equation}
\chi_j(x)=\frac{x^j}{1-1/x}+\frac{x^{-j}}{1-x}=\frac{x^{j+\frac12}-x^{-j-\frac12}}{x^{\frac12}-x^{-\frac12}}.
\end{equation}
At the unitarity bound \eqref{eq:unitaritybound} there are shortening conditions. For the scalar with dimension $\frac{d-2}2=1$ we define the character
\begin{equation}
\chi_{\phi}(q,x,y)=q(1-q^2)P_4(q,x,y).
\end{equation}
For spin $\ell=(\ell/2,\ell/2)$ conserved current we have the character
\begin{equation}
\chi_{\mathcal J_\ell}(q,x,y)=q^{2+\ell}\left(\chi_{\ell/2}(x)\chi_{\ell/2}(y)-q\chi_{\ell/2-1/2}(x)\chi_{\ell/2-1/2}(y)\right)P_4(q,x,y).
\end{equation}

The generating function for the spectrum in the case of a single four-dimensional free scalar field $\phi$ is
\begin{equation}
\label{eq:Zcharacters}
Z=\mathrm{PE}\left(\lambda\,\chi_{\phi}(q,x,y)\right),
\end{equation}
where we introduced the plethystic exponential
\begin{equation}
\mathrm{PE}(f(\lambda,q,x,y))=\exp\left(\sum_{k=1}^\infty\frac1k f\left(\lambda^k,q^k,x^k,y^k\right)\right),
\end{equation}
and introduced $\lambda$ as an additional index which counts the number of $\phi$'s in the resulting multiplets. The spectrum of conformal primary operators in the free four-dimensional theory is then given by decomposing $Z$ into the conformal characters. The operator counting of the interacting theory in the $\eps$-expansion agrees with that of the free theory, if one removes characters corresponding to the following multiplets (this follows from multiplet recombination, which we discussed in section~\ref{sec:multipletrecombination}):
\begin{itemize}
\item $\phi^3$, which becomes a descendant of $\phi$: $\chi_{[1,0,0]}=\chi_\phi+\chi_{[3,0,0]}$.
\item For each $\ell=4,6,\ldots$, one state of the form $\de^{\ell-1}\phi^4$, which becomes a descendant of $\mathcal J_\ell$: $\chi_{[\ell+2,\ell/2,\ell/2]}=\chi_{\mathcal J_\ell}+\chi_{[\ell+3,\ell/2-1/2,\ell/2-1/2]}$.
\end{itemize}
Note that the two components in the recombined multiplet have different number of fields in the free theory description, meaning that we expect $\lambda\chi_\phi+\lambda^3\chi_{[3,0,0]}$ for the long multiplet representing $\phi$, and likewise for $\lambda^2\chi_{\mathcal J_\ell}+\lambda^4\chi_{[\ell+3,\frac{\ell-1}2,\frac{\ell-1}2]}$ for the broken currents.

The expansion of $Z$ up to order $q^8$ reads
\begin{align}
Z&=1+q\lambda\chi_\phi+q^2\lambda^2\chi_{[0,0]}+q^3\lambda^3\chi_{[0,0]}+q^4\left(\lambda^2\chi_{\mathcal J_2}+
\lambda^4\chi_{[0,0]}
\right)
\nonumber\\&\quad +
q^5\left(\lambda^3\chi_{[1,1]}+\lambda^5\chi_{[0,0]}\right)+q^6\left(\lambda^2\chi_{\mathcal J_4}+\lambda^3\chi_{[\frac32,\frac32]}+\lambda^4\chi_{[1,1]}+\lambda^6\chi_{[0,0]}\right)
\nonumber\\&\quad +
q^7\left(\lambda^3(\chi_{[2,0]\oplus[0,2]}+\chi_{[2,2]})+\lambda^4\chi_{[\frac32,\frac32]}+\lambda^5\chi_{[1,1]}+\lambda^7\chi_{[0,0]}\right)
\nonumber\\&\quad +
q^8\big(\lambda^2\chi_{\mathcal J_6}+\lambda^3(\chi_{[\frac52,\frac52]}+\chi_{[\frac52,\frac32]\oplus[\frac32,\frac52]})+\lambda^4(2\chi_{[2,2]}+\chi_{[1,1]}+\chi_{[2,0]\oplus[0,2]}+\chi_{[0,0]})
\nonumber\\&\quad\qquad +\lambda^5\chi_{[\frac32,\frac32]}+\lambda^6\chi_{[1,1]}+\lambda^8\chi_{[0,0]}
\big)+O(q^9),
\end{align}
where we wrote $\chi_{[\Delta,j_1,j_2]}$ as $q^{\Delta}\chi_{[j_1,j_2]}$ for extra clarity.
We read off the operator content: Up to dimension $3$ the operators are $\1$, $\phi$, $\phi^2$, $\phi^3$, where the last operator is a descendant in the interacting theory by multiplet recombination. At dimension $4$ we note the primary operators $\mathcal J_2=T^{\mu\nu}$ containing two fields, and $\phi^4$ containing four fields, and so on. All the subsequent terms represent primaries, until the next descendant from multiplet recombination appears, which happens at dimension $7$ and spin $3$: $q^7\lambda^4\chi_{[\frac32,\frac32]}$. In the interacting theory this state becomes a descendant of the broken current $\mathcal J_4$.
We can compare the counting of scalar operators with the generating formulas for the cases up to eight fields is given in table~1 of \cite{Henning:2017fpj}, and we have found full agreement. 

The $\mathrm{SO}(4)$ characters described above are not sensitive to the parity of the operator. For instance, to distinguish a pseudovector, corresponding to the three-row Young tableau $y_{1,1,1}$, from a vector, one has to consider additional conditions. One option is to use the characters for $\mathrm{SO}(d)$ \cite{Dolan:2005wy} for sufficiently large $d$.

\subsection[Character decomposition for general $N$]{Character decomposition for general $\boldsymbol N$}
\label{app:charactergenN}

The characters for the global $\ON$ symmetry depend on the value of $N$; here we consider $N$ large enough so that no finite-$N$ effects arise, such as the identical vanishing of certain representations.
The character for a representation $([\Delta,j,\bar j],R)$ of $\mathrm{SO}(5,1)\times \ON$ is the product
\begin{equation}
\chi_{[\Delta,j,\bar j],R}(q,x,y,u)=\chi_{[\Delta,j,\bar j]}(q,x,y)\chi_{R}(u),
\end{equation}
and the generating function for the spectrum is
\begin{equation}
\label{eq:ZcharactersON}
Z=\mathrm{PE}\left(\lambda\,\chi_{\phi}(q,x,y)\chi_V(u)\right),
\end{equation}
where the plethystic exponential now treats the additional variable $u$ in the same way as the other variables.

\begin{table}
\centering
\caption{Characters for the $\ON$ global symmetry, expressed in terms of the $\mathrm S_N$ character $\hchi(u):=\hchi_{\square}(u)$.
}\label{tab:ONcharacters}
{\small
\renewcommand{\arraystretch}{1.55}
\begin{tabular}{|cllc|}
\hline 
\multicolumn{3}{|c}{Irrep}&$\chi_{R}(u)$
 \\
\hline
$S$  & \raisebox{-0pt}{$\bullet$}&\texttt{S} & $1$
\\
\hline
$V$ &  \raisebox{-0pt}{\tiny\yng(1)}&\texttt{V}& $\hchi(u)$
\\
\hline
$A$ & \raisebox{-3pt}{\tiny \yng(1,1)} &\texttt{A}& $\frac12\left(\hchi(u)^2-\hchi(u^2)\right)$
\\
$T$&\raisebox{-0pt}{\tiny \yng(2)}&\texttt{T}  &  $\frac12\left(\hchi(u)^2+\hchi(u^2)\right)-1$
\\
\hline
$A_3$&\raisebox{-3pt}{\tiny \yng(1,1,1)}&\texttt{A3}&  $\frac16\left(\hchi(u)^3-3\hchi(u)\hchi(u^2)+2\hchi(u^3)\right)$
\\
$H_3$&\raisebox{-3pt}{\tiny \yng(2,1)}&\texttt{Hm[3]} & $\frac13\left(\hchi(u)^3-\hchi(u^3)\right)-\chi_V(u)$
\\
$T_3$&\raisebox{-0pt}{\tiny \yng(3)} &\texttt{Tm[3]}& $\frac16\left(\hchi(u)^3+3\hchi(u)\hchi(u^2)+2\hchi(u^3)\right)--\chi_V(u)$
\\
\hline
$A_4$& \raisebox{-3pt}{\tiny \yng(1,1,1,1)}&---   & $\frac1{24}\left(\hchi(u)^4-6\hchi(u)^2\hchi(u^2)+3\hchi(u^2)^2+8\hchi(u)\hchi(u^3)-6\hchi(u^4)\right)$
\\
$Y_{2,1,1}$ & \raisebox{-3pt}{\tiny \yng(2,1,1)}& --- & $\frac18\left(\hchi(u)^4-2\hchi(u)^2\hchi(u^2)-\hchi(u^2)^2+2\hchi(u^4)\right)-\chi_A(u)$
\\
$B_4$ &  \raisebox{-3pt}{\tiny \yng(2,2)}&\texttt{B4} & $\frac1{12}\left(\hchi(u)^4+3\hchi(u^2)^2-4\hchi(u)\hchi(u^3)\right)-\chi_T(u)-1$
\\
$H_4$ &  \raisebox{-3pt}{\tiny \yng(3,1)} &\texttt{Hm[4]}& $\frac18\left(\hchi(u)^4+2\hchi(u)^2\hchi(u^2)-\hchi(u^2)^2-2\hchi(u^4)\right)-\chi_A(u)-\chi_T(u)$
\\
$T_4$& \raisebox{-0pt}{\tiny \yng(4)}&\texttt{Tm[4]}  & $\frac1{24}\left(\hchi(u)^4+6\hchi(u)^2\hchi(u^2)+3\hchi(u^2)^2+8\hchi(u)\hchi(u^3)+6\hchi(u^4)\right)-\chi_T(u)-1$
\\
\hline
\end{tabular}
}
\end{table}

A convenient way to treat the characters of the $\ON$ irreps is to express them in terms of characters $\hchi(u)$ of the permutation group $\mathrm S_N$. In table~\ref{tab:ONcharacters} we give the characters for the $\ON$ representations relevant for our computations.
The expressions can be found by the following method. Let $\hchi(u)$ denote the character of the fundamental representation {\tiny\yng(1)} of $\mathrm S_N$, $\hchi(u):=\hchi_{{\tiny\yng(1)}}(u)$, which we keep unevaluated. The character of the $\mathrm S_N$ representation with one row of $n$ boxes ${\tiny\yng(1)}\cdots{\tiny\yng(1)}$ is given by the order $\lambda^n$ part of the plethystic exponential of $\hchi(u)$:
\begin{equation}
\label{eq:characterSNTS}
\hchi_{({\tiny\yng(1)}\cdots{\tiny\yng(1)})_n}=\mathrm{PE}(\lambda\hchi(u))|_{\lambda^n}.
\end{equation}
The $\mathrm S_N$ characters of other representations of $\mathrm S_N$ can be found by matching with the tensor products. For instance, \eqref{eq:characterSNTS} gives immediately that $\hchi_{{\tiny\yng(2)}}(u)=\frac12\left(\hchi(u)^2+\hchi(u^2)\right)$, and then
\begin{align}
\hchi_{{\tiny\yng(1,1)}}(u)&=\hchi_{{\tiny\yng(1)}\otimes{\tiny\yng(1)}}(u)-\hchi_{{\tiny\yng(2)}}(u),
\\&=\frac12\left(\hchi(u)^2-\hchi(u^2)\right),
\end{align}
where we used that $\hchi_{R_1\otimes R_2}(u)=\hchi_{R_1}(u)\hchi_{R_2}(u)$. Continuing in this fashion one can determine the characters for $\mathrm{S}_N$ irreps of arbitrary Young tableaux. 

Next, the characters for the $\ON$ irreps are given by the removal and insertion of traces compared to the $\mathrm S_N$ irreps. For instance
\begin{equation}\label{eq:plethysm}
\hchi_{{\tiny\yng(2)}}(u)=\chi_T(u)+\chi_S(u).
\end{equation}
Solving for $\chi_T(u)$ gives the value in table~\ref{tab:ONcharacters}.	
The equation \eqref{eq:plethysm} is an example of the plethysm from $\mathrm{S}_N$ to $\ON$, and has been implemented in the computer in \texttt{Lie} \cite{Lie}, accessible with Mathematica through \texttt{LieLink} \cite{LieLink}.

The spectrum in the interacting theory is given by decomposing $Z$ of \eqref{eq:ZcharactersON} into characters, and subtracting representations that become descendants according to:
\begin{itemize}
\item $\varphi_V^3$ becomes a descendant of $\varphi$.
\item For $\ell=4,6,\ldots$, one state of the form $\de^{\ell-1}\varphi^4_S$ becomes a descendant of $\mathcal J_{S,\ell}$.
\item For $\ell=2,4,\ldots$, one state of the form $\de^{\ell-1}\varphi^4_T$ becomes a descendant of $\mathcal J_{T,\ell}$.
\item For $\ell=3,5,\ldots$, one state of the form $\de^{\ell-1}\varphi^4_A$ becomes a descendant of $\mathcal J_{A,\ell}$ .
\end{itemize}
The expansion to order $q^6$ reads
\begin{align}
Z&=1+q\lambda\chi_\phi\chi_V+q^2\lambda^2\chi_{[0,0]}(\chi_S+\chi_T)+q^3(\lambda^2\chi_{\mathcal J_1}\chi_A+\lambda^3\chi_{[0,0]}(\chi_V+\chi_{T_3}))
\nonumber\\&\quad +
q^4(\lambda^2\chi_{\mathcal J_2}(\chi_S+\chi_T)+\lambda^3\chi_{[\frac12,\frac12]}(\chi_V+\chi_{H_3})+\lambda^4\chi_{[0,0]}(\chi_S+\chi_T+\chi_{T_4}))
\nonumber\\&\quad +
q^5(\lambda^2\chi_{\mathcal J_3}\chi_A+\lambda^3(\chi_{[1,1]}(2\chi_V+\chi_{H_3}+\chi_{T_3})+\chi_{[1,0]\oplus[0,1]}\chi_{A_3})
\nonumber\\&\quad\qquad
+\lambda^4\chi_{[\frac12,\frac12]}(\chi_T+\chi_A+\chi_{H_4})+\lambda^5\chi_{[0,0]}(\chi_V+\chi_{T_3}+\chi_{T_5}))
\nonumber\\&\quad +
q^6(\lambda^2\chi_{\mathcal J_4}(\chi_S+\chi_T)+\lambda^3(\chi_{[\frac32,\frac32]}(2\chi_V+\chi_{A_3}+\chi_{H_3}+\chi_{T_3})+\chi_{[\frac32,\frac12]\oplus[\frac12,\frac32]}(\chi_V+\chi_{H_3}))
\nonumber\\&\quad\qquad
+\lambda^4(\chi_{[1,1]}(2\chi_S+3\chi_T+\chi_A+\chi_{T_4}+\chi_{B_4}+\chi_{H_4})+\chi_{[1,0]\oplus[0,1]}(\chi_A+\chi_{Y_{2,1,1}})+
\nonumber\\&\quad\qquad\qquad+\chi_{[0,0]}(\chi_{S}+\chi_T+\chi_{B_4}))
\nonumber\\&\quad\qquad
+\lambda^5\chi_{[\frac12,\frac12]}(\chi_V+\chi_{H_3}+\chi_{T_3}+\chi_{H_5})
+\lambda^6\chi_{[0,0]}(\chi_S+\chi_T+\chi_{T_4}+\chi_{T_6}))+O(q^7).
\end{align}
The multiplets in this list that correspond to descendants form the character sum $\lambda^3q^3\chi_{[0,0]}\chi_V+\lambda^4(q^5\chi_{[\frac12,\frac12]}\chi_T+q^6\chi_{[1,1]}\chi_A)$.

\subsection{Some explicit generating formulas}
 
Some explicit formulas for the generating function have been worked out. In the case $N=1$ for operators with no contracted derivatives at twist $k$, i.e.\ operators of the type $\de^\ell\phi^k$, we have \cite{Kehrein:1994ff,Roumpedakis:2016qcg}
\begin{equation}
q^k\prod_{p=2}^{k}\frac1{1-q^p}.
\end{equation}
For $k=4$, this counting includes the descendants at spins $3,5,\ldots$.

For operators with all derivatives contracted and $k=4,5,6,7,8$ fields, explicit expressions are given in table~1 of \cite{Henning:2017fpj}. We reproduce the only the case $k=4$, which reads
\begin{equation}
q^4\frac1{(1-q^4)(1-q^6)}=q^4+q^6+q^8+q^{10}+2q^{12}+q^{14}+2q^{16}+\ldots.
\end{equation}
Additional examples of counting of operators can be found in \cite{Kehrein:1994ff,Liendo:2017wsn,Derkachov:1995wg,Roumpedakis:2016qcg}.

\section{Presentation of data file}
\label{app:datafile}
Attached to this report is an ancillary data file in the form of a Mathematica package  \texttt{ONdata.m}. It contains the conformal data in computer-readable format. An important part of the data is the scaling dimensions for a large number of operators, and in tables~\ref{tab:implemented} and \ref{tab:implemented2} we give precise details of which operators have been implemented.

\begin{table}[ht]
\centering
\caption{Operators implemented.}\label{tab:implemented}
{\small
\renewcommand{\arraystretch}{1.25}
\begin{tabular}{|c|c|c|c|c|c|c|c|c|c|c|c|c|c|}
\hline
 $R$& $\ell$ & $0$ & $1$ & $2$ & $3$ & $4$ & $5$ & $6$ & $7$ & $8$ & $9$ & $10$--$20$ & Tables
\\\hline\hline
\multirow{2}{*}{\texttt{E}}  &  $\Delta^{\mathrm{4d}}_{\mathrm{max}}$   &   $12$  &   $11$  &   $10$  &   $11$  &   $10$  &   $11$ &   $10$  &   $11$  &   $12$  &  $13$  & $\ell+2$ (even)  & \ref{tab:evenscalars}, \ref{tab:evenspintwoising}, \ref{tab:evenspinfourising}, 
\\
&$i_{\mathrm{max}}$  & $13$  & $1$  & $8$  & $1$  & $7$  & $3$  & $4$  & $1$  & $5$  & $2$ & $1$  & \ref{tab:evenHSising}, \ref{tab:evenOpoddspinIsing}
\\\hline
\multirow{2}{*}{\texttt{O}}&  $\Delta^{\mathrm{4d}}_{\mathrm{max}}$ 
& $11$ & $10$ & $9$ & $10$ & $9$ & $10$ & $9$ & $10$ & $11$ & $12$ & \multirow{2}{*}{\graycell} 
& \multirow{2}{*}{\ref{tab:Z2oddscalars}, \ref{tab:Z2oddspinning}}
\\ & $i_{\mathrm{max}}$
& $8$ & $1$ & $4$ & $4$ & $3$ & $2$ & $2$ & $1$ & $2$ & $2$ & \graycell &
\\\hline\hline
\multirow{2}{*}{\texttt{S}}&  $\Delta^{\mathrm{4d}}_{\mathrm{max}}$ 
& $10$ & $9$ & $8$ & \multirow{2}{*}{\graycell}  & $8$ & $9$ & $10$ & \multirow{2}{*}{\graycell}  & $10$ & \multirow{2}{*}{\graycell}  & $\ell+2$ (even)
& \multirow{2}{*}{\ref{tab:singletscalars}, \ref{tab:singletspinning}}
\\ & $i_{\mathrm{max}}$
& $16$ & $1$ & $8$ & \graycell  & $5$ & $1$ & $7$ & \graycell & $1$ &\graycell  & $1$ &
\\\hline
\multirow{2}{*}{\texttt{V}}&  $\Delta^{\mathrm{4d}}_{\mathrm{max}}$ 
& $7$ & $6$ & $7$ & $6$ & $7$ & $8$ & $9$ & $10$ & \multirow{2}{*}{\graycell}& \multirow{2}{*}{\graycell}  & \multirow{2}{*}{\graycell} 
& \multirow{2}{*}{\ref{tab:ONV}}
\\ & $i_{\mathrm{max}}$
& $4$ & $2$ & $5$ & $2$ & $3$ & $3$ & $4$ & $4$  &\graycell&\graycell&\graycell &
\\\hline
\multirow{2}{*}{\texttt{T}}&  $\Delta^{\mathrm{4d}}_{\mathrm{max}}$ 
& $8$ & $7$ & $6$ & $7$ & $6$ & \multirow{2}{*}{\graycell}  & $8$ & \multirow{2}{*}{\graycell}  & $10$ & \multirow{2}{*}{\graycell}  & $\ell+2$ (even)
& \multirow{2}{*}{\ref{tab:Toperators}}
\\ & $i_{\mathrm{max}}$
& $8$ & $3$ & $4$ & $4$ & $3$ &\graycell & $1$ &\graycell & $1$ & \graycell & $1$ &  
\\\hline
\multirow{2}{*}{\texttt{A}}&  $\Delta^{\mathrm{4d}}_{\mathrm{max}}$ 
&  \multirow{2}{*}{\graycell}& $7$ & \multirow{2}{*}{\graycell} & $7$  & \multirow{2}{*}{\graycell} & $7$ & \multirow{2}{*}{\graycell} & $9$ &  \multirow{2}{*}{\graycell} & $11$ &  $\ell+2$ (odd)
& \multirow{2}{*}{\ref{tab:Aoperators}}
\\ & $i_{\mathrm{max}}$
& \graycell & $5$ &\graycell  & $4$ &  \graycell  & $1$ &\graycell  & $1$ & \graycell & $1$ & $1$ &
\\\hline
\multirow{2}{*}{\texttt{Tm[3]}}&  $\Delta^{\mathrm{4d}}_{\mathrm{max}}$ 
& $5$ & $6$ & $5$ & $6$ & $7$ & $8$ & $9$ & $10$ & \multirow{2}{*}{\graycell} & \multirow{2}{*}{\graycell} & \multirow{2}{*}{\graycell} 
& \multirow{2}{*}{\ref{tab:T3}}
\\ & $i_{\mathrm{max}}$
& $2$ & $1$  & $1$ & $1$ & $1$ & $1$ & $2$ & $1$ &  \graycell& \graycell &\graycell &
\\\hline
\multirow{2}{*}{\texttt{Tm[4]}}&  $\Delta^{\mathrm{4d}}_{\mathrm{max}}$ 
& $6$ & \multirow{2}{*}{\graycell} & $6$ & $6$ & $8$ & $9$ & $10$ & \multirow{2}{*}{\graycell} & \multirow{2}{*}{\graycell} & \multirow{2}{*}{\graycell} & \multirow{2}{*}{\graycell} 
& \multirow{2}{*}{\ref{tab:T4}}
\\ & $i_{\mathrm{max}}$
& $2$ &\graycell  & $1$ & $1$ & $2$ & $1$ & $3$ & \graycell &\graycell   &\graycell   &\graycell   &
\\\hline
\end{tabular}
}
\end{table}

\begin{table}[ht]
\centering
\caption{Operators implemented continued.}\label{tab:implemented2}
{\small
\renewcommand{\arraystretch}{1.25}
\begin{tabular}{|c|c|c|c|c|}
\hline
$R$  &$\Delta^{\mathrm{4d}}_{\mathrm{max}}$ & $\ell$ & Tables
\\\hline\hline
\texttt{Hm[3]} & $ 3+\ell+2\delta_{\ell1}$ & $1,2,3,4,5,6$ &  \ref{tab:otherONspinning}
\\\hline
\texttt{A3} & $ 3+\ell$ & $3,5,6$ &  \ref{tab:otherONspinning}
\\\hline
\texttt{Hm[4]} & $ 4+\ell$ & $1,2$ &  \ref{tab:otherONspinning}
\\\hline
\texttt{B4} & $6$ & $0,2$&  \ref{tab:otherONoperators}, \ref{tab:otherONspinning}
\\\hline
\texttt{Tm[5]} & $5$ &  $0$ & \ref{tab:otherONoperators}
\\\hline
\texttt{Hm[5]} & $6$ &  $1$ & \ref{tab:otherONspinning}
\\\hline
\texttt{Tm[6]} & $6$ &  $0$ & \ref{tab:otherONoperators}
\\\hline
\end{tabular}
}
\end{table}

\subsection{Installation}

Place the file \texttt{ONdata.m} in the Applications folder used by Mathematica and load it using the command
\begin{equation}
\label{packageloading}
\texttt{<}\texttt{<ONdata`}
\end{equation}
Alternatively, place it in an arbitrary folder and change the working directory to that folder using \texttt{SetDirectory}. Then load the package using \eqref{packageloading}. 

\subsection{Commands}

The Mathematica package uses the following symbols.
\begin{description}
\item[\texttt{e}] denotes $\eps$ in the $d=4-\eps$ expansion, and $\teps$ in the non-linear sigma model in the $d=2+\teps$ expansion.
\item[\texttt{n}] denotes the parameter \texttt n of the $O(\texttt n)$ global symmetry.
\item[\texttt{mu}] denotes $d/2$ for the spacetime dimension $d$.
\item[\texttt{ord}.] The term \texttt{ord x$^k$} denotes not computed results at $O (\texttt x^k)$.
\item[\texttt{const[i]}] is a placeholder for a numerical constant, to be described below.
\item[\texttt{constVal[i]}] gives the numerical value of \texttt{const[i]}.
\item[\texttt{z[i]}] denotes Riemann's zeta values \texttt{Zeta[i]}.
\item[\texttt{mz[i,j,\ldots]}] denotes the multiple zeta values. Use \texttt{multiZetaSub} to replace by numerical values.
\item[\texttt{P711}] denotes the constant $P_{7,11}$, also implemented as \texttt{const[4]}.
\item[\texttt{eta3[mu]}] denotes the order $N^{-3}$ correction to the critical exponent $\eta$, only implemented for particular values of \texttt{mu}.
\item[\texttt{eta1half}] denotes the order $1/N$ anomalous dimension of the operator $\texttt{Op[V,0,1]}=\varphi$ in the large $N$ expansion. This equals half of the order $1/N$ 
term in the critical exponent $\eta$.
\end{description}
The commands implemented are as follows
\begin{description}
\item[\texttt{Op[R,l,i]}] denotes the operator in the representation \texttt R and spin \texttt l with \texttt i$^{\text{th}}$ lowest scaling dimension.
\item[\texttt{IsingF$\langle i\rangle$[$\langle params\rangle$]}] denotes the operator at the given parameter values in family $i$ for $N=1$. $i=1,\ldots,13$.
\item[\texttt{ONF$\langle i\rangle$[$\langle params\rangle$]}] denotes the operator at the given parameter values in family $i$ for general $N$. $i=1,\ldots,11$.
\item[\texttt{DeltaE[Op]}] gives the scaling dimension of \texttt{Op} in the $d = 4 - \texttt e$ expansion.
\item[\texttt{DeltaN[Op]}] gives the scaling dimension of \texttt{Op} in the large $N$ expansion for $\texttt{mu} = d/2$.
\item[\texttt{DeltaZ[Op]}] gives the scaling dimension of \texttt{Op} in the non-linear sigma model in the $2+\texttt e$ expansion. This is only implemented for a limited set of operators.
\item[\texttt{ValueE[q]}] gives the value of the quantity \texttt q in the $d = 4 -\texttt e$ expansion.
\item[\texttt{ValueN[q]}] gives the value of the quantity \texttt q in the large $N$ expansion for $\texttt{mu} = d/2$. 
\item[\texttt{ValueZ[q]}] gives the value of the quantity \texttt q in the non-linear sigma model in the $2+\texttt e$ expansion. 
\item[\texttt{OpeE}.] \texttt{OpeE[Op1,Op2,Op3]} gives the squared OPE coefficient $\lambda^2_{\texttt{Op1},\texttt{Op2},\texttt{Op3}}$ in the $d = 4 - \texttt e$ expansion. \texttt{OpeE[Op]} gives the squared OPE coefficient $\lambda^2_{\phi,\phi,\texttt{Op}}$ of \texttt{Op} in the $\phi\times\phi$ OPE, in the $d = 4 - \texttt e$ expansion.
\item[\texttt{OpeN}.] \texttt{OpeN[Op1,Op2,Op3]} gives the squared OPE coefficient $\lambda^2_{\texttt{Op1},\texttt{Op2},\texttt{Op3}}$ in the large $N$ expansion for $\texttt{mu}=d/2$. \texttt{OpeN[Op]} gives the squared OPE coefficient $\lambda^2_{\varphi,\varphi,\texttt{Op}}$ of \texttt{Op} in the $\varphi\times\varphi$ OPE, in the large $N$ expansion for $\texttt{mu}=d/2$.
\item[\texttt{crossingM[R$_1$,R$_2$,R$_3$,R$_4$]}] gives basis of exchanged irreps and the corresponding crossing matrix.
\end{description}
The symbols introduced for irreps are \texttt E, \texttt O, \texttt S, \texttt V, \texttt T, \texttt A, \texttt{Tm[$m$]}, \texttt{A3}, \texttt{Hm[$m$]}, \texttt{B4}, and the general \texttt{YT[\{$r_1$,$r_2$,$\ldots$\}]}.

The symbols introduced for quantities are \texttt{CT}, \texttt{CJ}, \texttt{alpha}, \texttt{beta}, \texttt{gamma}, \texttt{delta}, \texttt{eta}, \texttt{nu}, \texttt{phic}, \texttt{omega}, \texttt{intercept[R]}, \texttt{RGbeta[g]}, \texttt{RGgamma[g]}, \texttt{RGm2[g]}, and \texttt{criticalCoupling}.

The constants implemented are
\begin{description}
\item[\texttt{constVal[1]}] $=0.03770767\ldots$ denotes the numerical value of the multiple zeta value $\zeta_{5,3}$.
\item[\texttt{constVal[2]}] $=0.008419668\ldots$ denotes the numerical value of the multiple zeta value $\zeta_{7,3}$.
\item[\texttt{constVal[3]}] $=0.0007892098\ldots$ denotes the numerical value of the multiple zeta value $\zeta_{5,3,3}$.
\item[\texttt{constVal[4]}] $= 200.3575\ldots$ denotes the numerical value of the period $P_{7,11}$.
\item[\texttt{constVal[5]}] $=-15.830116$ denotes a constant appearing in the OPE coefficient $\lambda^2_{\phi\phi\phi^2}$, computed in \cite{Carmi:2020ekr}.
\end{description}
For the definition of the multiple zeta values, see \eqref{eq:multizetavalues} in the main text.

\subsection{Possible issues}

\begin{itemize}
\item The package abuses the pre-installed symbols \texttt E (the natural logarithm) and \texttt O (the ``big-O'' symbol) as labels for the representations. This is not likely to cause any problem.
\item The commonly used one-letter symbols \texttt e, \texttt n, \texttt z, as well as many multi-letter symbols, have been defined within the \texttt{ONdata`} environment. This may cause problems when, without loading the package, copying the output from a file that was created with the package loaded.
\item For some of the values implemented in the large $N$ expansion, a simple substitution $\texttt{/.\,\{\,mu}\rightarrow\texttt{3/2\,\}}$ returns \texttt{Indeterminate}. Instead one may want to apply the command \texttt{//Normal[Series[{\color{teal}\#},\{mu,3/2,0\}]]\&}.
\end{itemize}

\subsection{Disclaimer}

The aim of this report has been to give a complete presentation of conformal data known to the date of the latest Arxiv submission, but future developments of new techniques will ultimately lead to it being superseded. While the most careful checks of the accuracy of the ancillary data file have been undertaken, its correctness cannot be guaranteed, and the author is grateful for comments and corrections.

\bibliographystyle{JHEP}
\bibliography{bibl}

\end{document}